\documentclass[12pt,a4paper,openany]{report}
\usepackage{graphicx}
\usepackage{acronym}
\usepackage[latin1]{inputenc}
\usepackage[normalem]{ulem}
\usepackage{amssymb}
\usepackage{amsfonts}
\usepackage{amstext}
\usepackage{dsfont}
\usepackage{setspace}
\usepackage[margin=0.9in]{geometry}
\usepackage{multicol}
\usepackage[main=british, french, german, latin]{babel}
\usepackage{graphicx}
\usepackage{subcaption}
\usepackage{fancyhdr}
\usepackage{afterpage}
\usepackage{setspace}
\usepackage{nomencl}
\usepackage{framed}
\usepackage{placeins}

\makenomenclature

\newcommand\blankpage{%
	\null
	\thispagestyle{empty}%
	\addtocounter{page}{-1}%
	\newpage}

\title{Categorising current-voltage curves in single-molecule junctions and their comparison to Single-Level Model}
\author{Giovanna Angelis Schmidt}

\begin{document}

	%----------------------------------------------------------------------------------------
	%	TITLE PAGE
	%----------------------------------------------------------------------------------------
	
	\begin{titlepage} 
		\newcommand{\HRule}{\rule{\linewidth}{0.5mm}} 
		
		\center 
		
		%------------------------------------------------
		%	Headings
		%------------------------------------------------
		
		\textsc{\LARGE Technische Universit\"at Dresden}\\[1.5cm] 
		
		\textsc{\Large Master in Organic Molecular Electronics}\\[0.5cm] 
		
		\textsc{\large Faculty of Physics}\\[0.5cm] 
		
		\textsc{\large Chair of Molecular Electronics}\\[0.5cm]
		
		%------------------------------------------------
		%	Title
		%------------------------------------------------
		
		\HRule\\[0.4cm]
		
		{\huge\bfseries Master Thesis\\ \hfill \\Categorizing current-voltage curves
			in single-molecule junctions and their comparison
			to Single-Level Model}\\[0.4cm] % Title of the document
		
		\HRule\\[1.5cm]
		
		%------------------------------------------------
		%	Author(s)
		%------------------------------------------------
		
		\begin{minipage}{0.4\textwidth}
			\begin{flushleft}
				\large
				\textit{Author}\\
				M.Eng. G.A. \textsc{Schmidt}\\
				
			\end{flushleft}
		\end{minipage}
		%~
		\begin{minipage}{0.5\textwidth}
			\begin{flushright}
				\large
				Supervision\\ 
				Professor Dr. A.  \textsc{Erbe}\\ 
				Professor Dr. F.  \textsc{Moresco}\\ 
			\end{flushright}
		\end{minipage}\\[6cm]
	
		{\large to achieve the academic degree}\\[0.4cm]
		{\large Master of Science (M.Sc.)}\\
		{\large in Organic Molecular Electronics}\\

		%------------------------------------------------
		%	Date
		%------------------------------------------------
		
		\vfill\vfill\vfill\vfill
		%{\large Date of submission:}
		\vfill\vfill\vfill\vfill
		
		{\large Dresden\\ \today}
		
		%------------------------------------------------
		%	Logo
		%------------------------------------------------
		
		\vfill\vfill
		%	\includegraphics[width=0.2\textwidth]{Logo_TU_Chemnitz.png}\\[1cm] 
		
		%----------------------------------------------------------------------------------------
		
		\vfill % Push the date up 1/4 of the remaining page
		
	\end{titlepage}
	\pagenumbering{gobble} 
	\afterpage{\blankpage}	
	\maketitle
	\chapter*{}
\begin{minipage}{0.9\textwidth}
\setstretch{1}
\begin{framed}
TU Dresden\\
Faculty of Physics\\
Schmidt, Giovanna\\
\\
\indent ``Categorizing current-voltage curves in single-molecule junctions and their comparison to Single-Level Model''\\
\\
Include bibliographical references\\
Master Thesis
\end{framed}
\end{minipage}
\vfill
\setstretch{1.5}
\begin{minipage}[b]{0.9\textwidth}
\begin{framed}
\textcopyright Copyright by TU Dresden\\

\noindent Without written permission of the promoters and the authors it is forbidden to reproduce or adapt in any form or by any means any part of this publication. Requests for obtaining the right to reproduce or utilize parts of this publication should be addressed to Technische Universit\"{a}t Dresden, Organic and Molecular Electronics (OME), Faculty of Physics 01062 Dresden.

\noindent A written permission of the promoter is also required to use the methods, products, schematics and programs described in this work for industrial or commercial use, and for submitting this publication in scientific contests.
\end{framed}
\end{minipage}

	\chapter*{}
\begin{minipage}{0.9\textwidth}
\end{minipage}
\vfill
\hfill
\begin{tabular}{l}
	\begin{minipage}{0.3\textwidth}
		\textit{I would like to dedicate this study to my husband, Matthias Schmidt, and my son, Wahriman Andrade de Ara\'{u}jo.}
	\end{minipage}
\end{tabular}
\vfill
\begin{minipage}[b]{0.9\textwidth}
\end{minipage}
\clearpage
	\setstretch{2}
	\chapter*{Statement of authorship}

I hereby certify that I am the solo author of this titled ``Categorizing current-voltage curves in single-molecule junctions and their comparison to Single-Level Model'' independently and without undue assistance from third parties. No other than the resources and references indicated in this document have been used. I have marked both literal and accordingly adopted quotations as such. There were no additional persons involved in the intellectual preparation of the present document. I am aware that violations of this declaration may lead to subsequent withdrawal of the academic degree.
\\
\vfill
\begin{minipage}{1\textwidth}
	\centering
	\begin{minipage}{.9\textwidth}
		\raggedleft
		 Giovanna Schmidt \\
		 Dresden, \today \\		 
	\end{minipage}
\end{minipage}\\ 
\vfill

\clearpage

	\setstretch{1.5}
	\chapter*{Acknowledgement}
\begin{minipage}{.8\textwidth}
``\textit{Science is God's language. The honour to witness its discoveries is the most splendid deed in our short existence.}''
\end{minipage}
\\
\vfill
%\begin{minipage}[b]{.9\textwidth}
\noindent I would like to express my gratitude to the Organic Molecular Electronic course and TU Dresden for giving me the opportunity to study with them and conduct my master's thesis research at the ``Helmholtz-Zentrum Dresden-Rossendorf''.

\noindent I want to thank Prof. Dr. Mannsfeld, my final course project's supervisor, who recommended me to Prof. Dr. Erbe. Completing this Master's Degree would not have been possible without the direct support and expertise of Prof. Dr Artur Erbe. He accepted and allowed me to stay in his group for almost one year, working with the state-of-the-art transport theory and the model developed by himself together with his former collaborators. I am very grateful for his patience, dedication to teaching, and willingness to share his knowledge with me. Every lesson he taught me helped me to archive my goals as scientist.

\noindent I am also grateful to Prof. Dr. Moresco, with her expertise on experimental chemistry and physics, for accept to review my thesis. 

\noindent I also want to extend my thanks to Claudia Neisser for preparing my samples, Holger Lager for introducing me to the measurement equipment, and PD. Dr Peter Zahn for aiding me with administrative questions. Furthermore, I am grateful to my colleagues at FWIO-T and the administrative staff of HZDR.

\noindent Finally, I would like to express my gratitude towards Matthias Schmidt, Wahriman andrade de Ara\'{u}jo and Debdutta Chakraborty for their invaluable support and feedback on my thesis. 
%\end{minipage}
\clearpage
	\pagenumbering{roman}
	\tableofcontents
	\listoffigures
	\addcontentsline{toc}{chapter}{\listfigurename}	
	\listoftables
	\newpage
	\addcontentsline{toc}{chapter}{\listtablename}
	\chapter*{Acronyms}
\markboth{ACRONYMS}{}

%\begin{multicols}{2}	
	% if multicol is used --> limit the wide of acronyms through value in brackets
	\begin{acronym}[6LoWPAN]
		\acro{AR}{Absolute Residuals}
		\acro{ESD}{Electrostatic Discharge}
		\acro{EBL}{Electron Beam Lithography}
		\acro{ET}{ElectronTransfer}
		\acro{GOF}{Good of Fitness criteria}
		\acro{HOMO}{Hihest Unoccupied Molecular Orbital}
		\acro{IV}{current-voltage}
		\acro{LED}{Light Emitter Diode}
		\acro{LOWESS}{Locally Weighted Scatterplot Smoothing}
		\acro{LUMO}{Lowest Unoccupied Molecular Orbital}
		\acro{LT}{Low Temperature}
		\acro{LM}{Standart Linear Model}
		\acro{MAR}{Median of Absolute Residuals}
		\acro{MCBJ}{Molecular Controllable Breaking Junctions}
		\acro{RT}{Room Temperature}
		\acro{SAM}{Self-Assembled Monolayer}
		\acro{SLM}{Single-Level Model}
		\acro{UFF}{Universal Force Field}
		\acro{VB PES}{Valence Band Photoemission Spectroscopy}
	\end{acronym}
%\end{multicols}

	\addcontentsline{toc}{chapter}{Acronyms}
	\chapter*{Terminology}

\begin{enumerate}
	\item \textbf{Coupling}: In the context of this text, it likely refers to the interaction between the molecule and the electrodes in a molecular junction.
		
	\item \textbf{Conductance Measurements}: Measurements of the electrical conductance, which is the ability of an object to conduct electric current.

	\item \textbf{de Broglie Wavelength}: A scale that describes the wave-like behaviour of particles at the quantum mechanical level.
	
	\item \textbf{Electronic Conductivities}: A measure of a material's ability to conduct an electric current.
	
	\item \textbf{Electron-Electron 'Forward' Scattering}: A specific type of scattering event involving electrons, influential in the context of electron transport.
	
	\item \textbf{Fermi Level}: The chemical potential for electrons (or the top of the distribution of electrons at absolute zero temperature) in a solid.
	
	\item \textbf{Hysteresis}: The dependence of the state of a system on its history, for example, the previous electrical current or voltage applied to the system.
	
	\item \textbf{I-V Characteristic}: Current-voltage characteristic, a graphical representation of the current through a circuit as a function of the voltage across it.
	
	\item \textbf{Molecular Electronics}: A branch of nanotechnology that uses single molecules or nanoscale collections of single molecules as electronic components.
	
	\item \textbf{MCBJ (Mechanically Controllable Break Junction)}: A technique used to precisely control the spacing between two metallic leads at the nanoscale, often used in molecular electronics studies.
	
	\item \textbf{Phonon Signatures}: Refers to the influence of lattice vibrations (phonons) in the material on the transport properties.
	
	\item \textbf{Resonance}: In this context, it likely refers to the condition when the energy level of the molecule aligns with the Fermi level of the electrodes, allowing efficient electron transport.
	
	\item \textbf{Single-Molecule Transport}: A term referring to the study of electrical conduction through individual molecules.
	
	\item \textbf{Strain/Stress}: Terms from physics and engineering referring to the forces applied to objects; in the context of molecules, this refers to changes in molecular structure due to external forces.
	
	\item \textbf{SLM (Single-Level Model)}: A simplified model used to describe electron transport through a single energy level, often used in molecular electronics.
\end{enumerate}

	\addcontentsline{toc}{chapter}{Terminology}
	
\chapter*{List of Symbols}
\begin{table} [h]

	\begin{tabular}{  l  l }
		I-V & Current-voltage\\ 
		&\\
		$\mathrm{G_0}$ &  Quantum of conductance\\
		&\\
		eV & Electron Volt \\
		&\\
		T & Transmission \\
		&\\
		$\mathrm{\AA}$ & Angstrom \\
		&\\
		$\mathrm{\phi}$ & Work function \\
		&\\
		$\lambda_{dB}$ & de Broglie Wavelength\\
		&\\
		$\mathrm{\alpha}$ & attenuation\\
		&\\
		$\mathrm{\hbar}$ & Planck's constant	
	\end{tabular}
\end{table}

	\addcontentsline{toc}{chapter}{Symbols}
	\newpage
	\addcontentsline{toc}{chapter}{Abstract}
	\chapter*{}
\markboth{ABSTRACT}{}
\section*{Abstract}

\noindent This thesis investigates the mechanically controlled break junctions, with a particular emphasis on elucidating the behaviour of molecular currents at room temperature. The core of this experimental investigation involves a detailed analysis of conductance, examining how it varies over time and with changes in the gap between electrodes. Additionally, this study thoroughly evaluates transmission properties, coupling effects, and current characteristics.
A pivotal aspect of the research was the meticulous current measurement, followed by carefully selecting optimal data sets. This process set the stage for an in-depth analysis of resonant tunnelling phenomena observed through a single channel. Notably, these experiments were conducted under open atmospheric conditions at room temperature. A significant finding from this study is the recognition that our current model requires refinement. This adjustment is necessary to encapsulate a broader spectrum of molecular transport mechanisms more accurately.
Furthermore, this work significantly advances our comprehension of quantum effects in single-molecule junctions, particularly concerning similar molecules to Corannulene extending to some organometallics. One of the essential disclosures is the identification of deviations in the transport model, primarily attributable to electron-electron interactions. This insight is crucial as it paves the way for developing a more comprehensive and precise model, enhancing our understanding of molecular-scale electronic transport.

\clearpage

\chapter*{}

\section*{Abstrakt}
\hfill
%\begin{minipage}{0.7\textwidth} 
%	``\textit{Science is God's language. The honour to witness its discoveries is the most splendid deed in our short existence.}''
%\end{minipage}\\ 

\noindent In dieser Arbeit wurde der molekulare Transport an mechanisch kontrollierten Bruchstellen untersucht, insbesondere bei der Beschreibung des molekularen Stroms bei Raumtemperatur.  Die Analyse der Leitf\"ahigkeit in Bezug auf Messzeiten und Abstand zwischen Elektroden wird in dieser experimentellen Studie umfasst sowie die Bewertung von \"Ubertragung, Kopplung und Strom. Einen theoretischen Hintergrund zur aktuellen Evaluation des Single-Level-Modells wird hier angeboten. Wir f\"uhren ihre Messungen in einer offenen Atmosph\"are bei Raumtemperatur durch und stellen fest, dass das Modell verbessert werden muss, um einen breiteren Bereich des Molek\"ultransports genau darzustellen.
\noindent Diese Arbeit tr\"agt dazu bei, unser Verst\"andnis von Quanteneffekten in Einzelmolek\"ulverbindungen zu verbessern. Hier wurde gezeigt, dass das Transportmodell durch Elektron-Elektron-Effekten divergiert; dieser Beweis kann zu Fortschritten in der molekularen Elektronik beitragen und ein besseres Transportmodell entwickeln.

%\newpage

	\clearpage
	\pagenumbering{arabic}
	\chapter{Introduction}

\section{Motivation and Objectives}

The motivation for choosing this topic was the challenge of understanding mesoscopic transport and quantum mechanics. It is to exploit quantum phenomena and dive deep into soft matter, one of the less exploited topics nowadays. Moreover, there is a need to learn about molecular measurements, even at room temperature, which can reveal the nanoworld's secrets.

\noindent This work aims to give an insight into mesoscopic transport and deals with phenomena signatures observed in experiments which are analysed statistically. The objective is to identify discrepancies between the molecular current transport single-level model (SLM) and its measurements and propose hypotheses for the phenomena causing these discrepancies.

\noindent We present our findings through histograms and current-voltage (IV) curves. Furthermore, we conclude that the SLM needs improvement to represent a broader range of molecules' transport accurately. Additionally, we explain the reasons why the single-level model does not fit the data at $V = 0\, \mathrm{V}$ and $V = \pm 1\, \mathrm{V}$. 

\noindent Finally,This work contributes to improving our understanding of quantum effects in single-molecule junctions, particularly for molecules like Corannulene. It presents evidence that the transport model deviates due to electron-electron interactions;
contributing to a more robust and accurate model.

\section{Molecular Electronics Background}

\noindent Modelling is the art of describing the real world. Despite all models being intrinsically incomplete, scientists emphasise the importance of accurately describing nature and avoiding excessive parametrisation when observing a phenomenon \cite{Box1976}.

\noindent Robert Mulliken and Albert Szent-Gyorgi conducted in $1940$ the first theoretical studies regarding charge-and-energy-transfer in molecules. They dealt with the so-called donor-acceptor systems. However, the tools required for experimental verification had not yet been invented at that time. The scanning probe microscope, for instance, was not invented until $1981$. While lithography had been around since $1820$, it was not commonly used for circuit printing until $1952$.
Additionally, the methods of connecting molecules to electrodes have been discovered recently. Further, Aviram and Ratner evaluated unimolecular quantum transport in 1979, electron dynamics and other transport mechanisms through molecules. Their work led to the Nobel Prize in Chemistry for Rudolph A. Marcus in 1992 \cite{CunibertiFagasRichter2005}.

\noindent In the past few decades, many newly engineered materials have emerged. These materials must be characterised structurally and electrically according to their intended applications. Experiments are conducted to obtain the electrical characterisation of these new materials. They are tested in the first approach in order to understand their behaviour.

\noindent The persistent development of our everyday modern electronic equipment gradually changes the requirements of new materials. New branches in the electronic industry arise from the fabrication of new sensing devices based on organic electronics, and these devices depend on the profound understanding of transport processes on the quantum scale. 

\noindent A simplified comprehension of the transport mechanism and the electronic structure of the molecular bonds are necessary for using them as new materials as circuit devices. In addition, we need to look deeper into the molecular orbitals for a better understanding. At the sub-atomic level, tunnelling may occur, and the transitions of electrons between the molecular orbitals can result in the emission or absorption of phonons of several energy ranges. For example, infrared photons can be involved in transitions that provoke changes in molecular vibrational modes, and transitions between two distant orbitals may involve photons of visible light. Furthermore, electrical transport occurs in the lowest unoccupied molecular orbital (LUMO) or highest unoccupied molecular orbital (HOMO). Depending on whether the energy level of the molecular channel is closer to the Fermi distribution of the contacts or which channels are participating in the transport. 
Understanding this complex transport mechanism requires a complete understanding of level spacing, charging, tunnel effect, couplings, energy exchange, vibrational states, and correlation of energies. Furthermore, researchers can increasingly control and tune experimentally \cite{PascalGehring2019} and use it to describe molecular transport and differentiate through the competition between different transport modes, which leads to diverse processes.

\noindent Molecular and organic electronics aim to understand molecules' transport mechanisms and use them to build new low-power components for commercial and biomedical applications. With this goal is realised the study of the self-assembled monolayers or single molecules by trapping them between electrodes, using anchor groups. 

\noindent A piezo actuator or a pure mechanical setup is used to open a gap in a thin metal wire to capture the molecules for measurement. We use either a liquid cell or a dry environment to conduct the respective measurements. The so-called mechanically controlled break junctions (MCBJ) have seen several changes during the past few years. The first problem was connecting the molecules in the electrodes, which was solved by including anchor groups. Likewise, there is a necessity to study their interference with molecular transport. Our measurements have been conducted at room temperature, leading to questions such as which is the extension of the low temperature (LT) models and when they are valid compared to models for room temperature (RT) experiments. This question has been addressed in this master's thesis, along with new propositions for the existing models.

\noindent Besides, the SLM is usually applied to model the transport of molecules trapped by molecular controllable breaking junctions (MCBJs). This method was developed and first reported \cite{Zotti2010} at room temperature. The model was born as the result of collaborations from Universidad Aut\'{o}noma de Madrid, University of Konstanz, Forschungszentrum Dresden-Rossendorf, Karlsruhe Institute of Technology, University of Zurich and ETH Zurich \cite{Zotti2010}.

\noindent Finally, the statistical evaluation permits understanding the RT measurements' transport in MCBJ through histograms. Furthermore, for the current, the SLM model was used to fit the current and extract the merit figures, such as transmission, coupling, and conductance. We focus on increasing our understanding of the phenomena behind molecular transport. Opening various possibilities for fabricating nanodevices can be exploited with nanotechnology advancements. 

\section{Related Work, the State of Art}

While silicon technology dominates mainstream applications with its unbeatable performance and reliability, organic molecular electronics carve out a significant niche, primarily due to their cost-effectiveness and simplified production processes. However, the scope of molecular electronics extends beyond just being an economical alternative. It offers a complementary approach to traditional silicon-based technology, opening new avenues for exploring mesoscopic physics at room temperature (RT).

\noindent Furthermore, molecular electronics are instrumental in advancing methodologies that enable detecting and analysing subtle phenomena often obscured by environmental noise in conventional settings. These advanced techniques, emerging from molecular electronics, are not merely incremental improvements but could herald a paradigm shift in our understanding of mesoscopic systems.

\noindent Inspired by the development of novel devices, scientists were developing molecules that can act as nanoscale wires, conducting electricity; Nozaki (2010) looked for a mechanism to control the conductances of these wires introducing defects\cite{Nozaki_2010}. The challenge is to control the conductance levels and understand the transport mechanisms at such a small scale. Mu and Zhongcan (2005) \cite{mu2005electronic} studied short DNA wires to create ultra-thin but efficient electrical wires.

\noindent Recent advancements in molecular electronics are predominantly oriented towards uncovering distinctive signatures at room temperature (RT). These efforts propose to bridge the gap between theoretical predictions and experimental observations, especially in the interaction between molecules and gold leads, as explored in the study by Lokamani (2023) \cite{Lokamani2023}. This line of research is critical for deepening our understanding of molecular-scale phenomena and enhancing the accuracy of theoretical models in molecular electronics.

\noindent Mended (2009) found out that these are molecules that can switch between different states (on/off, for example) in response to external stimuli (like electric fields) \cite{Meded_2009}. It is like a light switch but at the molecular level. Recent advancements have focused on improving the stability and reproducibility of these switches, which is crucial for practical applications.

\noindent In actuality, there is a growing focus on the potential applications of molecular semiconductors, as highlighted in the work of Buscemi (2022) \cite{Buscemi2022}. Innovative examples in this domain include the development of circuits that utilise Chlorophylls to mimic semiconductors for transistors, demonstrating the feasibility of employing organic molecules in practical electronic devices. Additionally, the exploration of `functional' molecules, particularly those with a metallic core, is gaining visibility in the literature. Studies like that of Kilibarda (2021) \cite{Kilibarda2021} investigate molecules with metal ion cores, evaluating its characteristics.

\noindent In recent years, there has been a significant shift in the experimental focus within molecular electronics. While earlier research predominantly concentrated on low-temperature (LT) experiments, there is now an increasing emphasis on conducting studies at room temperature (RT). The practical relevance of RT conditions for everyday molecular electronics applications drives this transition.

\noindent The choice of temperature at which experiments are conducted plays a crucial role in influencing the transport phenomena within molecular channels. At RT, specific mechanisms and behaviours can become either more prominent, emerging distinctly from or less prominent, obscured by the background noise. This temperature-dependent variability is primordial in understanding the intricate dynamics of molecular transport.

\noindent However, a topic of ongoing research and debate is the consistency of these phenomena across different temperatures. Researchers are actively exploring whether the mechanisms observed at RT are analogous to those at LT and what distinct transport mechanisms might uniquely influence the system under varying thermal conditions. This inquiry is crucial for developing a comprehensive understanding of molecular electronics that is applicable across a wide range of operational environments.

\section{Structure of the work}

\noindent In our study, we conducted a detailed experimental analysis of the SLM applied to molecular bridges composed of molecules attached to gold leads, specifically when these are immersed in a toluene solution. Our primary objective was to investigate the conductance and current properties of these molecular bridges. To achieve this, we carefully measured current voltage sweeps, time-dependent conductance and precisely quantified the conductances values in the molecule between the gold electrodes. Furthermore, our study extended to evaluating parameters such as transmission, coupling with leads, and the electrical current characteristics within these molecular structures. This comprehensive analysis provide deeper insights into the transport behaviour of molecule attached to MCBJ bridges in a toluene medium.

\noindent To present our study, we follow: 

\begin{itemize}
	\item The first chapter covers the motivation behind the interest in studying state-of-the-art molecular transport and the goals aimed to be archived.
	\item The second chapter is about the methodology employed and sample preparation, the procedures for the measurements, and the criteria of choice for the data included in this thesis.
	\item The third chapter explains the relevant theoretical background for explaining the data obtained in the measurements and current evaluation in the MCBJ. We discuss the criteria behind selecting molecules for the measurements, how they adhere to the MCBJ and the transport regimes in metals and molecules.
	\item In The fourth chapter, the SLM will be used to evaluate the current, and this model was developed for metal-molecule-metal junctions at room temperature, being appropriate for our data. Furthermore, we evaluate the difference between the measured data and the models' predicted curves. 
	\item In the last chapter, the conclusion summarises all topics presented in the thesis.	
\end{itemize}

\noindent The thesis is organised to give fluidity to the reader, so the measurements come before the theory and present the phenomenology behind the transport mechanism. Now, the reader is invited to our second chapter.

	\chapter{Methods and Implementations}
\label{sec:msetup}
Our research analyses molecules by examining their linear conductance and current-voltage characteristics. While there are other methods for characterisation, they typically require complex setups that must be adapted from low temperature to room temperature to provide significant measurements. However, these are beyond the scope of our work because these setups are more complex than ours. 
 
\noindent This work focuses on how carriers propagate from a gold single metal contact into a molecule and characterises their transport at room temperature (RT) according to:

\begin{itemize}
	\item Linear Conductance, and
	\item Current-Voltage Characteristics.
\end{itemize}

\noindent Some molecules are introduced into the bridge after being diluted in toluene because they are sensitive to oxygen or form clusters \cite{Kilibarda2021}. It uses a glass pipette (not sealed) on a mechanically controllable break junction (MCBJ). The linear conductance is obtained through a static configuration, where a slight voltage bias is applied to the molecule. We applied $100$ mV between the electrodes, during the opening of the bridge. Concerning the current-voltage characteristics is necessary do above this constant voltage (DC level) a sweep between $[-1 \, \mathrm{V}:1 \, \mathrm{V}]$ to obtain a non-linear current characteristic.
Additionally, the measurement range depends on the excitation energy of the molecule between the single-atom contact. Furthermore, the minimum resolution of these measurements depends on the noise background, and whether the measurement setup is adapted to changes from small to large background noises, enabling the proper signal measurement.
Although our measurement setup and acquisition of data at room temperature is very stable, an important note is that the MCBJ must be protected against voltage spikes, which may destroy the sample. The sample may be susceptible to any voltage or current spikes generated by sources in the setup. \cite{Erbe082020}.

\noindent The data was statistically evaluated and comparing the measurements to the model. Our measurements had no temperature control and were done for several students in different weather conditions over the years. In the same way `artefacts' can disappear in the statistics of a large number of data (min. 30 measurements), the influences caused by those conditions will not affect the results.

\section{Mechanically Controlled Break Junctions Principle}

The idea is to fabricate nanogaps on a nanowire on a flexible substrate (polyimide with a thickness of 0.5mm) by controlled bending of the substrate until the gold wire breaks. These substrates are stable and passive, and have a high resistivity (plastic/polymer).
The main parameter of such systems is the attenuation relative to the electrode displacement and substrate bending, which both depend on the sample geometry. Using this parameter, we can convert the z-axis displacement of the rod in the opening displacement of the bridge.

\begin{figure}[!htb]
	\centering
	\begin{minipage}{0.4\textwidth}
		\includegraphics[width=1\textwidth]{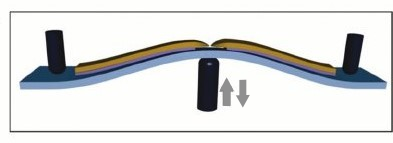}
		\subcaption[]{MCBJ work principle.}
		\label{fig:mcbj_principle}
	\end{minipage}
	\vfill
	\begin{minipage}{0.7\textwidth}
		\includegraphics[width=1\textwidth]{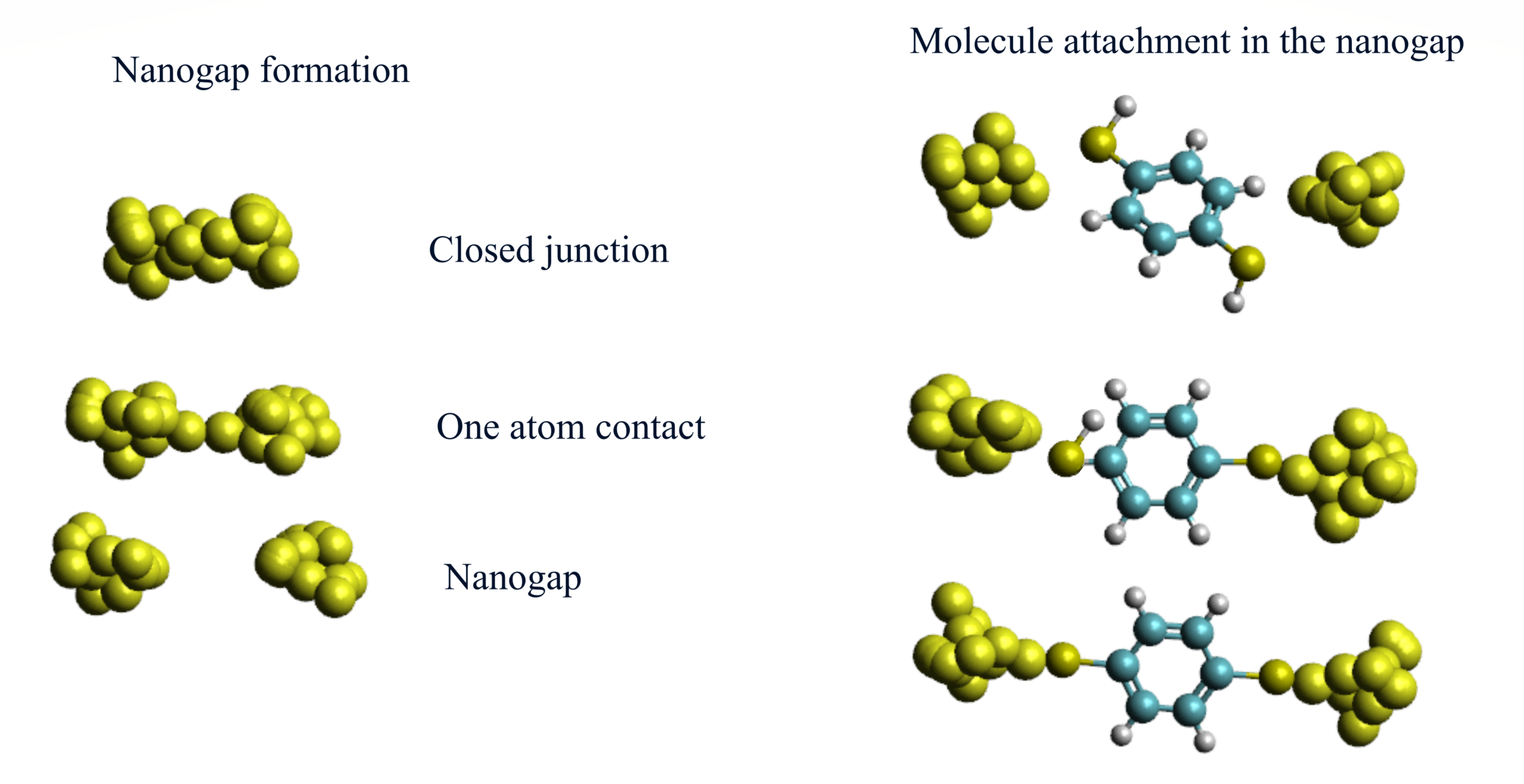}
		\subcaption[]{MCBJ nanogap opening and attachment of the molecule.}
		\label{fig:mcbj_tearing}
	\end{minipage}
	\caption{Conceptual work principle and lithographic image of a real MCBJ; (a) the rod which moves up and down; (b) and representation of a bridge opening and attaching a molecule.}
	\label{fig:bridges}
\end{figure}

\FloatBarrier

\noindent Figure \ref{fig:bridges} shows two pictures, one with  the sample and the parameters used to perform the controlled bending and pulled (\ref{fig:mcbj_principle}) and the other with an illustration of the process in (\ref{fig:mcbj_tearing}) showing the opening of the nanogap. 
\noindent A DC motor rotates, leading to an up- and down motion of the rod to open or close the bridge. The conversion of this radial movement is done with precise steps in horizontal displacement to open the bridge.

\noindent In Figures (\ref{fig:mcbj_principle})  \cite{Xiang2011}, the rod will tear the contacts apart to form the one-point contact. This procedure will continue until the bridge opens and the nanogap is created. This is used in both cases: calibration and molecule attachment.
The calibration will be carrying out for two cases: bridge dry and bridge with solvent. The opening and closing will occur several times until the material reaches a state of fatigue and, typically, breaks in the same place. We realise $100$ opening/closing cycles to calibrate the dry sample. The same procedure is repeated with the bridge immersed in solvent without the molecule.
In Figure (\ref{fig:mcbj_tearing}) \cite{Calame}, the process of capturing and contacting the molecule is illustrated. Once the molecule is caught and contacted the measurement can start.

%\clearpage

\subsection{Setups for MCBJ}

There are several setups possible for this measurement, for example, with either mechanical or piezoelectric actuator. They can also be supervised with a microscope in real-time. 

\begin{figure}[!htb]
	\centering
	\begin{minipage}{0.45\textwidth}
		\includegraphics[width=1\textwidth]{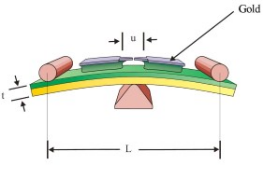}
		\caption{MCBJ parameters \cite{Calame}. In this setup, where $u$ is the nanowire length, $t$ is thickness, $L$ is the distance between rods, and  $\Delta z$ is vertical displacement: $L=16\, \mathrm{mm}$, $u=170\, \mathrm{nm}$, $t=0.5\, \mathrm{mm}$, and $\Delta z =0.1\, \mathrm{mm}/ \mathrm{rotation}$.}
		\label{fig:mcbj_parameters}
	\end{minipage}
	\hfill
	\begin{minipage}{0.45\textwidth}
		\includegraphics[width=1\textwidth]{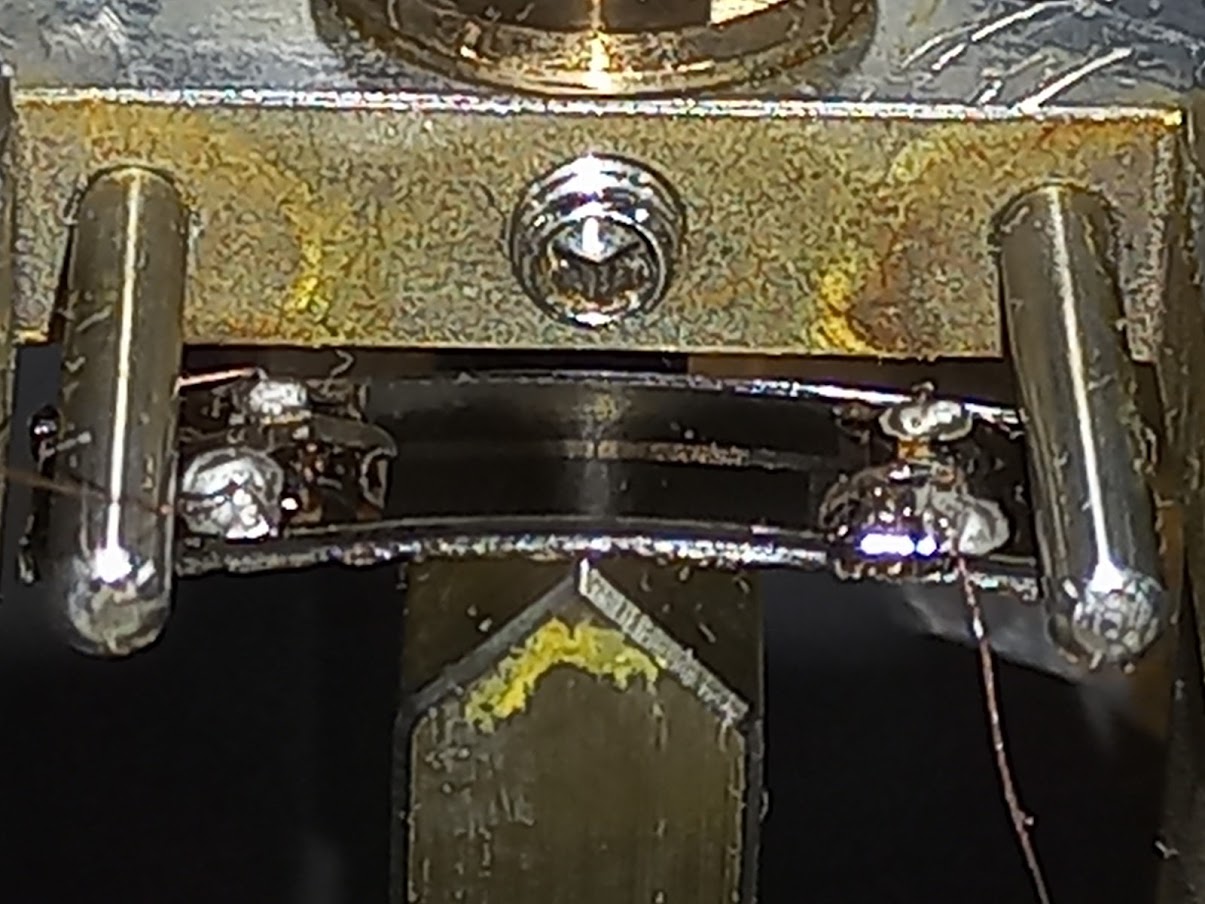}
		\caption{sample placed for measurement.}
		\label{fig:mcbj_sample}
	\end{minipage}
\end{figure}

\FloatBarrier

\noindent Nevertheless, all those setups have the same thing in common: the sample's geometry defines the attenuation factor which opens the bridge, which is expressed as:

\begin{equation}
	\alpha = \frac{6 u t}{L^2}, \label{attenuation}
\end{equation}

\noindent We show the parameters in Figure \ref{fig:mcbj_parameters}. The $\alpha$  in the Eq. \ref{attenuation}  is  the attenuation, which represents the variations in horizontal displacement (x-axis or length-axis) per vertical displacement (z-axis or thickness-axis) $\alpha = \Delta x / \Delta z$ \cite{Vrouwe2005}. Figure \ref{fig:mcbj_sample} shows the sample mounted on the setup, which can be directly compared to the illustration on the left hand-side in Fig. \ref{fig:mcbj_parameters}. Using the attenuation equation for the x-axis displacement, we obtained  $\Delta x = 150\,\mathrm{pm}/\mathrm{rotation}$ for our setup, what it is according with \cite{Vrouwe2005}.

\clearpage

\begin{figure}[!htb]
	\centering
	\begin{minipage}{0.8\textwidth}
		\includegraphics[width=1\textwidth]{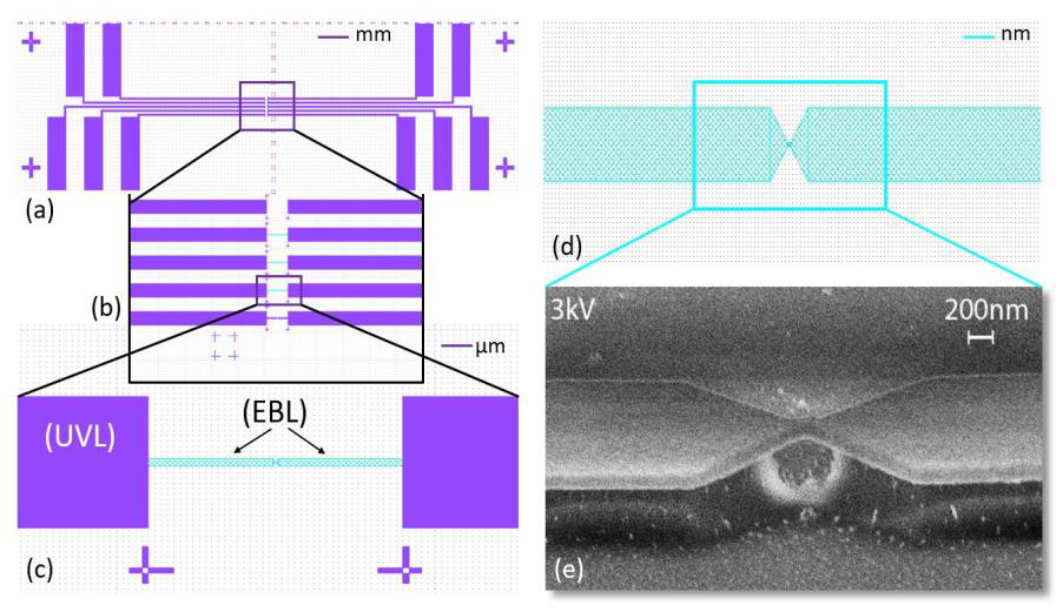}
		\caption{(a) Schematic;( b) Zoom on the  $L =135$ nm nanowire; (c) Zoom on one MCBJ; (d) The $Width=100$ nm contact's schematic; (e) The bridge's lithography. \cite{Sondhi2022}}
		\label{fig:mcbj_junction}
	\end{minipage}
\end{figure}

\FloatBarrier

\noindent Figure \ref{fig:mcbj_junction} shows a MBCJ sample prepared with ultra violet lithography (pads) and Electron Beam Lithography (ELB) technology (nanowires).

\begin{figure}[!htb]
	\centering
	\begin{minipage}{.8\textwidth}
		\includegraphics[width=1\textwidth]{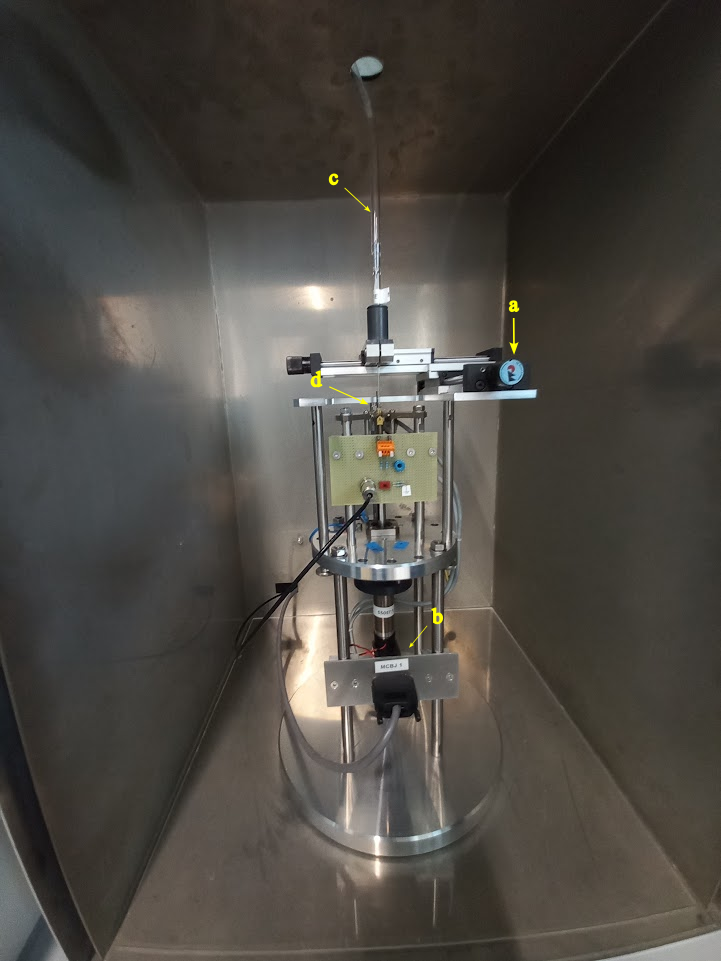}
		\caption{Photograph of the experimental setup. a.b. are positioning screws and only in b. is powered by a motor 1; c. plastic tube, which is attached to the pipette to feed the sample with toluene solutions, d. glass pipette and sample.}
		\label{fig:mcbj_setup}
	\end{minipage}
\end{figure}

\noindent The breaking process is sensed via an ampere-meter, which detects the  dropping current when the sample's bridge opens. During the breaking process, the current is measured, and the conductance is evaluated as multiple of the quantum conductance. This mechanical setup was chosen instead of a setup with a piezoelectric actuator, despite the higher speed of the latter, because piezo setups need extra protection against noise, and anti-vibration tables necessary to give extra protection to the system. A photograph of the setup is presented in Fig. \ref{fig:mcbj_setup}.

\clearpage

\subsection{Measurement Protocol}

In this section the sample preparation and measurements will be in detail explained.
\noindent The Solution with $0.01\,\mathrm{g}$ molecules in $10\,\mathrm{ml}$ toluene and the samples were prepared and measured by the group \footnote[2]{Nanoelectronics - Helmholtz Zentrum Dresden Rossendorf}, refer to Sodhi and Kilibarda \cite{Sondhi2022}\cite{Kilibarda2021} to see how they prepared the samples: using ultra violet lithography (UV-lithography), for the contacts and electron beam lithography for the bridges combined with lift off process. For the measurement, the first step is to connect the electrostatic discharge (ESD) protection to avoid spikes of electrostatic voltages, which can destroy the sample. Using gloves for this procedure is required to keep the sample clean.

\begin{figure}[!htb]
	\centering
	\begin{minipage}{0.5\textwidth}
		\includegraphics[width=1\textwidth]{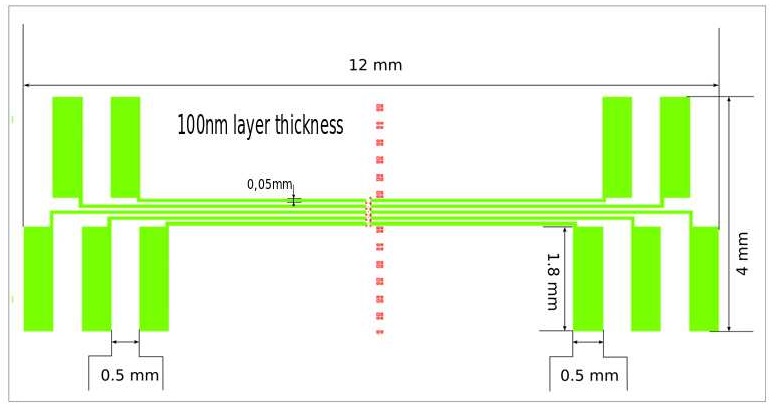}
		\caption{Sample schema, the nanowires between the contacts have a length of $135\, \mathrm{nm}$, and EBL did the break junctions electrodes with 40 nm width.}
		\label{fig:sample}
	\end{minipage}
\end{figure}

\noindent Figure \ref{fig:sample} contains the dimensions of our sample, the sample is done in a two-step process (pads) with ultra violet lithography, and ion-beams (nanowire).

\noindent Protocol:

\begin{enumerate}
	\item To prepare the samples, use tweezers and gloves, and connect the electrostatic discharge (ESD) protection.
	\item Place one sample on the prototype board and weld copper wires using a tiny droplet of silver paste, deposited on the contacts with a toothpick.
	\item Dry the highly conductive silver paste on sample contacts with a heat gun for 6 minutes.
	\item Prepare the epoxy glue by mixing two droplets of each glue part until the mix is homogeneous, to obtain a stable junction.
	\item Cover each weld contact with a small epoxy droplet and let it dry for approximately 15 minutes.
	\item Clean the sample with a nitrogen gun before transport.
	\item Put the sample in a plastic box or any box that can protect it during transport. Despite it does shield the MCBJ against ESD discharges, the box was touched while the ESD protection was connected in the arms, to discharge.
	\item Before fixing the sample on the setup, connect the ESD protection.
	\item Check if the measurement setup is grounded and connected to the sample. Turn on the low noise source and measurement setup.
	\item To set up the measurement, configure the resistance on the software to the sample resistance, our samples are usually 50 or 424 Ohms, depending on the sample contact-contact resistance; this value will control the opening and closing of the bridge; a fully closed bridge will correspond to the gold wire together with contacts resistance.
\end{enumerate}
\noindent Start the measurement and continue until the sample shows the first opening with the current measurement, and the position on the software is set to zero manually, so the bridge is calibrated to open and close from that point on. The measurement starts, and 100 openings and closings in dry conditions are realized. These measurements tend to break the bridge at the same position, after the same 100 openings and closings. A second calibration with toluene to clean the junction from particles, is done in the same manner as dry calibration. Further, the solution with the molecule can be used, and full measurements will be performed until the junctions breaks. We monitor the sample in toluene every hour, depending on the time it takes for the container with the molecule to dry out, since our bridge should not run dry without molecules. These setups are stable and take weeks until they destroy the MCBJ completely.
\clearpage
\subsection{Electrical Diagram of the Measurement}
This is a typical setup for molecules in a MCBJ measurement:

\begin{figure}[!htb]
	\centering
	\begin{minipage}{0.6\textwidth}
		\includegraphics[width=1\textwidth]{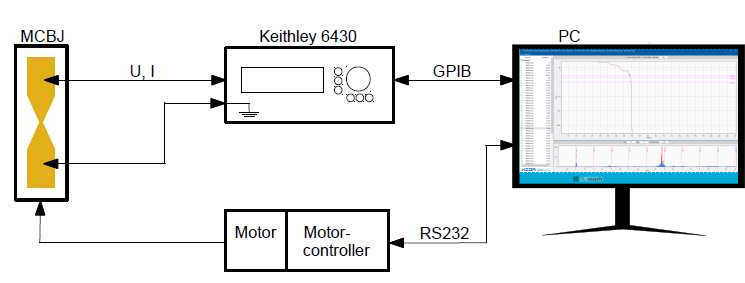}
		\caption{Circuit diagram.}
		\label{fig:mcbj_schema}
	\end{minipage}
\end{figure}

\noindent Figure \ref{fig:mcbj_schema} shows the measurement of a MCBJ in the probe station by a source meter. The measurement's interface is connected to the general propose input/outputs bus (GPIOB) on the computer, which controls the motion of the motor through the serial input/output (RS232), the cables are triax coaxial cables.
The control and supervision are done by proprietary software written in Python, which substitutes its former version in Labview\textsuperscript{\textregistered}.

\subsection{Criteria to Select the Data}

The curves used in this thesis were chosen according to several criteria. 
\noindent The data for the conductance graphics was selected based on the clean and synchronized curves around $0\, \mathrm{V}, 1\, \mathrm{G_0}$. The data selection was carried out manually, with a focus on selecting stable measurements either for opening or current. 

\noindent Furthemore, some measurements were chosen based on the opening after forming an one atom contact. This means they open the bridge at exactly one conductance quantum $1\, G_0$. For the molecule's histogram, curves were chosen based on a higher probability of having molecules present. This was achieved by examining the 1D histogram, with the conductance values ranging from $10^{-6}$ to $1\, G_0$, which are the documented conductance values for the molecules used in this thesis.

\noindent The criterion for selecting the current dataset was how noiseless and stable the absolute value of the current on the up and down sweep and how well the current fit with the SLM, using the goodness of fit (GOF) merit figure. The dataset shows $GOF > 90\%$ and without a high number of discontinuities. Thus the current was categorized and studied using histograms and I-V curves for fitting. 

\noindent Let's proceed to the next chapter and interpret the data. We used over 2000 measurements in total and selected the best 100 measurements whenever possible for each of the graphs. The current files contained c.a. 20,000 points, while the conductance files after normalisation contained c.a. 800 points. Hence we had to normalize the histograms by the number of files and bins to ensure the information presented is consistent, and can be compared with each other.

\section{Experiment Realisation}

\noindent The measurements were taken in an open atmosphere at RT, using a low-noise source. There was no control over the humidity in the laboratory where the measurements were conducted. The samples were fabricated by the group\footnote[1]{Nanoelectronics - Helmholtz Zentrum Dresden Rossendorf} with electron beam lithography of gold on polyimide a plastic substrate. 

\noindent We employed a sub-femtometer source and a pre-amplifier to measure the conductance of the opening and closing curves. The sample is placed in the setup Fig. \ref{fig:mcbj_setup}, and the sample Fig. \ref{fig:mcbj_sample} features a flexible substrate and one dc motor plus reducer with a rotation ratio of 55057:1 (value obtained in the data files), connected by a transmission mechanism. The transmission mechanism pushes the substrate in the z-axis direction with a rod while the computer monitors and records voltage, current, time, motor position, and speed. 
\noindent To collect the molecules dissolved in toluene, a pipette is placed above the sample to deposit the solution continuously, and these molecules can be caught by the junction after it is broken.

\begin{figure}[!htb]
	\centering
	\begin{minipage}{1\textwidth}
		\centering
		\includegraphics[width=.8\textwidth]{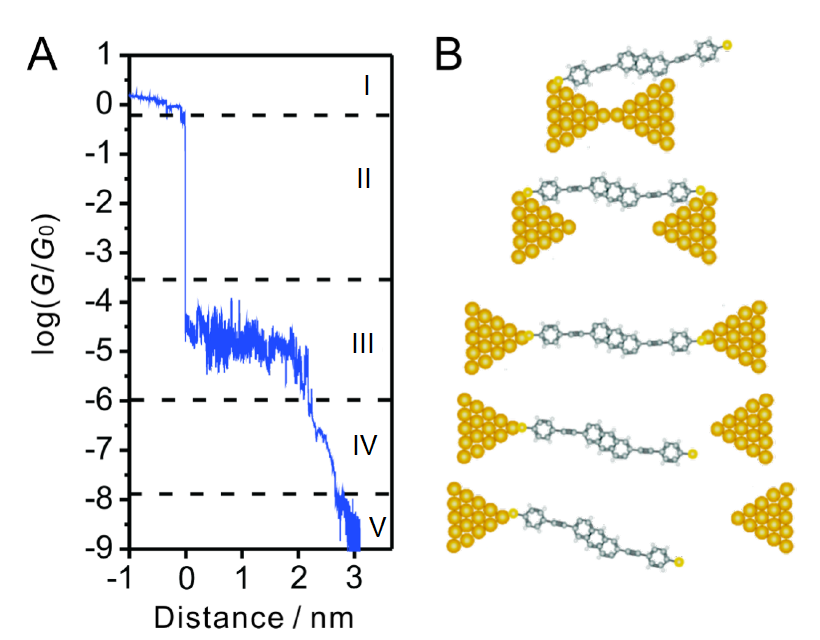}
		\subcaption[]{A. I is one atom contact, II tunnelling region, III Molecule; I-V Tunnelling region with a lower current than II; V Noise region. B. molecule possible positions for the respective numbers \cite{Hong2011}.}
		\label{fig:opa}	
	\end{minipage}
	\begin{minipage}{1\textwidth}
		\centering
		\includegraphics[width=.8\textwidth]{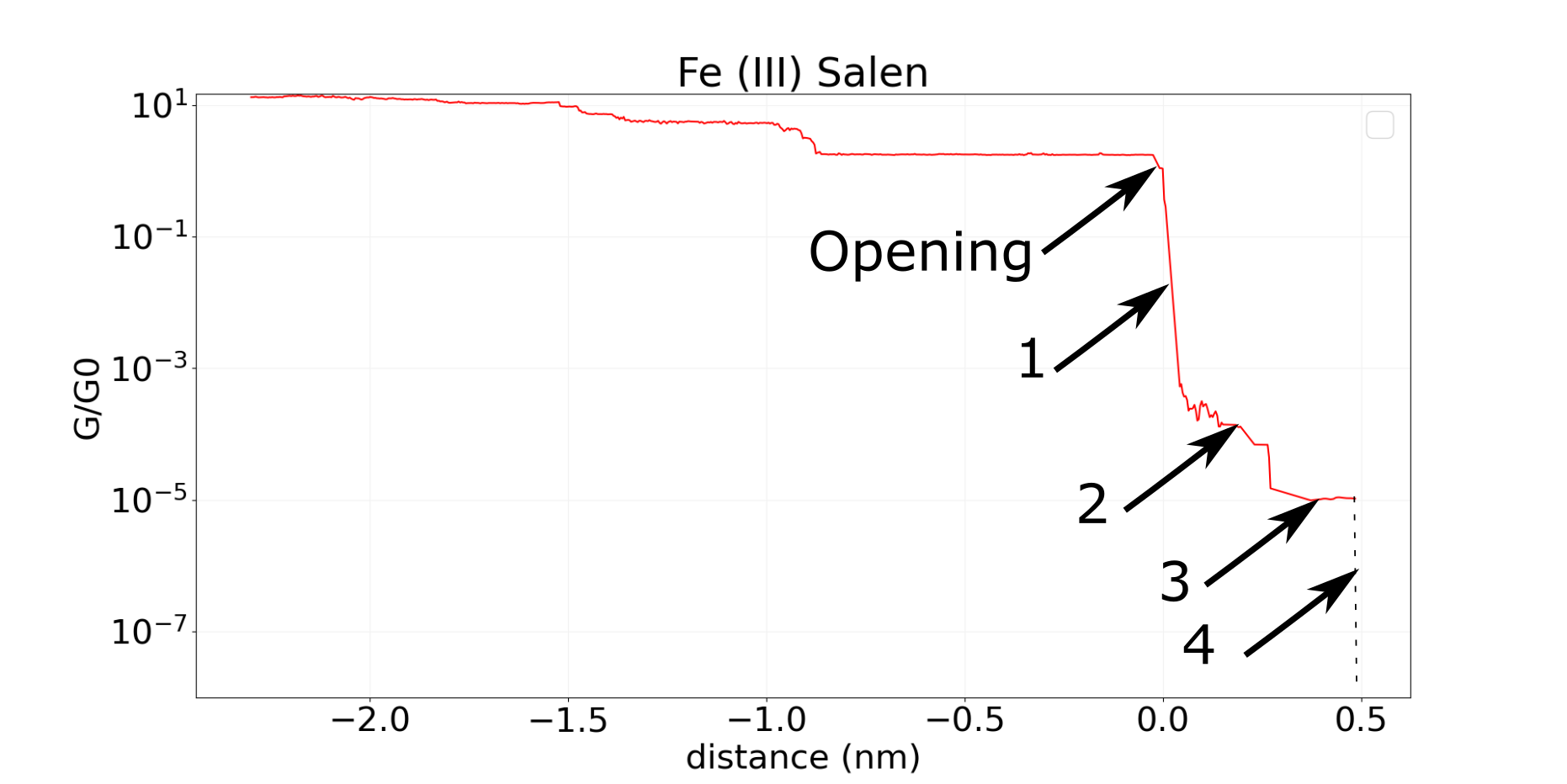}
		\subcaption[]{Bridge opening from a solution with Fe$^{+3}$ Salen.}
		\label{fig:opb}
	\end{minipage}
	\caption{The Figure in (a) is a model of the breaking process of the anthracene-based linearly conjugated molecule (AC) in the Au|AC|Au junction, realised by Hong \cite{Hong2011}; (b) Opening curve, the contact makes opening and closing cycles from $20\, G_0$ until the bridge breaks, the numbers indicates short time intervals where measurements are possible and the corresponding current regime, comparing Figures (a) and (b): 1. is the tunnelling just after the contact-to-contact opens (II), lower tunnelling currents are 2.and 3. in the molecule channel position (III), with different coupling modes each, 4. if any measurement happens in four will register as a low tunnelling (I-V), the grey dashes represent the conductances, and they are in a different colour from the red line, because at this point, the measurement starts the closing after the measurement of the molecule, and this range for the opening was included with graphics tools after the measurement of the data points.}
	\label{fig:op}
\end{figure}

\clearpage

\noindent In Figure (\ref{fig:opb}), the curve represents data from a single measurement of Fe$^{+3}$ Salen. The curve shows that the sample opened before the one-atom contact. The numbers 1 to 4 in the graph indicate potential measurement positions. Whether a measurement occurs at position $1$ under $1\, \mathrm{G_0}$, the tunnel current will be very high, reaching maximum absolute values of around $80 \, \mathrm{\mu A}$, but retaining the 'S' shape, which shows that this current is not a shot circuit limited by the compliance, which is typically linear. The positions $2$ and $3$ indicate one single molecule's measurements and it is presented in most of this thesis. In this graph, there is an actual measurement at position $4$, where a shallow tunnelling current is measured of a very low value of c.a. $0.1\, \mathrm{pA}$. During the measurement, the motor stops, and the I-V measurement is carried out at a constant motor position. That is the reason the conductance stays constant. The motor moves again after the sweep is finished, and the bridge starts to close. All the following values are stored in the next file. The next chapter will discuss further details on these measurements and their fitting, specifically in section \ref{sec:current_measfit}. 

\noindent Additionally, Fig. (\ref{fig:opb}) displays two possible tunnelling positions $2$ and $3$ in the range of the molecule, which changes depending on how the molecule attaches to the MCBJ.

\noindent A further remarks is that these experiments are stable and can run for weeks before the sample gets destroyed, making possible enough data to be measured to be statistically evaluated.

\begin{figure}[!htb]
	\centering
	\begin{minipage}{0.7\textwidth}
		\centering
		\includegraphics[width=1\textwidth]{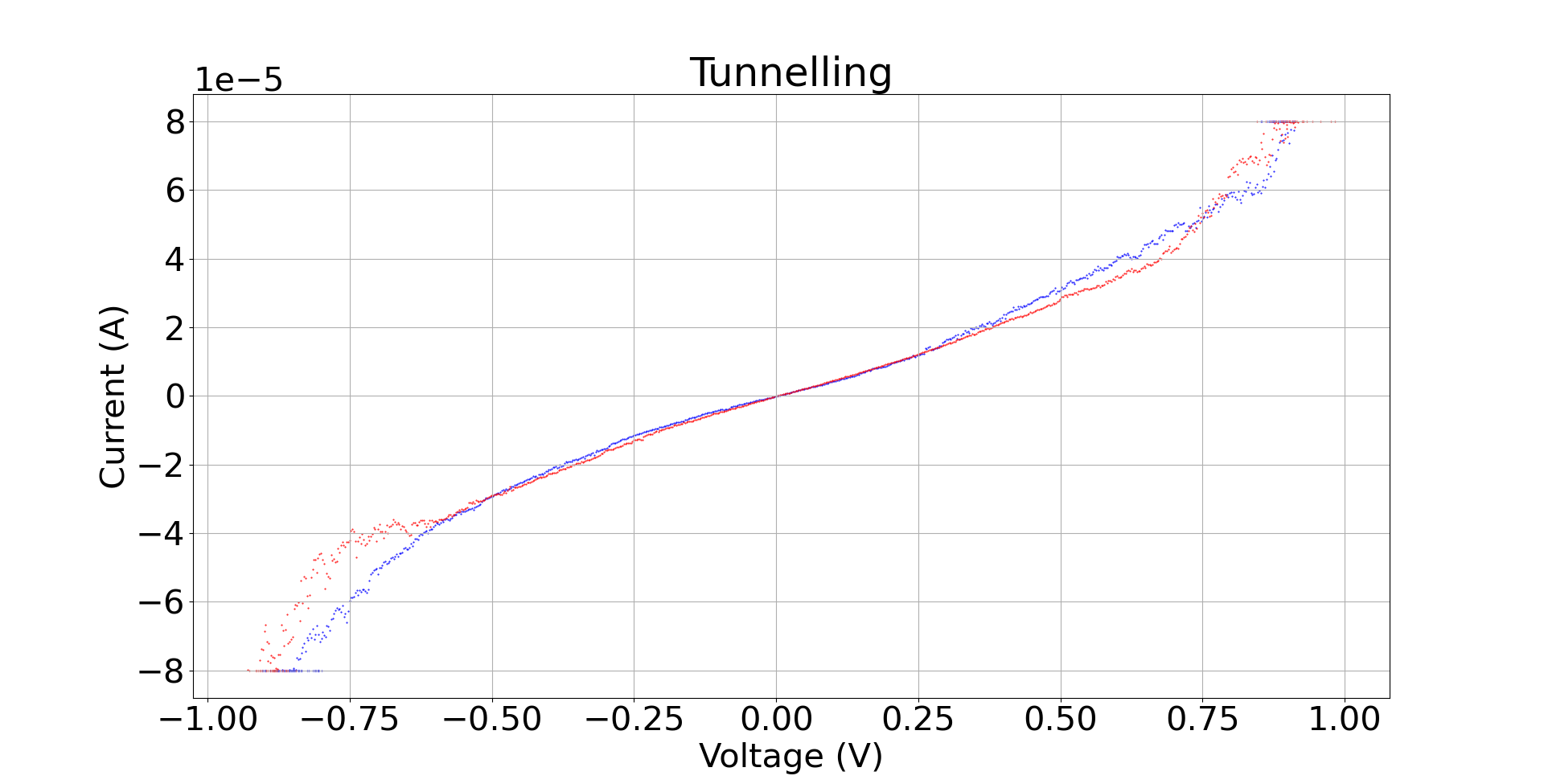}
		\subcaption[]{High tunnel current.}	
		\label{fig:tunn1}
	\end{minipage}
	%\hfill
	\begin{minipage}{0.7\textwidth}
		\centering
		\includegraphics[width=1\textwidth]{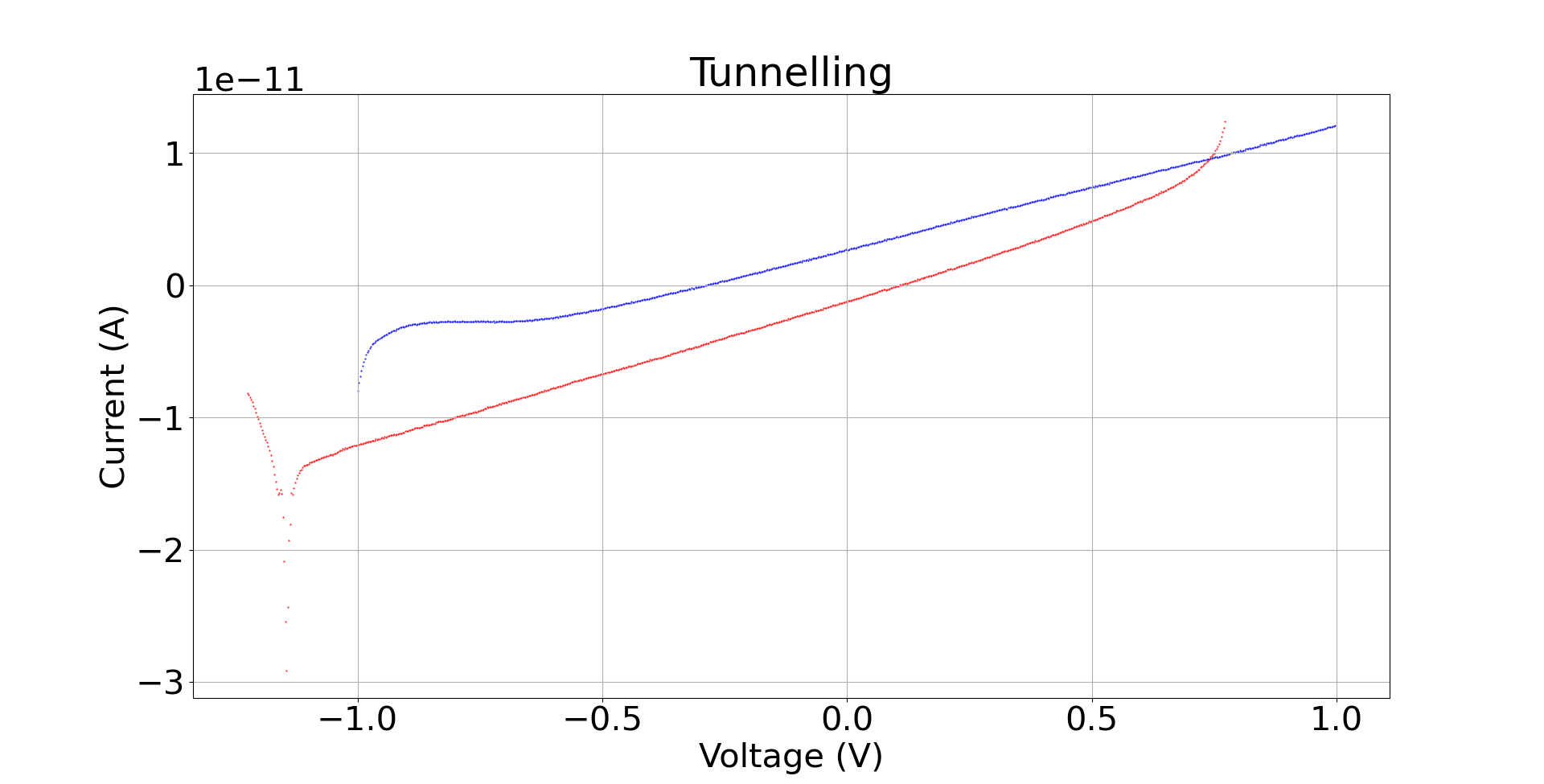}
		\subcaption[]{Low tunnel current.}
		\label{fig:tunn2}	
	\end{minipage}
	\caption{Tunnelling currents. In Figure (a), we observe comparable high tunnel currents, with up to $10\, \mathrm{\mu A}$, which is a typical tunnelling value if the leads are close as in the position 1 in Fig. \ref{fig:op}, and in (b) shows a case when the distance is lager, and the currents are smaller by a factor of $\approx 10^{-6}$ compared with (a), as in the position 4 in Fig. \ref{fig:op}. In some cases the distance become so large that the molecule can be considered detached from the leads.}
	\label{fig:tunnelling}
\end{figure}

\FloatBarrier

\noindent In the statistical evaluation of these measurements, we must synchronize them, which is done at a specific value below 1 G/G0 (at $0\, \mathrm{V}$). The data is synchronized and (time) shifted so that one can see the behaviour of all data in the same graph.

\noindent In Fig. \ref{fig:tunnelling}, the ``butterfly'' sweeps are from  $0\, \mathrm{V} \longrightarrow 1\, \mathrm{V} \longrightarrow -1\, \mathrm{V} \longrightarrow 0\, \mathrm{V}$, and depending on the molecule; the range is different, as for example, Co$^{+3}$ Salen $0\, \mathrm{V} \longrightarrow 0.8\, \mathrm{V} \longrightarrow -0.8\, \mathrm{V} \longrightarrow 0\, \mathrm{V}$. After the measurement finishes the bridge continues to open. If the plateau persists another measurement is realised. This proceed is repeated until the junction opens, and then the MCBJ closes until $20 \, G_0$.

\noindent In the forthcoming chapter, we present theoretical considerations and models pertinent to our data. This deliberate inversion aims to acquaint the reader with the essential background knowledge required for engaging with our discussion chapter effectively.
	\chapter{Molecules and Transport}

\section{Molecules in the Scope of this Thesis}

We utilised the organometallic compound Fe$^{+3}$ Salen in our initial experiments. However, we observed unstable and irregular current patterns characterised by noticeable gaps. In an effort to refine our current measurements, we introduced Corannulene at a late stage. This decision was driven by Corannulene's promising characteristics, which made it particularly suitable for studying current within the Single-Level Model (SLM) framework. Additionally, our research extended to include other organometallic complexes derived from Salen, such as with inclusion of cobalt and manganese atoms. These were incorporated into our analysis later based on the extensive data we had accumulated for them.

\noindent The primary focus of our study, and consequently this thesis, has been moved from Fe$^{+3}$ Salen to Corannulene. As such, the most comprehensive and detailed information about current measurement and SLM pertains to this molecule. While we have endeavoured to present a thorough analysis, including additional molecules, particularly at a later stage, may have introduced some complexity in the narrative. We acknowledge this potential for untidiness in our presentation and apologise for any inconvenience it may cause for reading.

\subsection{Fixation of pi-Conjugated Molecules on 	Gold Surfaces via Thiol Bond}
The electronic current going into the molecular orbitals is influenced by how the molecules are bound and connected to electrodes and the location of HOMO and LUMO levels, which will conduct current after a resonance between molecular channels and leads contact to be archived. This resonance between the molecular channel and Fermi level in the gold leads significantly impacts molecular transport. In addition, adhesion between the molecule and the surface is crucial to controlling its electronic properties, which atom in the lead the attachment is realised, and if both leads are attached or only one to the molecule. The MCBJ technique involves creating two electrodes on an insulating substrate using lithography. These electrodes have narrow conductive gold wires, and their openings can be adjusted to match the length of the molecule by bending the substrate. However, the exact configuration of the molecule on the electrode tips is currently unknown. Which get attached to the electrodes usually through thiols group $(R-SH)$%, sulphide $(R-S-R)$, %disulphide $(R-S-S-R)$,  thiocyanate $(R-SCN)$, alcohols $(R-OH)$, amine $(R-NH_2)$, alkene $(R-CH=CH_2)$ or silane $(SiCl_3, Si(OMe)_3, Si(OEt)_3)$ 
 \cite{Ulman1991}. The relative positioning between the molecule and electrode also determined by the presence of the surrounding molecules. In our case, the solvent and the dissolved molecules belong to the surroundings and can slow down or prevent reactions to catch the atom between the leads, and this phenomenon is also known as a steric hindrance. The reaction is between the gold and anchor group, and the contact will be stable between the electrode gap, and to several distances of the electrodes the molecule still binds \cite{Briechle2012}.

\section{Ballistic Transport}

Ballistic transport can be studied as classical or quantum ballistic transport. For molecular electronics, this depends only to the size of the studied molecule. Whether the elastic length in the molecule is bigger than the mean free path, which is bigger or has the size in the same order of magnitude as the molecule, the transport regime will be quantum ballistic. While a metal conductor has dimensions much larger than the mean free path, it exhibits Ohmic behaviour. When these dimensions become of the order of the mean free path, the conductor exhibits other transport effects. In the case of the system dimensions are much smaller than the mean free path, the carriers will move ballistically through the conductor:

\begin{eqnarray}
	&& L_m >> L_X, 
\end{eqnarray}  

\noindent $L_X$ system dimension, $L_m$ the mean free path \cite{Datta2005}.

\noindent A conductor is referred to as ballistic, if the conductor is too small related to the mean free path of the charge carriers moving inside it. In this case, the conductance is a multiple of the quantum conductance \cite{aldea2016}, and their carriers will not lose energy, momentum or phase because they will not suffer collisions. The electrons will be close to the Fermi level without enough energy to scatter because there are no free states available for them to scatter into, which depends on the intrinsic properties of the conductor.

\subsection{Tunnelling}

%\subsubsection{Direct tunnelling}	
Based on the shape of the energy barrier, it can be classified into direct tunnelling and Fowler-Nordheim (FN) tunnelling. In direct tunnelling, electrons move within a trapezoidal barrier and are confined within it, never escaping. Conversely, in FN tunnelling, the triangular shape of the barrier allows electrons, which initially travel within the barrier and then out, eventually ``escaping'' the energy barrier.

\begin{equation}
	I \sim \sinh\left( \frac{eV\, d}{\hslash}\sqrt{\frac{ m_e}{2\varphi}}\right) \label{eq:dist} ,
\end{equation}

\noindent $m_e$, electron mass, $\varphi = E_F-E_{HOMO}$ or $\varphi = E_{LUMO} - E_F$ is the high of the barrier \cite{Huisman2009}. The direct tunnelling is explained by Simmons' model modified for low fields.

\subsubsection{Simmons Asymmetrical Barrier Model}

\noindent The model for the tunnelling occurring between the leads and molecule depends on the position of the single channel relative comparison to the Fermi level in the leads. Whether this channel has energy much higher than the Fermi level of the leads $E_0 \geq E_F$, the barrier will be asymmetrical, and the Simmons' model for asymmetrical tunnelling barrier (or trapezoidal barrier \cite{Simmons1964}) will describe the transport more inaccurate than the SLM. The reason is that the SLM description is based on the resonance between the channel and leads.

\subsubsection{Resonant Tunnelling}
A quantum well is a structure that confines electrons in a specific space. Per the laws of quantum mechanics, the energy of confined electrons is fixed and can only exist at specific levels. However, the electrons can still move in the perpendicular direction, creating quasi-bound states in a two-dimensional electron gas \cite{Kosik2004}.

\noindent Resonant tunneling describes a phenomenon in which the electron transmission coefficient through a structure exhibits sharp peaks at specific energy levels. To understand this tunnelling effect, one can imagine the scenario as being bounded by infinite walls. The energy levels within a quantum well, regardless of its shape, are determined by solving an eigenvalue problem. When the energy of an electron precisely aligns with the quantum well's virtual resonant energy level, the transmission coefficient approaches unity. Under these conditions, the electron can tunnel through the barrier with minimal reflection \cite{Kosik2004}.

\section{Single Level Model (SLM)}
\label{sec:SLM}

The Zotti-Erbe\footnote[3]{Zotti-Kirchner-Cuevas-Pauly-Huhn-Scheer-Erbe} SLM is based on the resonant-tunnelling model, i.e. the Fermi level of contacts and channel are align. However, the current in the resonant condition is hard to detect, and the IV characteristic has an `S' shape. It emerged for the first time for room temperature to describe the single molecule current-voltage characteristic \cite{Zotti2010}:

\begin{itemize}
	\item the transport is phase coherent (tunnelling mechanism);
	\item the current is dominated by a single channel resonance in the entire voltage range explored in the experiments \cite{Zotti2010}.
\end{itemize}

\label{sec:another_app}
\noindent We write down the equations resulting from the Landauer formalism \cite{Zotti2010}:

\begin{equation}
	I(U) = \frac{2e}{h}\int_{-\infty}^{\infty} T(E,U) [f_L(E-eU)-f_R(E)] dE, \label{currentAp}
\end{equation}

\begin{equation}
	T(E,U) = \frac{4\Gamma_1\Gamma_2}{(E-E_0)^2 + (\Gamma_1+\Gamma_2)^2}, \label{eq:Breit-Wiegner}
	\label{TransmissionAp}
\end{equation}

\begin{equation}
	E_0(U) = E_0 + \frac{eU}{2}\frac{(\Gamma_1-\Gamma_2)}{(\Gamma_1+\Gamma_2)}.	\label{E0Ap}
\end{equation}

\begin{figure}[!htb]
	\centering
	\begin{minipage}{0.48\textwidth}
		\centering
		\includegraphics[width=1\textwidth]{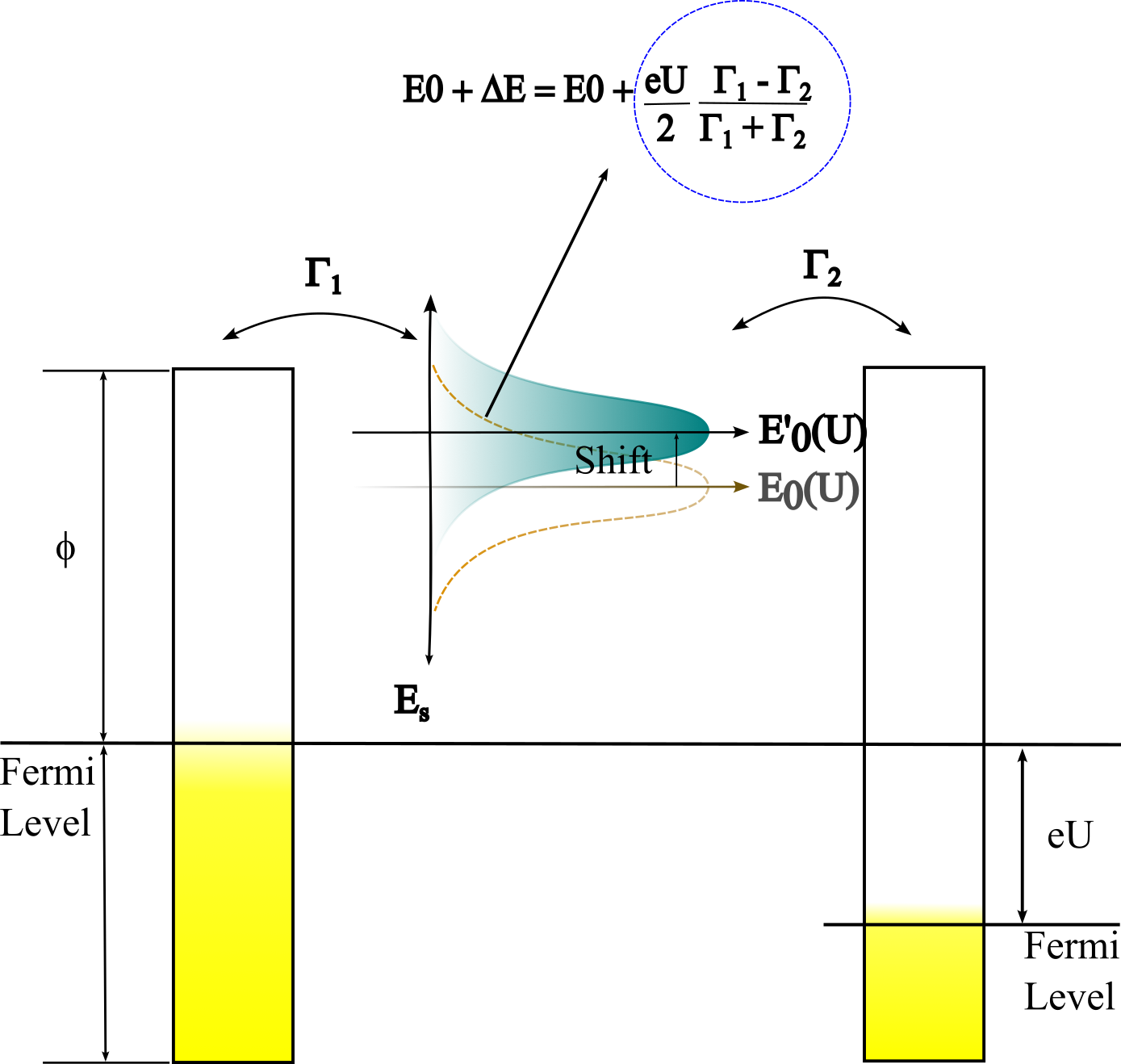}
	\end{minipage}
	\label{fig:shift}
	\caption{This figure represents the equation \ref{E0Ap}, and $\phi$ in this figure is the work function.}
\end{figure}
\noindent The expressions Eq. \ref{currentAp} and \ref{eq:Breit-Wiegner} are the SLM formalism. Additionally, Eq.\ref{E0Ap} is an adjustment for the channel position \cite{Zotti2010}.
\noindent The energetic position of the molecular orbital, concerning the Fermi energies of the metals, depends on the charge transfer between metal and molecule \cite{Zotti2010}, which changes with coupling and can also be influenced by the bias applied at the source and drain. Whether both electrodes have the same coupling as the molecule does not matter, but in the case of an asymmetric junction, the coupling of the channel will be affected by the bias at one electrode more than at the other. Hence, the implementation of the factor for adjustment is necessary.

\noindent The potential difference between the one-atom contact junction, with potential $U$, is $\Delta U$. The current sweep is applied to only one lead, and the other is grounded relatively to $\Delta U$, so $\Delta U = U_n-U_{n-1}$, and $\Delta E = E_n-E_{n-1}$, and `n' is sub-energy states $n\geq 1$ inside the channel. Electrons are submitted to the energy potential $\Delta E = Ue$ between the leads, which will move the channel up and down, allowing proximity with the gold lead Fermi level producing the resonance with molecular orbital. The total energy necessary to cross the channel is $E_\mathrm{channel} = Ue - E_0$, $E_0 = E_F - E_\mathrm{HOMO/LUMO}$ \cite{Briechle2012}.

\subsection{Chemical Nature of the Molecular Channels}

The combination of on-site and transition energies contributes to the transport proprieties of the conductive channel. For proper calculation, we need to consider the energy of the whole system and account for the structural importance of the atomic orbitals to the transport.
Metals with `$s$' orbital in the outermost shell have better-quantified conductance. Therefore, their conductance values will be closer to the integer-numbered levels. This quantification can be observed in alkaline and noble metals. Another family of metals have `$sp$-like' orbitals, such as transition metals with `$sd$'-hybridised orbitals. The geometry of the molecular orbitals will influence the transport and the geometry of the leads because of their influence, first on the wave function and second on the coupling between the atoms. Also, they must be electrically neutral to avoid losing electrons \cite{Erbe082020}.

\noindent The number of participating orbitals gives the dimension of the matrix of coupling elements. This matrix is used for the evaluation of the transmission coefficients. Therefore, the dimension reduces to one when only the s-orbital participates in transport. 
Each different orbital `$s, p, d$'  has a different transmission coefficient and hybridisation. However, when the orbitals are consecutively occupied with electrons, the transmission coefficient of the whole atom is also influenced by the electron-electron interaction \cite{Erbe082020}. In the case of a one-electron system, the channel conductance adds up to

\begin{equation}
	|Channels\rangle = T_s|s\rangle  + T_p|p\rangle + T_d|d\rangle + T_h|h\rangle. \label{channel}
\end{equation}

\noindent Equation \ref{channel} uses $T_s$, $T_p$, $T_d$ as transmission coefficients or probabilities for the orbitals, and $T_h$ is for their hybridization. When the density of states of these orbitals greatly differs from the Fermi energy of the contact metal, the conductance matrix's trace elements will not be integer multiples of the conductance quantum.

\noindent Therefore, the junction experiments at low temperatures can prove that channels role in single-atom charge transport. This transport will occur via tunnelling between the tips and Cooper pairs at low temperatures close to $0$ K, and it is possible to see that bulk materials have different proprieties from single atom contact \cite{Erbe082020}.

\section{Transport Mechanisms in Molecules attached to MCBJ}
The conductance of a molecule is highly dependent on how it is attached to the junction. So, if the molecule itself is a good conductor, but the anchor group that attaches the molecule to the gold junction is bad, then we have a poor conductor overall. In this case, the coupling dominates the conductance. Additionally, the transport also depends on whether the molecule's single channel is close or distant from the contacts' Fermi level, the whole system must be always taken into account. When connecting molecules to electrodes, there are two types of bonds: chemical and physical; chemical or covalent bonding results in solid coupling to the electrodes and good conductance.
\noindent On the other hand, physical bonding relies on weaker van der Waals forces or polarisation, resulting in poor coupling to the metal. When a molecule is placed between two metal electrodes, its molecular orbitals moves trying to align with the band structure of the electrodes and getting closer to them. The reason is that the electronic states in the metal are occupied up to the Fermi edge. If no potential is present at the contact, the Fermi levels are situated between the HOMO and LUMO in the molecule \cite{Eberlein2008}. Hence, we apply a small potential to the electrodes. As a result, the Fermi level on the electrodes will align with molecular orbitals, and the current can flow.
The conductive properties of a molecular system depend on its lowest unoccupied molecular orbital (LUMO), its highest occupied molecular orbital (HOMO), and its vibration modes. The vibrational modes may couple with the electron during a tunnelling event, in case of a weak coupling or off-resonance tunnelling, because when an electron is scattered, releasing or absorbing a phonon, and its energy may change. The current will lead to a strong coupling if the conductance $G(G_0)$ is sufficiently high. In this case, electron-phonon-scattering or inelastic electron-electron-scattering can reduce the current because the electron will lose energy instead. The analysis of energy transmission diagrams can distinguish the type of coupling. In such a diagram, the respective effects will be separated into very distinguishable regions \cite{Erbe2020}.

\noindent Nevertheless, for incoherent tunnelling, the electron tunnels via a series of sites, which are characterised by potentials well; if the electron stays for long in one of them, it will suffer phase perturbations. This kind of tunnelling is, in principle, temperature independent; in case these mentioned ``resistances'' among sites are more prominent than quantum resistance, Coulomb charging and blockade effects may occur. During its journey, electrons play an essential role in exciting vibronic modes of the molecule. This excitation happens if the energy of an electron exceeds the threshold for excitation $V >\hslash\omega/e$ \cite{aldea2016} or their multiples. However, these blockades will not be activated at RT. Also, the vibronic modes are exited at RT, making it very hard to leave signatures in those measurements.

\noindent In molecular transport, quantum interference can occur when the size of the molecule is same magnitude to the electronic phase coherence length. This similarity happens when the electron wave reaches a point where it can travel through different paths that eventually intersect. It can either enhance or diminish the conductance, depending on the kind of interference if it is constructive or destructive \cite{Guedon2012}, \cite{Aradhya2012}, \cite{Ballmann2012}. These paths should be smaller than the electronic phase coherence length. This path length explains, for instance, that the conductance of a cross-conjugated molecule can be significantly lower than that of a linearly conjugated one due to an anti-resonance in the transmission function \cite{Guedon2012}. However, these effects are only observable at LT, as the background noise at RT masks their signatures. Furthermore, any changes in their population are difficult to detect in RT measurements.

\noindent Despite the perfect conductivity of nanoscale devices, resistance is still present. The trace of the channel's transmission matrix is simplified in the linear regime to one element \cite{Erbe} \cite{Erbe2020}. Typically, the process involves the interaction of molecules with electrodes and coupling with vibrations and external factors like light and thermal gradients. It is crucial to consider the molecule's transport, the spectral densities involved in the coupling, and the impact of interactions between molecules and electrodes, the broadening and shifting effects, as they provide valuable information when conducting conductance calculations \cite{CunibertiFagasRichter2005}

\noindent Another point to be considered is the conductance of a molecule is composed by 

\begin{equation}
	G =  G_{Ballistic}+G_{Elastic}+G_{Inelastic}+G_{Asymmetric}, \label{eq:cond}
\end{equation}

\noindent contributions of the ballistic, elastic, inelastic and asymmetric conductances. The elastic term is from elastic scattering events such as electron-electron scattering. The inelastic term is from the weak coupling between electron-phonon, so when a high bias is applied, it might excite vibrational modes, and then electrons can tunnel from left to right electrode but inelastically absorbing or emitting a quantum of energy $\hslash \omega$\cite{Sondhi2021}.

	\chapter{Results and Discussions}

The attachment of molecules in electrodes is not trivial and makes mechanically controlled break junctions (MCBJ) bridges a complex topic depending on the molecule. For instance, the literature consistently documented quantum effects like electron tunnelling at room temperature across various molecules, forming the foundation of Single-Molecule Level (SLM) studies. On the other hand, some molecules, such as Corannulene, present diverges of the model, other as Fe$^{+3}$ Salen shows a flat curve for low bias, and hysteresis, making it harder to evaluate the discrepancies between model and transport. Several sections of this chapter exploit the model divergence and attempt to explain it based on the literature. 

\noindent Our primary objective is to present our statistical evaluation, aligning them with theoretical models to identify and understand any discrepancies. In this work, single-measurement graphs are used, which effectively highlight key trends and anomalies in the data. However, for the statistical analysis, we maintain the minimum number suggested by the literature of at least 30 samples. This threshold is widely acknowledged as sufficient for ensuring the reliability and robustness of statistical conclusions, thereby enhancing the credibility of our findings.

\noindent In the upcoming sections, we compare the SLM, and our observed measurements. Our primary aim is to dissect and elucidate the reasons behind the model's inability to accurately represent the data, particularly at the origin point and in the tail regions. We provide substantive evidence supporting our hypothesis that electron-electron interactions are the predominant cause of the observed discrepancies at the origin. In the tails, the divergences are caused by channel saturation, which is also not well represented in the model. The SLM model's inability to accurately fit the data in this regime could be attributed to its oversimplified handling of these shifts. Additionally, channel saturation indicates that the model might not adequately account for factors like the density of states in the electrodes and molecule, molecular orbital broadening or deformations, and the dynamic response of the molecule's orbitals to the applied electric field. Additionally, while this thesis primarily focuses on these aspects, we briefly touch upon other potential factors influencing current transport, acknowledging their presence but noting that they fall beyond the scope of our current study.

\noindent Each phenomenon requires a deep understanding, and since low-temperature (LT) measurements are not within the scope of this work, we will only rely on the phenomena observed at room temperature (RT), which will allow a shallow evaluation. However, we will find comprehensive and enough explanations for the `artefacts' that appear throughout the measurements. Our Corannulene's measurements are more suitable to study resonant tunnelling in a single channel than Fe$^{+3}$ Salen, because our measurements for this molecule were more smooth with fewer `artefacts'. In these measurements, Fe$^{+3}$ Salen's current characteristics show more discontinuities, kinks and plateaus than Corannulene. Additionally, a significant level of hysteresis in these measurements tends to obscure the phenomena we aim to observe, making Corannulene a more suitable subject for our investigation.

\section{Discussions}
\label{chapter:Measurements}
\subsection{Opening Curves}
\noindent The samples mounted on the mechanically controllable breaking junctions (MCBJs) presented here have been measured at room temperature, 
and in a non-controlled environment (the humidity, pressure, and temperature were not controlled).
The measurements were analysed using a Python script\footnote[3]{Author/programmer Holger Langer, Nanoelectronics - Helmholtz Zentrum Dresden Rossendorf}.
The script synchronizes the measurements, in an interval of $10$, between $[-5, 5]$ i.e., each cycle from $20\, G_0$ to $G_0/N\sim10^{-9}$, where `N' is chosen by us to represent the opened state. This procedure will uniformly span the same duration for each measurement, from the moment the bridge is fully closed until it is completely open. Given that the DC motor operates at a constant speed during the opening process, we can synchronize the measurement times across x-axis. This synchronization allows us to use the gathered data effectively in creating histograms, which are crucial for the statistical evaluation of our samples.

\noindent The opening curves in one-atom contact will show stairs; because at the nanoscale, the conductance will change from continuous line, observed in bulk materials, to steps. This result can be observed at room temperature, as it is already known at low-temperature \cite{CuevasScheer2010}.

\begin{figure}[!htb]
	\centering
	\begin{minipage}{0.8\textwidth}
		\centering
		\includegraphics[width=1\textwidth]{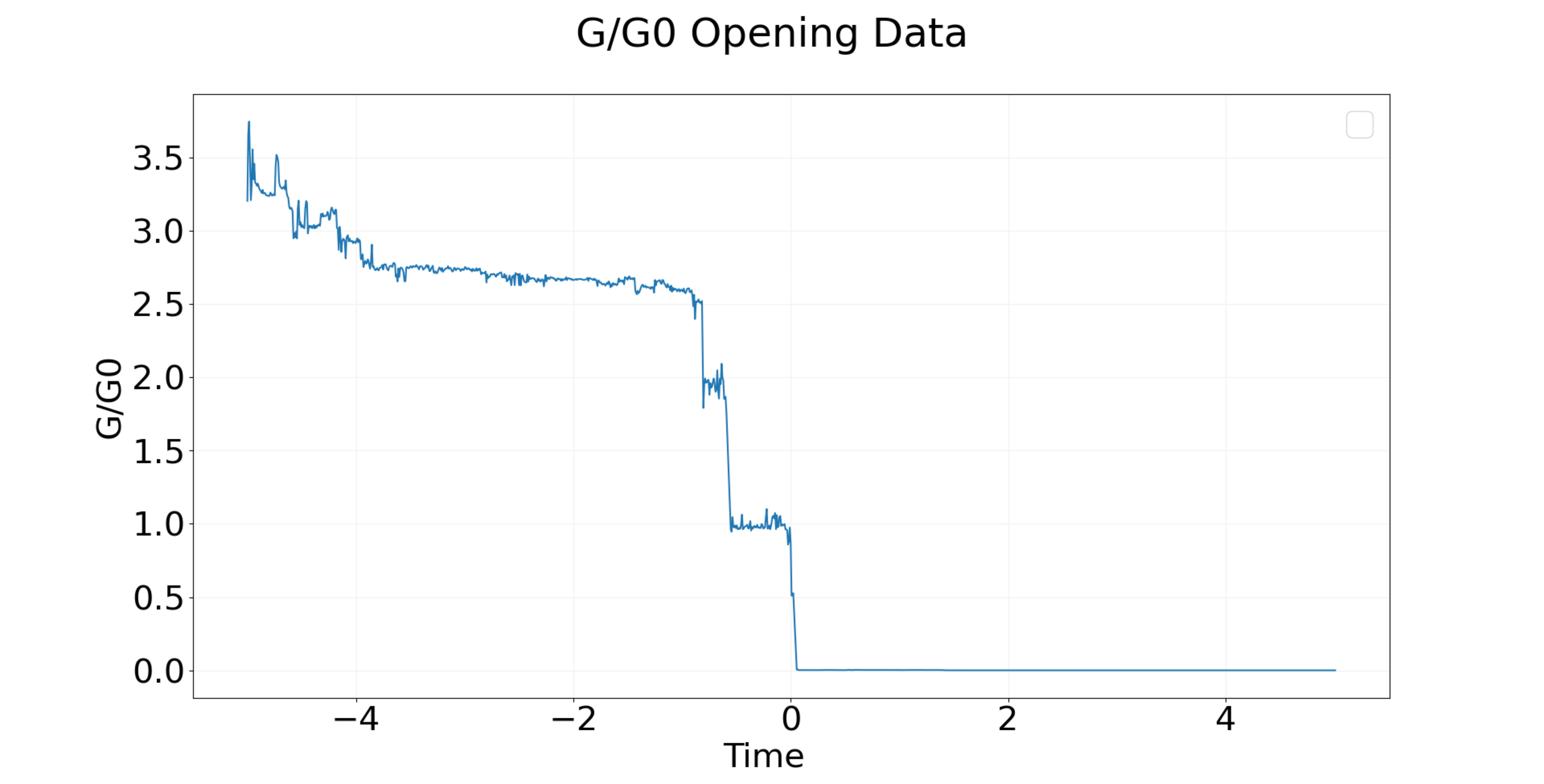}
		\subcaption[]{}
		\label{fig:opening2}
	\end{minipage}
	\caption{MCBJ opening curves for the dry sample measurement at RT. The synchronization (see Kilibarda \cite{Kilibarda2021}) change the time axis to the interval $[-5:5]$ for all measurements, the $x$ axis is in arbitrary units. In Fig. (b) the bridge opens at $t=0$, i.e. for $t>0$ , the bridge is opened $G/G0=0$, and for $t<0$ the bridge is closed. %Also the effect of the resistor from the MCBJ bridge's circuit should be subtracted once 
	The pronounced quantization steps observed in the graph correlate with the involvement of orbitals in gold's transport mechanism. This phenomenon occurs when the bridge, due to its limited size, no longer exhibits properties characteristic of gold in its bulk form. The subtle step noted at $2.5\, G_0$ just prior could be attributed to the atomic configuration at the tip of the bridge, right before the manifestation of quantum effects.  %and this residence is present in all current  measurement curve for the gold. This resistance does not affect the system after the bridge opens, it is just for circuit protection and does not interfere in our analysis.
	}
\end{figure}

\FloatBarrier

%\noindent After 100 measurements were obtained at RT. The handling of the data was done using Python, so the measurement results can be plotted in the same figure. It is possible to see the steps for each measurement.
\noindent The measurement in Figure \ref{fig:opening2}, shows the example of one curve with steps. We observe that the steps do not show integer multiples of $G_0$, as expected for metals. The closing curves will not be used to evaluate the current in the bridge because this process is less reliable, due to the behaviour of the sample's contacts.

%\clearpage

\subsection{Histogram from the Measurements}

This section details three distinct sets of measurements conducted on a gold bridge: in dry conditions, with pure toluene, and with a molecule dissolved in toluene. Initially, for all measurements, the sample is set up and measured inside a fume hood at room temperature. In this phase the dry bridge realises 100 cycles of opening and closing for calibration, alongside current and voltage measurements. Subsequently, the setup is modified by mounting a pipette filled with pure toluene. The measurement process is then repeated, extending over 100 cycles to enhance calibration accuracy. During this stage, the combination of the bridge and pure toluene serves to clean the sample. Following this, another 100 cycles of opening and closing are conducted, typically resulting in the sample breaking at a consistent position. The final step involves introducing a solution of toluene mixed with the molecule into the pipette for measurement.

\begin{figure}[h]
	\centering
	\begin{minipage}{1\textwidth}
		\centering
		\includegraphics[width=1\textwidth]{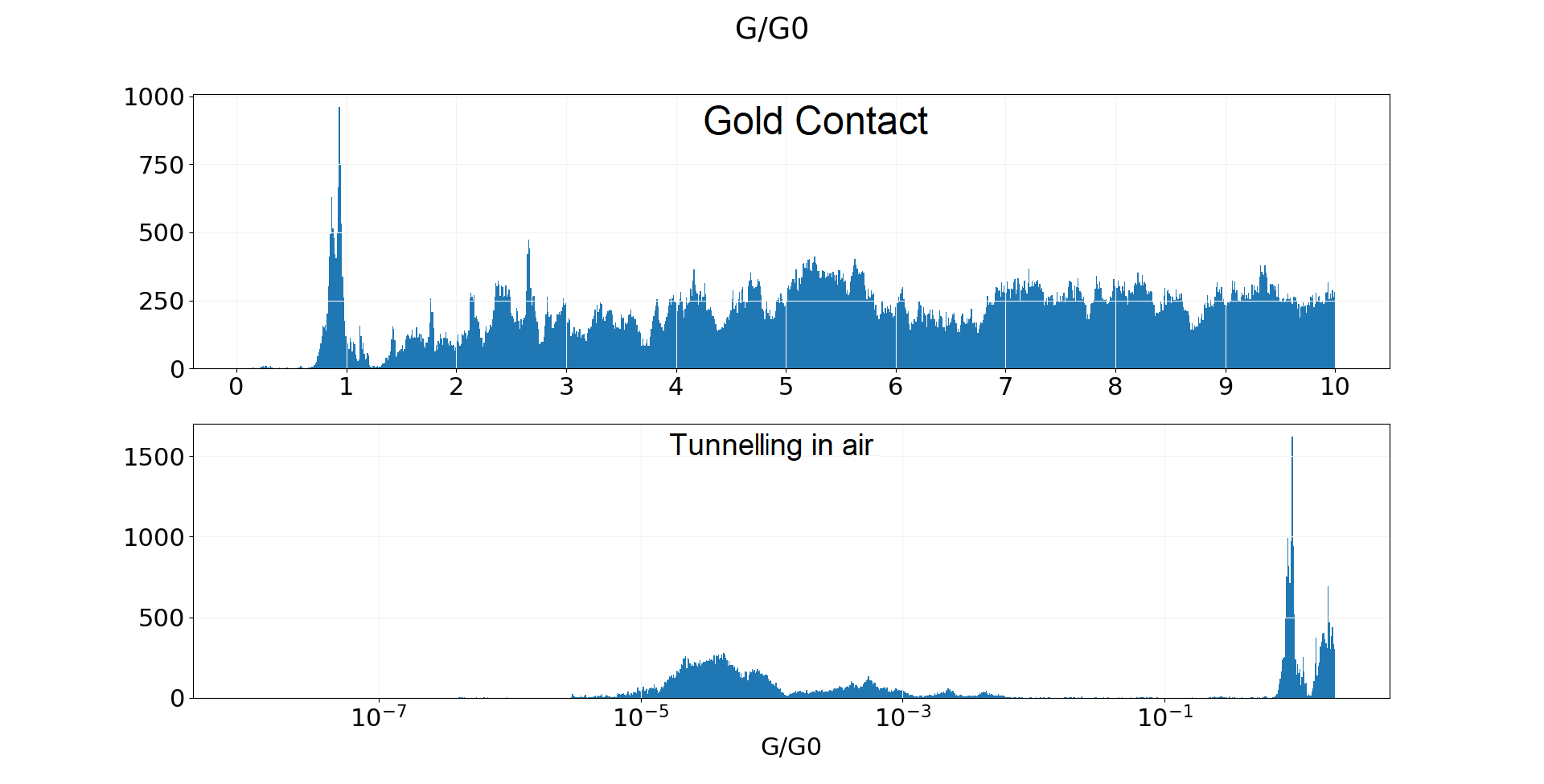}
		\subcaption{Bridge dry. Top: histogram of $G/G_0=0...10$. Bottom: $G/G_0=10^{-8}...10^0$.}
	\end{minipage}
	\begin{minipage}{1\textwidth}
		\centering
		\includegraphics[width=1\textwidth]{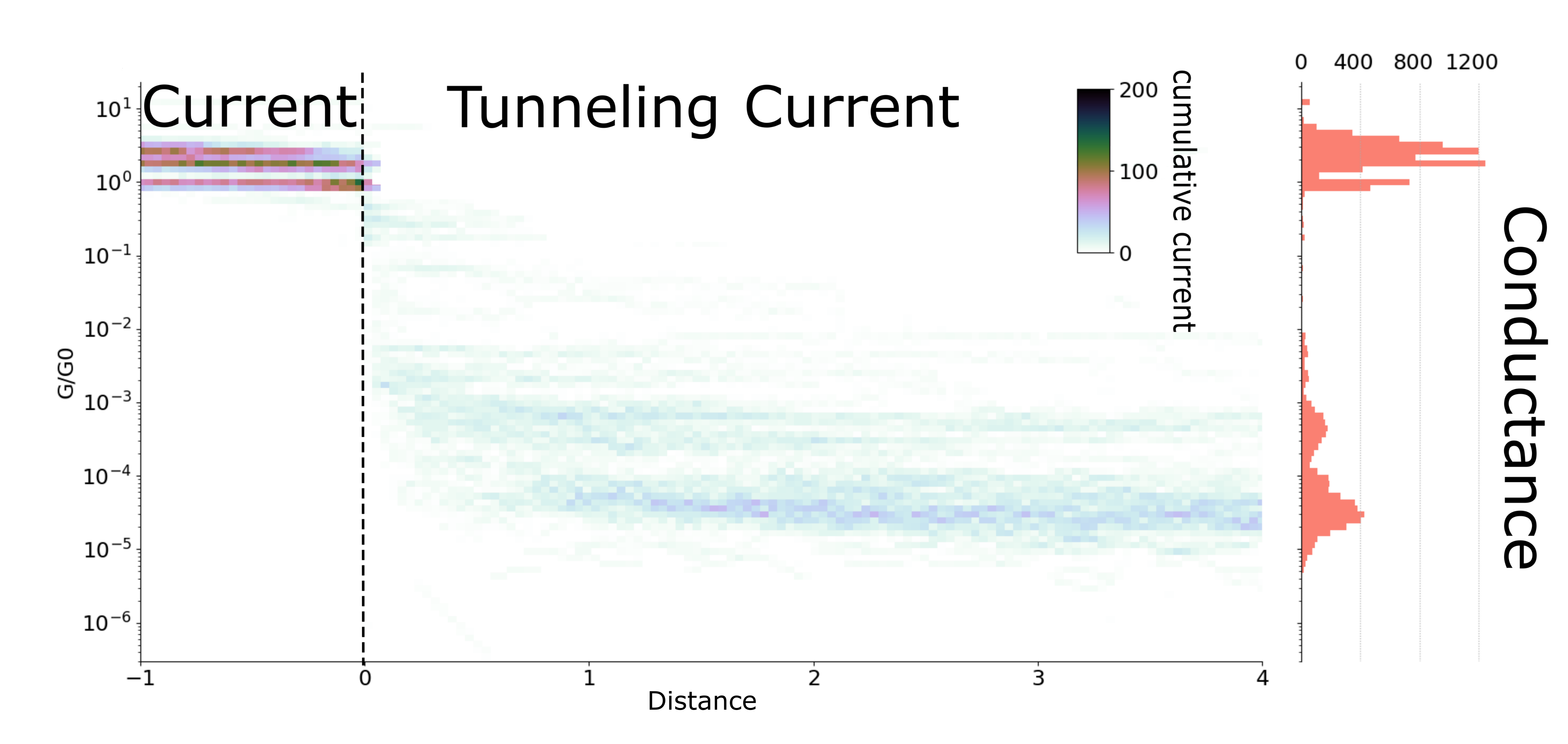}
		\subcaption{G/G0 vs Distance vs Counting: Dry sample one point contact.}
	\end{minipage}
	\caption{Histograms for a MCBJ dry bridge, c.a. 200 measurements. (a) shows a 1D histogram for the $G/G_0$ counts.  $G/G_0$ assumes values in the range of $1..10$, in case of closed bridge (gold contacts closed, top figure in (a)). In case of an open bridge the values are $G/G_0<<1$ and the tunnel current is visible (gold contacts opened, bottom figure in (a)). (b) shows a 2D Histograms: $G/G_0$ vs. the corresponding distance (dimensionless) (check ref. \cite{Kilibarda2021}) and proportional to time $distance\sim time$. The colour scale indicates the bin counts for pairs of [$distance,\, G/G0$] in the main plot. Above the main plot: 1D histogram of distance counts those distances, for which there is a current flow. On the right-hand side to the main plot: 1D histogram of $G/G_0$ counts. The counts on the colour bar are to report of the arbitrary value of files in this case c.a. $200$.}
	\label{fig:hist}
\end{figure}

\FloatBarrier

\noindent Figure \ref{fig:hist} is a result of dry measurement, in Figure \ref{fig:hist} a we can see the peak at $1\, G_0$ this peak is broadened, because the statists of each measurements and its different atoms arrangements, changing the conducting channel. 
The one-atom contact gold at room temperature is exposed to a wide noise spectrum such as thermal noise \cite{Konczakowska2018}. The leads are not isolated for the dry measurement; for example, humidity can cause an undesired effect in the measurements, which will appear in the measurement curve.

\begin{figure}[!htb]
	\centering
	\begin{minipage}{1\textwidth}
		\centering
		\includegraphics[width=1\textwidth]{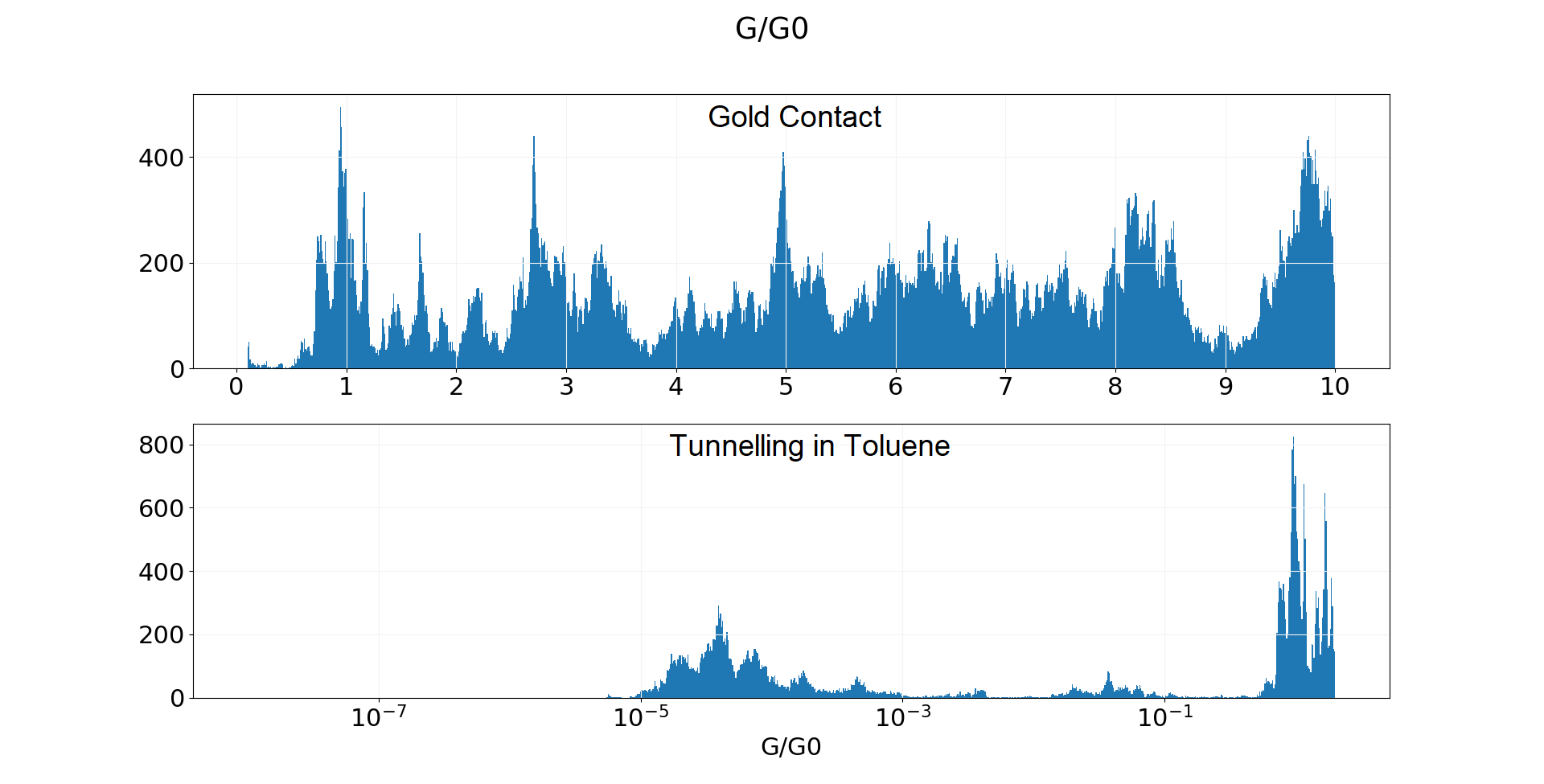}
		\subcaption{Bridge with toluene, histogram of $G/G_0$.}
	\end{minipage}
	\begin{minipage}{1\textwidth}
		\centering
		\includegraphics[width=1\textwidth]{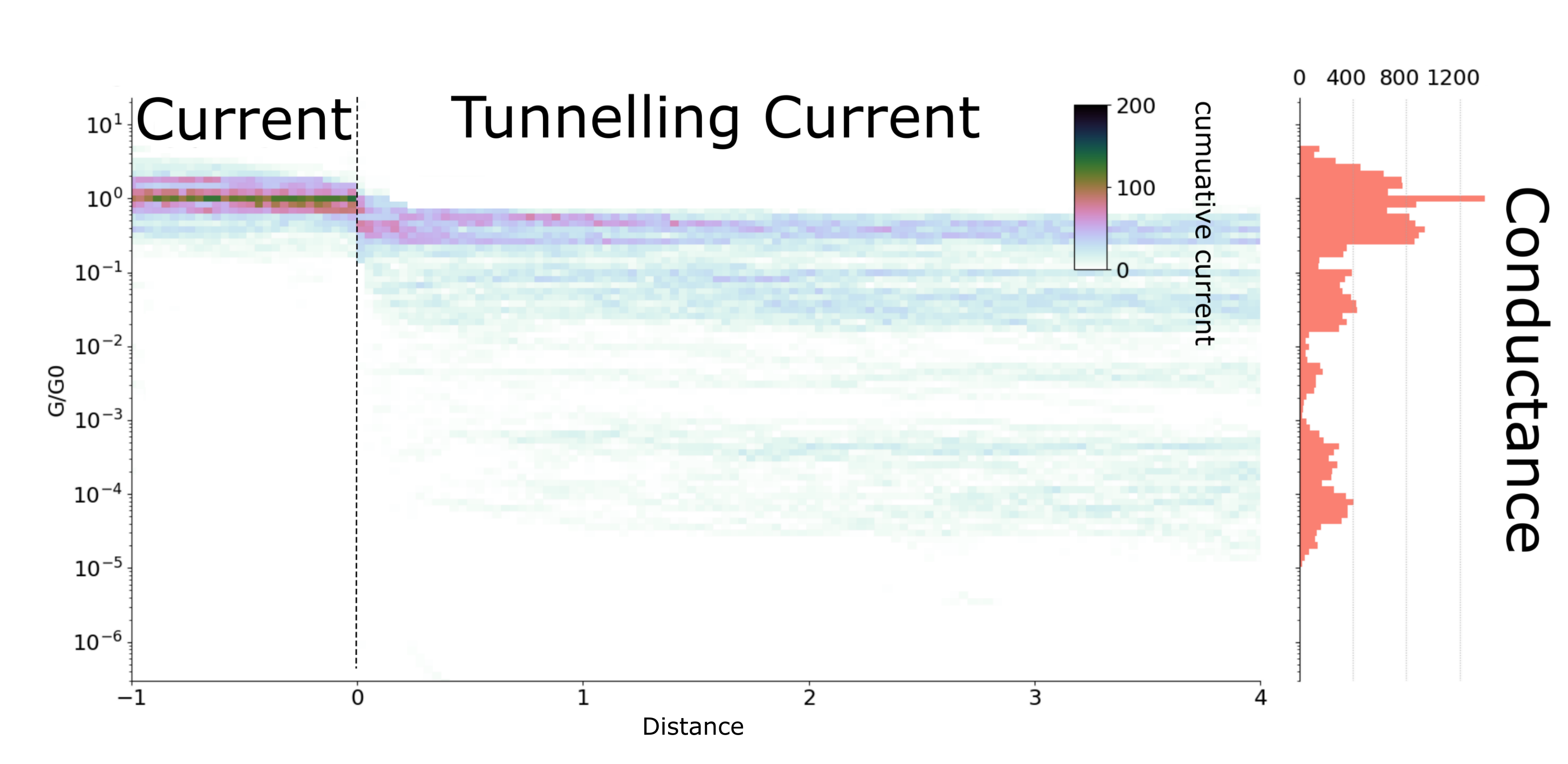}
		\subcaption{G/G0 vs Distance vs Counting: toluene and sample measurements,one-atom contact.}
	\end{minipage}
	\caption{Histograms for MCBJ toluene measurements, c.a. 200 measurements. Compare with\ref{fig:hist}. Our measurements with toluene do not seem to be very clean; it would be expected from this measurement that it has much less tunnelling than the dry one.}
	\label{fig:toluene}
\end{figure}

\FloatBarrier

\noindent In Figure \ref{fig:toluene}, it is represented both the counting in the histogram when the bridge is closed and open. The counts observed in the open bridge state are attributed to the tunnelling effect between contacts immersed in the solvent \cite{LukaGuth2016}. The noise levels do not change. 
Furthermore, the solvent plays a crucial role beyond just acting as a medium. It is used to dissolve the molecular powder, enveloping and protecting the molecules from environmental factors. This encapsulation not only enhances the stability of molecular junctions in our measurements but also significantly influences their conduction properties \cite{CuevasScheer2010}. Additionally, the solvent facilitates the diffusion of molecules towards the metal electrodes, thereby enabling the repeated formation of new and independent junctions \cite{LukaGuth2016}.

\noindent Also the distance in all graphs are dimensionless, the calculation of the distance on Eq. \ref{eq:dist} is repeated bellow:
\begin{equation}
	distance \sim \frac{\hslash}{eV} \left(                                                                                                                                                                                              \sqrt{\frac{2\phi}{m_e}}\right) \sinh^{-1}\left( I \right)  \label{Eq:SiTu}
\end{equation}

\noindent In our analysis, we employ Simmons' Model for low fields to elucidate the tunnelling events, specifically how they correlate with the distance between the leads \cite{Huisman2009}. A key aspect of our methodology is the assumption that the distance between the tips during the sample's bending is proportional to time ($d \sim t$), a condition made possible by our constant bending rate. This proportionality is crucial as it allows us to deduce the specific distance at which the tunnelling event occurs. Neither Eq. \ref{attenuation} was used to obtain the distance in this thesis nor \ref{Eq:SiTu}. Although Simmons' approach is considered a superior method in this context \cite{Vrouwe2005} it is also acknowledged that it does not provide a precise measurement of the distance for the tunnelling events.

\begin{table} [h]
	\caption{Relevant parameters for toluene \ref{fig:toluene}: $E_0$ the channel energy, $\Gamma$ coupling energy, $\mu$ the
		dipole moment, $\varepsilon_r$ the low frequency dielectric constant, and $\sigma$ the DC conductivity \cite{LukaGuth2016}.}
	\begin{tabular}{ | c | c | c | c | c | c | c |}
		\hline
		$\Gamma$ (eV) & $|E_0|$ (eV) & $\Phi$ (eV) & d (nm) & $\mu$ (D) & $\varepsilon_r$ (Fm$^{-1}$) & $\sigma$ (S cm$^{-1}$) \\
		\hline
		$0.83 \pm 0.55$ & $0.84 \pm 0.15$ & $1.14 \pm 0.39$ & $1.05 \pm 0.11$ & $0.375$ &  $2.4$ & $810^{-16}$ \\
		\hline
	\end{tabular}
	\label{tab:toluene}
\end{table}

\noindent Table \ref{tab:toluene}  \cite{LukaGuth2016} shows important values for this measurement, two of them are $E_0$ (channel energy) and $\Phi$ (work function), which have larger values for toluene compared to most functional molecules and for this reason will neither interfere in the measurements nor cause ambiguities to distingue between solvent and molecule. In combination with the lower toxicity of toluene compared to other solvents, toluene is better suitable for the measurements\cite{LukaGuth2016}.

%\clearpage

\begin{figure}[h]
	\centering
	\begin{minipage}{1\textwidth}
		\centering
		\includegraphics[width=1\textwidth]{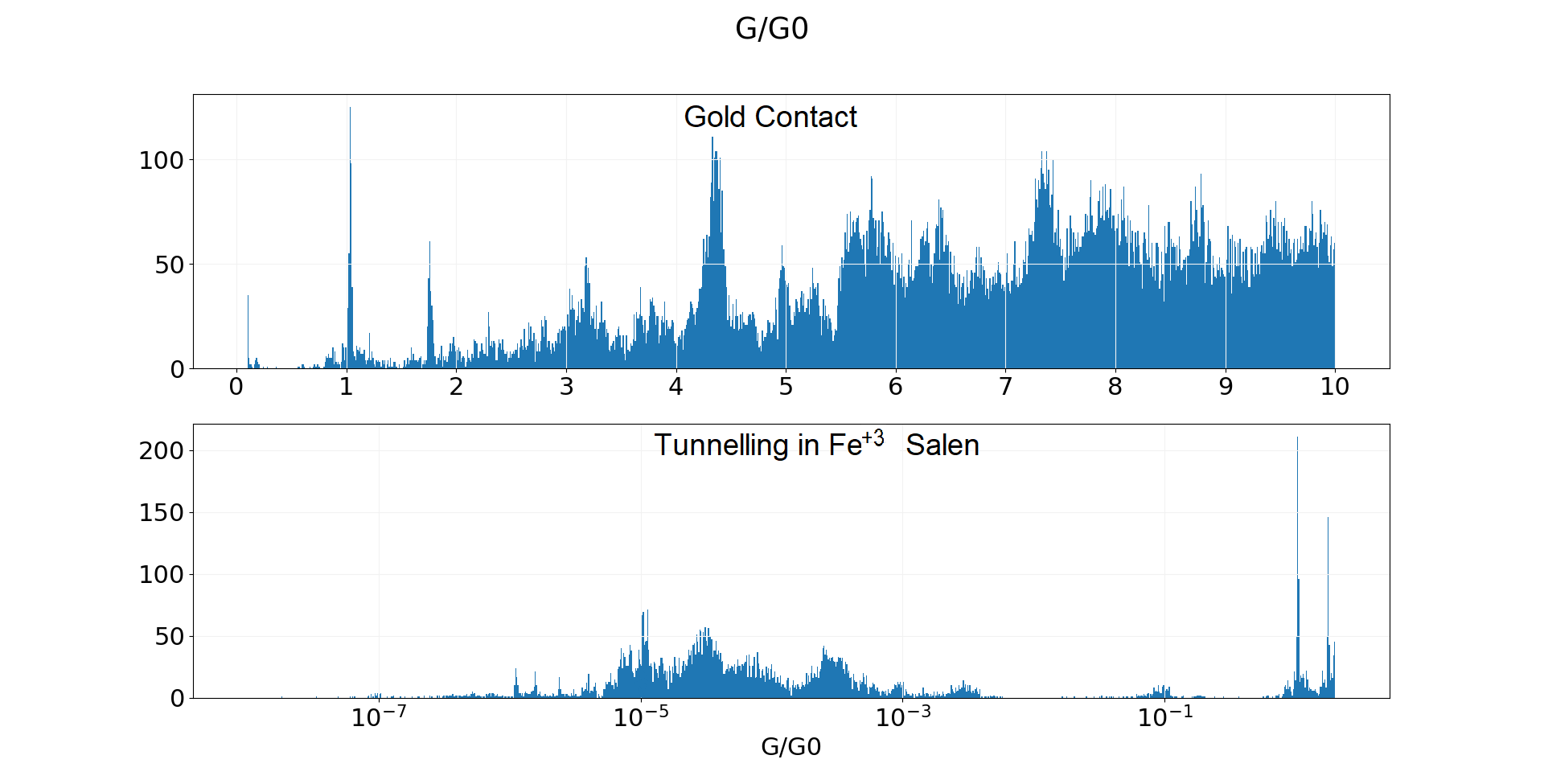}\\
		\subcaption{Molecule immersed in toluene, histogram of $G/G_0$.}
		\label{fig:hist_mola}
	\end{minipage}	
	\begin{minipage}{1\textwidth}
		\centering
		\includegraphics[width=1\textwidth]{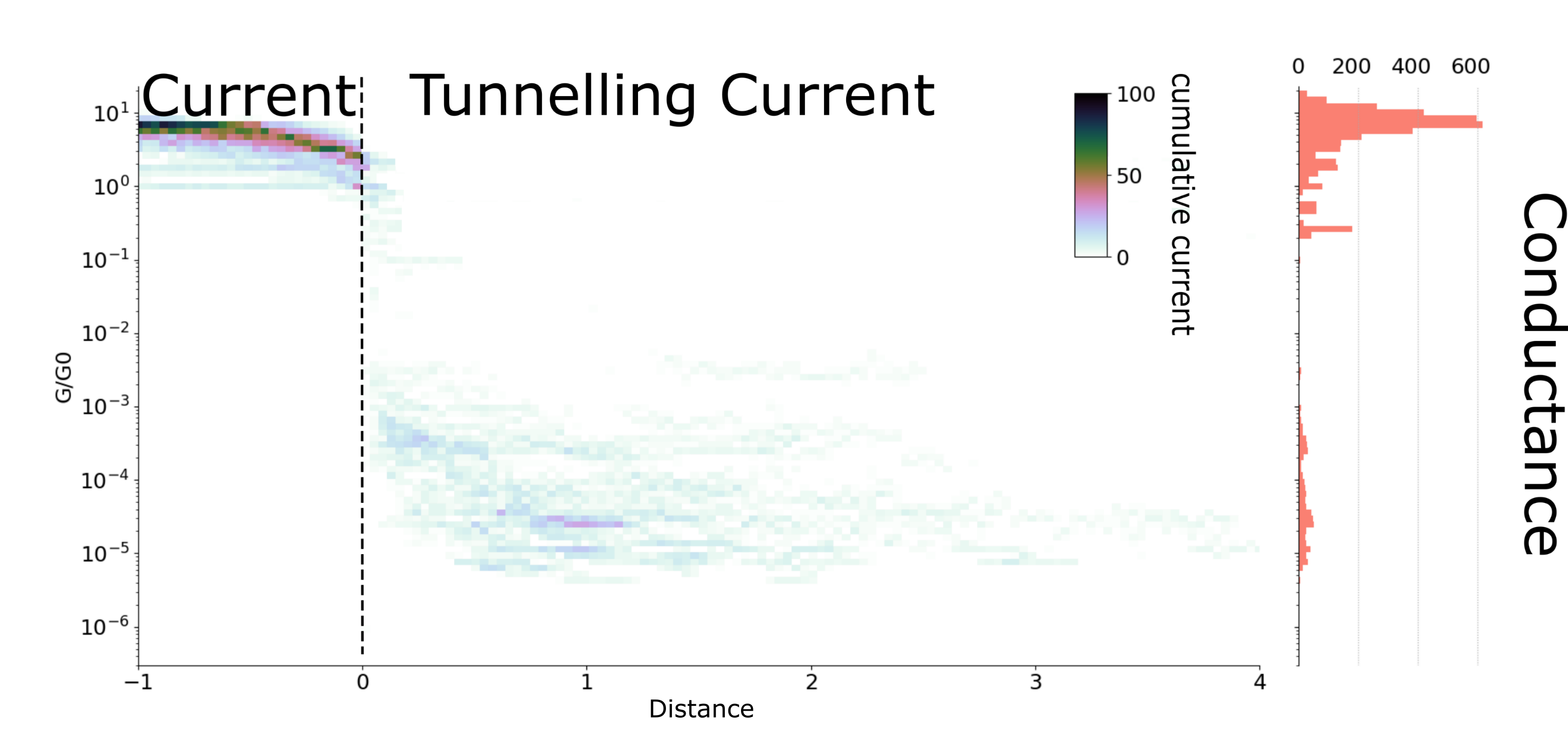}
		\subcaption{G/G0 vs Distance vs Counting: molecule immersed in toluene in the bridg.}
	\end{minipage}
	\caption{Histogram for MCBJ, molecule in toluene, 100 molecules measurements; compare with \ref{fig:hist}, and \ref{tab:toluene}.}
	\label{fig:hist_mol}
\end{figure}

\FloatBarrier

\noindent Figure \ref{fig:hist_mol} show  the histogram for Fe$^{+3}$ Salen. The conductance of the molecule can be estimated based on the graphs' values ranging from $3.10^{-5}\, \mathrm{G_0}$ to $2.10^{-2}\, \mathrm{G_0}$, the reason is, when there is molecule between the leads this values will often appear in the countings of the histogram. Tunneling transport mechanisms have been extensively modeled and studied in the literature, with various approaches being well-documented. One prominent method is the semi-classical Simmons Model \cite{Simmons1964}, \cite{Kilibarda2021}, \cite{Hartman1964}. Alternatively, the concept of resonant tunnelling is often favoured for investigating molecular transport. This approach has significantly contributed to the development of the Single-Molecule Level (SLM) model, which provides a more nuanced understanding of molecular transport phenomena, and facilitating efficient electron transport. 

\noindent Additionally, in Figure \ref{fig:hist_mol} the tunnelling of charge carriers (the colourful cells, between $[0,5]$ in \ref{fig:hist_mol}) from the solvent is still present but much less dominant (exponential decay with distance) because the transport occurs within the molecule channels mainly compared with the transport in the solvent. 

\clearpage

\section{Current}

\noindent The SLM model \cite{Zotti2010} considers exclusively resonant tunnelling and does not account for other mechanisms that may be present, explaining only the pronounced increase in the current at resonant conditions. Its exclusive focus on resonant tunnelling leads to imprecise predictions at a certain range of the current sweep. As a result, the fitting around $U=0\, \mathrm{V}$ and $U=1\, \mathrm{V}$ is not perfect. Around $U=0\, \mathrm{V}$, the resonance was not achieved, and around $U=1\, \mathrm{V}$, the channel saturates and the energy alignment that facilitated resonant tunnelling is lost. In molecular systems, other factors like electron-phonon interactions, electron-electron interaction, molecular conformation changes, and environmental effects can also influence tunnelling, whose might not be fully covered by the SLM model. The intricacies of these topics, including their implications in molecular systems, will be thoroughly examined and discussed in the forthcoming sections

\subsubsection{Fitting the Current with SLM}

\noindent The linear regression analysis performed in our study was rigorously assessed for its accuracy using established goodness-of-fit criteria. Goodness-of-fit often refers to how well the regression line approximates the data points. Analysing the residuals (differences between observed and predicted values) is one of the methods to indicate goodness-of-fit; in our fitting to the SLM, a smoothed curve was not for fitting but as a means of getting proper fitting weights \cite{Kilibarda2021}. 

\noindent In this process, outliers were not eliminated; instead, they were assigned smaller weights, which first generated a robust LOWESS (Locally Weighted Scatterplot Smoothing) curve. After the deviation of the original data points from this smoothed curve is measured, these distances are used as weights in the subsequent fitting of the SLM model. Ideally, the obtained residuals should be randomly dispersed around zero for a well-fitted model.

\noindent Our algorithm incorporated a two-cycle process. Initially, a `window of influence' spanning three data points was established, employing the tricubic weight function to determine the influence between neighbouring points. This function assigned whole influence to central points and minimal to lateral points, applied across all data points. However, the final adjustments were only made after all corrections were computed. In the second cycle, we penalised outliers based on the absolute residuals (AR) calculated from the difference between original and smoothed data points. After determining the median of these absolute residuals (MAR) over five cycles, we assigned new initial weights for the subsequent final fitting routine from GNU Scientific Library. The current numeric integrations equations used the CQUAD routine is robust to ``Inf'' and  ``NaN'' values.

\noindent For a detailed description of the fitting method refer to \cite{Kilibarda2021}.

\noindent In order to carry out this study, we need to concentrate on curves that fit accurately; these curves will show a typical smooth `S' shape, with a slight decrease in conductance at low bias; the SLM can model these curves. Unfortunately, these curves are rare, and even those resulting from molecules with a slightly more elaborate shape fail to fit in some plot regions.

\begin{figure}[!htb]
	\centering
	\begin{minipage}{.8\textwidth}
		\centering
		\includegraphics[width=1\textwidth]{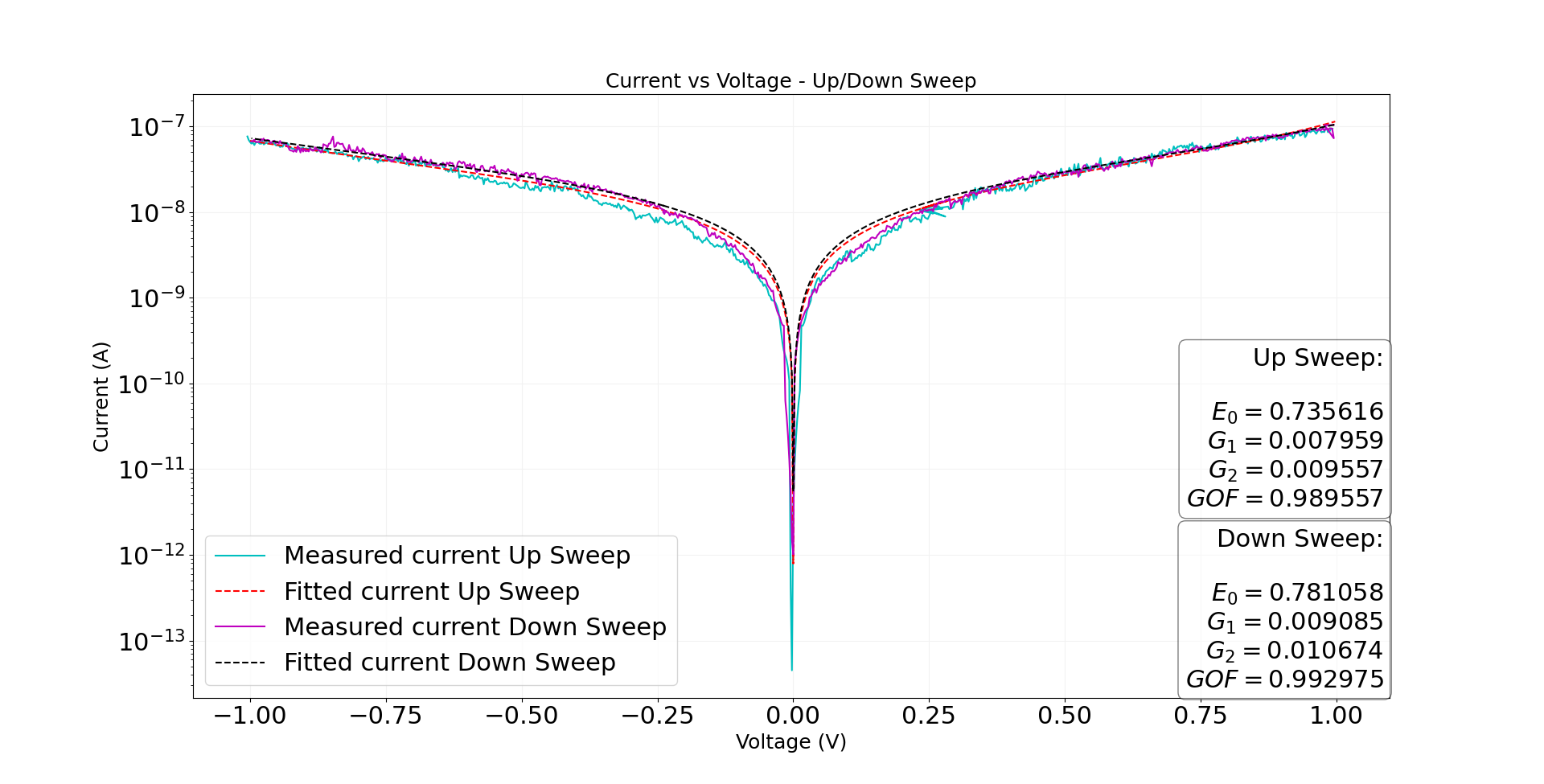}	
	\end{minipage}
	\caption{Example of a good data with good fit.}
	\label{fig:GoodFit}
\end{figure}

\FloatBarrier

\begin{figure}[!htb]
	\centering
	\begin{minipage}{.8\textwidth}
		\centering
		\includegraphics[width=1\textwidth]{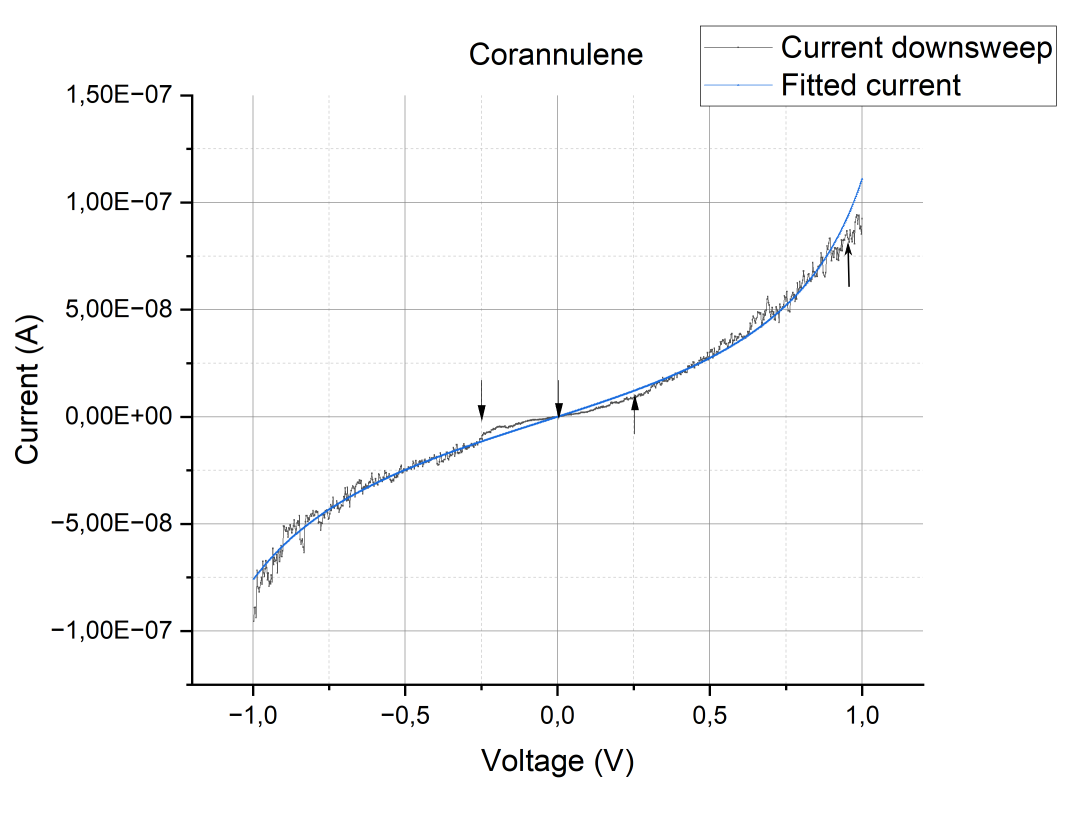}
	\end{minipage}
	\caption{In this figure, there are example with measurements where the fit fails. The arrows in the figures point out the region around the fitting fails.}
	\label{fig:pbFit}
\end{figure}

\noindent In this text, we is given preference to expose the data in linear scale, but sometimes it is necessary to move to a logarithmic scale for the current to highlight small differences as in the Fig. \ref{fig:pbFit}, where discrepancies between data and fitting are visible. Observe that the fit curves follow the data almost perfectly, failing only $[-0.30\, \mathrm{V},0.25\, \mathrm{V}]$. This divergence between SLM and data is the focus of the thesis: good data without outliers where the model did not fit when expected to be a perfect fit.

\FloatBarrier

\noindent  As observed in Fig. 4.6, bad-fitting are those around the origin, which we evaluate in this chapter and try to explain. Our analysis extends to examining the tails of the distribution as well.

\noindent To systematically address these issues, we will categorize the measurements and conduct a detailed analysis throughout this chapter. This structured approach allows us to understand the factors contributing to the divergences between model and measurement, and to propose potential explanations for these discrepancies.

\clearpage

\subsubsection{Categories for the fitting}

The Corannulene data was divided according to Fig. \ref{fig:schema_fit}, for evaluation of the SLM:

\begin{figure}[!htb]
	\centering	
	\begin{minipage}{1\textwidth}
		\centering
		\includegraphics[width=1\textwidth]{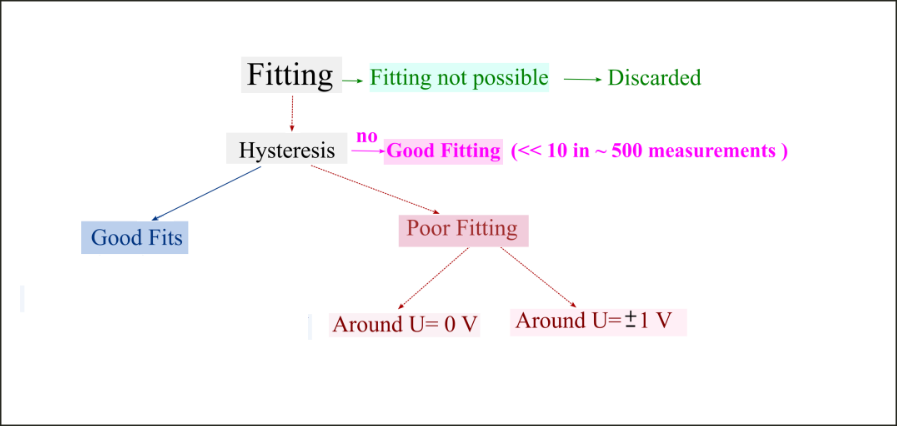}
	\end{minipage}
	\caption{Three different scenarios to consider when fitting: The case where the fit is not possible, when the fitting saturates too early; The case where the hysteresis absent, this were very few cases less than 10 within c.a. 500 measurements; In case of pronounced hysteresis, although the model fits partially for a specific voltage range. However, it does not accurately describe direct tunnelling around the origin ($U=0\, \mathrm{V}$), and it doesn't take into account the current saturation around the tails ($U=\pm 1 \, \mathrm{V}$).}
	\label{fig:schema_fit}
\end{figure}

\FloatBarrier

\noindent This thesis introduces a classification scheme that specifically accounts for the presence and extent of hysteresis in the measurements. Hysteresis is quantitatively assessed using the following formula: $\Delta H = (I_{upsweep}-I_{downsweep})/abs(I_{max})$, where $\Delta H$, represents the hysteresis factor. This factor is calculated by taking the difference between the up-sweep and down-sweep current values, and then dividing this by the absolute maximum current value.

Based on the calculated $\Delta H$ value, the hysteresis is categorized as follows:

\begin{itemize}
	\item For values $\Delta H < 0.1$, the measurement is classified as having `no hysteresis'.
	\item For values $0.1 < \Delta H < 0.2$ , it is categorized as `small hysteresis'.
	\item For $\Delta H > 0.2$ the classification is `large hysteresis'.
\end{itemize}

\noindent This categorization allows for a more nuanced understanding of the hysteresis phenomenon in the context of our measurements, providing a clear, quantifiable method to differentiate between varying degrees of hysteresis.

\FloatBarrier

\noindent From this point forward, the terms `no hysteresis', `small hysteresis', and `high hysteresis' will be employed within this document, each carrying the specific definitions outlined previously. This terminology will ensure precise and consistent communication of the hysteresis characteristics observed in our study.

%\clearpage

\subsubsection{Current Measurements and Fitting}
\label{sec:current_measfit}

\begin{figure}[!htb]
	\begin{minipage}{0,46\textwidth}
		\centering
		\includegraphics[width=1\textwidth]{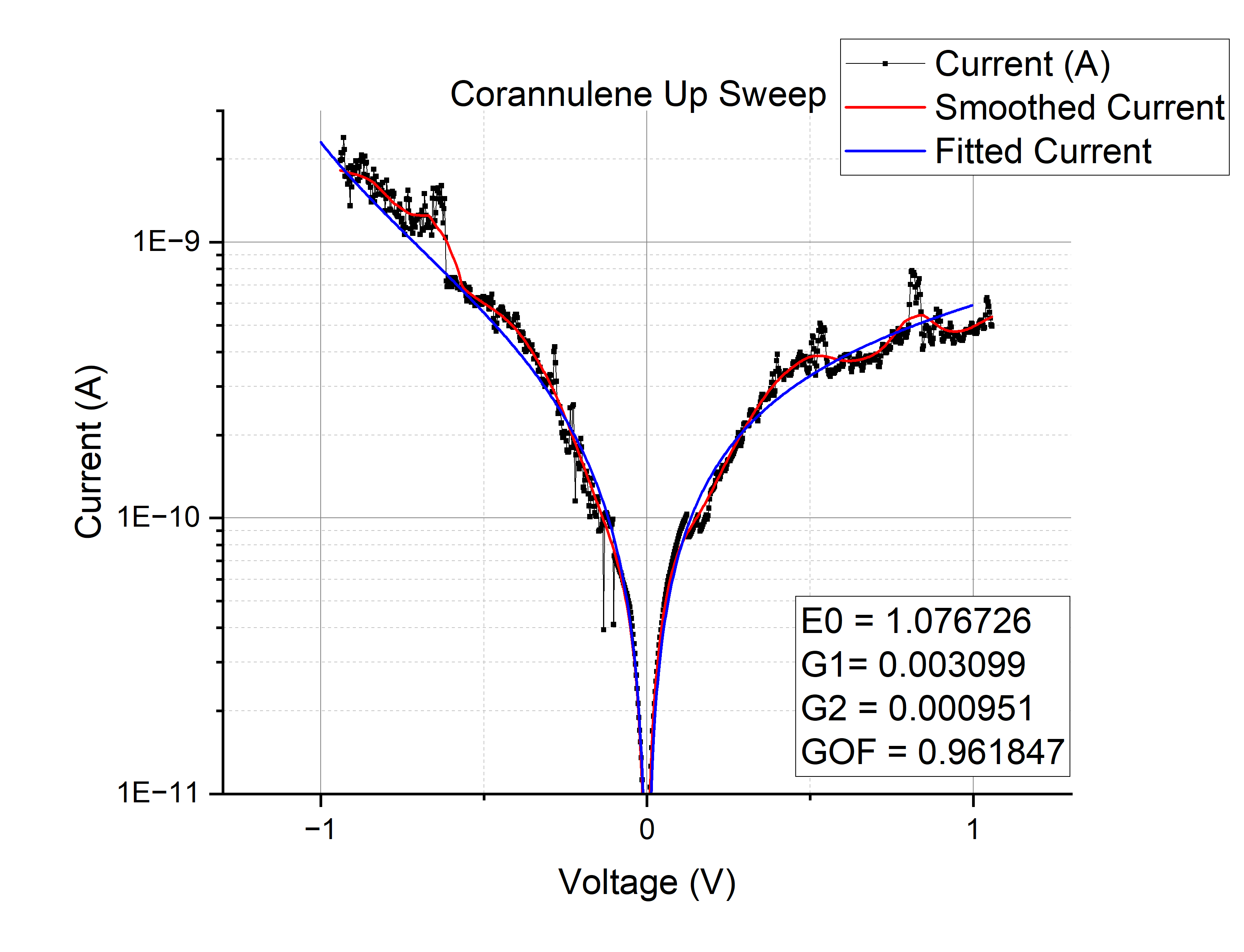}
		\subcaption{Fitting up sweep in logarithmic scale.}
		\label{fig:cta}
	\end{minipage}
	\begin{minipage}{0,46\textwidth}
		\centering
		\includegraphics[width=1\textwidth]{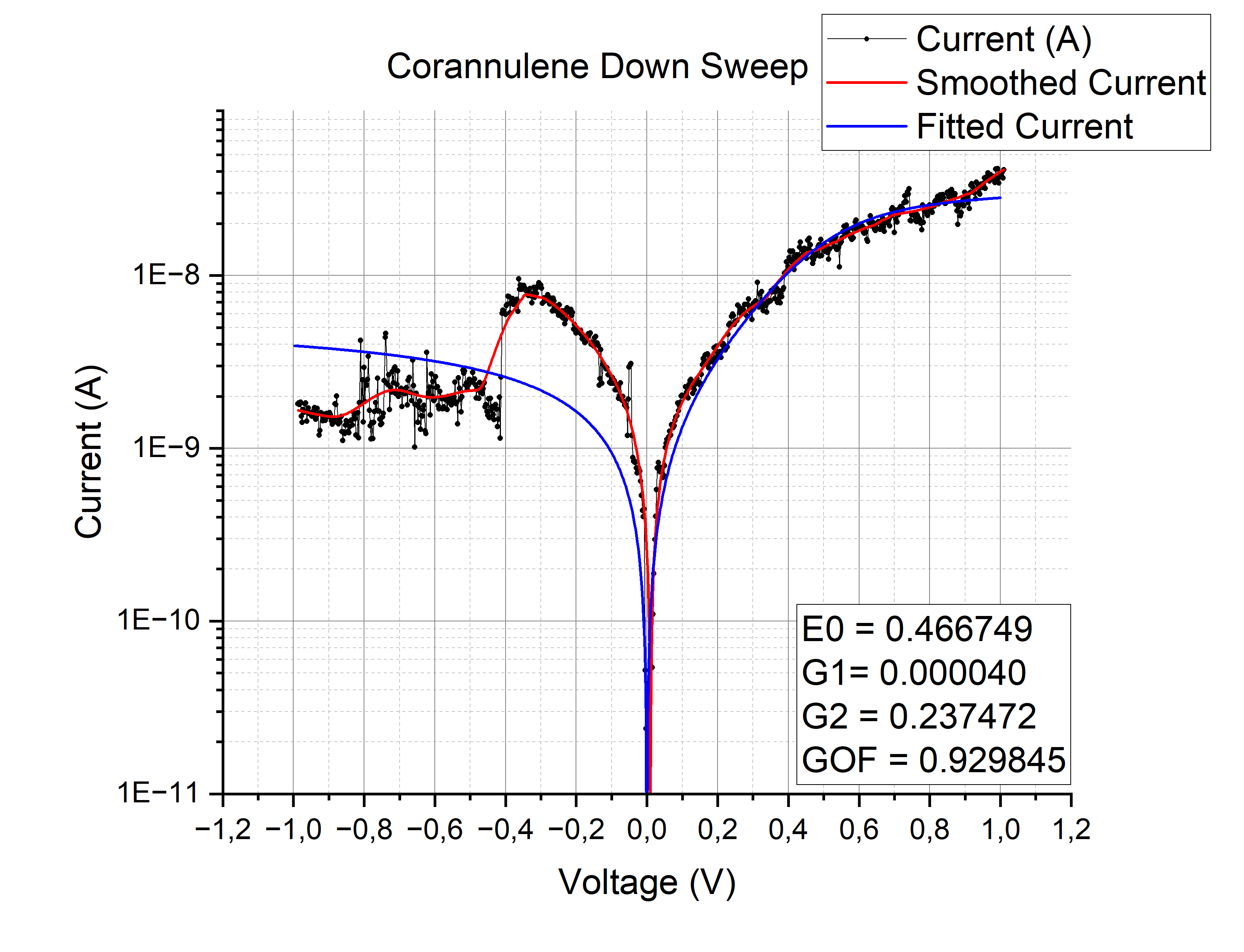}
		\subcaption{Current vs voltage: fitting down sweep in logarithmic scale.}
		\label{fig:ctb}	
	\end{minipage}
	\begin{minipage}{0.46\textwidth}
		\centering
		\includegraphics[width=1\textwidth]{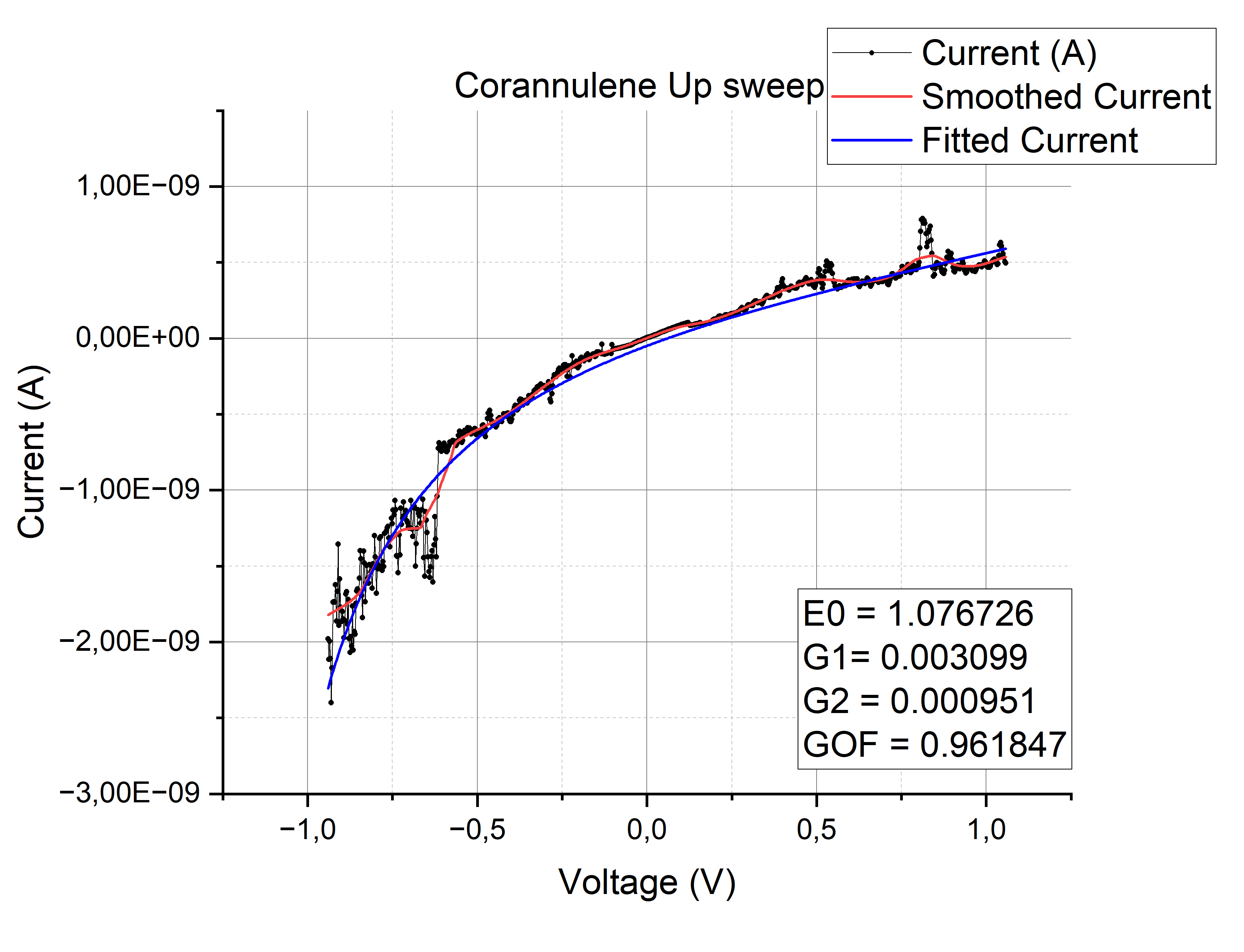}
		\subcaption{Current vs voltage: fitting up sweep in linear scale.}
		\label{fig:ctc}
	\end{minipage}
	\hfill
	\begin{minipage}{0.46\textwidth}
		\centering
		\includegraphics[width=1\textwidth]{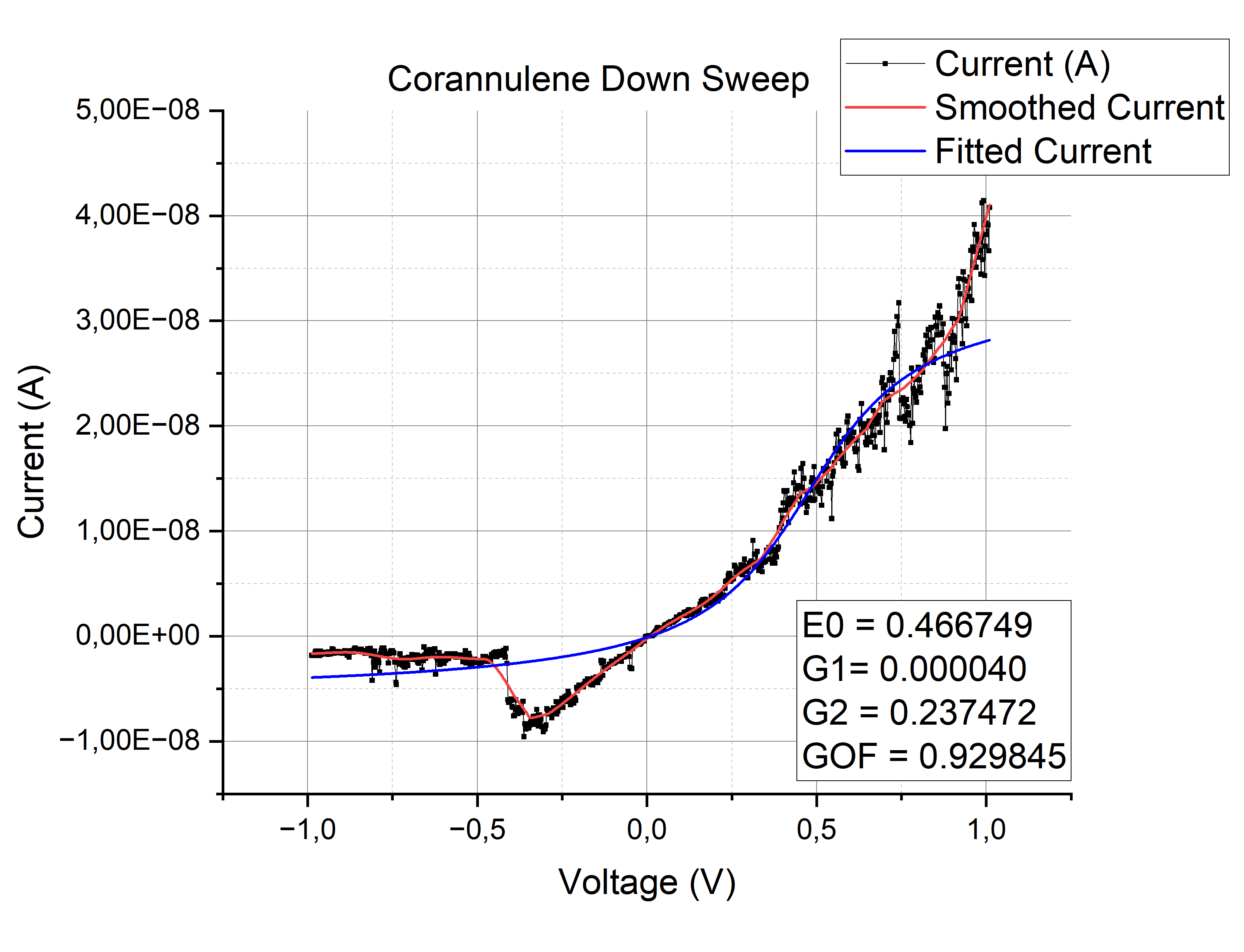}
		\subcaption{Current vs voltage: Fitting down sweep in linear scale.}
		\label{fig:ctd}
	\end{minipage}
	\caption{Current vs voltage sweeps in Fe$^{+3}$ Salen: measured current, smoothed and fitted curves. The boxes in the right corners in (a) (b) (c) (d) show the fit parameters. While (a)/(c) shows a good fit (higher GOF), the fit is less good in (b)/(d) due to the jump in the sweep at $V=0.4\, \mathrm{V}$, one reasonable explanation for this break of the current flow is because the gold electrodes changes (bindig energy) or a detachment, although our setup is very stable.}
	\label{fig:current}
\end{figure}

\FloatBarrier

\noindent Figure \ref{fig:current} shows the SLM fits, which gives us a fitting approach to obtain the current. Despite the model cannot predict the presence of the molecule without further analysis, we use the 1D histograms \ref{fig:hist_mola} to observe through the conductance's histogram whether a molecule exists between the contacts. Figure \ref{fig:cta} is from one measurement with good fitting;  the resonant tunnelling model is used in the fitting, which follows the data with the exception of origin and tails. %The reason for this mismatch will be considered in this chapter and it is the main point of this thesis.
The simplification in the current equation to the tunnel resonant model can cause these poor fitting in Figures \ref{fig:cta} to \ref{fig:ctd}, excluding those mismatches caused by noise, or conformation adjustments to the leads. However, only the resonant tunnelling transport mechanisms is used in the SLM. The disadvantage of remove the transport mechanisms or effects that are not predominant in the channels introduces systematic errors in the fitting by virtue of this omission. Compare the curves in Fig. \ref{fig:current} with the appendix \ref{app:somefits}; there exhibits more examples where the SLM fails.

\noindent One example of a phenomenon which is not included in the model is electron relaxation due to electron-phonon interactions. Although it can not be observed at RT in IV curves, it increases the phonon density because of their excitation during the relaxation. So, whether these measurements are to be realised at LT, it is pertinent to include those interactions in the model to make it general. It causes significant changes at LT \cite{CuevasScheer2010},\cite{Sondhi2022}, and the current increases by a small factor, highlighting the difference between model fit and observations. 

%\noindent Another phenomenon is the position of the molecule which can change during the current flow (conformation), leading to current jumps which can be observed at $U=0.4\, \mathrm{V}$ in Figure \ref{fig:ctb} and \ref{fig:ctc}, but there are other causes with might provoke similar changes and it is not possible differentiate them, as biding energy or a change of coupling. Nevertheless, one possible dominant phenomenon is electron-electron interactions at RT  \cite{Kreupl1998}, and this will be a central point of discussion in this thesis.

\noindent Another observed phenomenon involves the molecule's position, which may shift during current flow (conformation), leading to noticeable variations in current. These fluctuations are evident at $U=0.4\, \mathrm{V}$ in Figures \ref{fig:ctb} and \ref{fig:ctc}. However, similar changes could be triggered by other factors, making it challenging to distinguish among them, such as variations in binding energy or changes in coupling. Despite these complexities, one potential predominant factor is the electron-electron interactions at room temperature (RT), as highlighted by Kreupl 1998 \cite{Kreupl1998}. This aspect will form a focal point for discussion in this thesis.

%\clearpage

\subsection{Current in Toluene}

In this section, we investigate in details the toluene measurements. Despite Toluene is recognized as a non-polar solvent, it exhibits notable hysteresis in these measurements. A plausible explanation for this observed hysteresis could be the inadvertent contamination by water vapor. The presence of water vapour, which is soluble in toluene, may lead to the dissolution of its ions. These dissolved ions could potentially play a significant role in the measurements as hysteresis. Understanding this aspect will allow comprehending the measurements curves, including the factors that might influence its accuracy and determining which factors can be considered negligible.

\begin{figure}[!htb]
	\centering
	\begin{minipage}{1\textwidth}
		\centering
		\includegraphics[width=1\textwidth]{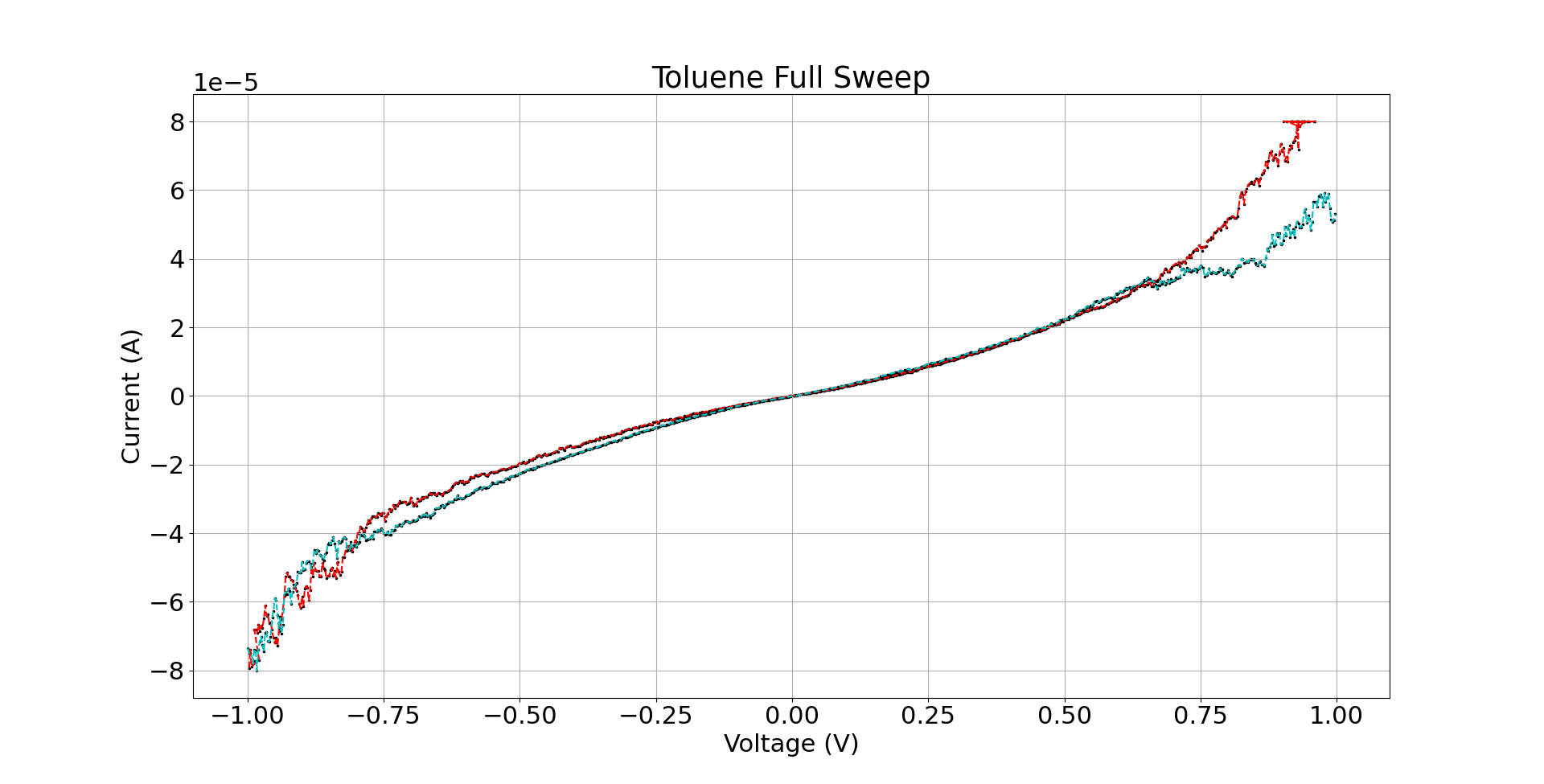}
		\subcaption[]{Measurement partially without hysteresis and with very small one on the tails mostly.}
		\label{subcap:tola}	
	\end{minipage}
	\hfill
	\begin{minipage}{1\textwidth}
		\centering
		\includegraphics[width=1\textwidth]{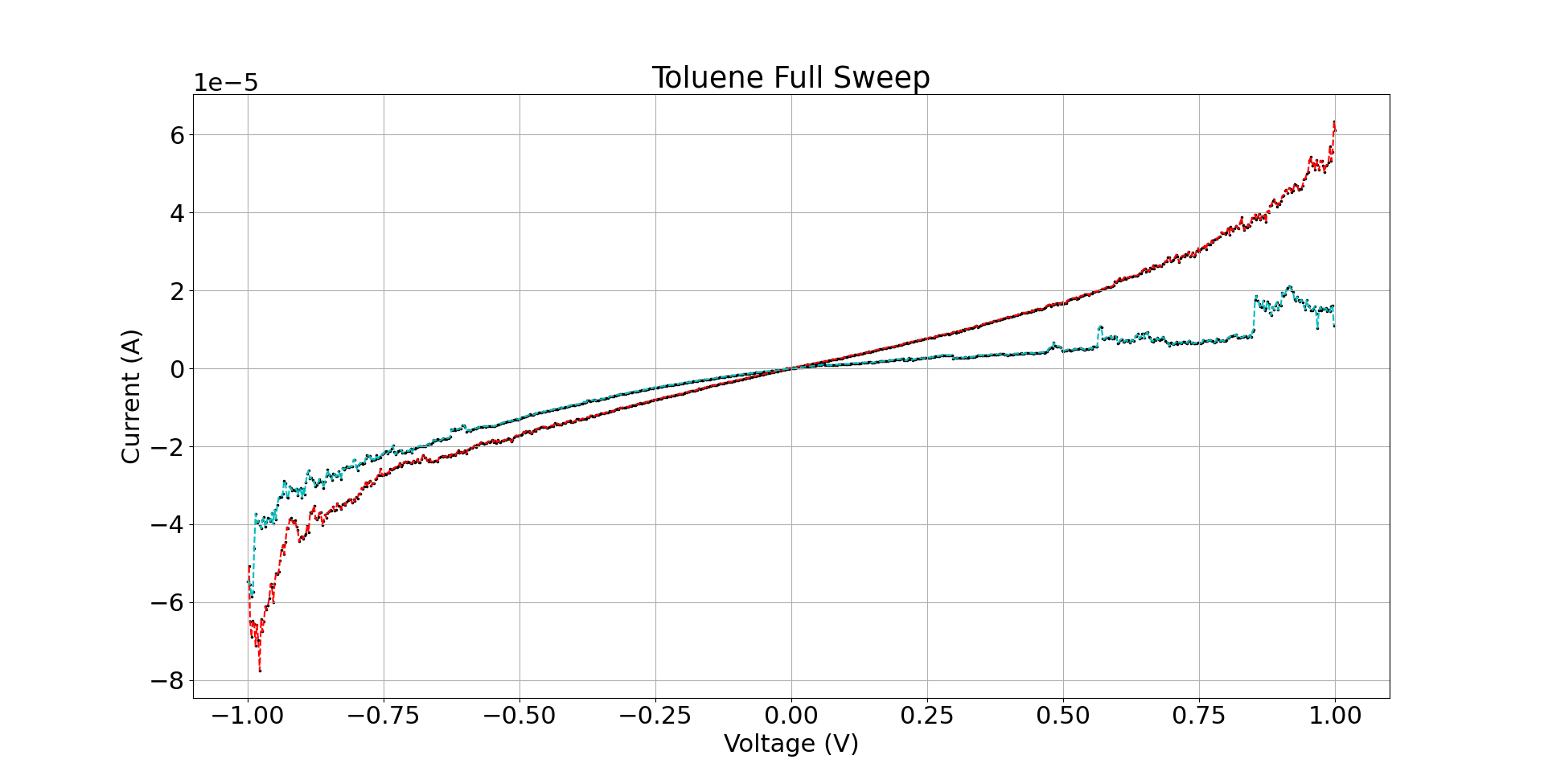}
		\subcaption[]{Small hysteresis, also open in both tails.}
		\label{subcap:tolb}		
	\end{minipage}
	\caption{Current vs voltage: Toluene. Measurements up and down sweep, (a) (b) show the same thing with different perspectives, the hysteresis present in the solvent's current measurement. As it is possible to see, their differences, since they change a lot. As previously discussed, one plausible explanation could be the absorption of water vapour by the solvent when exposed to the laboratory's atmospheric conditions, leading to its subsequent decomposition in ions, allowing the solvent store energy.}
	\label{fig:tol}
\end{figure}

\FloatBarrier

\noindent Observe that none of this hysteresis are stable, and each measurement has a different one, some being narrower than others. This phenomenon cannot always be neglected for low bias, as that seen in Figure \ref{subcap:tolb}. Also, on the right-hand side of the graph, it is possible to see the curves do not finish at the same point they started (at $\mathrm{U}=1\, \mathrm{V}$, the initial branch from $0$ to $1\, \mathrm{V}$ sweep is discarded), probably due to the capacitive effect, refer to section \ref{sec:msetup} to see how the measurement was realised: it is possible that water ions dissolved in the toluene are producing a Helmholtz double layer around the leads, causing the hysteresis' loop to open. 

%\clearpage 

\subsection{Current in Corannulene}

In these measurements conducted at room temperature (RT), the discrepancy observed between the model predictions and actual current measurements may stem from electron-electron interactions \cite{Eberlein2008}, \cite{walczak2005}. %This effect will be thoroughly explored throughout this chapter.

\begin{figure}[!htb]
	\centering
	\begin{minipage}{1\textwidth}
		\centering
		\includegraphics[width=1\textwidth]{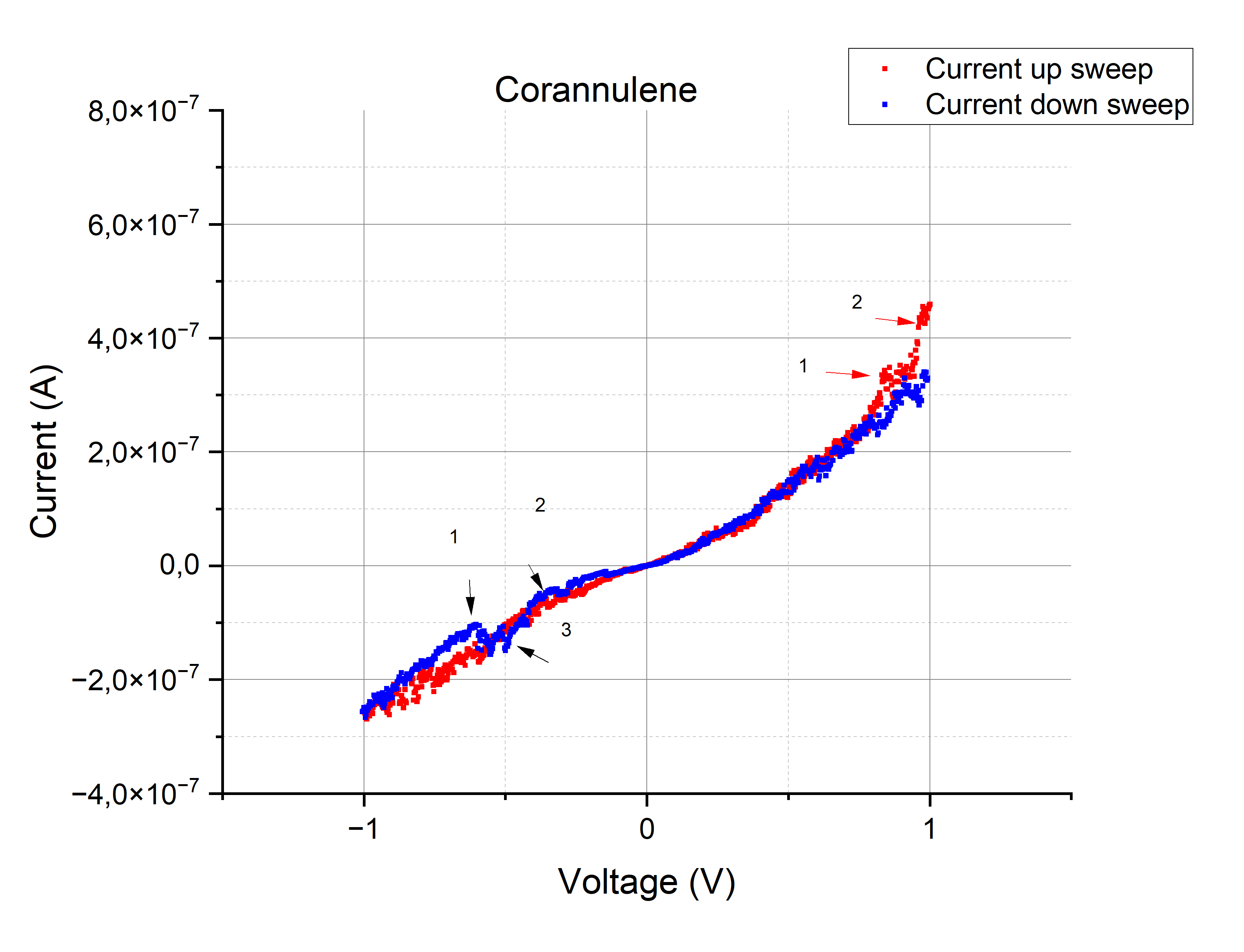}
		%\subcaption[]{}		
	\end{minipage}
	\caption{Current full sweep in Corannulene. Example of category 1: $T= 2.58 \, 10^{-3}$, for down sweep $T= 3.03 \, 10^{-3}$. Here, the colour of the arrows indicates they point out `artefacts' in different excursions of the data, (red) for the up sweep and (black) for the down sweep data. In this measurement, the hysteresis can be neglected, and considered hysteresis free.}
	\label{fig:sca}	
\end{figure}

\FloatBarrier

\noindent Current in Figure \ref{fig:sca} has small negligible hysteresis. Actually, there are always hysteresis in our measurements, some bigger than others. In this thesis, we consider the curves that the up and down sweep almost intercept each other as hysteresis free. We use the previous criteria to classification to consider or ignore the hysteresis, we explained them in the previous sections, in the chapter beginning. The arrows' colours distinguish to which curve they point out: (red) arrow to red data, (black) arrow to blue data.

\begin{figure}[!htb]
	\centering
	\begin{minipage}{1\textwidth}
		\centering
		\includegraphics[width=1\textwidth]{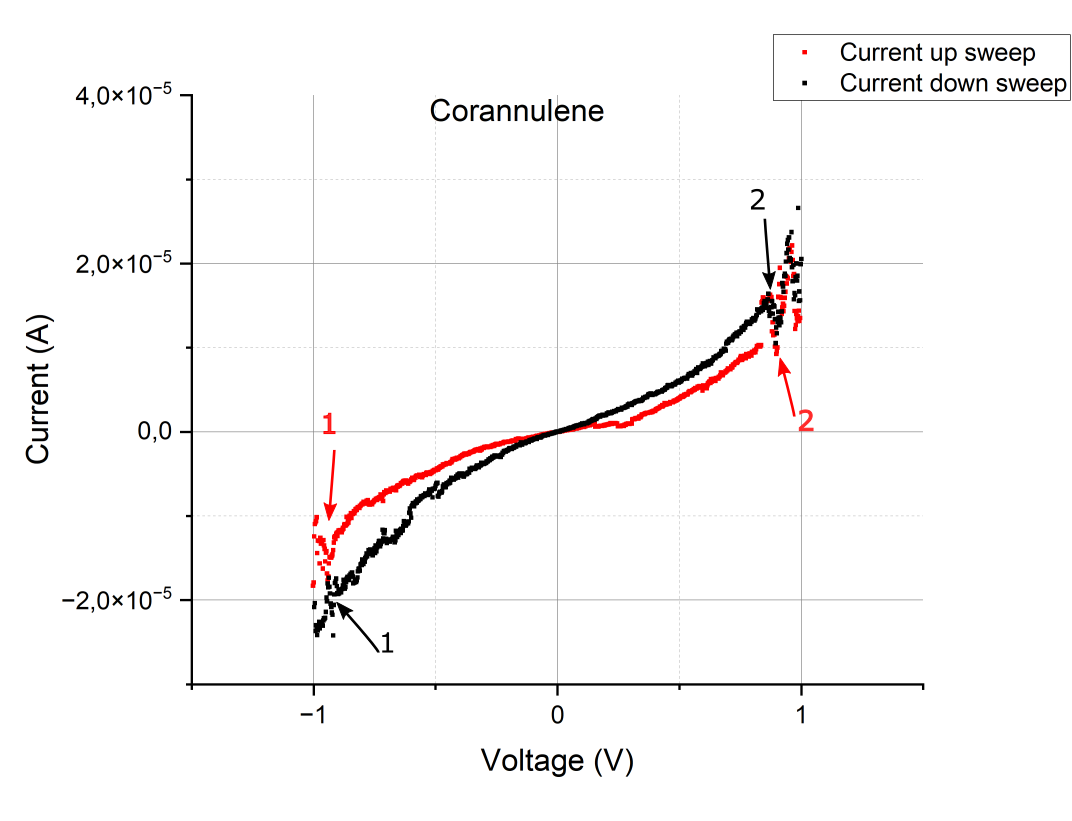}
		%\subcaption[]{Current with hysteresis}			
	\end{minipage}
	\caption{Current full sweep in Corannulene. Here we have a measurement with hysteresis and discontinuities at $U=\mp 1\, \mathrm{V}$, the (red) and (black) arrows (1,2) has the same function than in \ref{fig:sca}.}
	\label{fig:hys}
\end{figure}

\FloatBarrier

\noindent In this analysis, both molecule (Corannulene) and solvent (toluene) used are non-polar. Therefore, a probable cause of the hysteresis in Figure \ref{fig:hys}  is caused by water vapour absorption from the atmosphere in the solvent \cite{LukaGuth2016} and $\mathrm{H}_2\mathrm{O}$ dissociation. The spreading of the data in the tails is caused mainly by thermal noise.

\noindent We now shift our focus to the characteristics of the current sweep. The current initiates at $1\, \mathrm{V}$ and moves towards the origin (as indicated by the red curve), undergoing a noticeable change in shape. It becomes flatter as it crosses the origin, maintaining this flattened trajectory until another transition occurs, after which it progresses to $-1\, \mathrm{V}$. At this voltage, the sweep direction inverts, heading back towards zero. Around the zero point, the curve flattens out again (this effect is more pronounced in some graphs than others, with additional examples provided in Appendix \ref{app:somefits}). The sweep continues past the origin, remaining flat until another change is observed, leading to a faster increase in current back to $1, \mathrm{V}$. The difference around the origin is where the model fails to predict data; the SLM predict a ``linear'' behaviour to small voltages, which is in accord with the tunnelling resonant model, when there is an evident (red curve) flattening or less evident (black curve).

\begin{figure}[!htb]
	\centering
	\begin{minipage}{1\textwidth}
		\centering
		\includegraphics[width=1\textwidth]{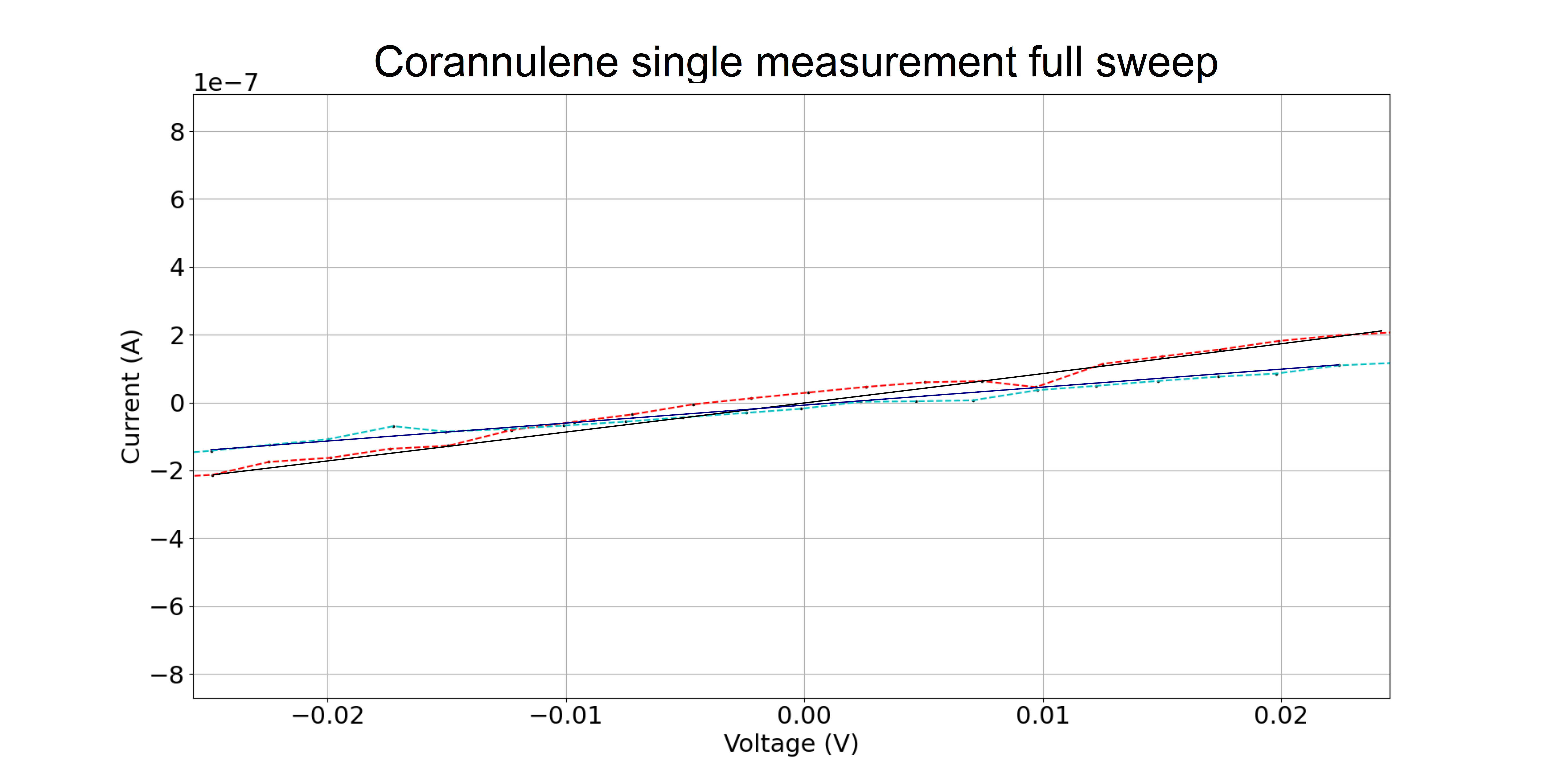}
		%\subcaption[]{}	
	\end{minipage}
	\caption{Zoom into a current sweep for small bias, the  hand-drawn lines (black) show the the curve different inclinations; single measurement.}
	\label{fig:zoo_med1}
\end{figure}

\begin{figure}[!htb]
	\begin{minipage}{1\textwidth}
		\centering
		\includegraphics[width=1\textwidth]{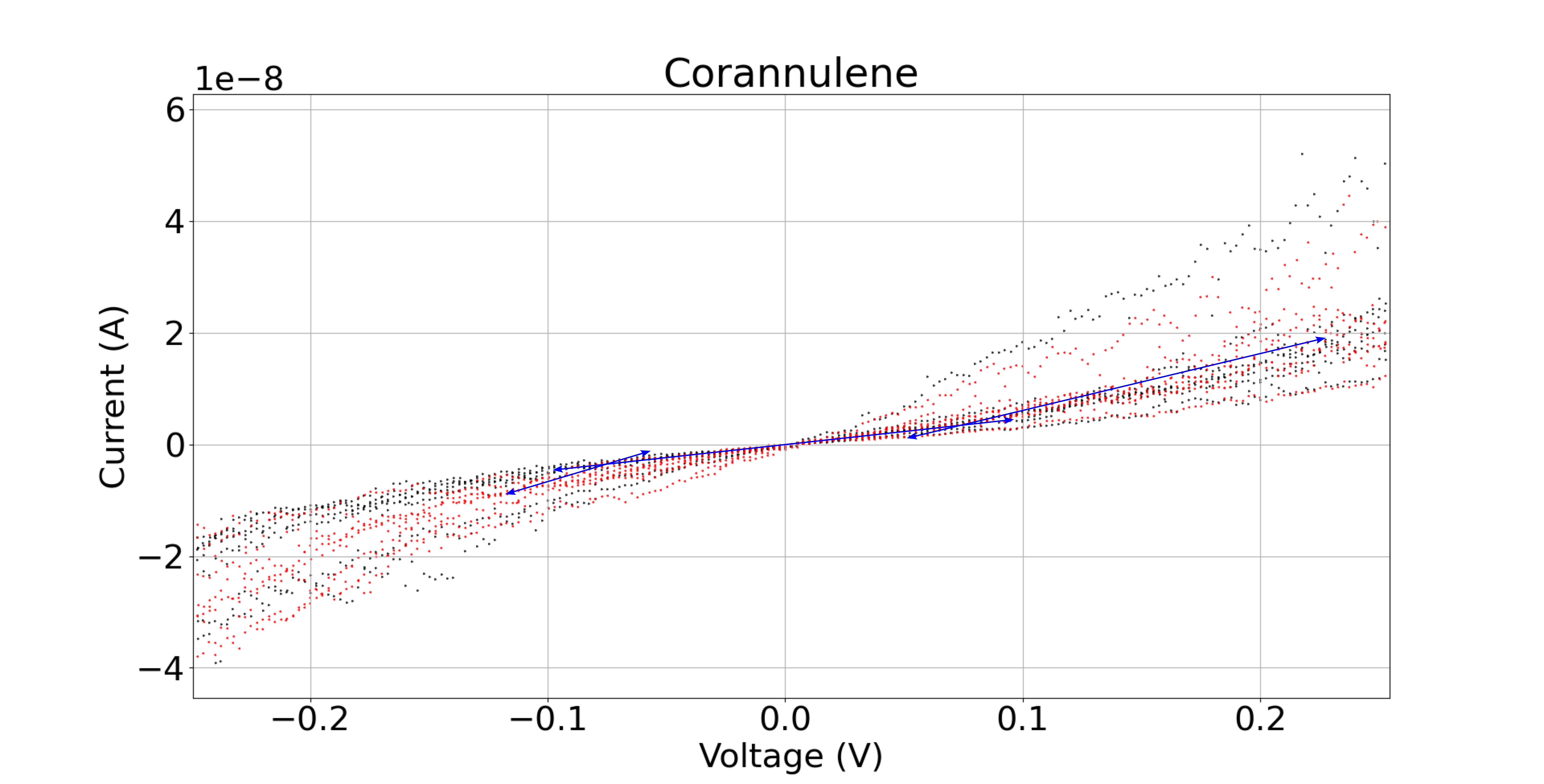}
		%\subcaption[]{}	
	\end{minipage}
	\caption{Zoom into a current sweep for small bias, the  hand-drawn lines show the the curve different inclinations; with this multiple measurements is possible to observe the tendency of the curves.}
	\label{fig:zoo_med}
\end{figure}

\FloatBarrier

\noindent After a detailed analysis of the measurements presented in Figures \ref{fig:zoo_med1} and \ref{fig:zoo_med}, a linear region is noticeable for small bias, indicative of direct tunnelling. Notably, the slope of the red current exhibits a change during the sweep, as highlighted by the inserted arrows and lines for enhanced visualization. To refine the fitting, it's crucial to identify the missing charge interaction and transport mechanism in the Standart Linear Model (LM), and a plausible explanation for these interactions is electron-electron interactions. As suggested by Datta (2005) is possible that scattering is playing a role \cite{Datta2005}. Nevertheless Datta works is more related to $3 \, \mathrm{D}$ and $2 \, \mathrm{D}$, he explains that in such a conductance range, transport is predominantly influenced by the electron localization effects and electron-electron forward scattering.

\noindent To further investigate this phenomenon, the experimental approach proposed by Laible (2020) \cite{Laible2020} can be employed, involving bending a bridge in a device setup. This setup includes positioning the bridge on a scanning stage, illuminated by a laser, with the resultant image captured by a detector. Determining whether the bending leads to scattering, or if a more intricate transport mechanism is at play, remains challenging. Ideally, breaking the bridge without bending it would offer clearer insights. Alternatively, a dark setup could be used to prevent photo-excitation, allowing for observation of photon generation, as discussed by Healy (2001) \cite{Healy2001}. The possibility of a scattering event can be explored by illuminating the molecule with laser light before and after the current sweep.

\begin{figure}[!htb]
	\centering
	\begin{minipage}{1\textwidth}
		\centering
		\includegraphics[width=0.8\textwidth]{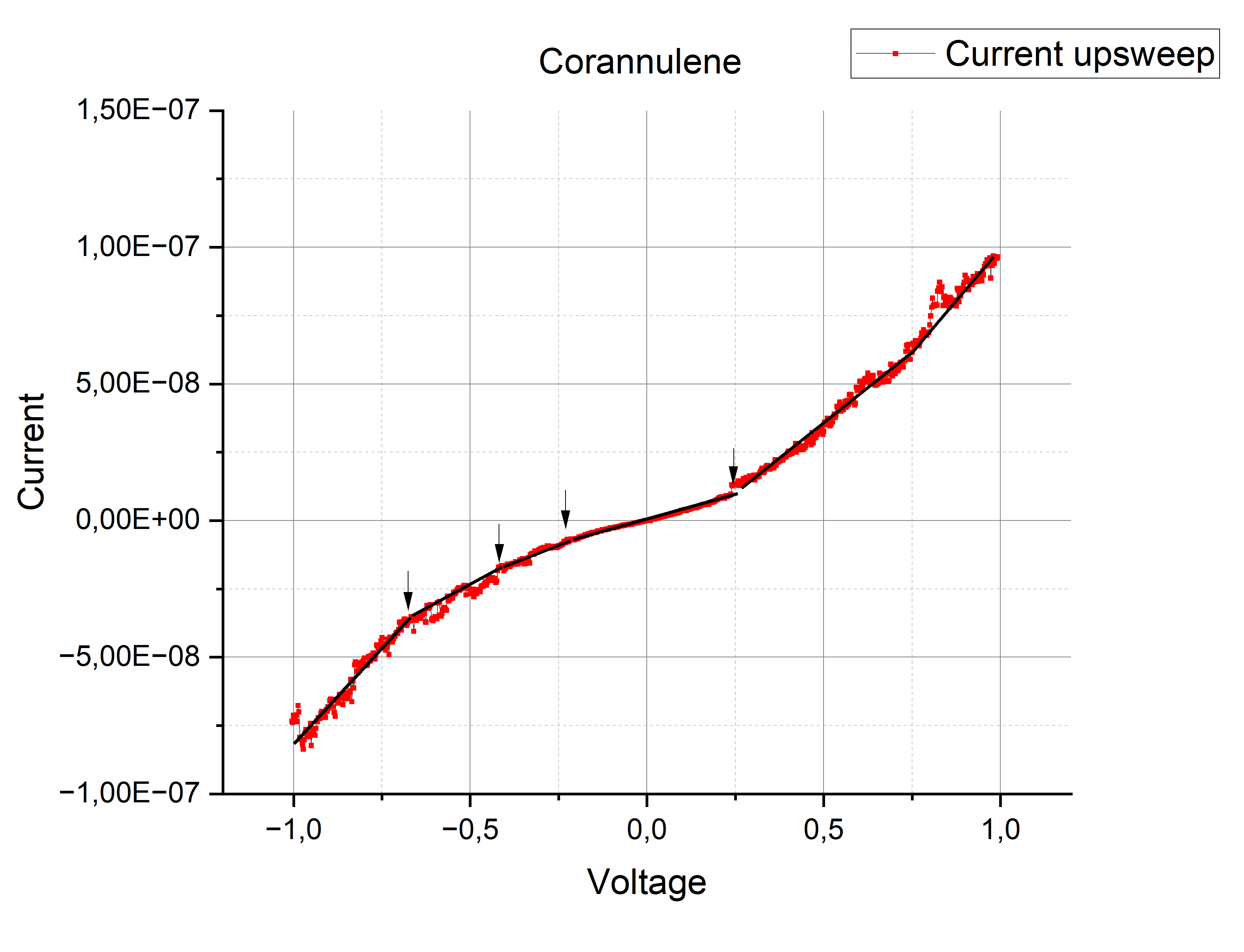}
		\subcaption[]{In this current graph, the arrows points out here, where the change in the curve happens.}	
	\end{minipage}
	\begin{minipage}{1\textwidth}
		\centering
		\includegraphics[width=0.8\textwidth]{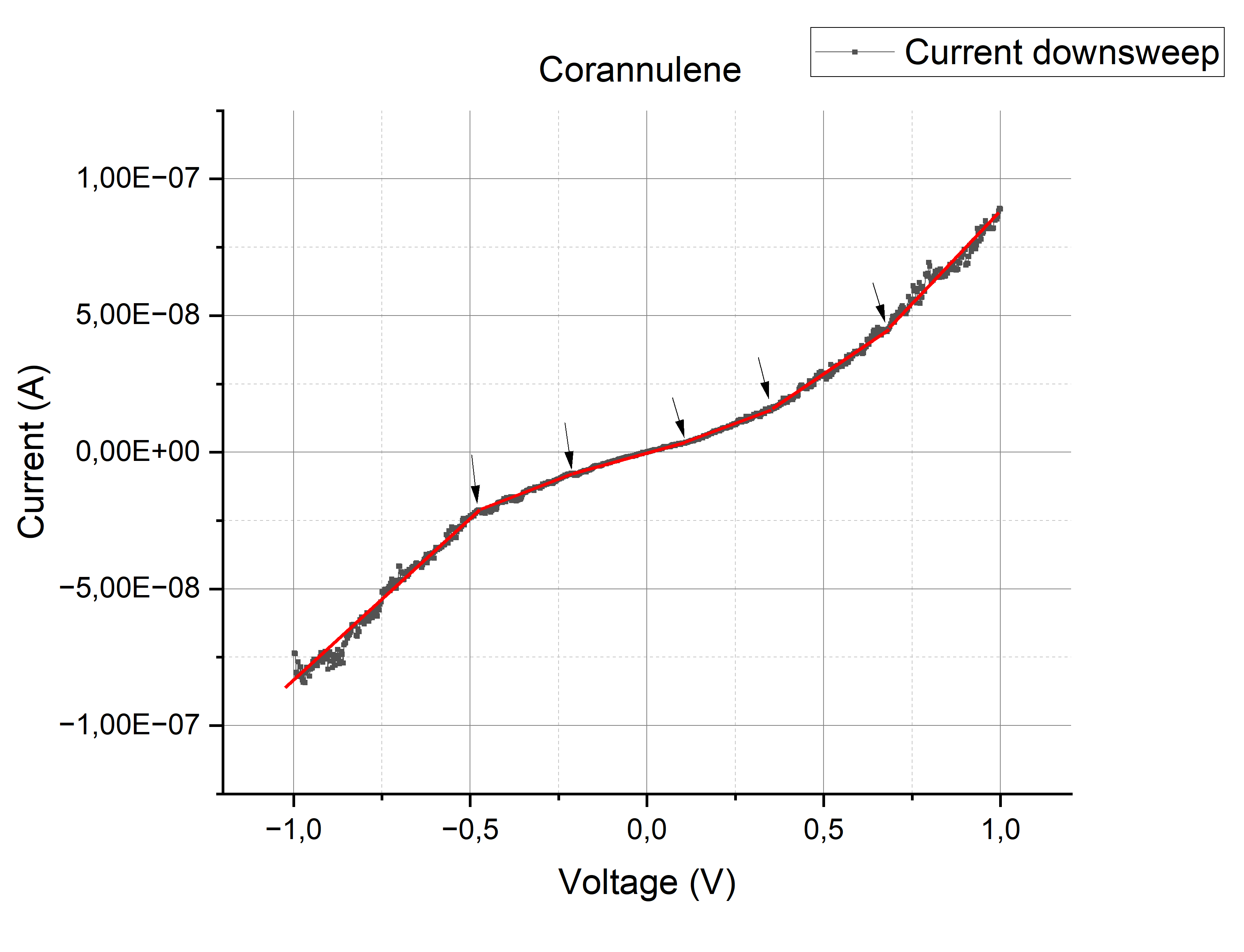}
		\subcaption[]{}	
	\end{minipage}
	\caption{Current vs voltage: The arrows indicate the change of slope in down- and up-sweep, from a stable measurement. The curves were drawn manually following the data. After it was placed on the measurement, to exhibits the change of curvature. The drawn curves has a contrasted colour making them distinguishable of their respective data. This changes are not described by the model, and these changes we want to investigated and understand.}
	\label{fig:slope}
\end{figure}

\FloatBarrier

\noindent Figure \ref{fig:slope} shows the change in the slope of the current. For down-sweep, this change occurs in the range $\pm 500\, \mathrm{mV}$; for up-sweep, it changes to $\pm 250\, \mathrm{mV}$. The arrows show a region where the flow of current has a visible change, and in the rage $[-0.25:0.25]$, a change in the inclination of the curve happens, followed by a modification in its growing rate; the current does not have an exponential change in this range, but linear with a tiny angle in relation to the x-axis, more examples can be seen in Appendix \ref{app:norm}.

%\clearpage

\subsection{Current in Fe$^{+3}$ Salen}

In this subsection, we dive into the intricacies of a particular experimental measurement, highlighting the various events that impact its outcomes. A key challenge encountered is the evaporation of toluene, a solvent used in the experiment. This evaporation process can introduce variables that affect the accuracy and reliability of our results. Additionally, ambient noise presents a significant obstacle, as it can obscure specific signatures within the current measurements. The noise interference, whether from electronic equipment or environmental factors, makes it difficult to discern subtle but crucial features in the data. For an accurate interpretation and analysing of the experimental results, it is necessary to understand and mitigate these issues with noise and environment interferences.

\begin{figure}[!htb]
	\centering
	\begin{minipage}{.8\textwidth}
		\centering
		\includegraphics[width=1\textwidth]{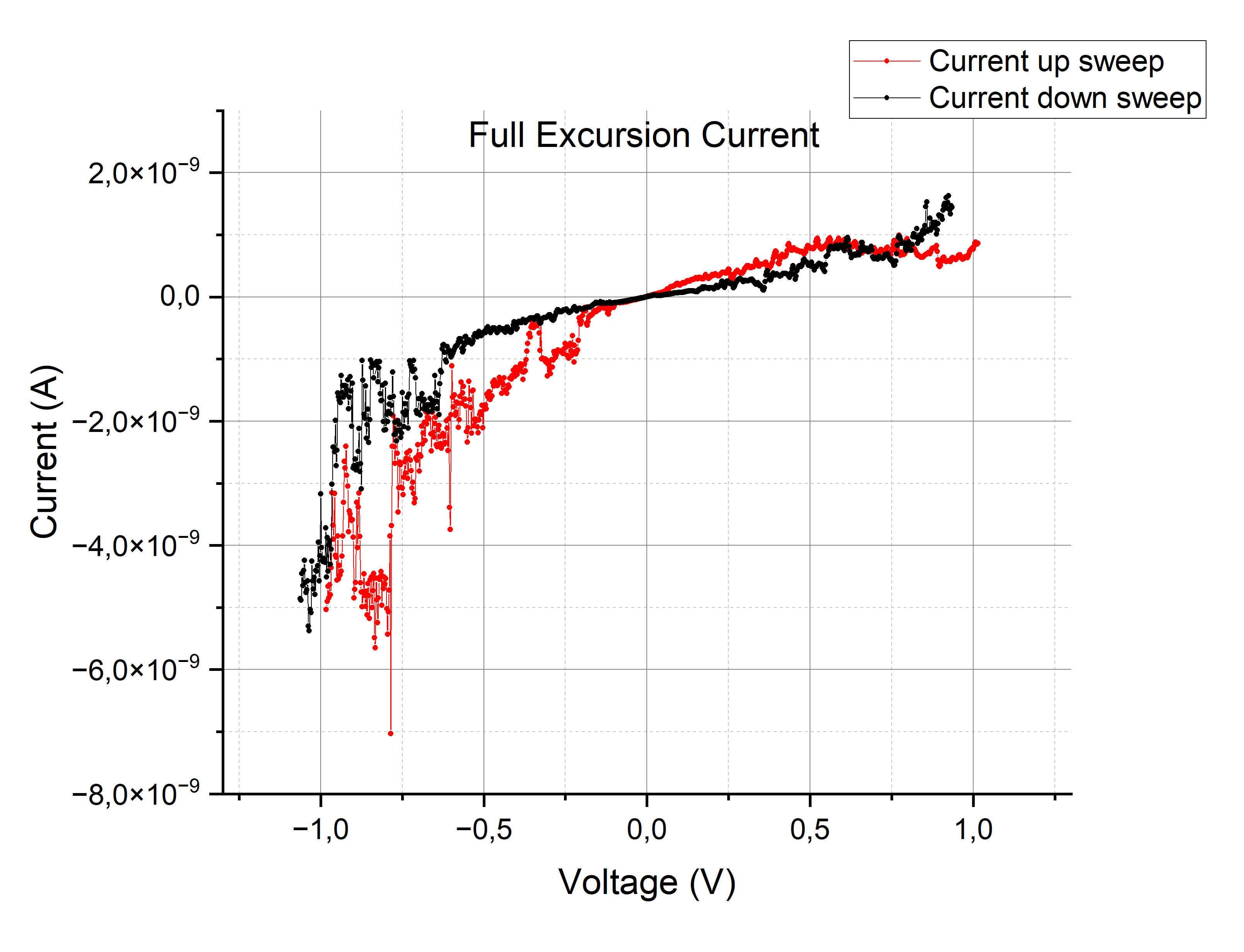}	
	\end{minipage}
	\caption{IV-curve for a measurement loop in Fe$^{+3}$ Salen. This graph exhibits a very characteristic shape compared to others measurements, it is a good example of measurement with hysteresis for Fe$^{+3}$ Salen.}
	\label{fig:data}
\end{figure}

\FloatBarrier

\noindent The initial observation from Fig. \ref{fig:data} is the hysteresis evident in the IV sweep. This hysteresis could be influenced or exacerbated by the presence of Fe$^{+3}$ ions. While literature such as Kato (2018) \cite{Kato2018} reports similar phenomena in other molecules, those instances typically involve molecules with bi-stable positions or, as Akiyoshi (2021) \cite{Akiyoshi2021} notes, molecules exhibiting ferroelectricity and spin-crossover in the presence of significant number of ions with numerous degrees of freedom. However, a recurring theme in molecular electronics literature is the attribution of hysteresis to environmental factors. In our experiment, this implies that the solvent itself may be playing a role, potentially storing energy during the sweep and thus contributing to the observed hysteresis.

\noindent Another contributing factor to the observed fluctuations in current during the sweep could be the molecule attached to the leads. The molecule's own vibrational modes - such as stretching, bending, and wagging - might alter the electronic properties of the system or behave as extra source of environmental noise, once they can not be distinguished at RT. These vibrational changes can affect the molecule's attachment position to the leads or even result in temporary loss of contact, leading to interruptions in current flow. This complex interplay of molecular dynamics and environmental factors, highlights the difficulty of analysing and interpreting data in molecular electronics.

%\clearpage

\subsection{Current Measurement after Consecutive Opening  -  Case Study: Fe$^{+3}$ Salen}

\begin{figure}[!htb]
	\centering
	\begin{minipage}{1\textwidth}
		\centering
		\includegraphics[width=.6\textwidth]{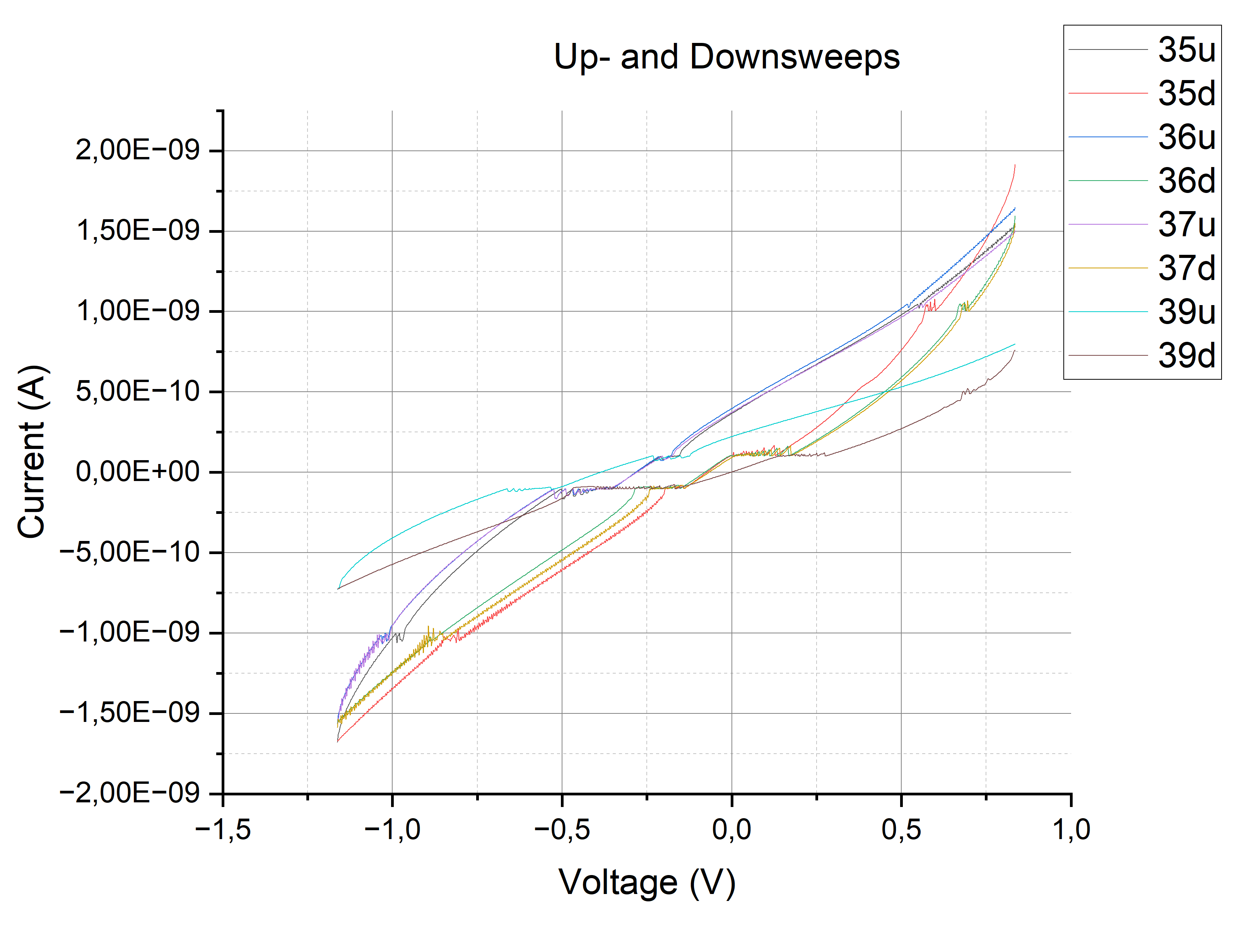}
		\subcaption[]{}	
	\end{minipage}
	\begin{minipage}{.46\textwidth}
		\centering
		\includegraphics[width=1\textwidth]{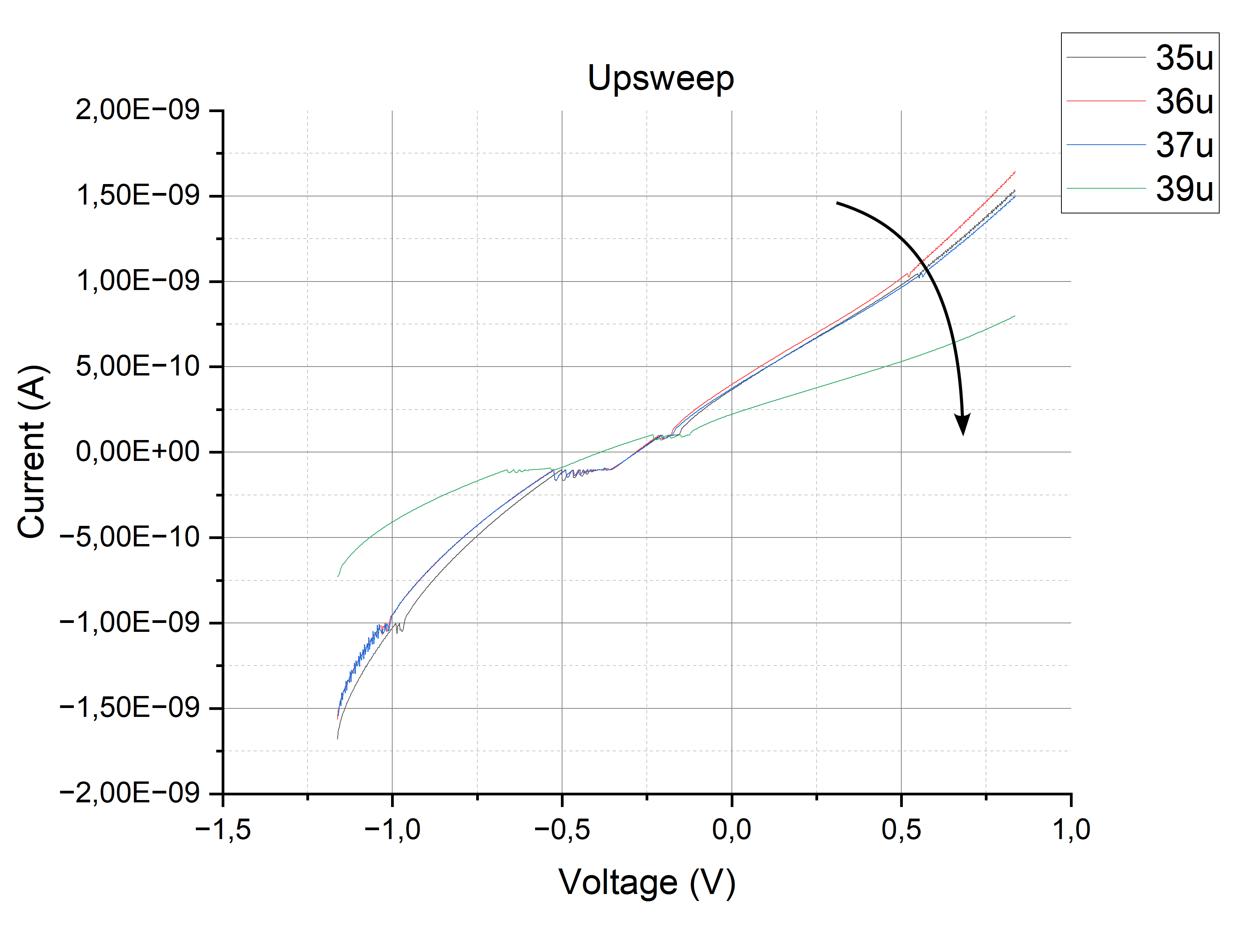}
		\subcaption[]{}	
	\end{minipage}
	\hfill
	\begin{minipage}{.46\textwidth}
		\centering
		\includegraphics[width=1\textwidth]{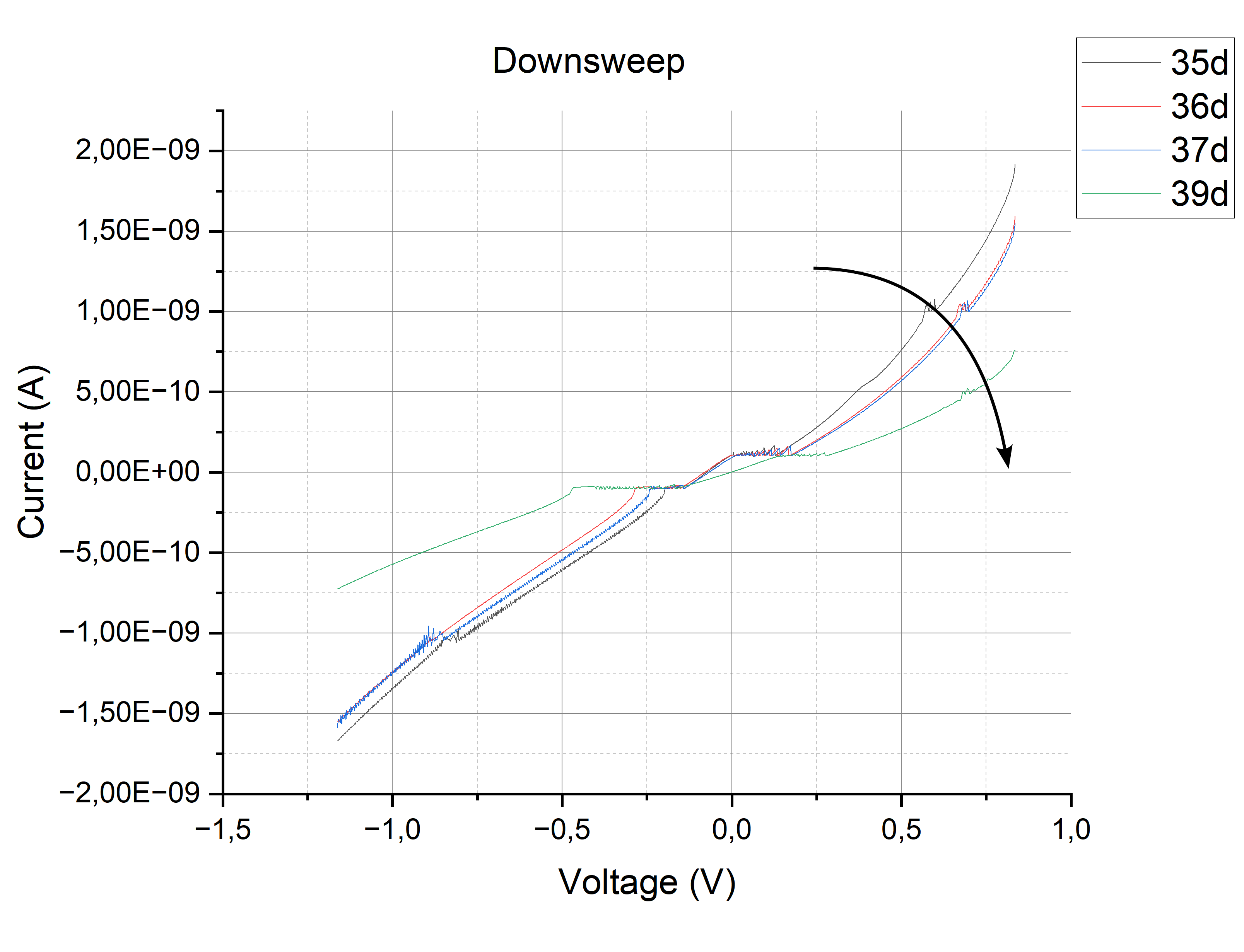}
		\subcaption[]{}	
	\end{minipage}
	\caption{IV sweeps for the bridge in four curves during the opening for Fe$^{+3}$ Salen. (a) shows several consecutive sweeps, labelled 35u (up), 35d (down) to 39u (up), 39d (down). The curves exhibit different current values for each sweep pair (up \& down). Especially 39u/d shows a very different inclination although the same shape than the others sweeps, due coupling change between two sweeps. (b) shows the up and (c) shows the down-branches of (a), to illustrate the effect. The arrows show the change of inclination in consecutive measurements. These measurements are done with with different distances between the contacts, and coupling.}
	\label{fig:opening}
\end{figure}

\noindent The curves in Fig. \ref{fig:opening} exhibit a strong hysteresis. Evaluating this phenomenon in Fe$^{+3}$ Salen requires further study. This molecule features an ion with the highest shell partially filled with five electrons unpaired, which can induce hysteresis. The literature also reports the presence of different hysteresis in organometallics regarding their oxidation states \cite{Akiyoshi2021} \cite{Kato2018}. However, there are several possible sources of hysteresis, as those caused by capacitive effects and those caused by electrochemical effects. These latter hysteresis types are usually constantly present and are not very pronounced large, usually caused by water contamination of the air humidity.

%\clearpage

\FloatBarrier

\noindent In Fig. \ref{fig:opening}, the series of consecutive opening measurements exhibit distinct characteristics, each corresponding to different opening positions within the same molecular conductance plateau (as identified in position III in \ref{fig:opa} and positions 2 or 3 in \ref{fig:opb}). 

\noindent Moreover, one particular measurement shows a marked change in the inclination of the IV sweeps. This pronounced shift could likely be a result of a change in the coupling between the molecule and the gold leads. Such a change in coupling can significantly influence the electronic properties of the system, thereby affecting the sweep characteristics. This observation underscores the sensitivity of molecular electronic measurements to the precise molecular configuration and the nature of its interaction with the electrodes, Lokamani (2023) \cite{Lokamani2023} exploits the effects caused by the geometry of the leads during the attachment and measurement.

%\clearpage

\subsection{Single Level Model -  Case Study: Corannulene}

In this section, we provide examples of when the fit gives poor results around zero ($U=0\, \mathrm{V}$) or at the tails ($U=\pm 1 \, \mathrm{V}$). These discrepancies are typically observed during one or both directions of the voltage sweep. A common scenario involves a channel being open ($U=\pm 1 \, \mathrm{V}$) and reaching saturation much earlier than predicted by SLM, thereby conducting the maximum current. In such cases, our model fails to accurately represent this saturation effect. 

\begin{figure}[!htb]
	\centering
	\begin{minipage}{.8\textwidth}
		\centering
		\includegraphics[width=1\textwidth]{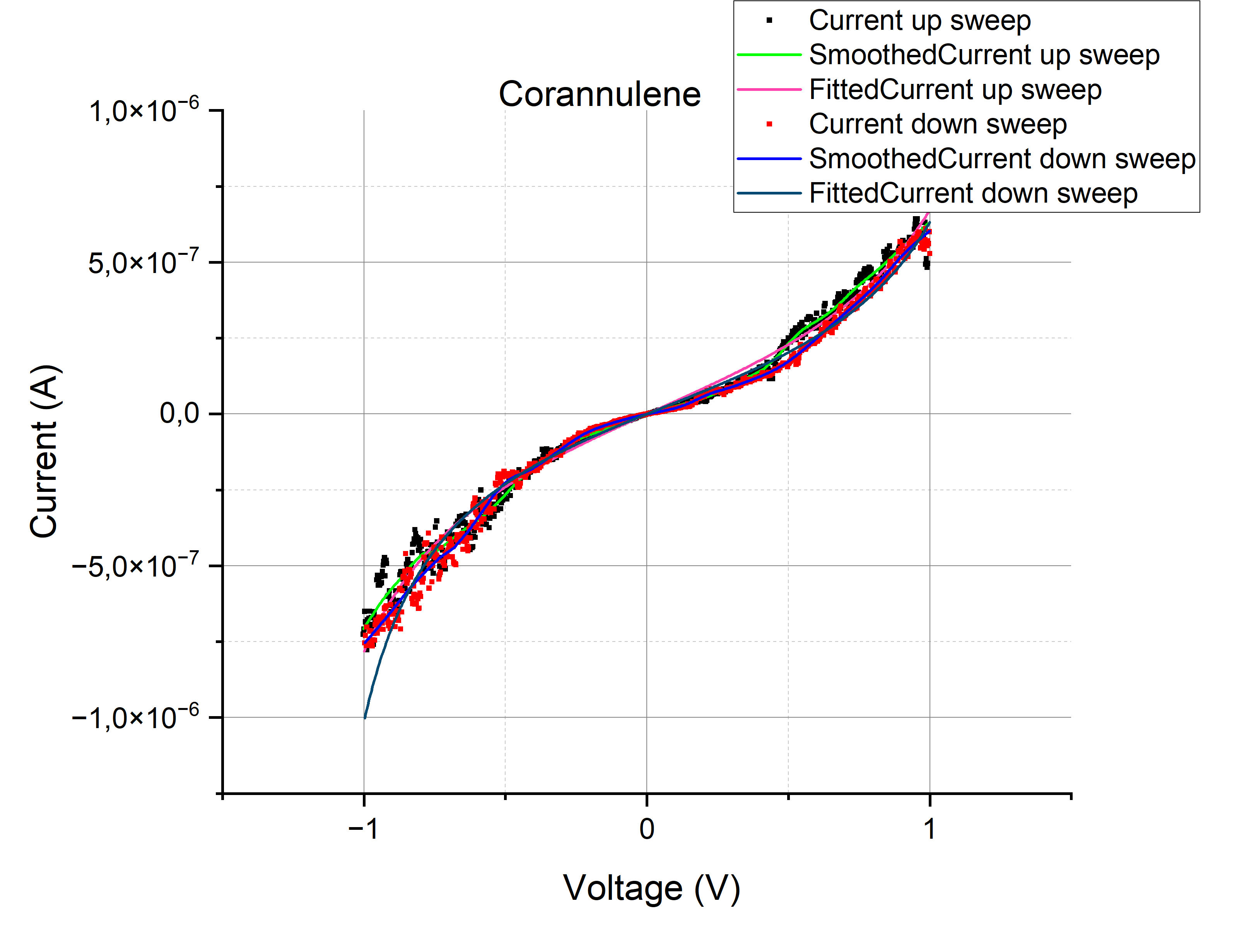}
		%\subcaption{}
		\label{fig:model3}	
	\end{minipage}
	\caption{Corannulene: full sweep with smoothened curve and fit curve, with respective values for transmission for the up-sweep $T= 5 \, 10^{-3}$, and for the down-sweep $T= 5.5 \, 10^{-3}$.}
\end{figure}

\begin{figure}[!htb]
	\centering
	\begin{minipage}{.8\textwidth}
		\centering
		\includegraphics[width=1\textwidth]{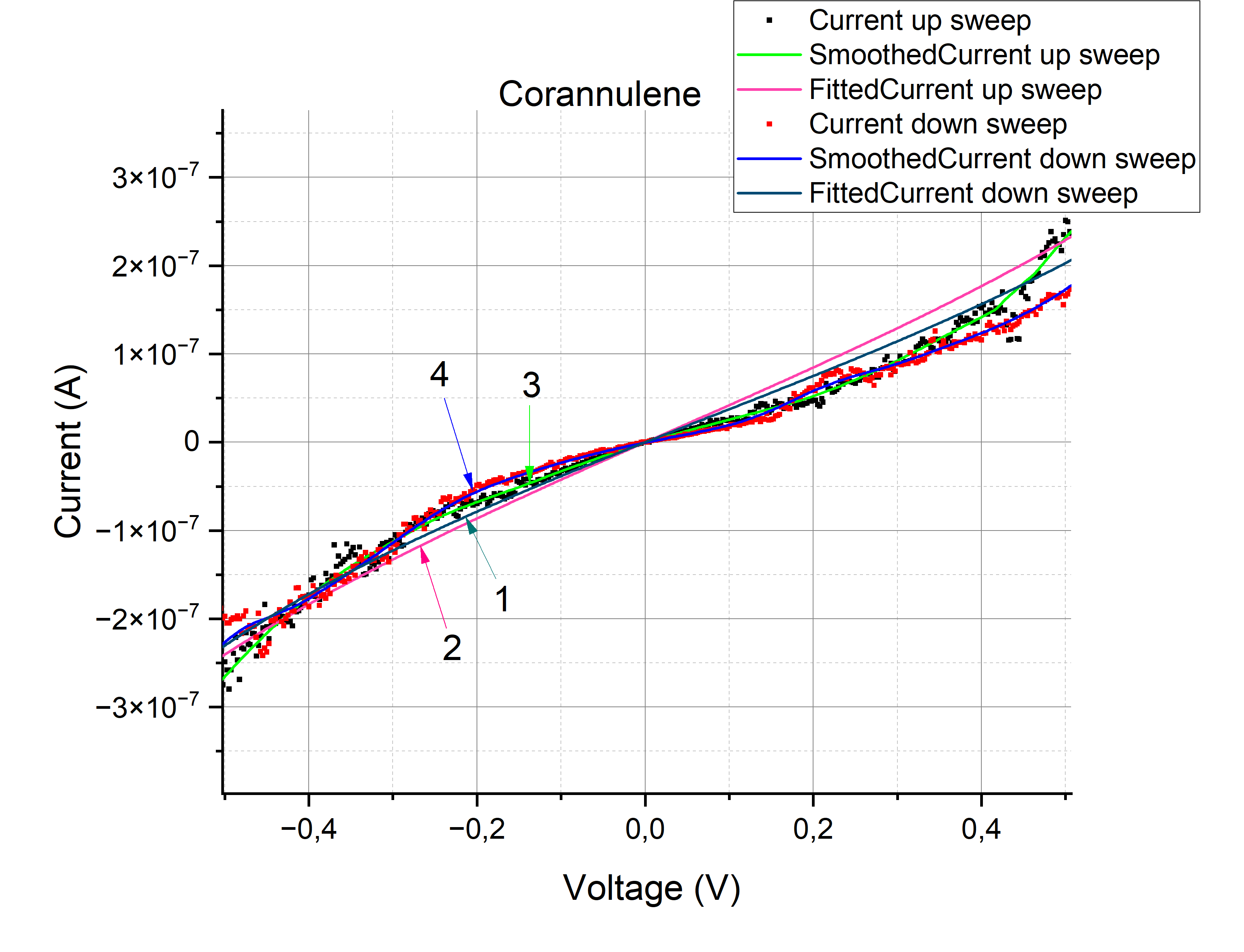}
		\subcaption{Corannulene, smoothening and fits.}
		\label{fig:model4}	
	\end{minipage}
	\hfill
	\begin{minipage}{.8\textwidth}
		\centering
		\includegraphics[width=1\textwidth]{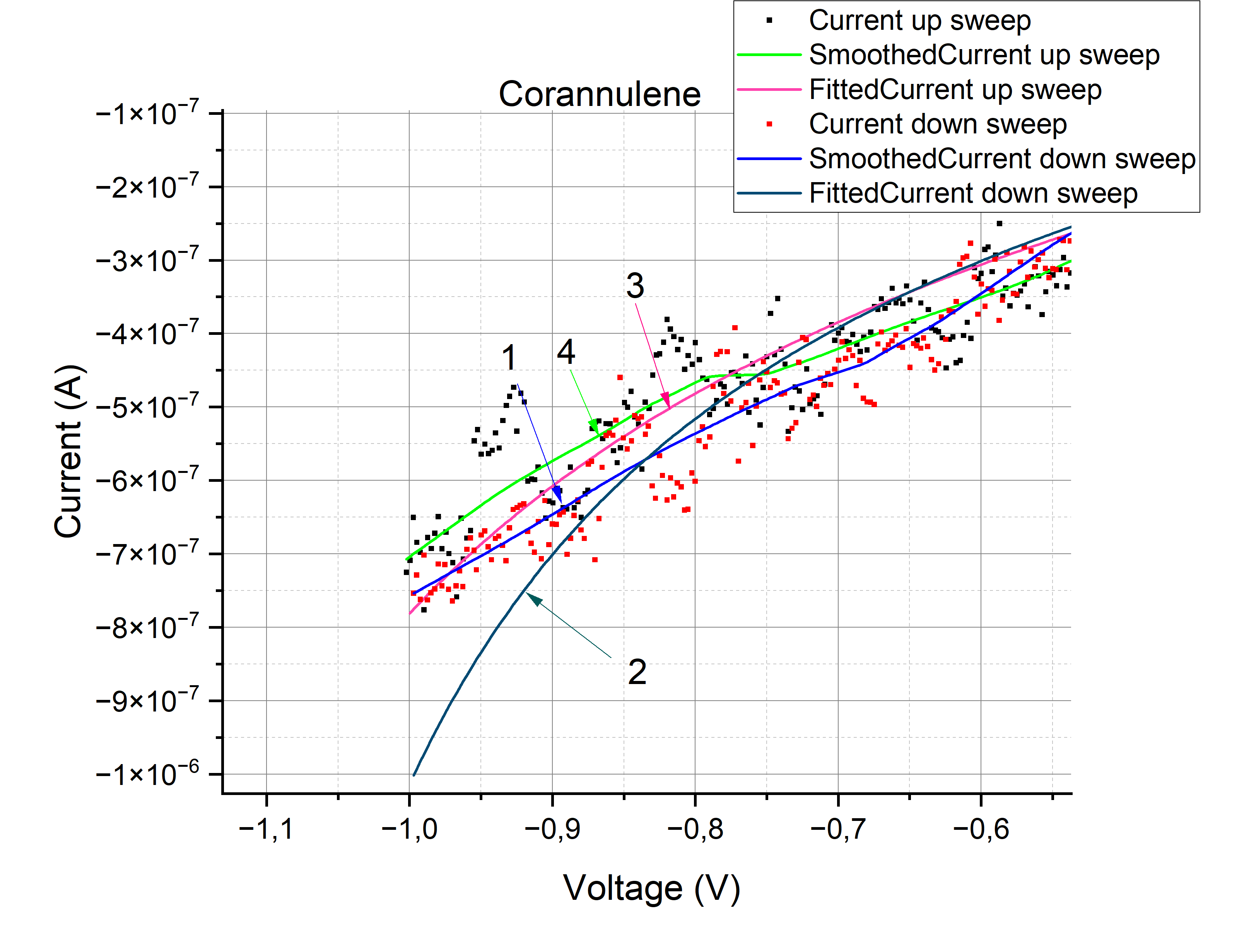}
		\subcaption{Corannulene measurement, and fits in the tail.}
		\label{fig:model5}		
	\end{minipage}
	\caption{Corannulene: Compare the zooms to previous figure; %(a) full sweep with smoothened curve and fit curve; 
		(a) zoom around zero the arrows numbered 1. 2. 3. and 4. show the difference between model and measurement, and (b) the zoom around the tail. The arrows indicate the differences between measurements and fits.}
\end{figure}

\FloatBarrier
%\clearpage

\noindent These figures show that the single level model fails to describe the full excursion of current. Observe the graph in the origin, Figure \ref{fig:model4}, the measured current is less than what the model predicts, and the current-voltage curves appear flatter around zero compared to the model's projections, as further illustrated also around the tails, Figure \ref{fig:model5}, an earlier saturation of the current.

\noindent This observation serves as a clear indicator that our current model fails to sufficiently account for several transport mechanisms which can be present. We hypothesize that electron-electron interactions are primarily responsible for the effects observed in the IV curves, especially considering that their signatures are discernible in current measurements conducted at room temperature. To incorporate these interactions into the SLM effectively, a deeper understanding of this phenomenon is necessary.

\noindent Furthermore, it is important to note that not all molecules exhibit the same behavior around the zero-voltage mark. For instance, Corannulene demonstrates a more linear response in this region as compared to the organometallic compounds discussed in this thesis. This differential behaviour can be further explored in Figure \ref{fig:scattering}, located in Section \ref{sec:singel_molecule}. The comparison highlights the unique electronic properties of different molecular structures and emphasize the need for a model that can adapt to these variances.

%\clearpage

\subsection{Lorentzian Distribution and Fitting in Salen organometallics and Corannulene}
\label{sec:singel_molecule}

\begin{figure}[!htb]
	\centering
	\begin{minipage}{0.48\textwidth}
		\centering
		\includegraphics[width=1\textwidth]{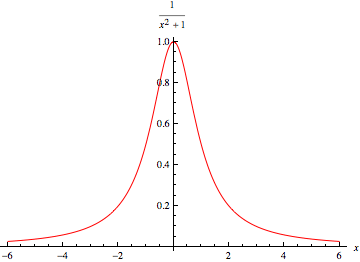}
	\end{minipage}	
	\caption{Typical shape od the transmission $T$ as function of energy can be described as an Lorentzian bell curve. However, for resonant tunnelling the peak may broaden and the curve can show a saturation plateau with maximum transmission $T=1$.}
	\label{fig:Laurentzian}
\end{figure}

\noindent The Figure \ref{fig:Laurentzian} illustrates the expected transmission shape when it does not reach saturation. The profile is symmetric, characterized by a smooth increase from 0 to 1 and a gradual decline back to zero. In this context, $T=1$ represents the 'maximum transmission' achievable for molecular channels modelled using the Lorentzian bell curve.

\noindent It is important to note that the rate and maximum transmission for different measurements approach to $T=1$ can vary. Some measurements exhibit a quicker ascent to the maximum transmission value. In such cases, the transmission is said to be 'saturated' at $T=1$, indicating that the current values are higher compared to instances of lower transmission values. This transmission saturation reflects different electronic behaviours of the channels under study (coupling), and this difference in the saturation's speed is necessary for understanding the dynamics of the system being analysed.

\begin{figure}[!htb]
	\centering
	\begin{minipage}{.48\textwidth}
		\centering
		\includegraphics[width=1\textwidth]{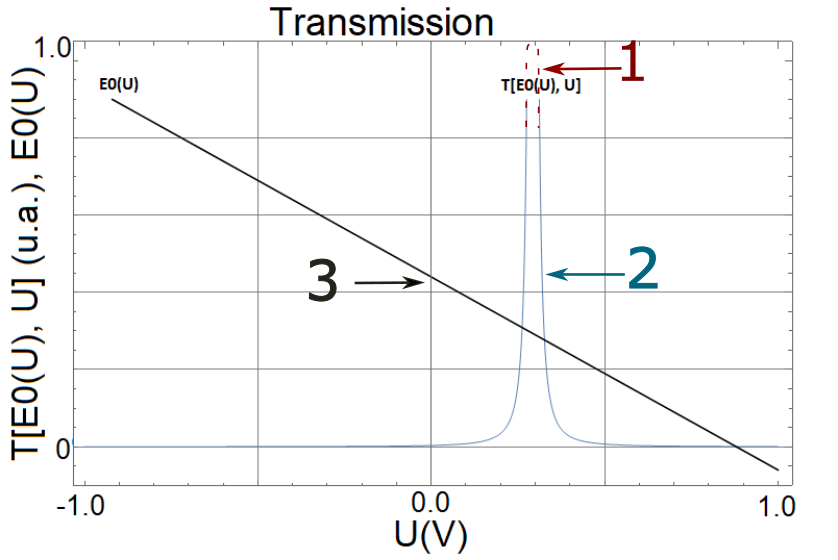}
		\subcaption{Transmission landscape for one channel, located at $E_0=0.44$, inside a single molecule channel during down-sweep; $E_0=0.44$ eV, $\Gamma_L=0.00004$, $\Gamma_R=0.23386$, $GOF=95$\%. The value in the discontinuity for the transmission is $T=1$, compare with \ref{fig:Laurentzian}.}
		\label{fig:transmission}	
	\end{minipage}
	\hfill
	\begin{minipage}{.45\textwidth}
		\centering
		\includegraphics[width=1\textwidth]{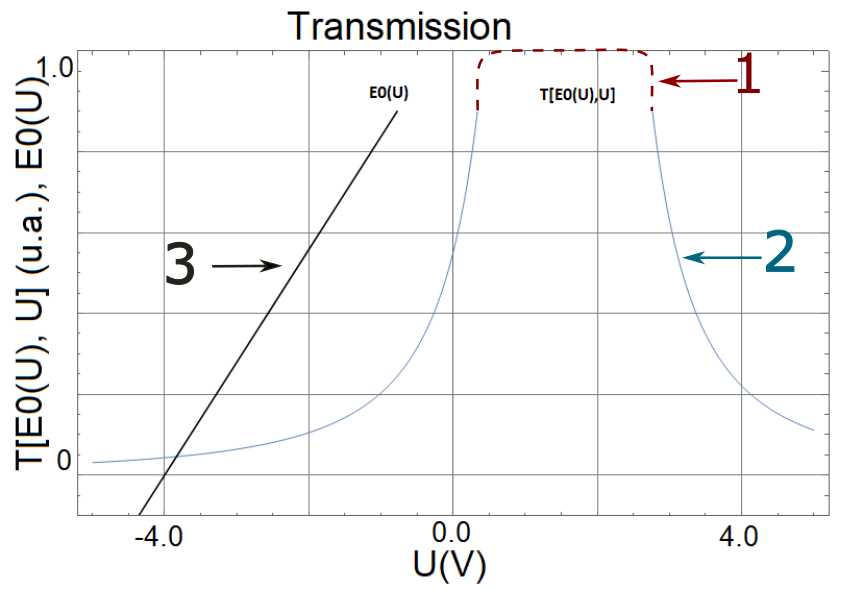}
		\subcaption{Transmission landscape for one channel, located at $E_0=1.117281$, inside a single molecule channel, during up-sweep; $E_0=1.117281$ eV, $\Gamma_L=0.003372$, $\Gamma_R=0.000955$, $GOF=98.9$\%. The value in the discontinuity for the transmission is $T=1$, compare with \ref{fig:Laurentzian}.}
		\label{fig:transmission_up}		
	\end{minipage}
	\caption{Transmission in Fe$^{+3}$ Salen. Figures a. and b. show the change in transmission from down-sweep to up-sweep. In a. and b., the numbers have the same meaning: (1) is the red dotted region and represents the convergence of the curve to $1$. The fast convergence can be comprehended by observation of the equation \ref{eq:Breit-Wiegner}, the transmission has an elevated growth rate for the measured values of $\Gamma$ around the channel $E_0$ and the transmission saturates in $1$; (2) belongs to the graphs calculations with the equation, and number (3) The change of position of the channel energy $E_0$ of the channel when a supply is applied between the terminals during the current sweep from $[-1:1]$. These graphs bring important information about how the sweep affects the channel and the current. Observe that $E_0$ changes and sometimes leaves the rage of maximum transmission. Also, in b. there is a broadening of the transmission around $E_0$, which brings the possibility of more states participating in the current.}
\end{figure}

\FloatBarrier

\noindent Figure \ref{fig:transmission} presents the calculated transmission $T(E_0,\Gamma_1, \Gamma_2)$ using the fit parameters previously determined for IV sweeps in Fe$^{+3}$ Salen. Notably, during the down-sweep, the transmission `T' reaches values close to $\approx 1$ for voltages around $U=0.295\,\mathrm{V}$, indicating a $100\%$ efficient charge carrier transmission through the molecule, a phenomenon known as channel resonance. For the up-sweep there is a broad voltage interval where resonance occurs, as predicted by the fitting model.

\noindent We realize that the model predicts exactly the current tunnelling through the molecule, despite a GOF of $95\,\%$. Also, $E(U)$ is affected by the current sweep during all measurements. Additionally, the resonant voltage occasionally falls outside the sweep range, what indicates that channel is not archiving resonance, once it is out side the sweep region. We need to emphasis at this point that these are sweeps from the same measurement sample. It is crucial to note that these variations are within the same measurement, indicating alterations in the molecule's conformation during the experiment, which could affect the coupling between molecule and leads. This could lead to discrepancies between the measured values and some of the expected behaviour predicted by the fitting curve.

\noindent For small $\Gamma$, the electron remains time long enough in the molecule and this can diphase the quantum state and can cause incoherent transport, i.e. $e-ph$ scattering and $e-e$ inelastic scattering. Additionally, $G< 0.5$, meaning that if $e-ph$ scattering happens will be $e-ph$ forward scattering \cite{Sondhi2021}, what means that after the scattering event the electron gains energy. For small molecules with one main ``redox'' site, transport evolves into a two-step hopping process and the analytical approximation of the expression 

\begin{equation}
	I(U) = \frac{2e}{h}\int_{-\infty}^{\infty} T(E,U) [f_L(E-eU)-f_R(E)] dE, \label{current}
\end{equation}

\noindent cannot be used to describe the current-voltage curves. Instead, either a two-barrier model is used in these cases, especially for small molecules with conjugate groups connected to a non-conjugate core \cite{PascalGehring2019} or include a ne term in the resonant tunnelling model. 
This is in agreement with the flat current curves around the origin results for Fe$^{+3}$ Salen, and its hysteresis only give a level DC to the measurements, still allowing to visualise its plateaus around zero-voltages. Although the most of these incoherence effects is observed at LT \cite{PascalGehring2019}. We discard this conservative approach, once the SLM can do good predictions of current also in organometallics. However, we decided focus on coranulenne which has no ion inside, and has lower hysteresis than Fe$^{+3}$ Salen. We want to explore whose of these incoherent phenomena are still observable at RT, or whether they are obscured by background noise. This explains mismatches between the SLM and regions with low bias and low transmission observed in several plots.
\noindent A final word about organometallics, they still can be modelled using the SLM.
The ion centres inherent in these compounds play a significant role in the overall electronic behaviour.They contribute to potential interactions with the current passing through the molecule. This introduces additional complexities, such as disturbances and enhanced electron-electron interactions. 

\begin{figure}[!htb]
	\centering
	\begin{minipage}{1\textwidth}
		\centering
		\includegraphics[width=.8\textwidth]{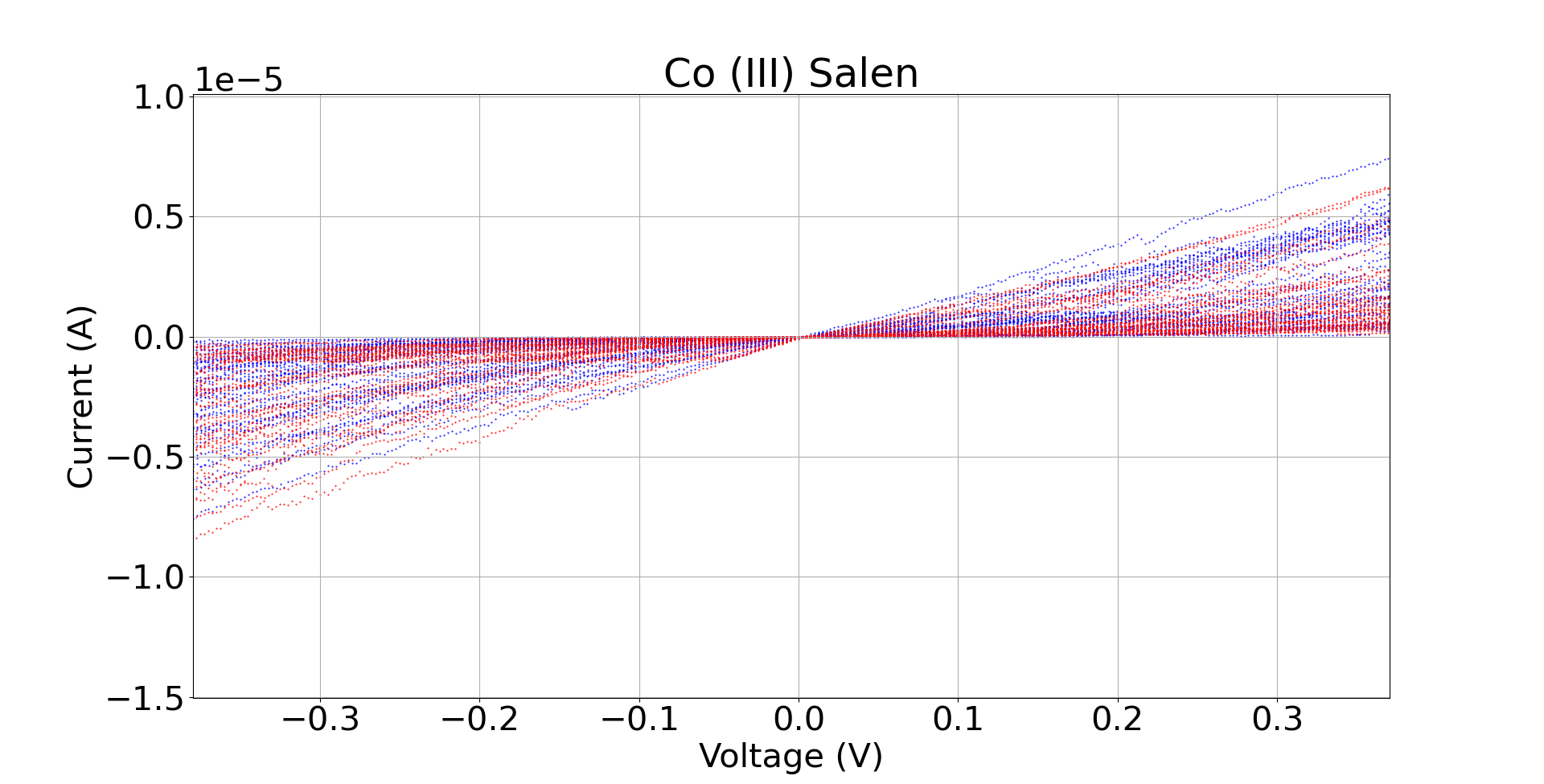}\\
		\subcaption{Current vs voltage: selection of ca. 100 down-sweeps.}
		\label{fig:highorder1}	
	\end{minipage}
	\hfill
	\begin{minipage}{.8\textwidth}
		\centering
		\includegraphics[width=1\textwidth]{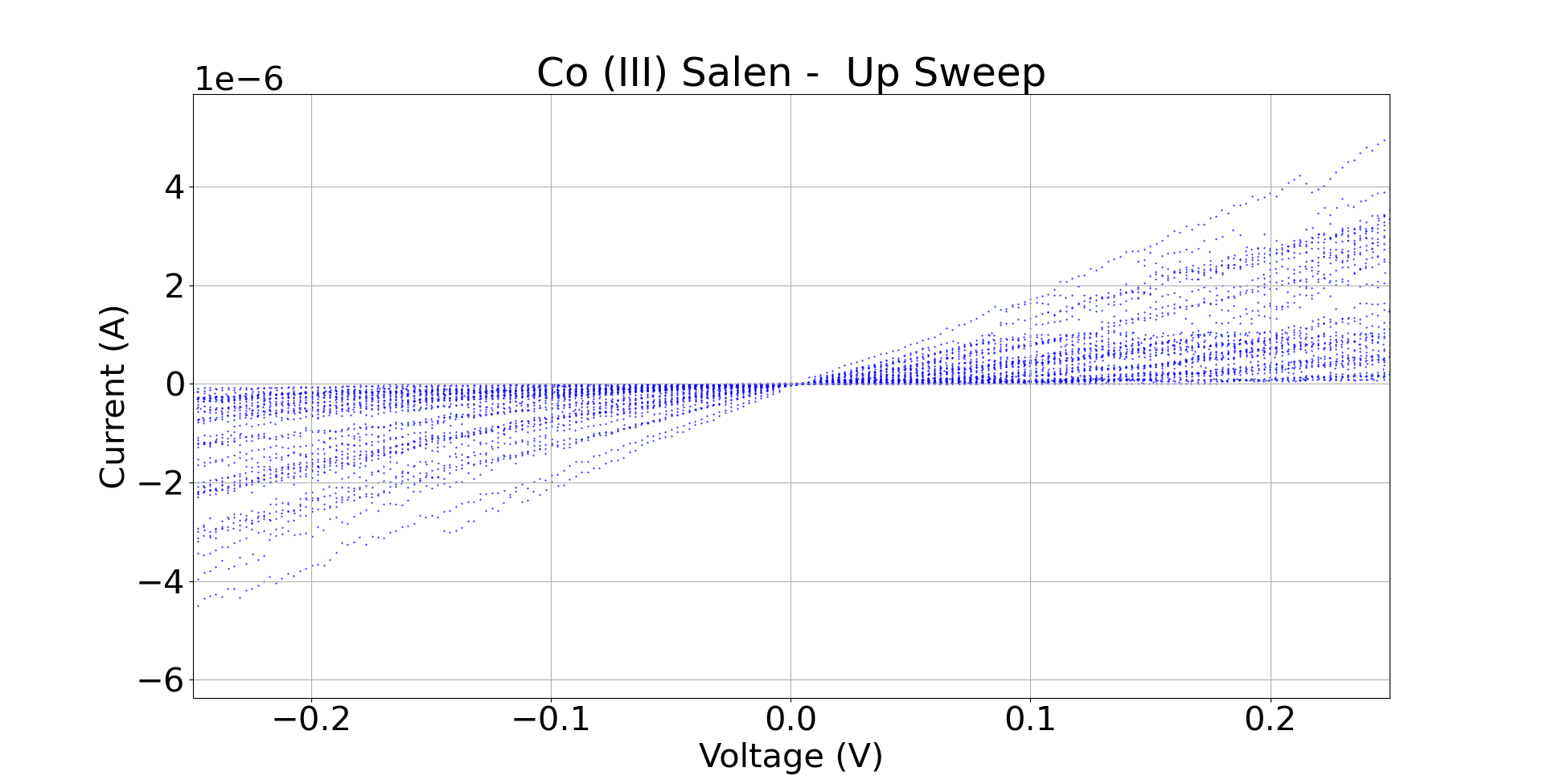}\\
		\subcaption{Current vs voltage: selection of ca. 100 up-sweeps.}
		\label{fig:highorder2}
	\end{minipage}
	\begin{minipage}{.8\textwidth}
		\centering
		\includegraphics[width=1\textwidth]{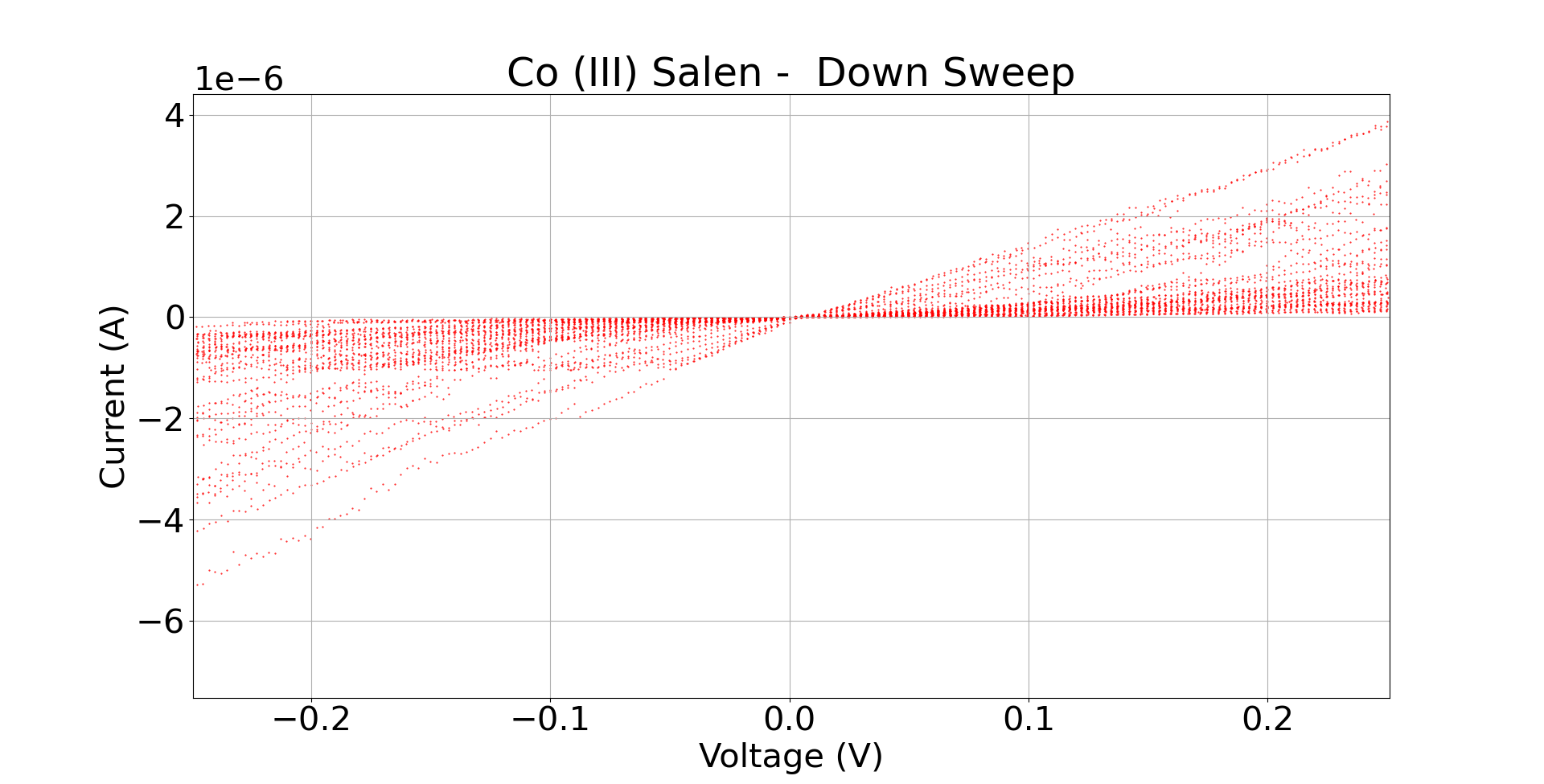}\\
		\subcaption{Current vs voltage: selection of ca. 100 up-sweeps.}
		\label{fig:highorder3}
	\end{minipage}
	\caption{Linearity during current measurement around zero for clusters of Co$^{+3}$ Salen. In Figure (a), we have a full sweep in full measurement range; in (b) and (c) we show a zoom around zero with the up and down-sweeps separated in each figure.}
\end{figure}

\FloatBarrier

\noindent Note that in Figure \ref{fig:highorder1}, the current shows a linear behaviour around zero, with a very small inclination. The transmission value of these samples is greater than $0.1$, and we assume here are probably clusters instead of single molecules, and more than one channel is participating in the charge transport. Previous studies on organometallics of Salen have reported single molecules with a transmission value between $10^{-6}$ and $10^{-3}$\cite{Sondhi2022}. However, because a value of transmission bigger than $0.1$, can not just be immediately considered as belonging to a cluster; As demonstrated by Kim et al. (2011) \cite{Kim2011}, this threshold does not necessarily indicate the presence of a cluster formation. Their research on single molecule benzene junctions reported transmission values exceeding $T>0,1$, accompanied by relatively high absolute currents, approximately $\sim 10^{-6}\mathrm{A}$. Their finding suggests that even individual molecules can exhibit significant transmission and current levels, challenging the assumption that high transmission values are exclusive to clustered molecular arrangements. Therefore, a more careful approach is necessary when analysing transmission data in molecular electronics, considering the potential for high transmission in single molecule scenarios.

\begin{figure}[!htb]
	\centering
	\begin{minipage}{1\textwidth}
		\centering
		\includegraphics[width=1\textwidth]{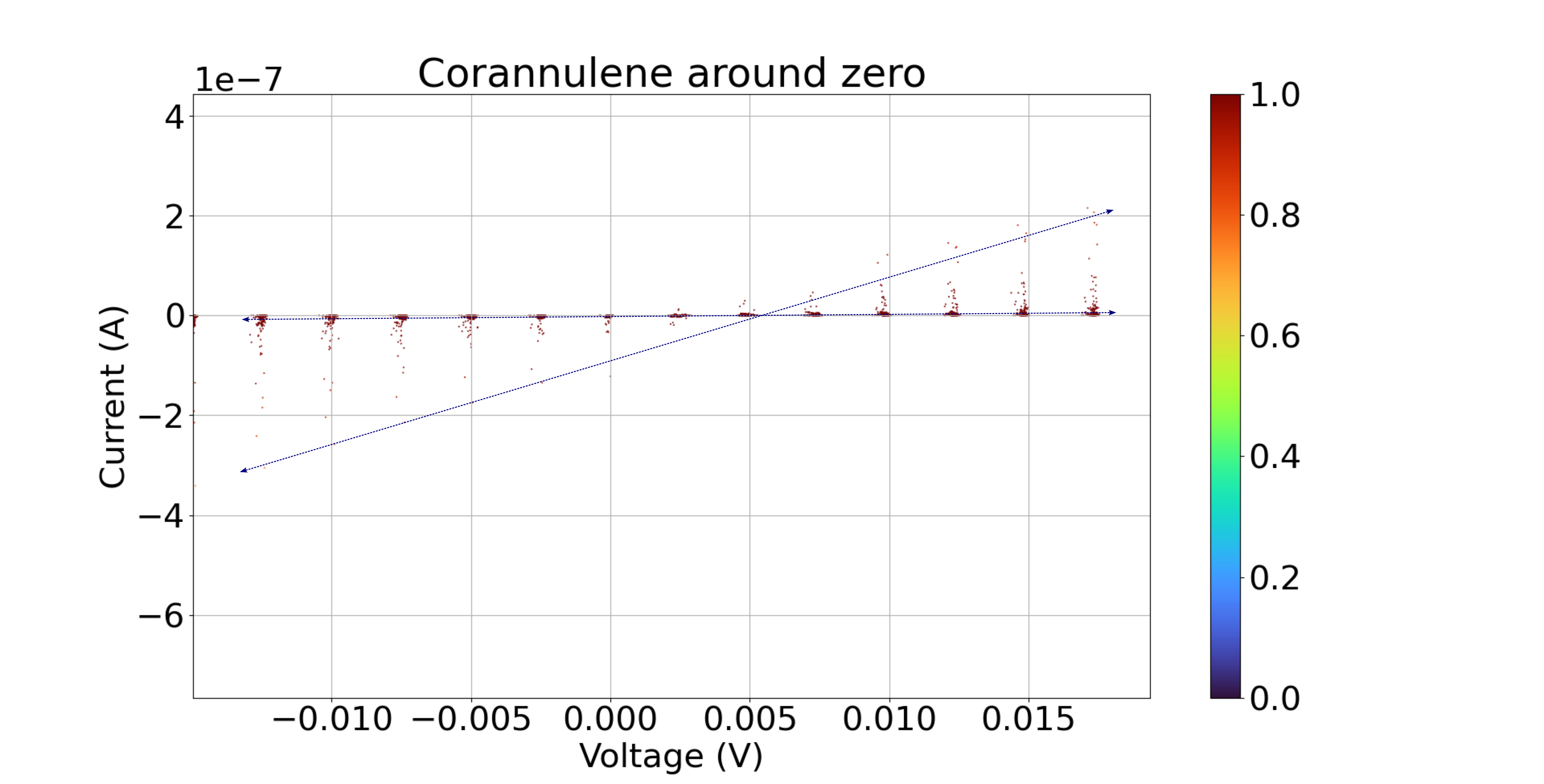}
		\subcaption[]{Corannulene measurement zoom around zero, with hand drown direction curve, to highlight the magnitude of the highest currents around the axis origin.}
		\label{fig:scatteringa}
	\end{minipage}
	\begin{minipage}{1\textwidth}
		\centering
		\includegraphics[width=1\textwidth]{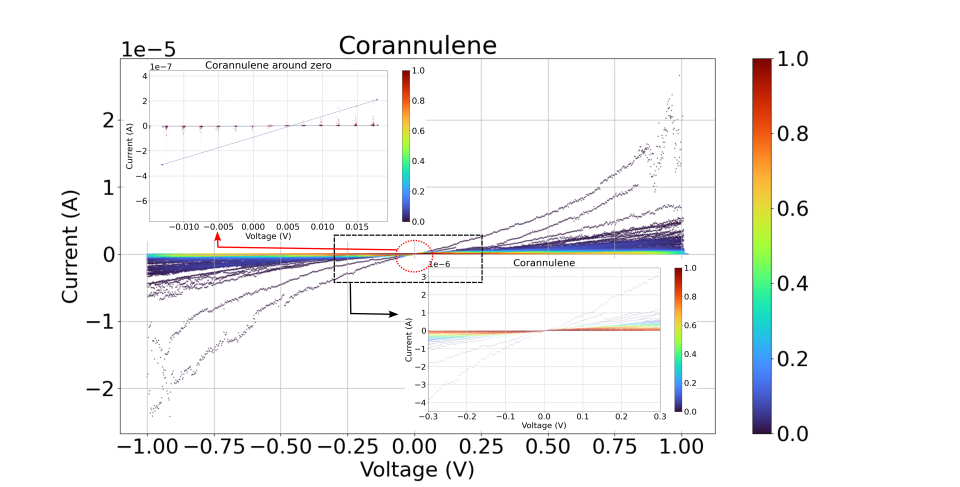}
		\subcaption[]{Corannulene measurement full range, ca. 100 measurements.}
		\label{fig:scatteringb}
	\end{minipage}
	\caption{Current vs voltage: scattering curves of Corannulene for data around zero in (a) and full range in (b). The color code represents the normalized current $I_n$ of c.a. 100 measurements, see also appendix \ref{app:norm} for Corannulene.}
	\label{fig:scattering}
\end{figure}

\FloatBarrier

\noindent In Figures \ref{fig:scatteringa} and \ref{fig:scatteringb}, we note a distinct sub-linear behavior (flattening) around zero voltage in the IV sweeps, a phenomenon not captured by our current fit model. We postulate that this flattening is likely due to electron-electron interactions, as suggested by the work of Kosik (2004) \cite{Kosik2004}. These interactions, notably absent in the standard implementation of the LM, could be a critical factor in explaining this observation. Furthermore, Kosik's study indicates that at low temperatures (LT), Coulomb blockades play a significant role in this voltage region, an influence that is not observed at room temperature (RT) \cite{Kosik2004}. This discrepancy between LT and RT measurements highlights the temperature-dependent nature of electron interactions in these systems.

\noindent Additionally, we provide a comprehensive analysis of this behaviour in the appendix \ref{app:norm}. Here, we present various graphs illustrating the linear nature of the slope around the origin. This linear trend does not contrast with the sub-linear flattening observed at zero voltage. Although this flat behaviour there are still current and it is linear. The emphasis on the complexity of electron behaviour in these measurements shows the potential limitations of the SLM in fully capturing these nuances.

%\FloatBarrier

\begin{figure}[!htb]
	\begin{minipage}{1\textwidth}
		\centering
		\includegraphics[width=1\textwidth]{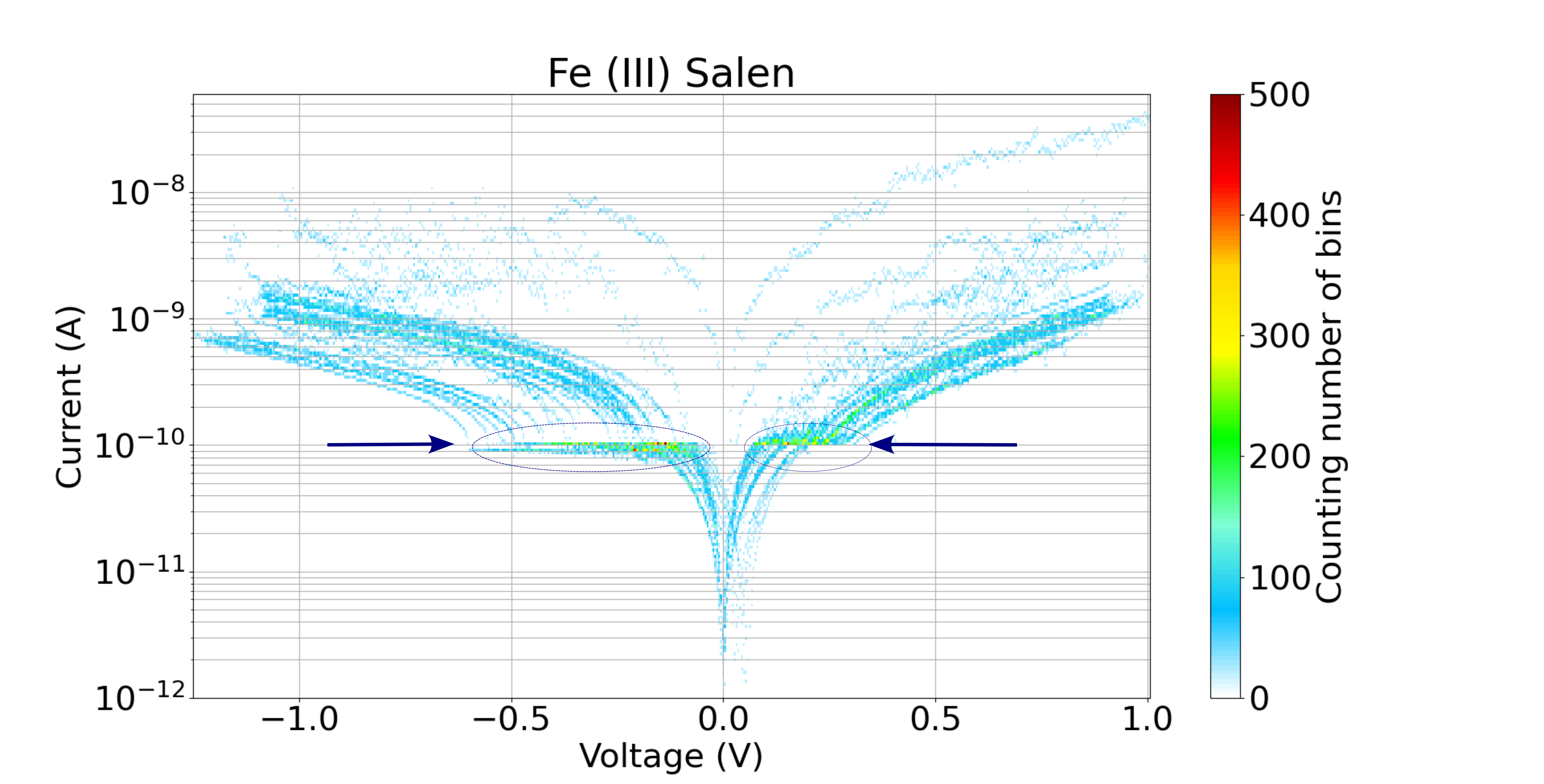}
		%\subcaption[]{}
	\end{minipage}
	\caption{Current vs voltage: The counting represents the number of bins for the pairs $[current, \, voltage]$ which superpose each other. This graph shows the presence of non-linearity in the current measurement of ca. 30 sweeps, indicated by the arrows. The color code of each dot represents the respective number of bins.}
	\label{fig:kinks_hist2d}
\end{figure}

\FloatBarrier

\begin{figure}[!htb]
	\centering
	\begin{minipage}{.8\textwidth}
		\centering
		\includegraphics[width=1\textwidth]{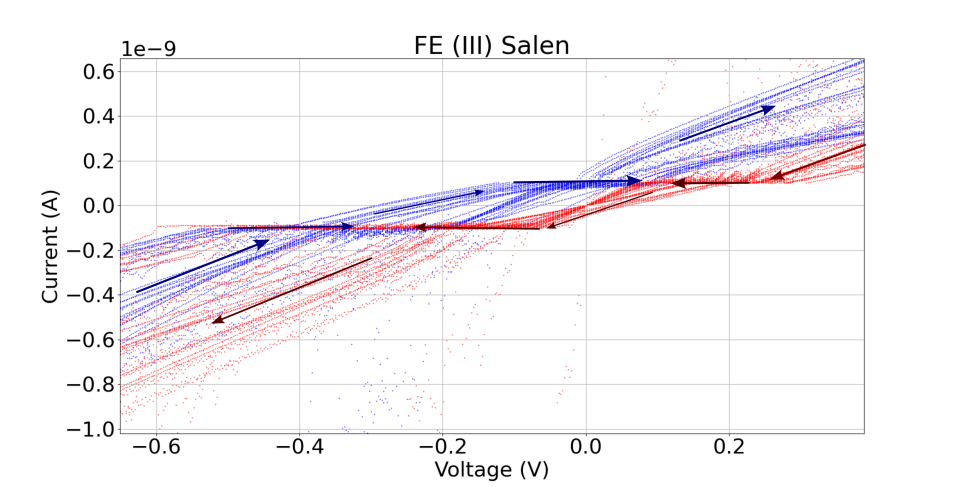}
		\subcaption[]{}
		\label{fig:zoom0}
	\end{minipage}
	\begin{minipage}{.8\textwidth}
		\centering
		\includegraphics[width=1\textwidth]{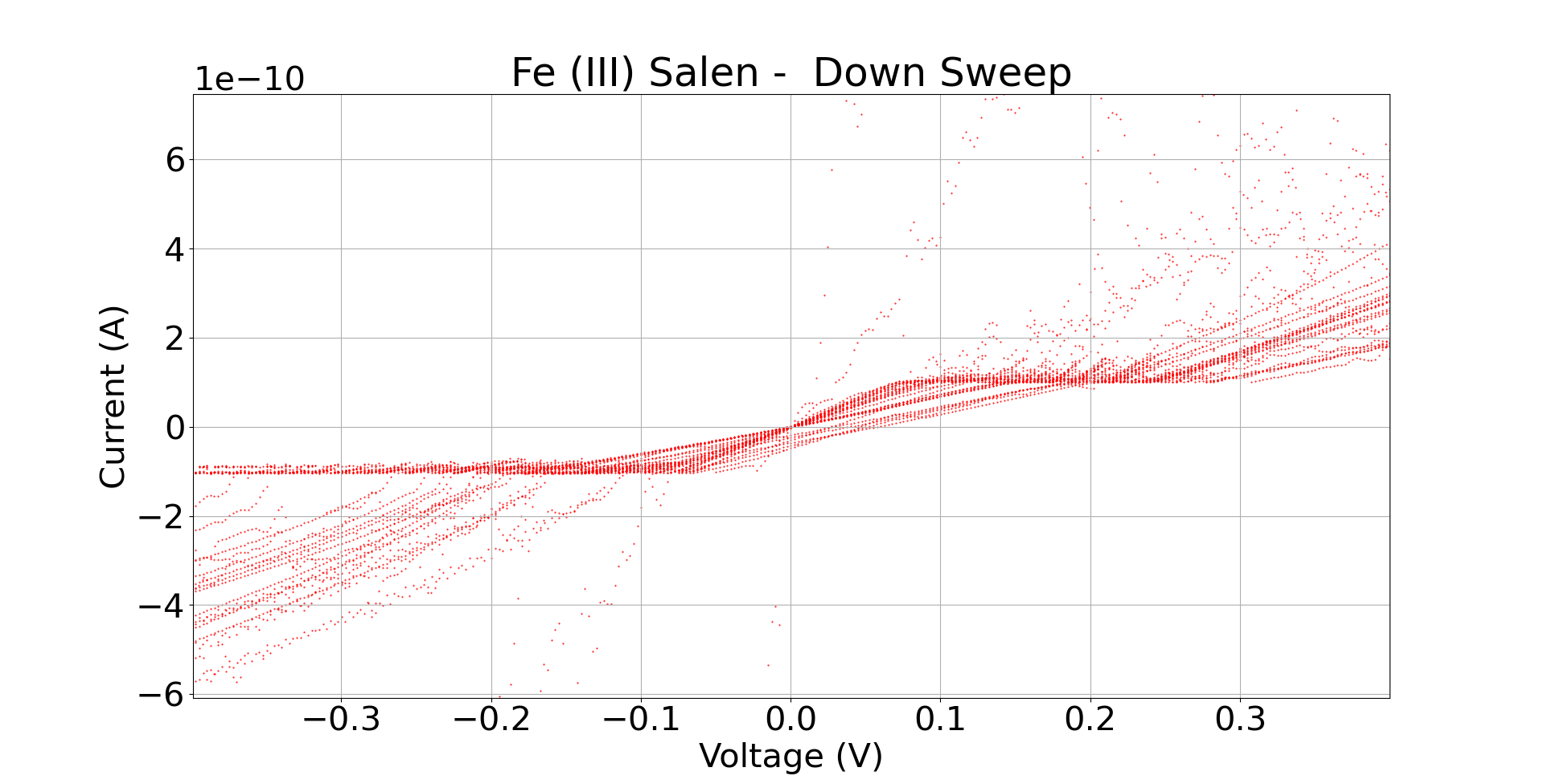}
		\subcaption[]{}
		\label{fig:zoom1}
	\end{minipage}
	\hfill
	\begin{minipage}{.8\textwidth}
		\centering
		\includegraphics[width=1\textwidth]{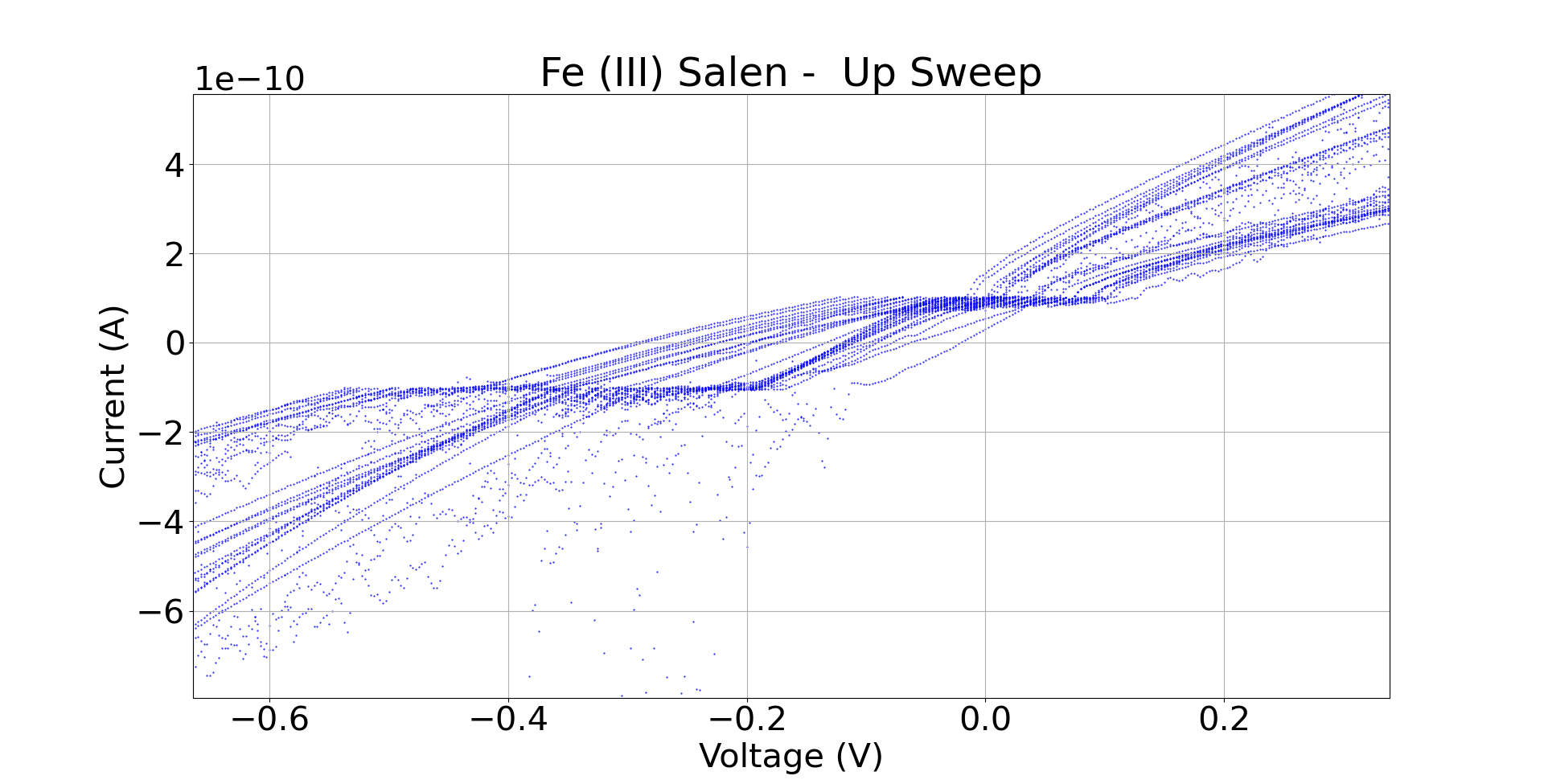}	
		\subcaption[]{}
		\label{fig:zoom2}
	\end{minipage}
	\caption{Current vs voltage: the presence of non-linearity in the current measurement of ca. 30 sweeps of Fe$^{+3}$ Salen single molecules. In Figure (a), we show full sweep and direction indicated by arrows; (b) and (c) are up- and down-sweep separated in each figure.}
\end{figure}

\noindent In the case of electron-phonon interaction, and the electronic metal-molecule coupling is weak, the physical description can be in terms of rate equations, which take into account vibronic effects in a non-perturbative manner \cite{CuevasScheer2010}. These effect signatures are typically studied at LT \cite{Sondhi2021}. However, the presence of these signatures in single molecules at RT has not been extensively studied. The main reason is that all relevant phonon modes are expected to be thermally excited at RT. Neither Figures \ref{fig:kinks_hist2d}, \ref{fig:zoom1}, and \ref{fig:zoom2} have any signatures of phonon excitations, nor do we observe any clear indications of phonon excitations. At LT, whether the sweep coincides with integer multiples of the vibrational energy $\hbar \omega$, can induce a vibrational inelastic current \cite{CuevasScheer2010}. This effect is molecule-dependent, and the number of these phonons is typically not significant enough to induce visible changes in the current graphs but in its differential measurements when they affect the transport of some molecules. At RT, even with an increased population of phonons, detecting their influence on current molecular transport is challenging with our current measurement setup. Consequently, the investigation into the role of phonons at RT, while intriguing, is beyond the scope of this current work.

\noindent On the other hand, changes in the measurements can be a result of electron-electron interactions, particularly since the SLM, primarily designed for resonant tunnelling scenarios, does not align perfectly with our measurements. This hypothesis is supported by Kosik (2004) \cite{Kosik2004}. The potential mismatch between the SLM predictions and actual measurements indicates the need for a more comprehensive model to encapsulate the full range of interactions, including electron-electron dynamics.

\FloatBarrier

\noindent Coulomb blockades: the molecules can act as charge traps at low bias. In case the electrons are exposed to a capacitive effect, and only high-energy electrons can pass through the molecule. The current flow will stop (LT) or be very low (RT) for electrons with low energies. Fig. \ref{fig:scatteringb} shows that even if they are deactivated, the coulomb blockade's effect should not be neglected at RT, as these figures show around zero the flattening caused by the remaining electrons iterations, which provokes an electron-electron iteration, which affects the current flow \cite{Kreupl1998}.

\noindent When the electron scatter interacts with other electrons and gains energy in this scattering crossing the molecule \cite{Datta2005}, we have to consider its effects. However, the SLM is based on electron coherence for resonant tunnelling; the molecule is weakly coupled, and in such cases, the charge centres within the molecule can influence the electrons, whether they are pre-existing or induced by deformations in the molecular structure due to its soft nature. Nevertheless, only the elastic electron-electron scattering is effective in the case of localisation effects, as Datta  2025  \cite{Datta2005} defined them. In Laible 2020 \cite{Laible2020} measurements, it is suggested that electron-electron inelastic scattering is present too. However, he worked with very different geometry, which, for this reason, has distinct effects.

\clearpage

\subsection{Single Level Model - Study of the case: Fe$^{+3}$ Salen}

In the realm of our measurements, a pertinent question emerges: Are we observing the behavior of single molecules or molecular clusters? This query is especially relevant considering the range of transmission and current values for Fe$^{+3}$ Salen reported in the literature, as noted by Sondhi (2022) \cite{Sondhi2022}. To address this, the current section is dedicated to analysing our fit results using the SLM and comparing these outcomes with existing literature. This comparative analysis aims to elucidate whether the phenomena we are observing can be attributed to individual molecules or to the collective behaviour of molecular clusters. Such a distinction is critical for accurately interpreting our data and aligning our findings with the broader understanding of molecular electronics.

\begin{figure}[!htb]
	\centering
	\begin{minipage}{.8\textwidth}
		\centering
		\includegraphics[width=1\textwidth]{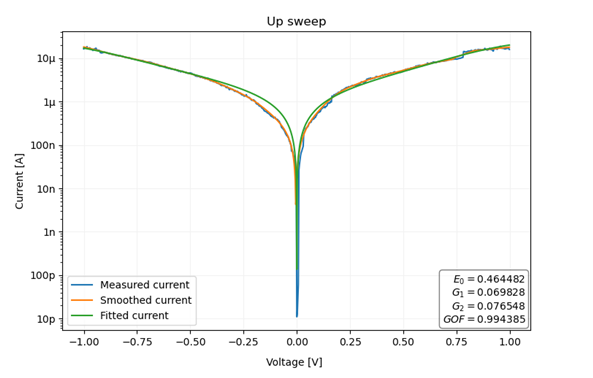}
		\subcaption[]{}
		\label{fig:clustervssingle1a}
	\end{minipage}
	\hfill
	\begin{minipage}{.8\textwidth}
		\centering
		\includegraphics[width=1\textwidth]{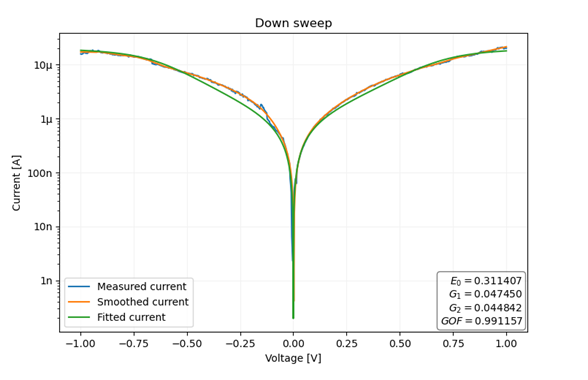}	
		\subcaption[]{}
		\label{fig:clustervssingle2b}
	\end{minipage}
	\caption{Current vs voltage: current measurement, the fit matches the transmission around $T\approx 10^{-5}$ suggesting we are dealing with a single molecule a. up-sweep b. down-sweep.}
	\label{fig:clstr}
\end{figure}

\FloatBarrier

\begin{figure}[!htb]
	\centering
	\begin{minipage}{.8\textwidth}
		\centering
		\includegraphics[width=1\textwidth]{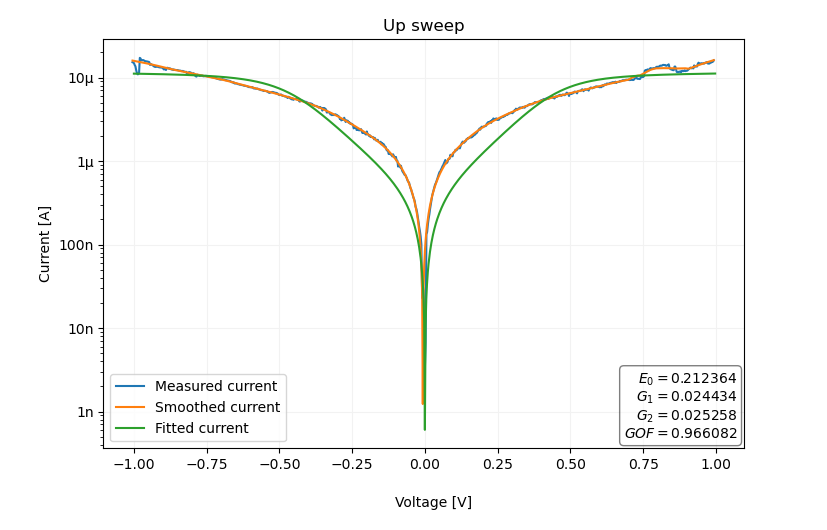}
		\subcaption[]{}
		\label{fig:clustervssingle1}
	\end{minipage}
	\hfill
	\begin{minipage}{.8\textwidth}
		\centering
		\includegraphics[width=1\textwidth]{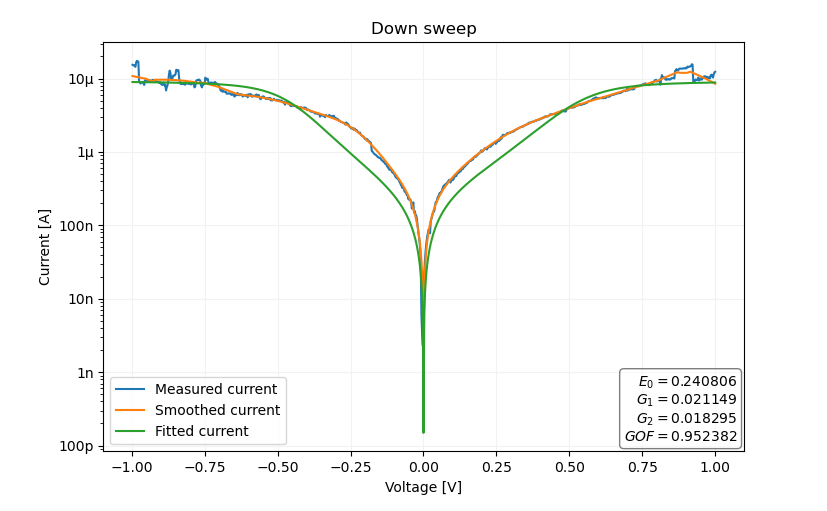}	
		\subcaption[]{}
		\label{fig:clustervssingle2}
	\end{minipage}
	\caption{Current vs voltage: current measurement, here the model does not fit, the values of the transmission are high $T>10^{-3}$, suggesting we are dealing with a molecule cluster (a) up-sweep (b) down-sweep; in log scale.}
\end{figure}

\FloatBarrier

\noindent In Fig. \ref{fig:clustervssingle1a} and \ref{fig:clustervssingle2b}, we observe that IV sweep fits better to the SLM than in Fig. \ref{fig:clustervssingle1} \ref{fig:clustervssingle2}, where the fit is poor. A good reason to explain the differences is that a single molecule and a cluster have been measured respectively, and the clusters respond differently to the transport through their channels. Clusters have larger currents because they feature a higher number of open channels. Those channels can reorganize in a more robust channel, which cannot be so sensible to the changes caused by the current sweep or coupling, enabling transmission values closer to one. 
\noindent The current jumps in the subsequent measurement is interpreted as a position change of the molecule/cluster, interrupting the current flow temporarily and/or changing the coupling.

\clearpage

\section{Transmission and Coupling}

\subsection{Transmission and Coupling -  Case Study: Fe$^{+3}$ Salen}

This analysis presents a series of graphs that investigate the transmission values in  Fe$^{+3}$ Salen, specifically in the context of current tunnelling processes. The choice of Fe$^{+3}$ Salen as the subject of this study is not driven by any specific attributes that distinguish it markedly from other molecules like Corannulene in terms of coupling properties; the coupling in Corannulene does not differ from Fe$ {+3}$ Salen. Besides its values, there are no peculiar or exceptional trends or points to be exploited.

\begin{figure}[!htb]
	\centering
	\begin{minipage}{1\textwidth}
		\centering
		\includegraphics[width=1\textwidth]{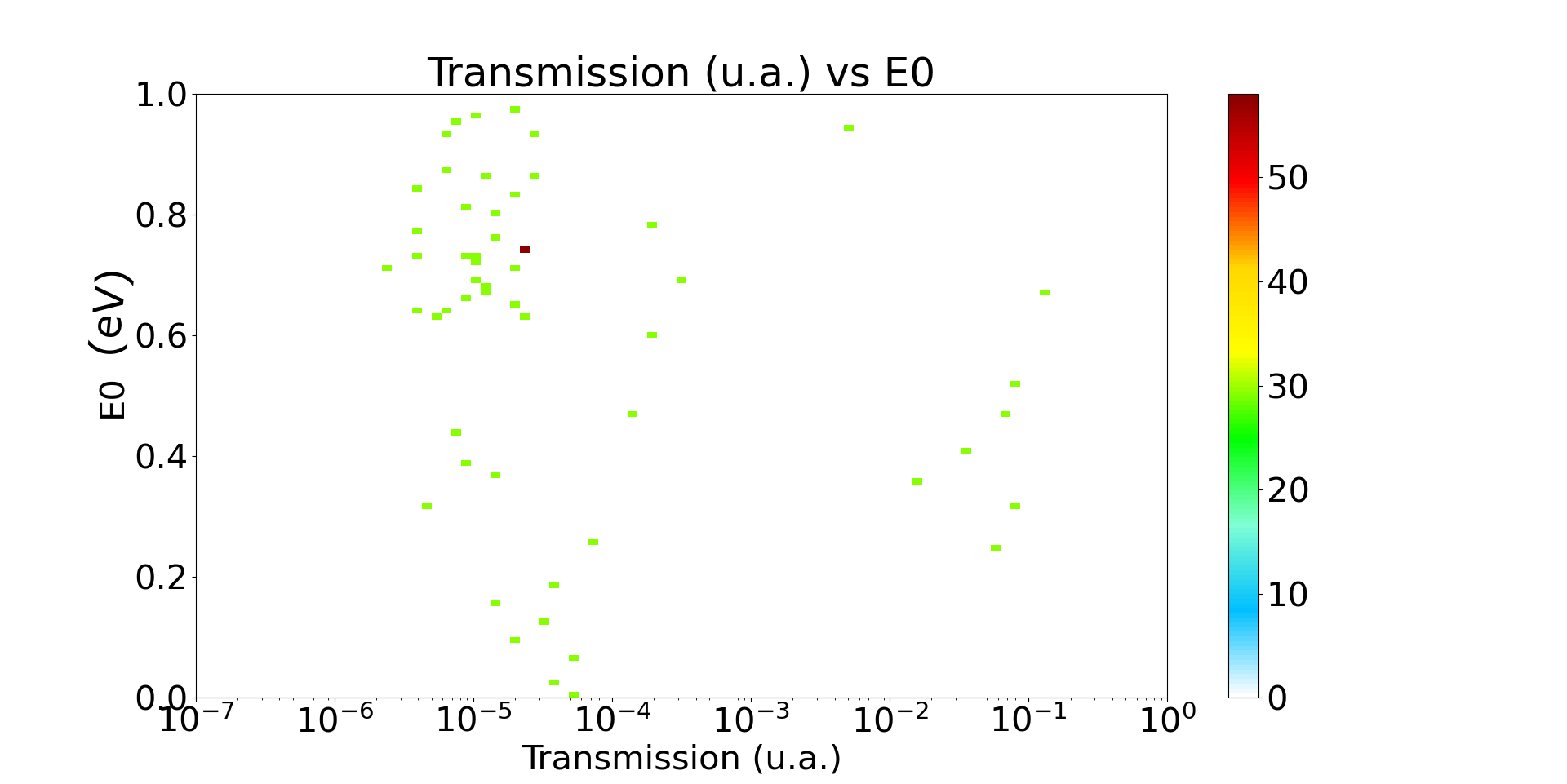}
	\end{minipage}
	\caption{$E_0$ vs transmission relation for an ensemble of ca. 30 measurements with molecules. The color code represents the number of counts for each data bin of the data pair $[T,E_0]$.}
	\label{fig:t1}
\end{figure}
\begin{figure}[!htb]
	\begin{minipage}{1\textwidth}
		\centering
		\includegraphics[width=1\textwidth]{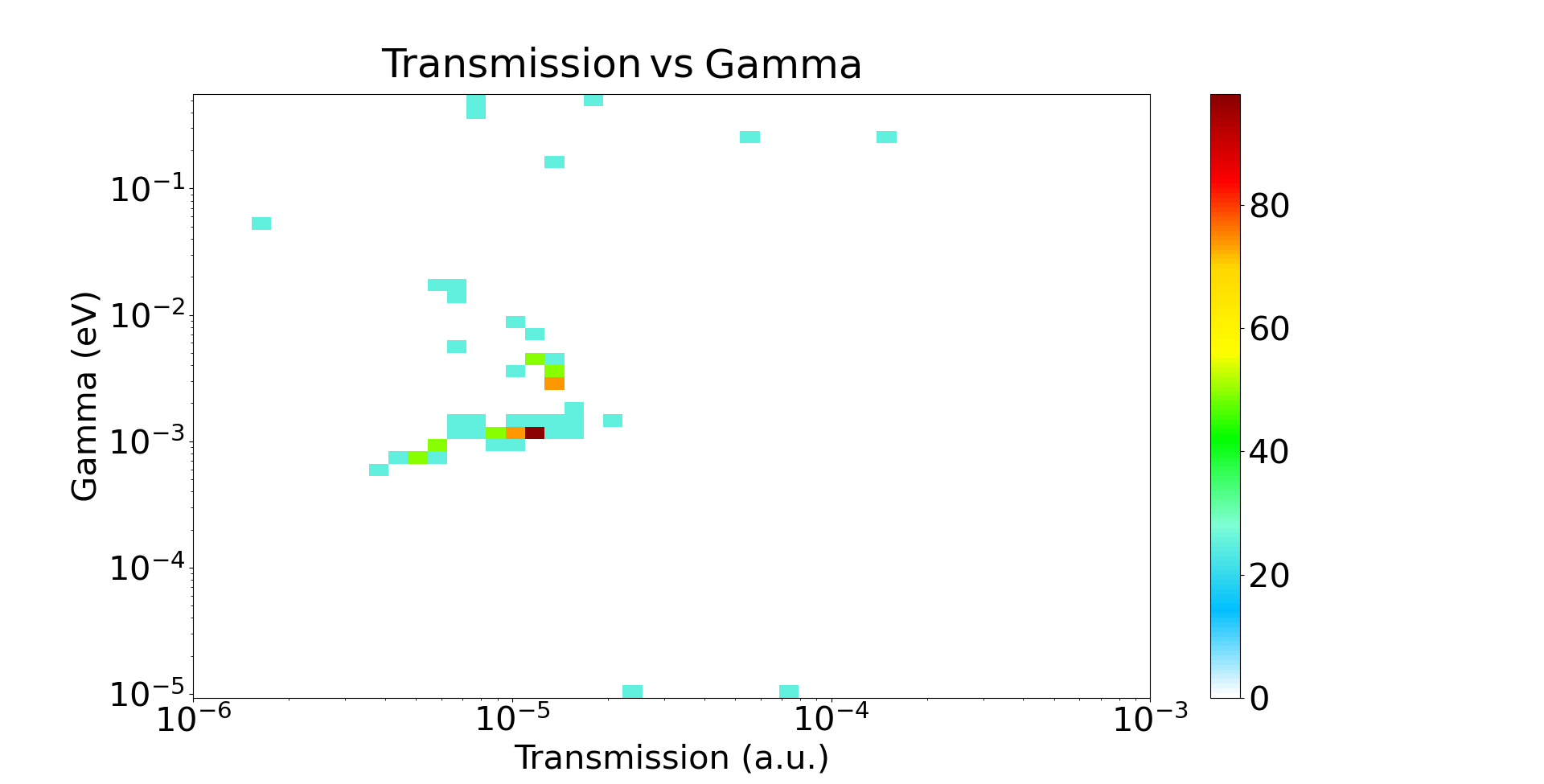}	
	\end{minipage}
	\caption{Coupling $\Gamma$ vs transmission for an ensemble of ca. 30 measurements with molecules.The colour code represents the number of counts for each data bin of the data pair $[T,\Gamma]$.}
	\label{fig:t2}
\end{figure}

\FloatBarrier

\noindent The data in Fig. \ref{fig:t1} and Fig. \ref{fig:t2} show data with molecules only, the colour scale is related to the number data points. They show that for the full sweep the molecule transmits, absolute $[0:1]$, and the coupling does not has much variance around $10^{-3}$.

\begin{figure}[!htb]
	\centering
	\begin{minipage}{1\textwidth}
		\centering
		\includegraphics[width=1\textwidth]{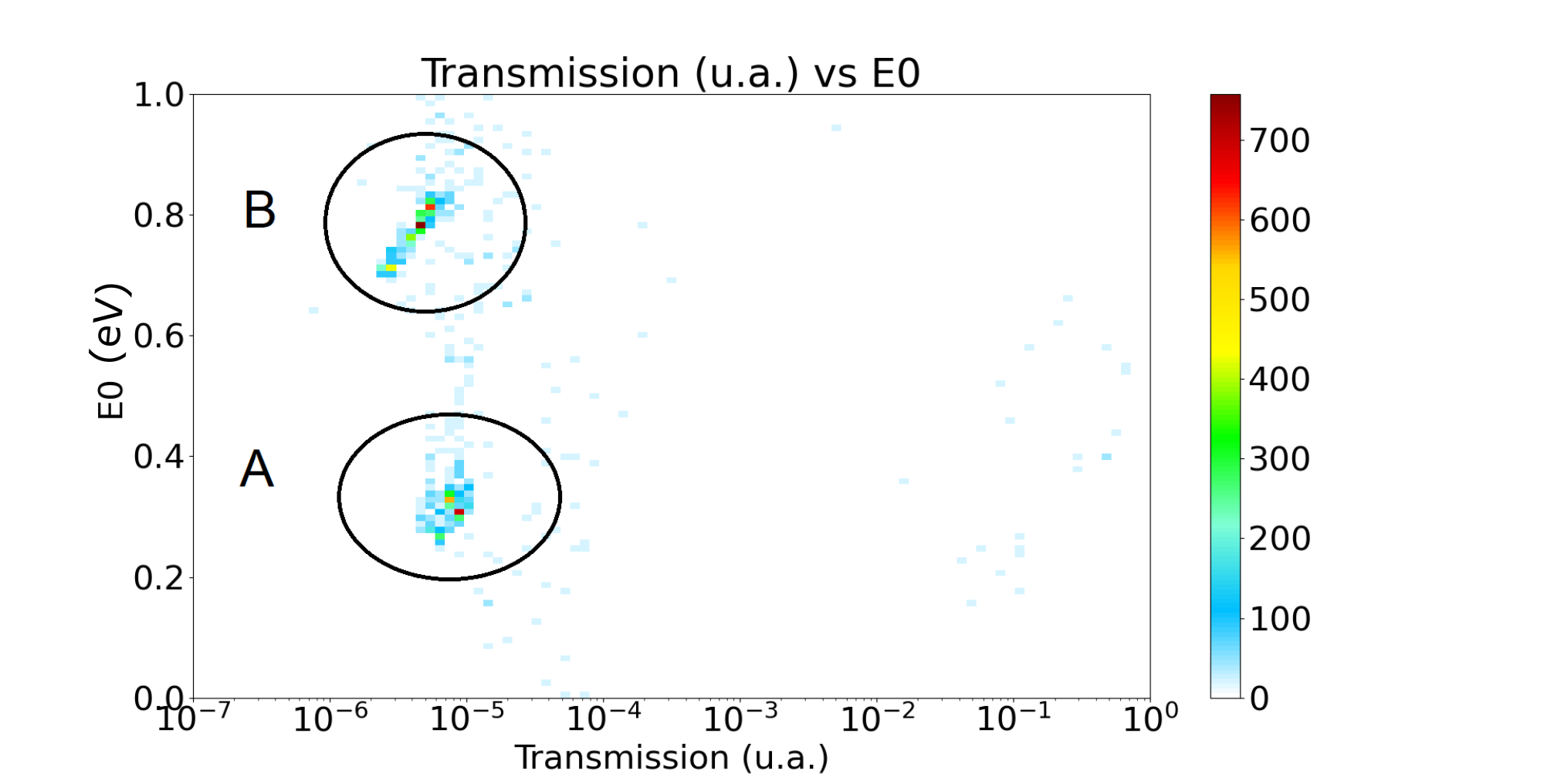}	
	\end{minipage}
	\caption{Channel energy $E_0$ vs transmission for data fits, for an ensemble of ca. 400 measurements with molecules. The colour code represents the number of counts for each data bin of the data pair $[T,E_0]$.}
	\label{fig:tt1}
\end{figure}

\FloatBarrier

\noindent Data point clusters A and B show the highest count of $[Transmission,Energy]$ data points, where A belongs to up-sweep and B to down-sweep measurements. We assume that the respective measurements belong to single molecules, because in contrast to clusters the measurements of the transmission values are more well defined around $10^{-5}..10^{-3})$, also it is clear two channels around $E_0=0.3$ and $E_0=0.8$.

%\clearpage

\begin{figure}[!htb]
	\centering
	\begin{minipage}{1\textwidth}
		\centering
		\includegraphics[width=1\textwidth]{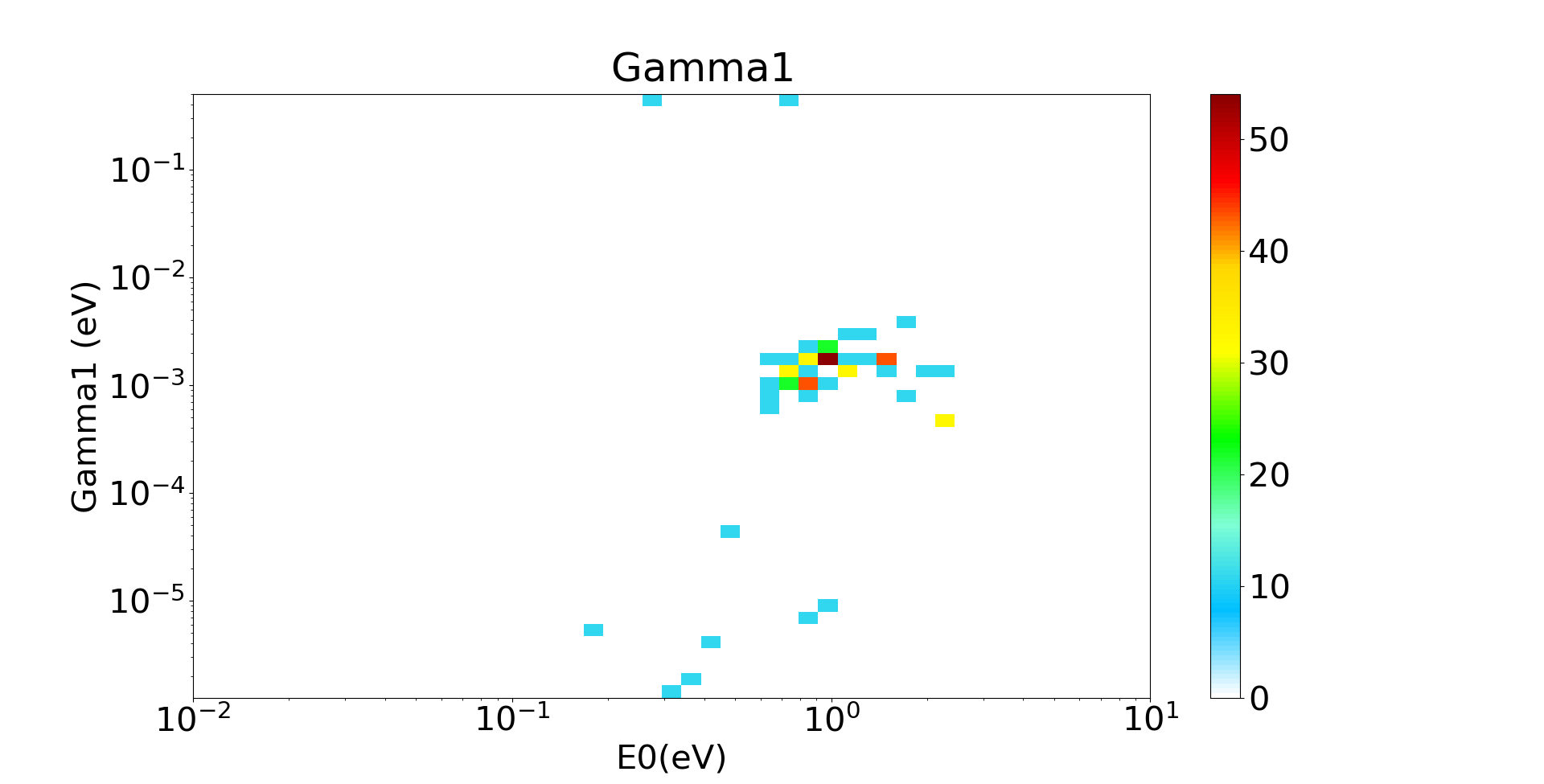}
		\subcaption[]{Couppling $\Gamma_1$ and transmission relation for ca. 30 measurements. The color code represents the number of counts for each data bin of the data pair $[E_0,\Gamma_1]$.}
		\label{fig:g1}
	\end{minipage}
	\hfill
	\begin{minipage}{1\textwidth}
		\centering
		\includegraphics[width=1\textwidth]{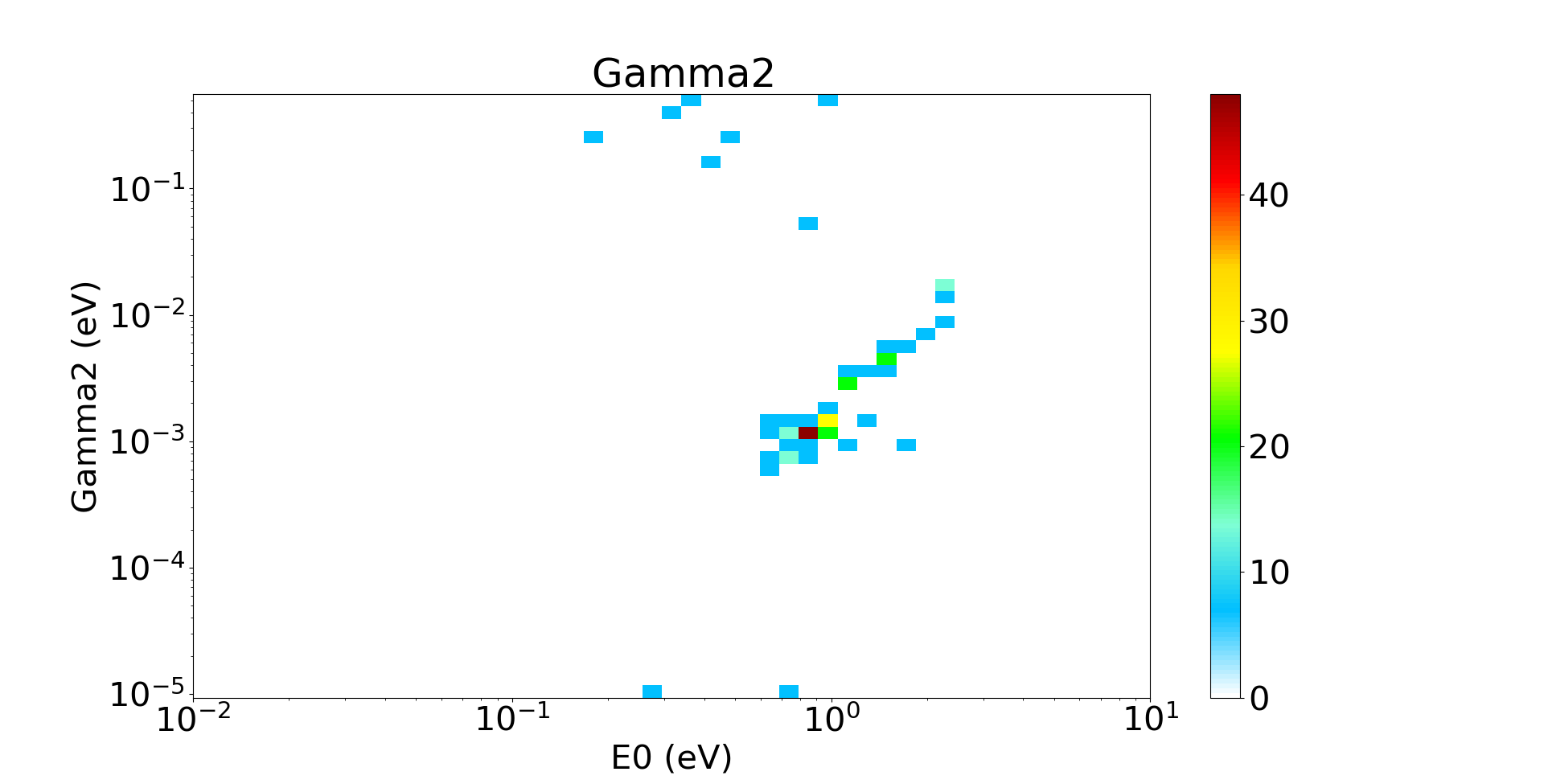}	
		\subcaption[]{Coupling $\Gamma_2$ and transmission relation for ca. 30 measurements. The color code represents the number of counts for each data bin of the data pair $[E_0,\Gamma_2]$.}
		\label{fig:g2}
	\end{minipage}
	\caption{The values in these figures result from the fitting. The ensemble consists of ca. 30 measurements with molecules.}
\end{figure}

\FloatBarrier

\noindent Fig. \ref{fig:g1} and \ref{fig:g2} show similar behaviour as in Fig. \ref{fig:tt1}, likely due to the repetition of the transmission for specific values of coupling. When both sides of the molecule are strongly coupled to the gold leads, on reason is a higher probability for the up- and down-sweep to show the same values for coupling, appearing twice in the counts, the authors Kilibarda 2021 \cite{Kilibarda2021}, Simmons 1964 \cite{Simmons1964}, Hartman 1964 \cite{Hartman1964} has more details about coupling.

\noindent Fig. \ref{fig:g1} and \ref{fig:g2} show the values and tendencies for $\Gamma_1$, $\Gamma_2$ and $E_0$ for the data. This is expected for one specific energy $E_0(U)$. In the plot with 30 fits, we observe that the count increases in the region of the transmission of the Fe$^{+3}$ Salen.

\begin{figure}[!htb]
	\centering
	\begin{minipage}{1\textwidth}
		\centering
		\includegraphics[width=1\textwidth]{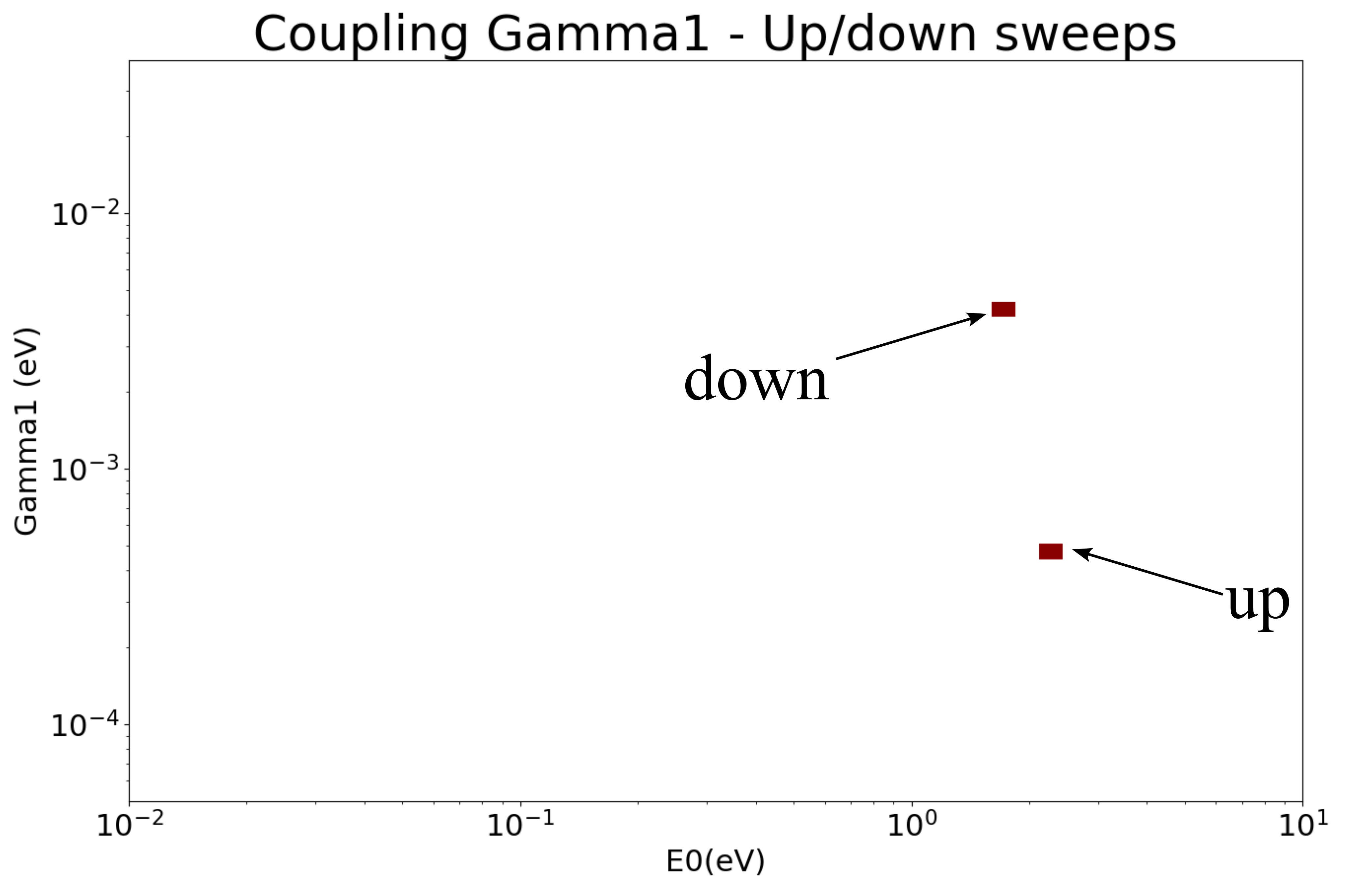}
	\end{minipage}
	\caption{The Figure shows the coupling constants for up- and down-sweep, they are distinctively different; up-sweep shows higher coupling constant   $\Gamma_1$ than the down-sweep. Observe that the channel is a bit shifted as expected because the effect caused by the sweeps in its the position.}
	\label{fig:gammas}
\end{figure}

\FloatBarrier

\begin{figure}[!htb]
	\centering
	\begin{minipage}{1\textwidth}
		\centering
		\includegraphics[width=1\textwidth]{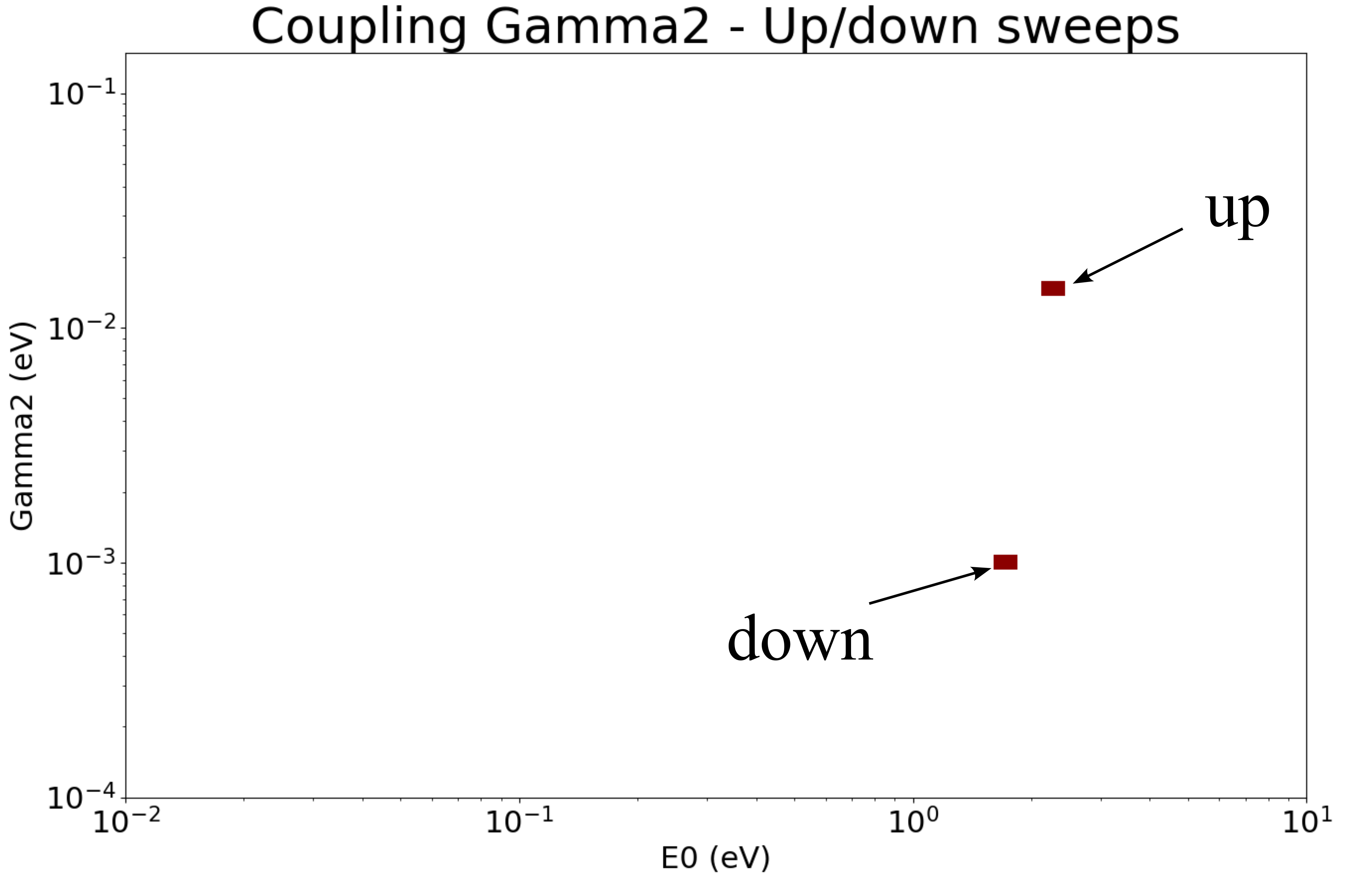}
	\end{minipage}
	\caption{The Figure shows coupling $\Gamma_2$ in the same fashion as \ref{fig:gammas}, and the same trend observation for up- and down-sweep, respectively. Compare to \ref{fig:gammas}.}
	\label{fig:gammas2}
\end{figure}

\FloatBarrier

\noindent In our study, we have chosen to focus on the coupling values, specifically $\Gamma_1$ and  $\Gamma_2$ of one measurement, derived from a singular measurement. A pertinent question in this context is whether the observed characteristics, particularly the coupling between molecule and leads, if they are inherently preferential or merely a consequence of the initial direction chosen for the sweep process, and how they can be either lost or become smaller after change the sweep direction.

\noindent It is important to note that our measurement methodology initiates at a zero point and progresses upwards to one. The measurement realisation is detailed in the chapter \ref{sec:msetup}, and this particular branch of current is subsequently disregarded, and removed. In these measurements, reversing the direction of the sweep at the one-volt appears, that the sweep have no observable influence in the coupling. However, the change of coupling for different directions of sweep will repeat in several measurements.

\noindent Considering the shifts in the channel energy, which is in agreement with the adjustment for $E_0$ in Eq. \ref{E0Ap} in section \ref{sec:SLM}, and the change in the sweep direction can affect the coupling strongly, once it changes the direction in which the charges move, or the electron distribution inside the molecule, making bonds stronger in one direction than in another.

\noindent According to \cite{Lokamani2023} in his thiol coupling study; there is a dependence of coupling of the thiol group and distance. Depending on the coupling position from tip to tip of the leads, edge to tip, or edge to edge, the coupling will change. It was observed that for planar configurations the coupling increased, the bending affects supporting the formations of optimal angles for the molecule in relation to the leads. Furthermore, at top sites the channel coupled strongly. For small distances, Lokamani 2023 \cite{Lokamani2023} verified  that the stronger coupling was at bridge sites and facets edges. We believe their result show the configuration of their molecule for direct tunnelling, which is extensible for other molecules too. The electronic coupling is stronger for when symmetrical geometries are more probable, this was observed in their tip to tip configuration \cite{Lokamani2023}.

\clearpage

\section{Conclusive Discussion}
In this chapter, we explained using measurements graphs as example of the possible phenomena present that are producing the mismatch between SLM and data, also the models describe well the direct tunnelling, but do not describe other effects around $U=0\,\mathrm{V}$ which affects the tunnelling. One assumption was ``electron-phonon scattering''. However, there is no definitive signatures for this current characteristic. Hence the mismatching from SLM and data for the IV characteristic could not be attributed to $e-ph$ scattering, once there as no means to evaluate the existence of these signatures at RT, because the limitation of our methods. There are other transport mechanisms which leave their signature in the IV curve, such as electron-electron interactions, which lead to scattering and according to \cite{Datta2005} in the range of the molecules' conductance will experience electron-electron scattering. Furthermore, the Coulomb blockades which can be measured in LT experiments are mostly not observable at RT \cite{Kreupl1998}, still the potential (of these `blockades') can lead to elastic scattering according to \cite{Kirchner202110}. Electron-electron scattering will be inelastic, causing decoherence.
The next section introduces a semi-classical approach to explain in a simplified way the hypothesis of electron-electron interaction in the results.

\subsection{Hypothesis of Scattering}
According to \cite{CuevasScheer2010,Sondhi2021,Datta2005}, considering only the conductance values $G<1\, G_0$ non-ballistic effects can be observed at LT and inelastic effects contribute $G<0.5\, G_0$. Here $e-ph$ forward scattering dominates once the electron gains energy while for $G>0.5\, G_0$ $e-ph$ backscattering dominates \cite{Sondhi2021}, and the electron loses energy; the conductance of Fe$^{+3}$ Salen is in the range of $3\ 10^{-6} \, \mathrm{S} < G < 10^{-3} \, \mathrm{S}$, so the molecule experience both kinds of transport phenomena. Thus in presence of a tunnel barriers, electron-electron interactions can be observed at RT in the channel\cite{Kreupl1998}. 
In this section, we consider the electron-electron elastic and inelastic scattering hypothesis.

\noindent Using the energy-time uncertainty approach to quantity the phenomenon, a rough estimate is obtained:

\begin{equation}
	\Delta E \cdot \Delta \tau \geq \frac{1}{2}\hbar \label{UPH}.
\end{equation}

\begin{table}[htb!]
	\centering
	\begin{tabular}{| c | c | c| c | c |}
		\hline
		Potential (V) & Energy (eV)& $\lambda_{dB}$(\AA) &  $\tau$ (fs) & $L_e$(\AA) \\
		\hline
		0.1 & 0.1 & $40$ & 3.3 & 6.17 \\
		0.275 & 0.275 & $23.4$ & 1.2 & 2.4 \\
		1 & 1 & $12$ & 0.37 & 2.06 \\
		\hline
	\end{tabular}
	\caption{de Broglie wavelength $\lambda_{dB}$ (Fermi wavelength is this wavelet at Fermi level), relaxation time $\tau$, elastic mean free path $L_e$, calculated for different scattering potential.}
	\label{tab:deBroglie}
\end{table}

\noindent Nevertheless, the values in Tab. \ref{tab:deBroglie} shows that the molecular dimensions of $24 \mathrm{\AA}$ and scattering length are on the same magnitude of some molecules, for example, Fe$^{+3}$ Salen (Corannulene is smaller), suggesting again that inelastic events are present too. The Table \ref{tab:charges} gives an estimate of a net change of charges within of the molecule, this suggest that there are probability for elastic scattering events.

\begin{table}[htb!]
	\centering
	\begin{tabular}{| c | c | c| c |}
		\hline
		Energy (eV) & Current (nA) & Time (fs) & Number of Apparent Charges\\
		&   &   & into the channel (yC or nA.fs)\\
		\hline
		0.1 & 0.001 & 3.3 &  3.3 $10^{-3}$\\
		0.275 & 0.1 & 1.2 &  120 $10^{-3}$ \\
		1 & 10 & 0.27 &  2.7\\
		\hline
	\end{tabular}
	\caption{Number of charges. Calculated from energy and current relaxation time \cite{Kirchner202110}.}
	\label{tab:charges}
\end{table}

\noindent The exact nature of the transport phenomenon and their iteration mechanism give the opportunity for further research and future experiments.

	\chapter{Conclusion and Further Work}

This study aimed to understand and learn more about the transport of single molecules, which can be seen as an advanced and extensively researched area of molecular electronics with much activity in the field. However, the transport characteristics of single-molecule junctions are not fully understood. These phenomena were evaluated experimentally using statistical methods because their `robustness' for experimental data, which makes possible infer conclusions from them.

\noindent It also engaged theoretical topics that are well advanced, e.g., the conductance in soft molecules in systems with coupling in a mechanically controllable break junction (MCBJ), current and single level model (SLM). Although the theory is well-developed, experimental confirmation is still required. Additional theoretical investigations are going on, and corrections between experimental and SLM are investigated. In this instance, we want to highlight a feature of the flat but still linear current increase at low bias in the sweep curves.

\noindent From our results, we conclude that the coupling affects the conductance measurements. This coupling is related to the binding between the molecules and leads. Furthermore, investigations were done on the electronic conductivities.

\noindent Let us subclassify the coupling in three regimes:
\begin{enumerate}
	\item The non-contact regime: the molecule is not attached to the lead in both side, but one of the sides is only close enough for a small amount of current to flow. In this arrangement, the junction has a non-symmetric behaviour. 
	\item The contact regime: The molecule contacts the leads on both sides, leading to binding positions concerning the junction. In this case, the coupling is almost symmetrical, corresponding to high coupling values. 
	\item The contact regime with molecule under stress: With the stress applied through the leads, the resulting deformation of the molecule will introduce a degree of asymmetry. Consequently, the values will be reduced to coupling below unity.
\end{enumerate}

\section{The Current Measurement}

\noindent From the obtained current curves, we could distinguish several features. Further investigations were needed to determine hat it as affecting the `S' shape. This evaluation allowed us to study some of the molecules' general electronic properties from the RT measurements.

\noindent Despite this, the central theme of this study was SLM. However, it was observed that the model did not accurately fit all the measured data, especially at the origin (around zero on the current-voltage (IV) characteristic) and tails (around 1V on the IV characteristic). There are strong indications that the deviation from the model around the origin of IV curve is caused by electron-electron interaction. On the other hand, the channel earlier saturation in the tails of current measurements stays open for further studies. The SLM's single-channel resonant tunnelling model approach does not cover this effect. 

\noindent For evaluation of the SLM, the measurements were divided into three groups according to the fitting for the Corannulene's molecule, the energy storage effect (hysteresis) was considered caused by water vapour dissolved in the solvent. Nonetheless, it does not seem to be the case for organometallics of Salen because it is an ion, but the SLM was being treated here, not hysteresis. For the SLM evaluation was chosen a non-polar molecule as Corannulene:

\begin{enumerate}
	\item Data fits: in a minimal number of cases, the fit was perfect in most of them, but there was still a substantially small difference around the origin;
	\item Data has a bad fitting: most of the data is in this group, some failing around the origin and others around the tails. For one side around the origin, the mismatching suggests either some potential is producing scattering or some Coulomb effect between the electrons is disrupting the current. On the other side, the current tails seem to suggest that once the channel is opened and occupied, there is no room for current increase, and the current saturates;
	\item Data do not fit: here, an infinite amount of reasons can be the cause, from the saturation of the fitting itself too early to data without molecules or extremely noisy insufficient data. These data were excluded as `trivial' cases.
\end{enumerate}

\noindent A solution to improve the model, it was proposed that these interactions should be included in the equations as additional terms in the Hamiltonian of the model.

\noindent Overall, these results indicate that the quantum transport in molecular junctions is a rich source of incentives for the advancement and development of state-of-the-art theoretical methods and applications. Interpreting and comprehending molecular architecture effects in a distinct nonequilibrium setting is an exciting opportunity for researchers of various fields across the scientific community.
\noindent The following section contains suggestions for further works, which are the open questions about SLM.

\section{Further Scope}

\noindent Each molecule can open a field of new observations. A promising avenue for future research involves identifying new setups able to present phonon evidence (signatures) at room temperature and the perfect molecule able to exhibit this phenomenon.

\noindent Incorporating electron-electron interactions into the SLM significantly improves modelling molecular electronics. These interactions are critical for accurately describing the behaviour of electrons crossing molecular spaces, potentially leading to more precise predictions of electronic properties and behaviours in molecular junctions.

\noindent Another critical area of research is the precise determination of the resonance energy ($E_0$) value of the single channel, another essential topic to be addressed, allowing the evaluation of the exact value of energy when resonance happens. This information identifies whether the channel exhibits resonance with the Femi level of the gold leads having the same de Broglie wavelength or another value (bigger/smaller) inside the molecule. It is critical to know which approaches can be used to evaluate the transport once it is expected that most molecules have a size in the range of quantum ballistic transport.

\noindent Additionally, it is essential to know the shape for modelling in this single-level channel. Suggestions for initial investigation include the Lorentzian shape, symmetric and asymmetric. However, it might be possible that a Gaussian shape should be more accurate. A suggestion at this point is to use the current integral in the Fourier domain, which can open a different inside of the energies participating in the channel.

\noindent Enhancing our understanding of molecular transport mechanisms promises to revolutionize the modelling of mesoscopic transport in soft materials. It is becoming increasingly clear that resonant tunnelling does not fully cover the complexities of electron transport through molecules. We can develop more comprehensive and accurate models by broadening our understanding beyond resonant tunnelling. These models will better reflect the intricate nature of molecular transport and pave the way for innovative applications in nanotechnology and materials science.	
	\newpage
	\pagenumbering{roman}
	\appendix
	\chapter{Current with Mean Normalization}
\label{app:norm}

\begin{equation}
	I_n= \frac{I-I_{mean}}{I_{max}-I_{min}}
\end{equation}

\begin{figure}[!htb]
	\centering
	\begin{minipage}{.7\textwidth}
		\centering
		\includegraphics[width=1\textwidth]{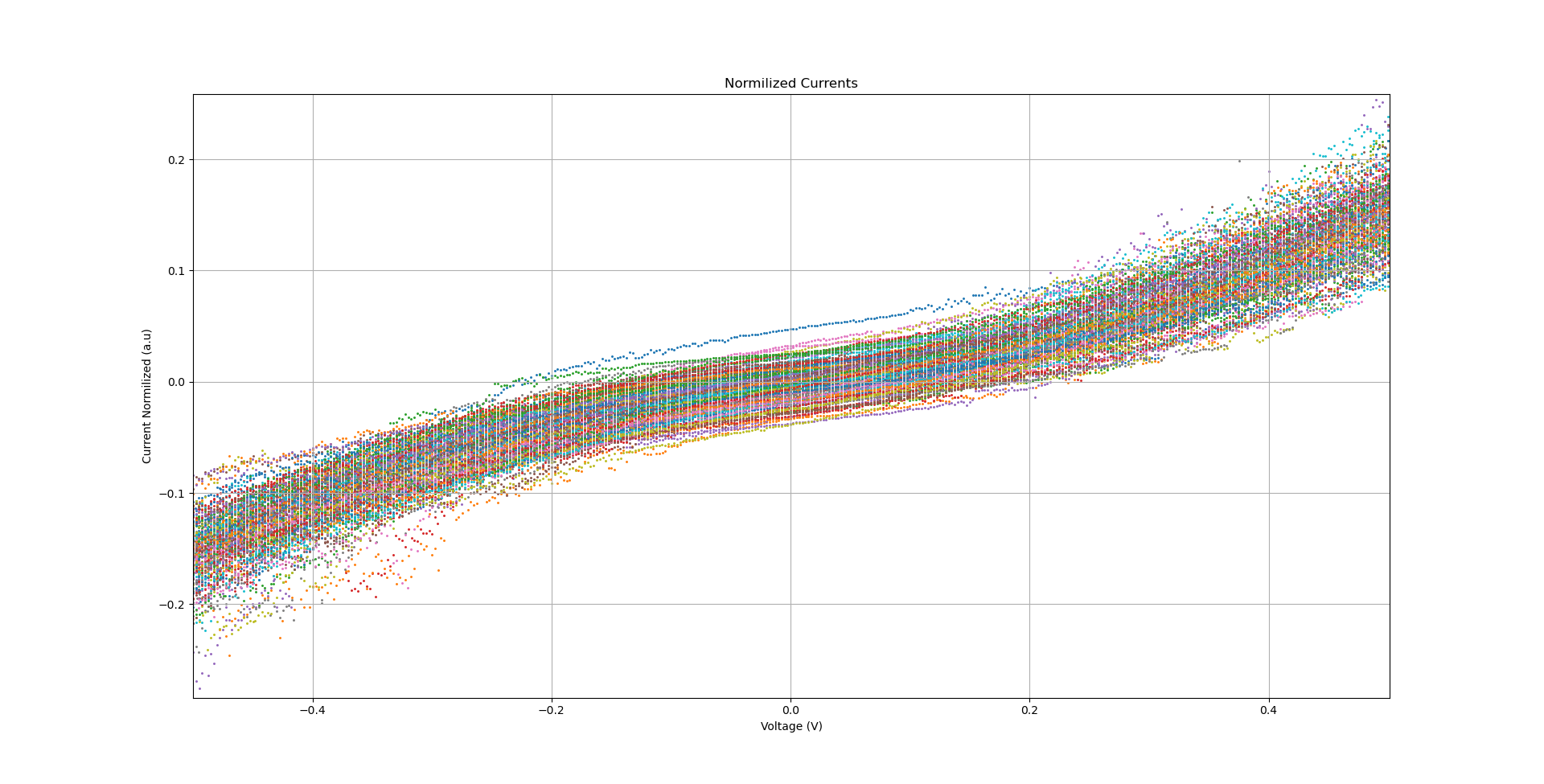}
		\subcaption[]{}
		\label{fig:norma}
	\end{minipage}
	\begin{minipage}{.7\textwidth}
		\centering
		\includegraphics[width=1\textwidth]{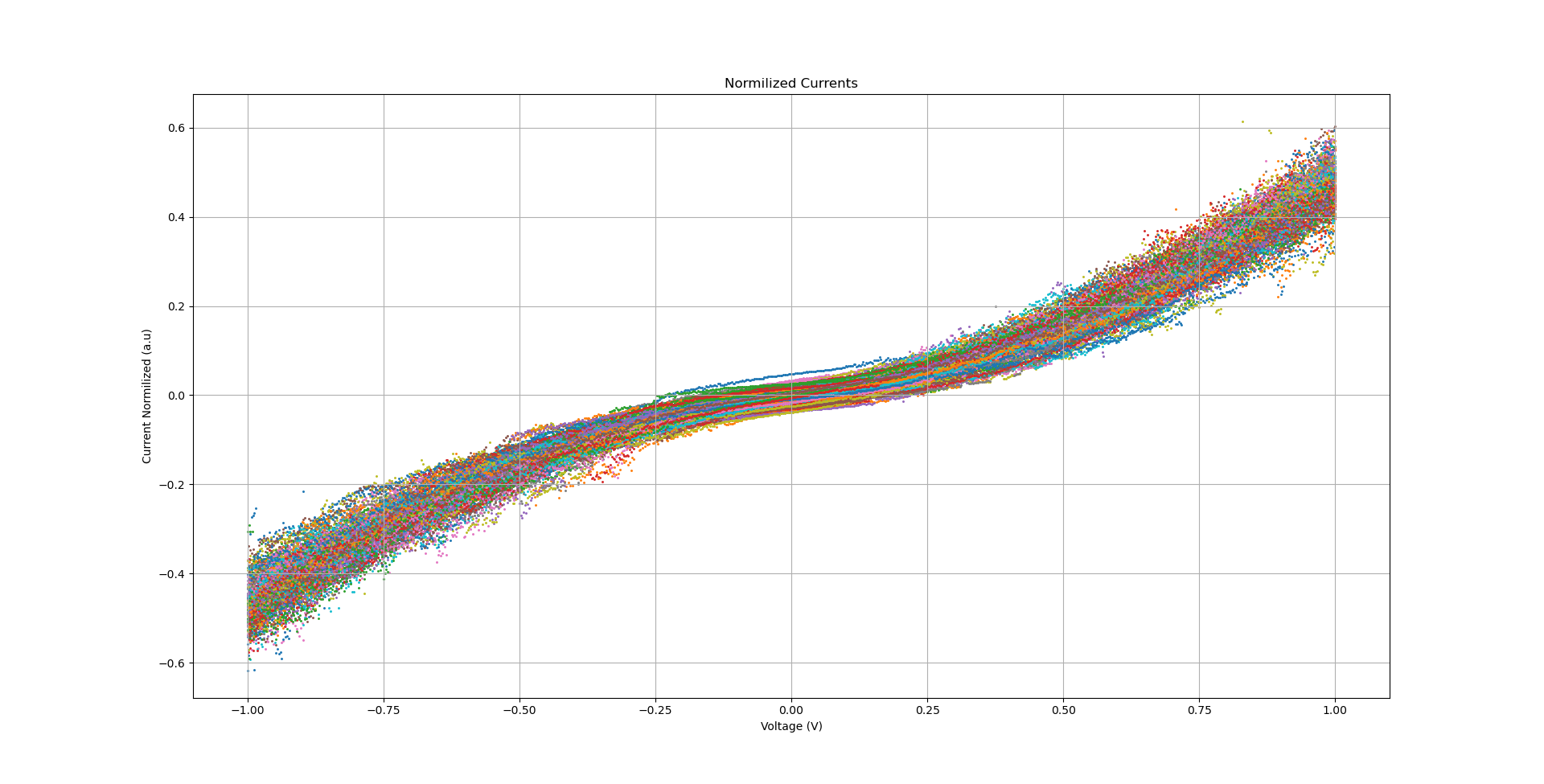}
		\subcaption[]{}
		\label{fig:normb}
	\end{minipage}
	\caption{The 'S' chape of the current a. shows a linear region around zero and b. shows the full curve.}
	\label{fig:norm}
\end{figure}

\section{Categories of Measurements}

\subsection{Measurements without hysteresis or very small}

\subsubsection{Measurements with good Fitting}

\begin{figure}[!htb]
	\centering
	\begin{minipage}{.7\textwidth}
		\centering
		\includegraphics[width=1\textwidth]{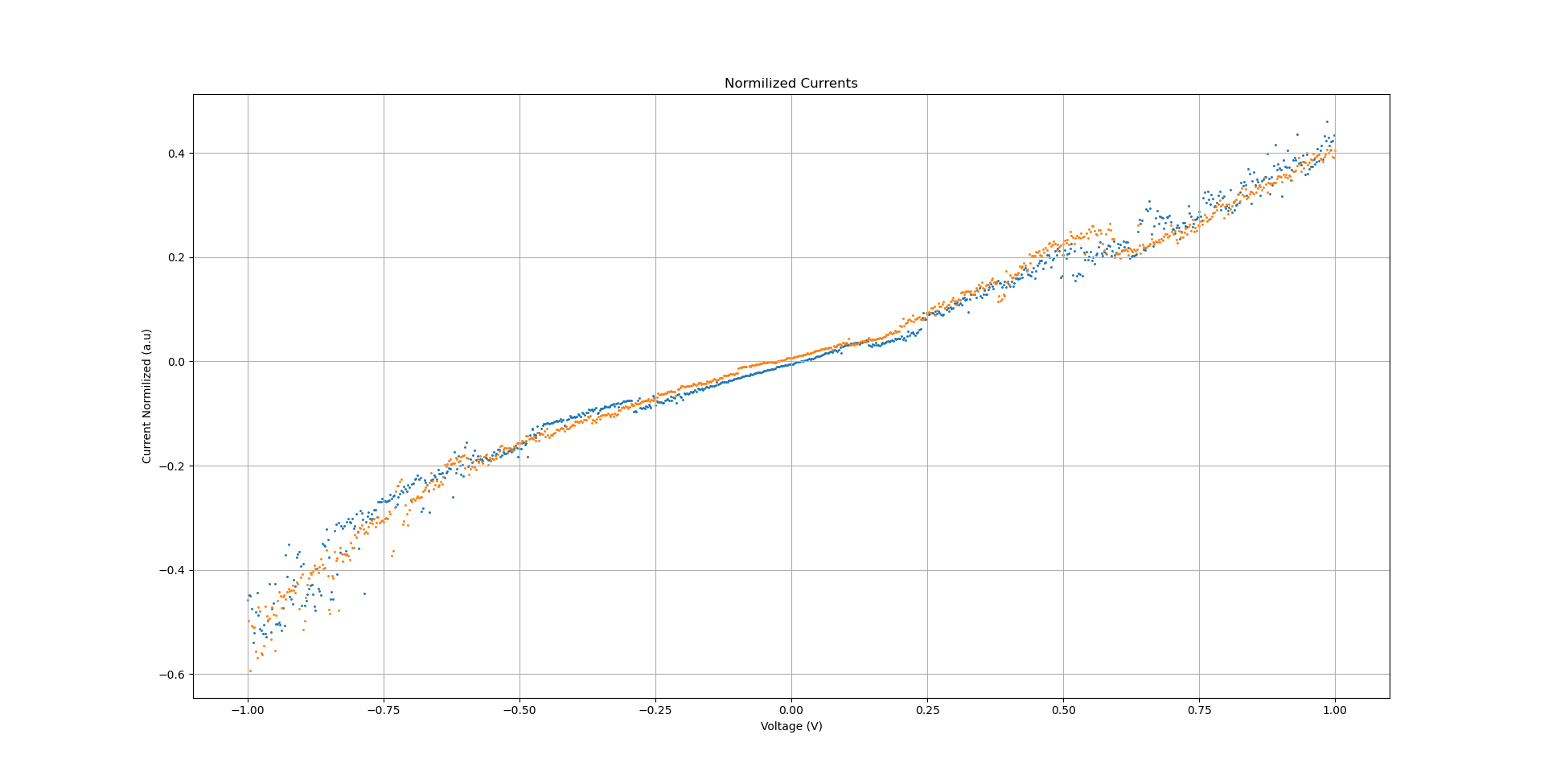}
		\subcaption[]{}
		\label{fig:cat1a}
	\end{minipage}
	\begin{minipage}{.7\textwidth}
		\centering
		\includegraphics[width=1\textwidth]{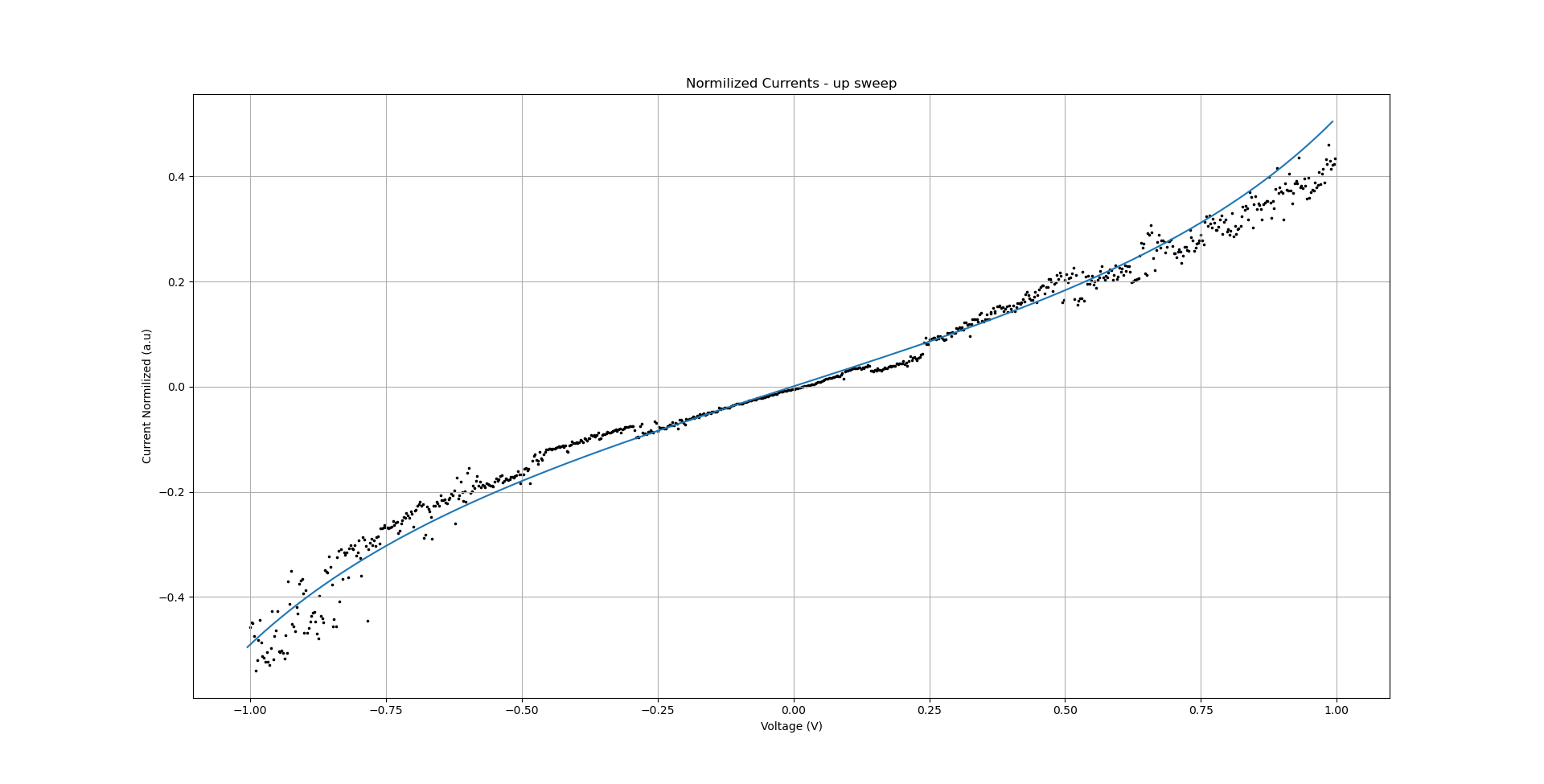}
		\subcaption[]{}
		\label{fig:cat1b}
	\end{minipage}
	\begin{minipage}{.7\textwidth}
	\centering
	\includegraphics[width=1\textwidth]{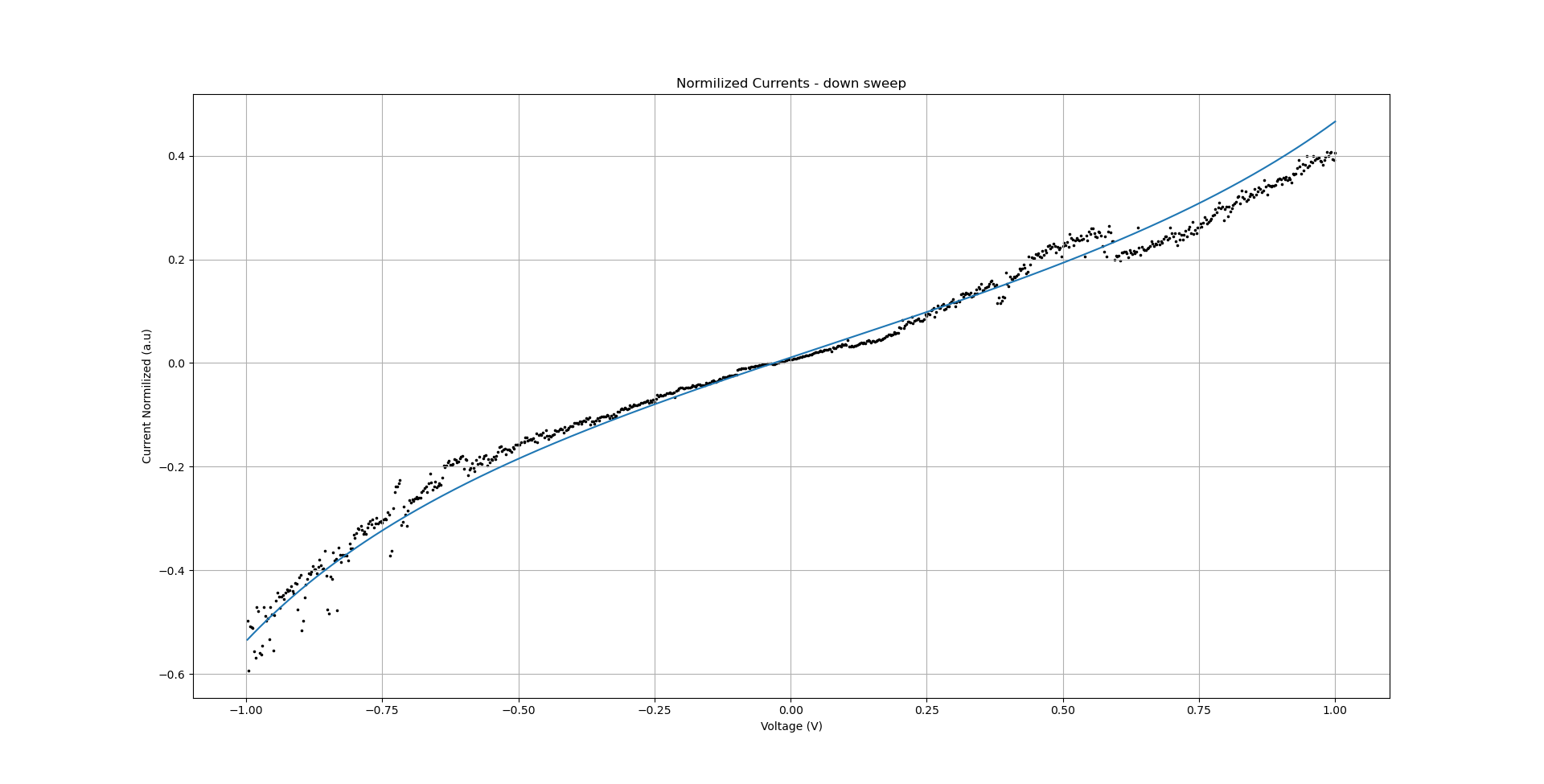}
	\subcaption[]{}
	\label{fig:cat1c}
	\end{minipage}
	\caption{}
	\label{fig:cat1}
\end{figure}

\clearpage

\subsubsection{Measurements with poor Fitting (origin and tails)}

\begin{figure}[!htb]
	\centering
	\begin{minipage}{.7\textwidth}
		\centering
		\includegraphics[width=1\textwidth]{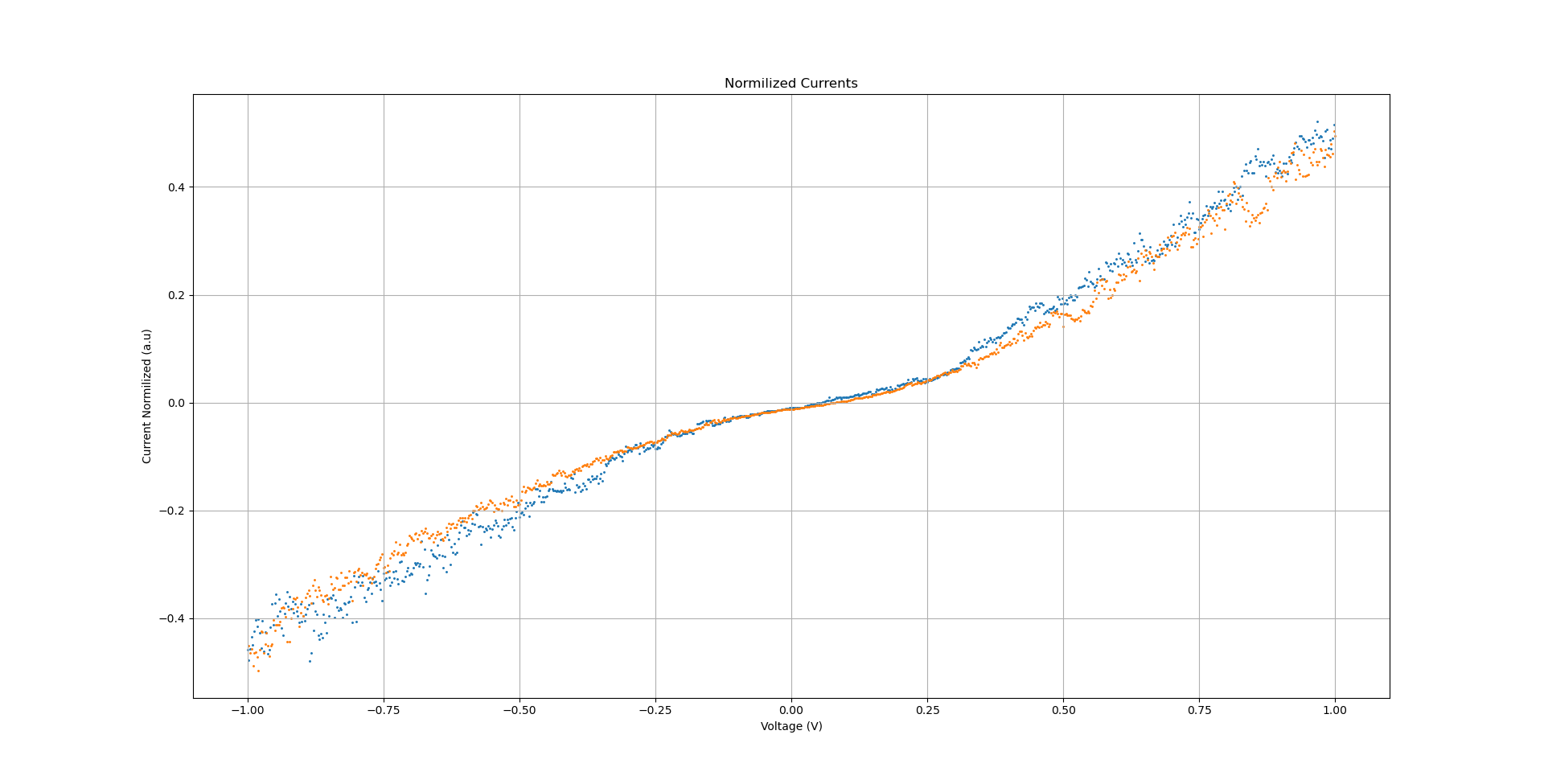}
		\subcaption[]{}
		\label{fig:cat3a}
	\end{minipage}
	\begin{minipage}{.7\textwidth}
		\centering
		\includegraphics[width=1\textwidth]{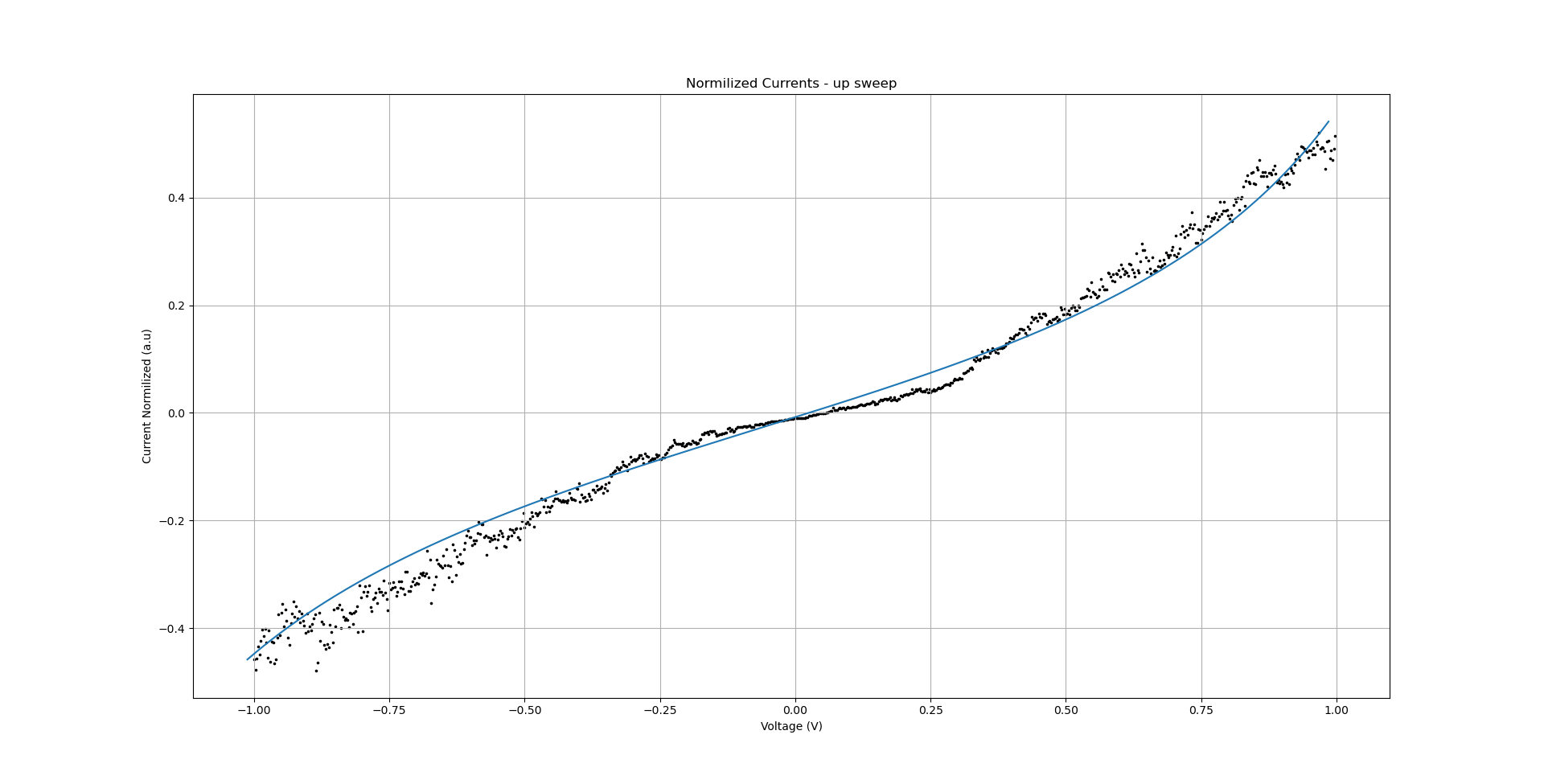}
		\subcaption[]{}
		\label{fig:cat3b}
	\end{minipage}
	\begin{minipage}{.7\textwidth}
		\centering
		\includegraphics[width=1\textwidth]{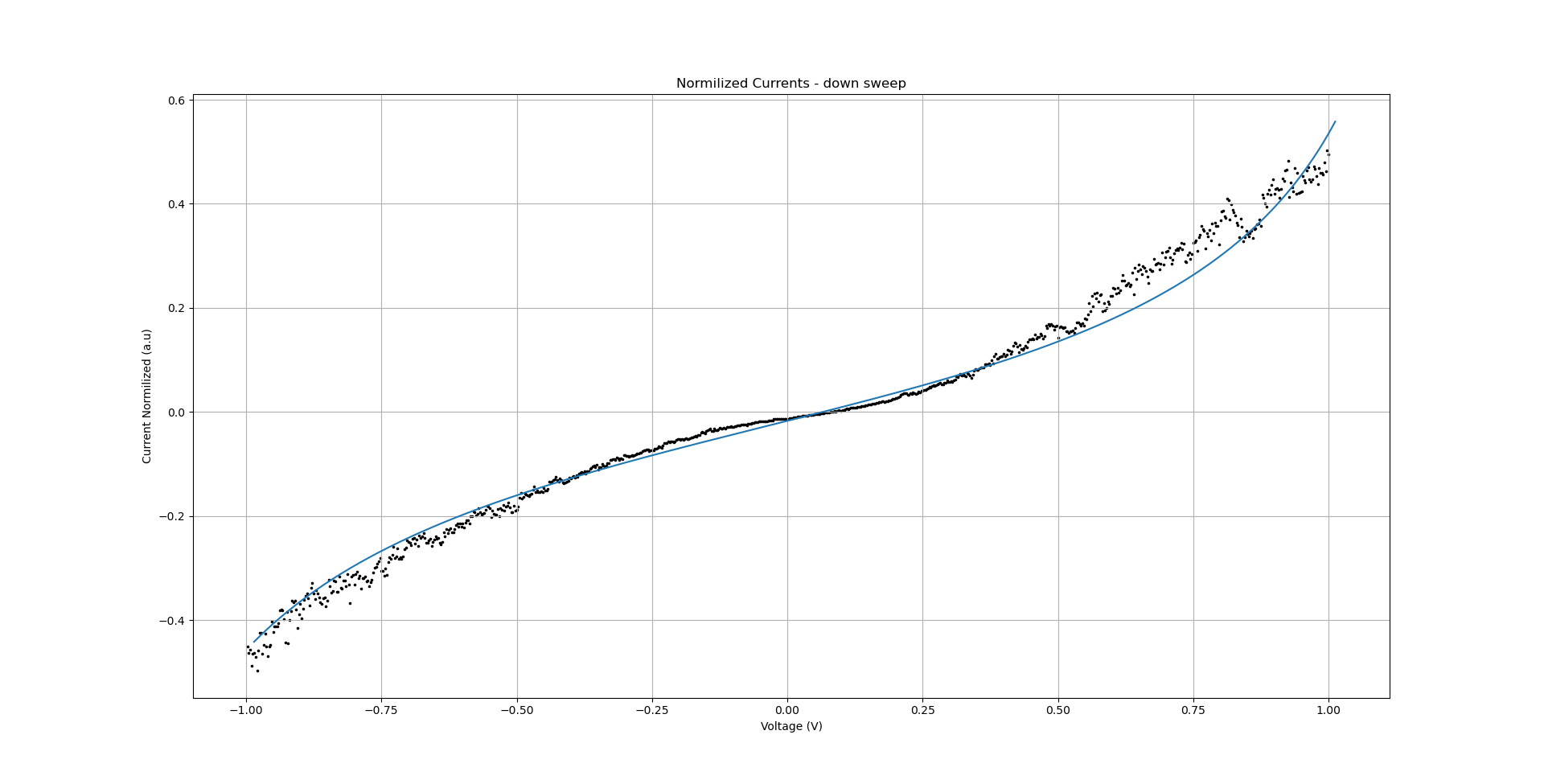}
		\subcaption[]{}
		\label{fig:cat3c}
	\end{minipage}
	\caption{}
	\label{fig:cat3}
\end{figure}

\clearpage

\subsubsection{Measurements with poor Fitting (origin)}

\begin{figure}[!htb]
	\centering
	\begin{minipage}{.7\textwidth}
		\centering
		\includegraphics[width=1\textwidth]{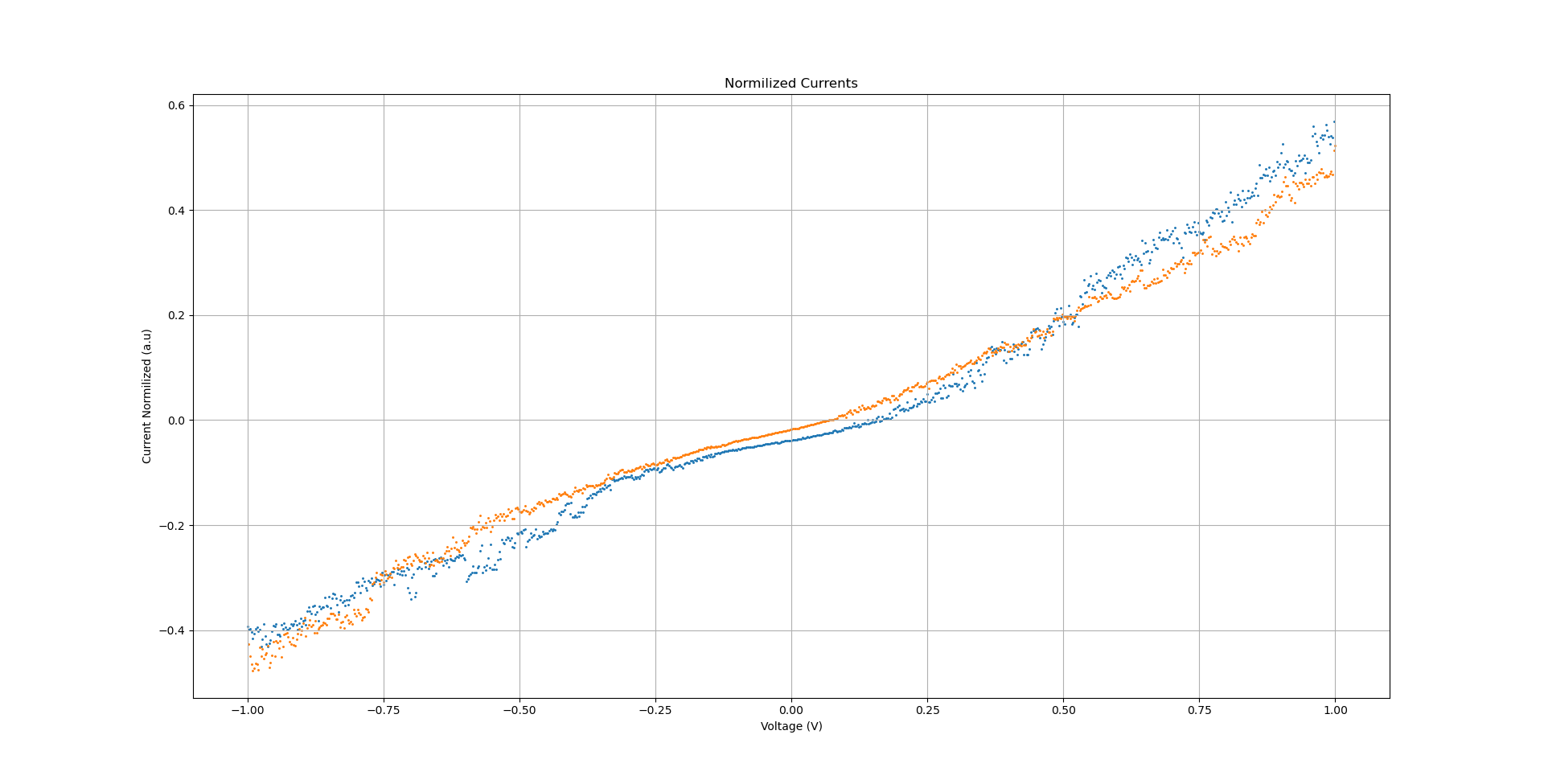}
		\subcaption[]{}
		\label{fig:cat6a}
	\end{minipage}
	\begin{minipage}{.7\textwidth}
		\centering
		\includegraphics[width=1\textwidth]{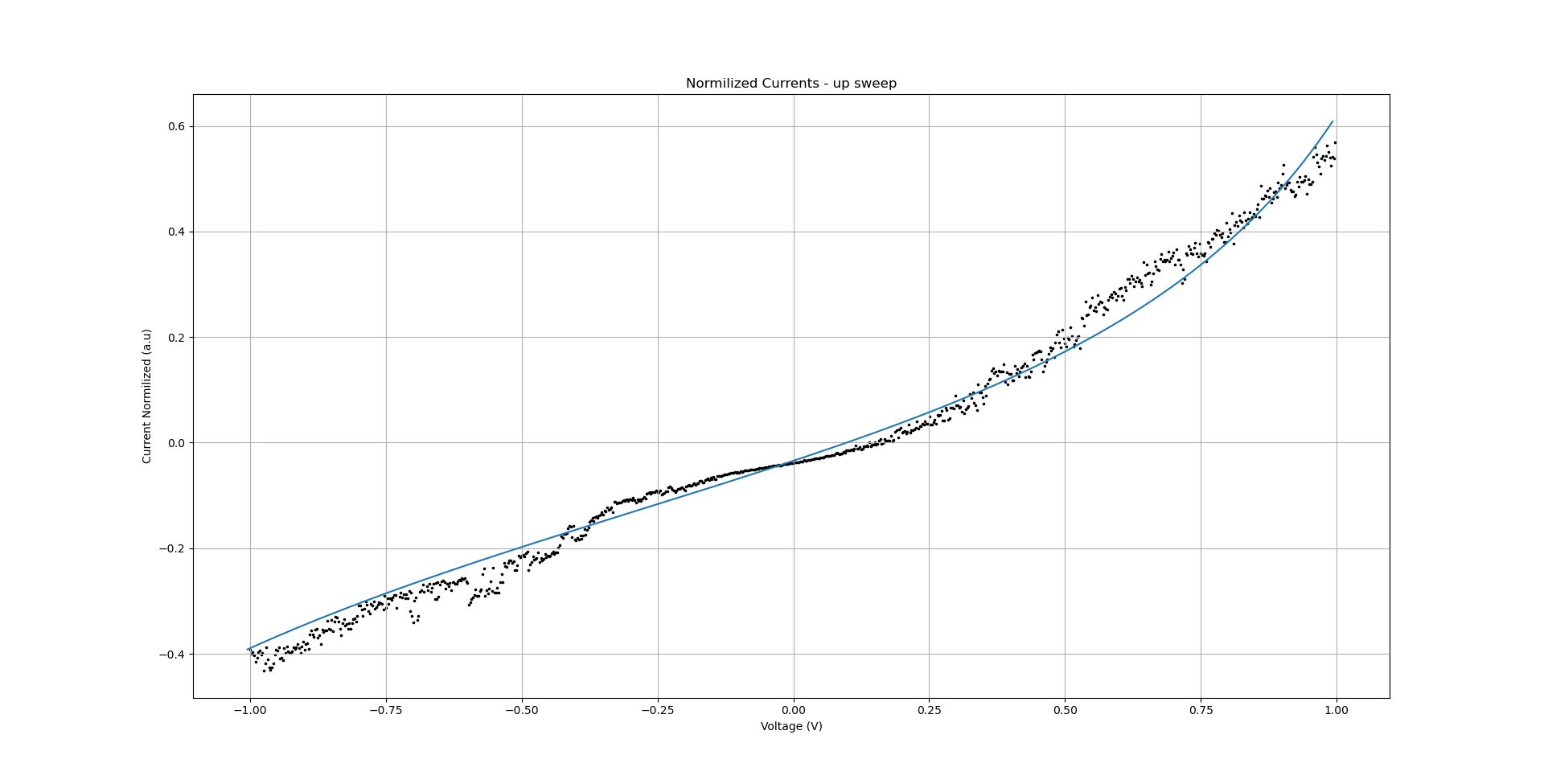}
		\subcaption[]{}
		\label{fig:cat6b}
	\end{minipage}
	\begin{minipage}{.7\textwidth}
		\centering
		\includegraphics[width=1\textwidth]{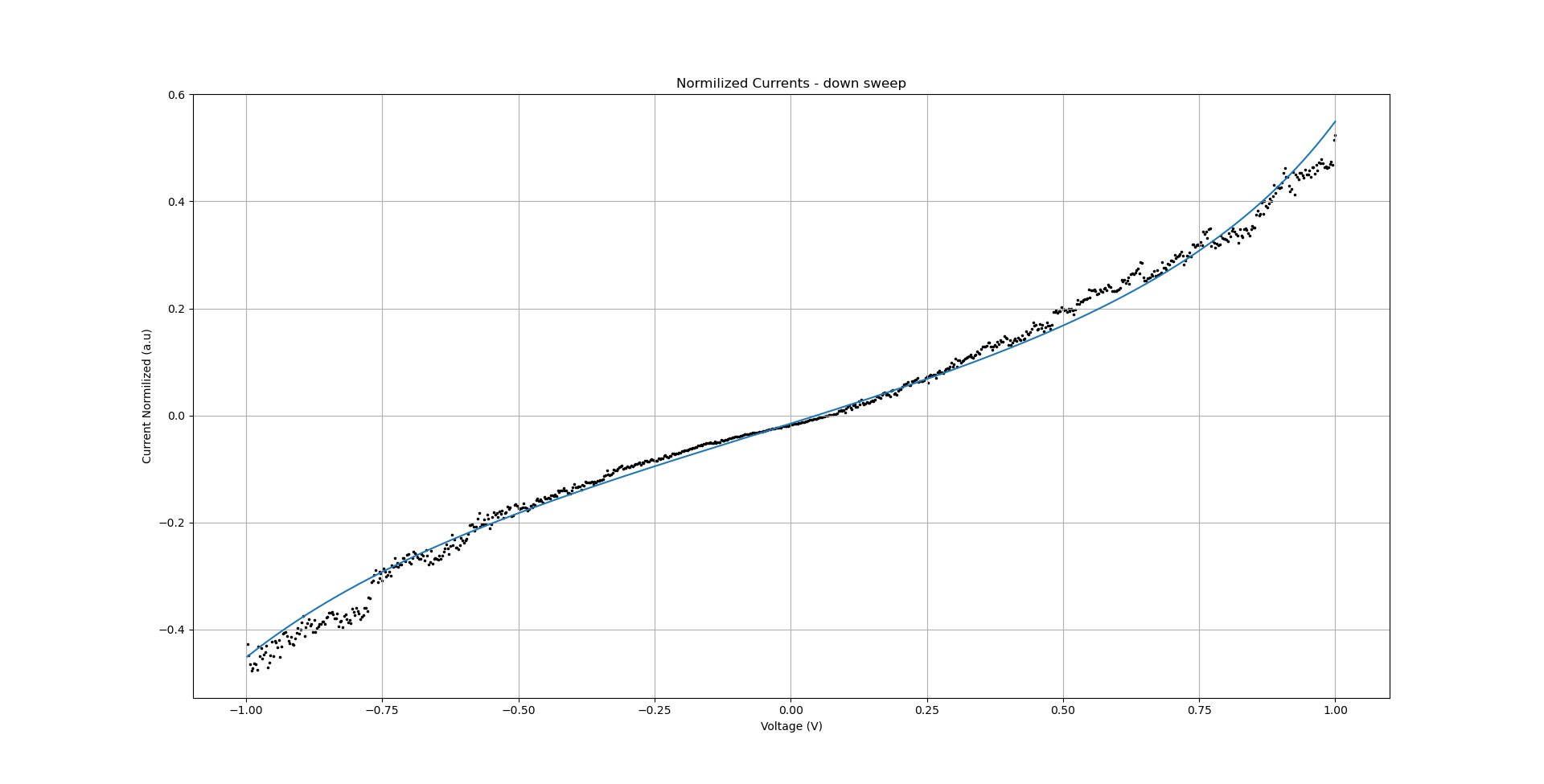}
		\subcaption[]{}
		\label{fig:cat6c}
	\end{minipage}
	\caption{}
	\label{fig:cat6}
\end{figure}

\clearpage

\subsubsection{Measurements with poor Fitting (tails)}

\begin{figure}[!htb]
	\centering
	\begin{minipage}{.7\textwidth}
		\centering
		\includegraphics[width=1\textwidth]{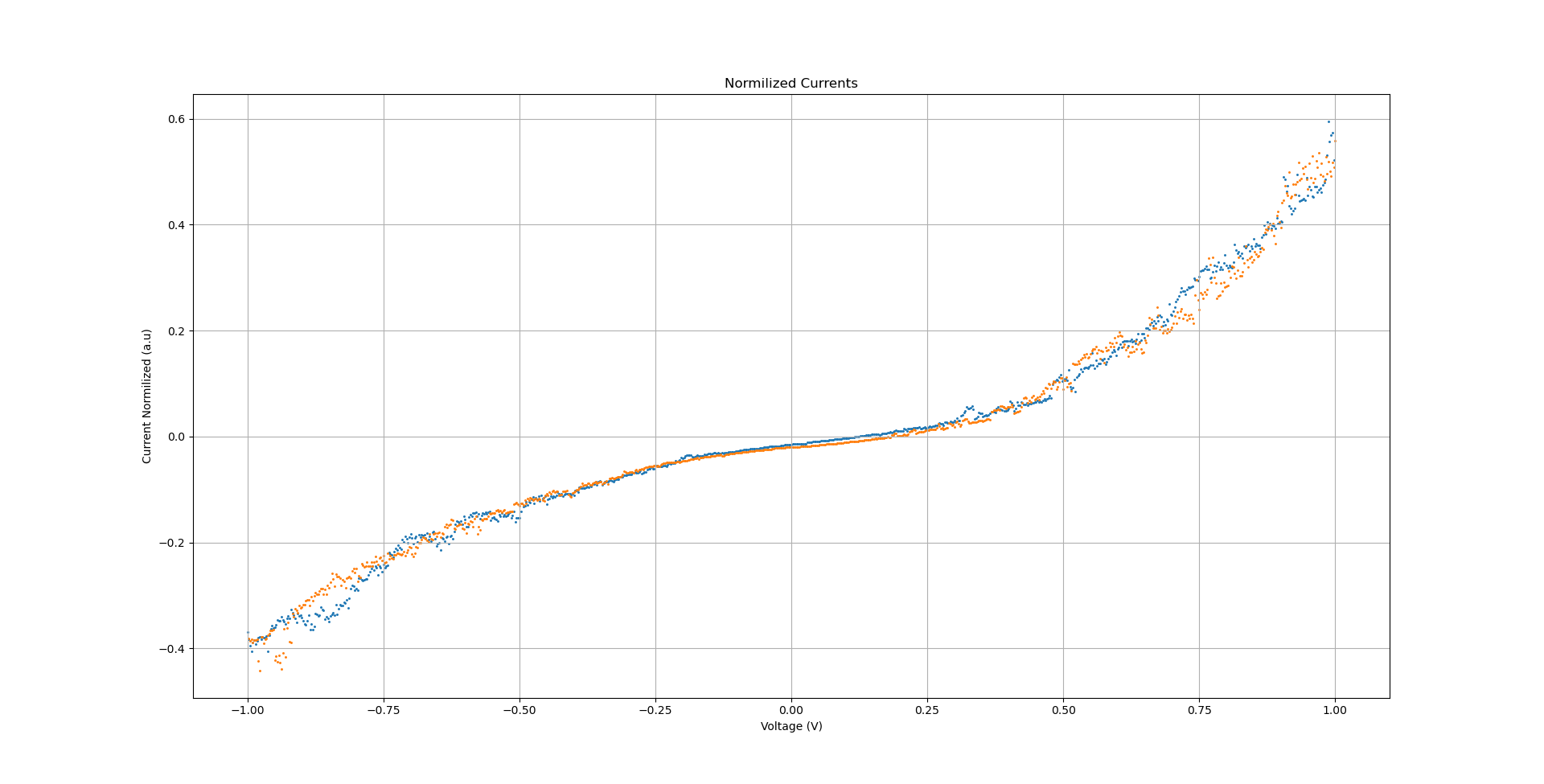}
		\subcaption[]{}
		\label{fig:cat5a}
	\end{minipage}
	\begin{minipage}{.7\textwidth}
		\centering
		\includegraphics[width=1\textwidth]{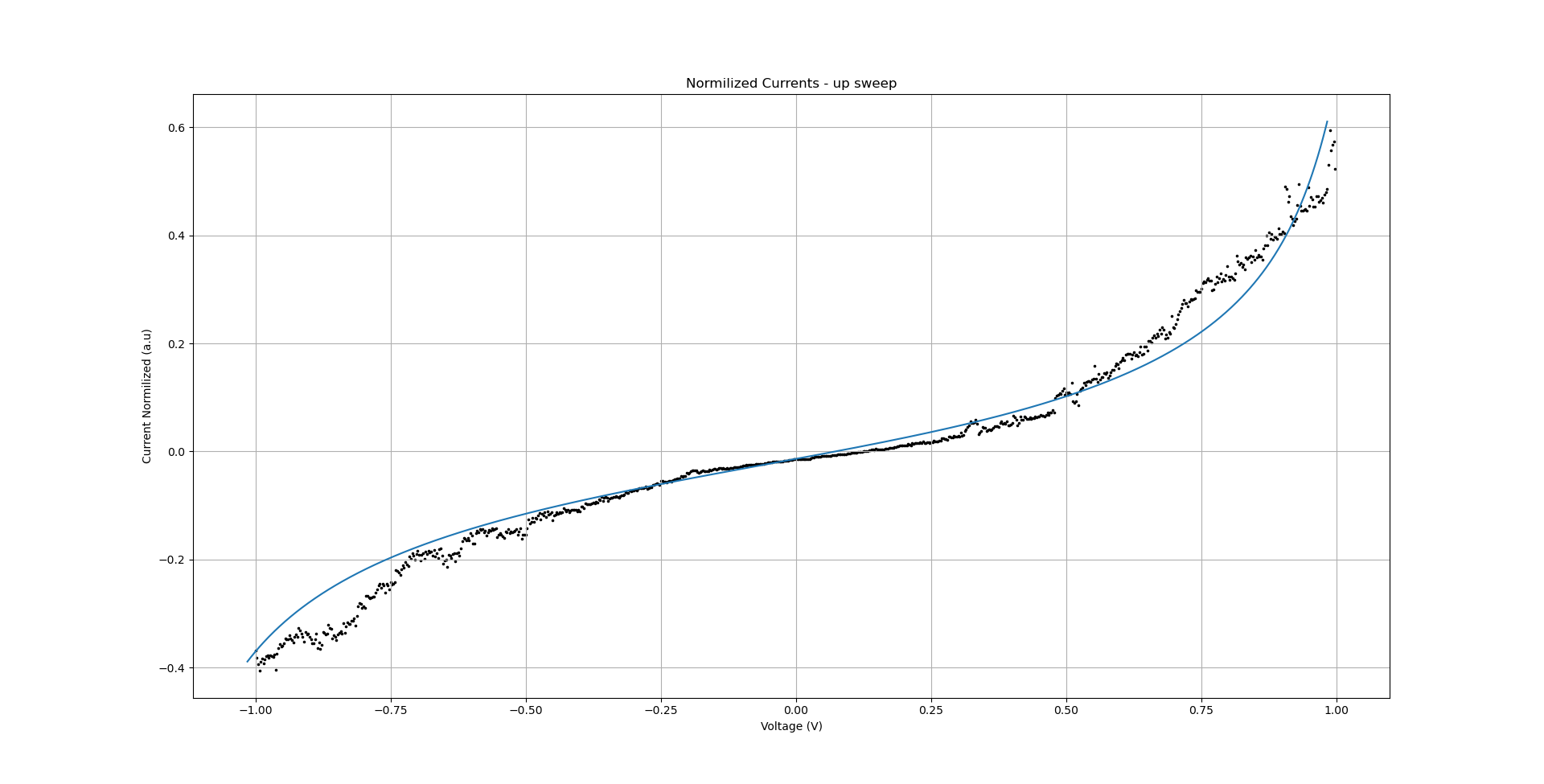}
		\subcaption[]{}
		\label{fig:cat5b}
	\end{minipage}
	\begin{minipage}{.7\textwidth}
		\centering
		\includegraphics[width=1\textwidth]{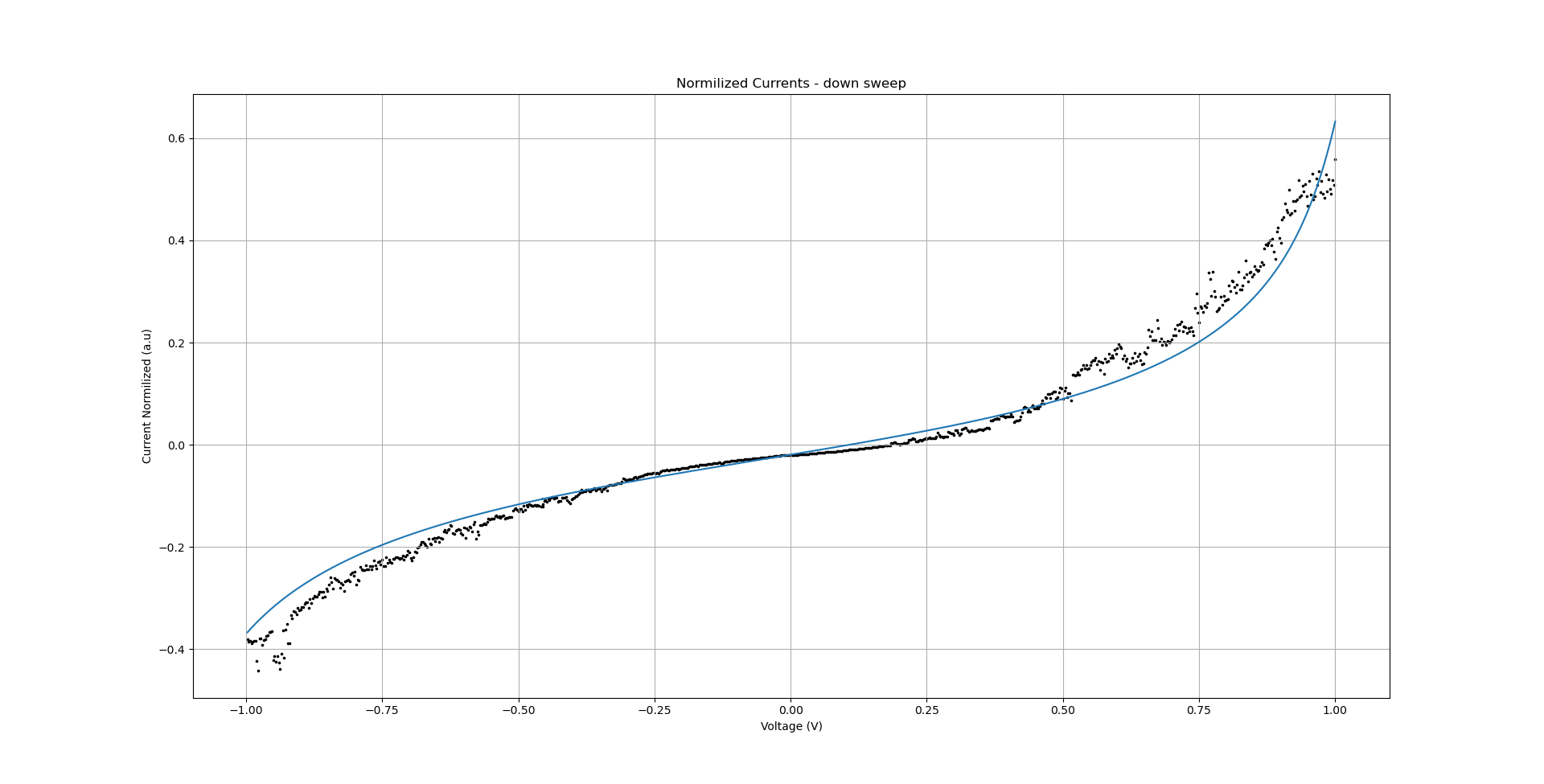}
		\subcaption[]{}
		\label{fig:cat5c}
	\end{minipage}
	\caption{}
	\label{fig:cat5}
\end{figure}

\clearpage

\subsection{Measurements with hysteresis}

\subsubsection{Measurements with good Fitting}

\begin{figure}[!htb]
	\centering
	\begin{minipage}{.7\textwidth}
		\centering
		\includegraphics[width=1\textwidth]{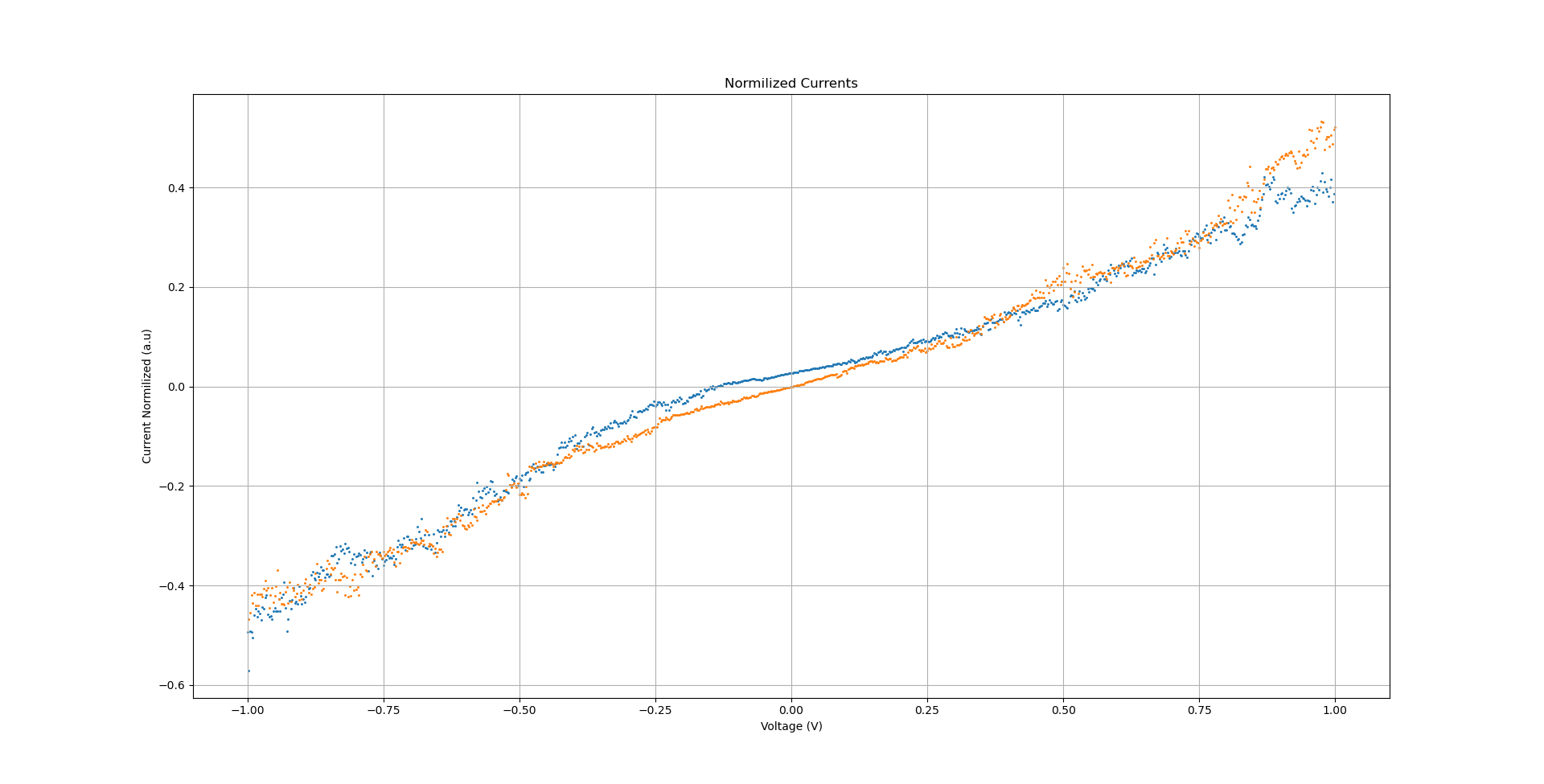}
		\subcaption[]{}
		\label{fig:cat2a}
	\end{minipage}
	\begin{minipage}{.7\textwidth}
		\centering
		\includegraphics[width=1\textwidth]{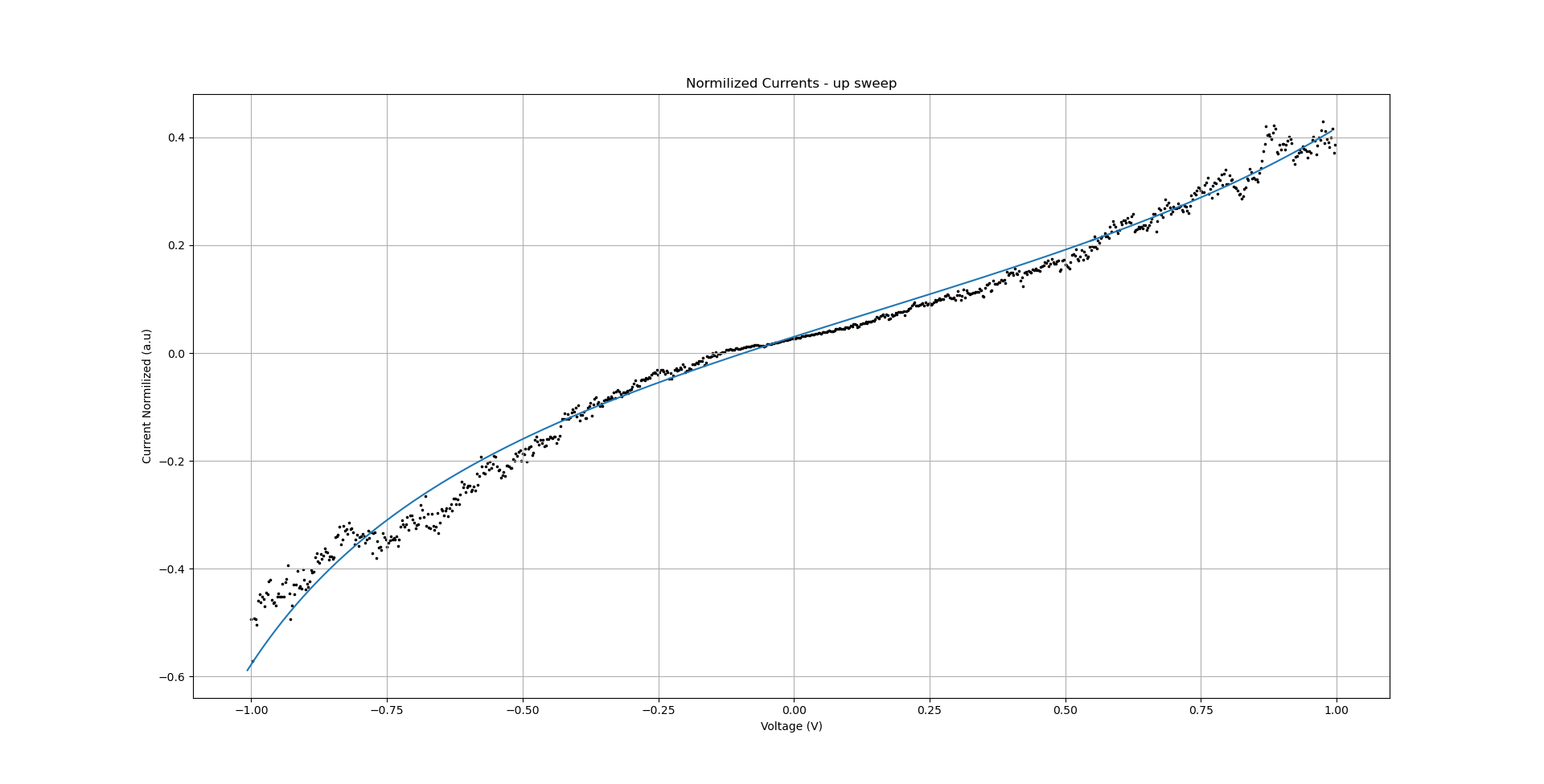}
		\subcaption[]{}
		\label{fig:cat2b}
	\end{minipage}
	\begin{minipage}{.7\textwidth}
		\centering
		\includegraphics[width=1\textwidth]{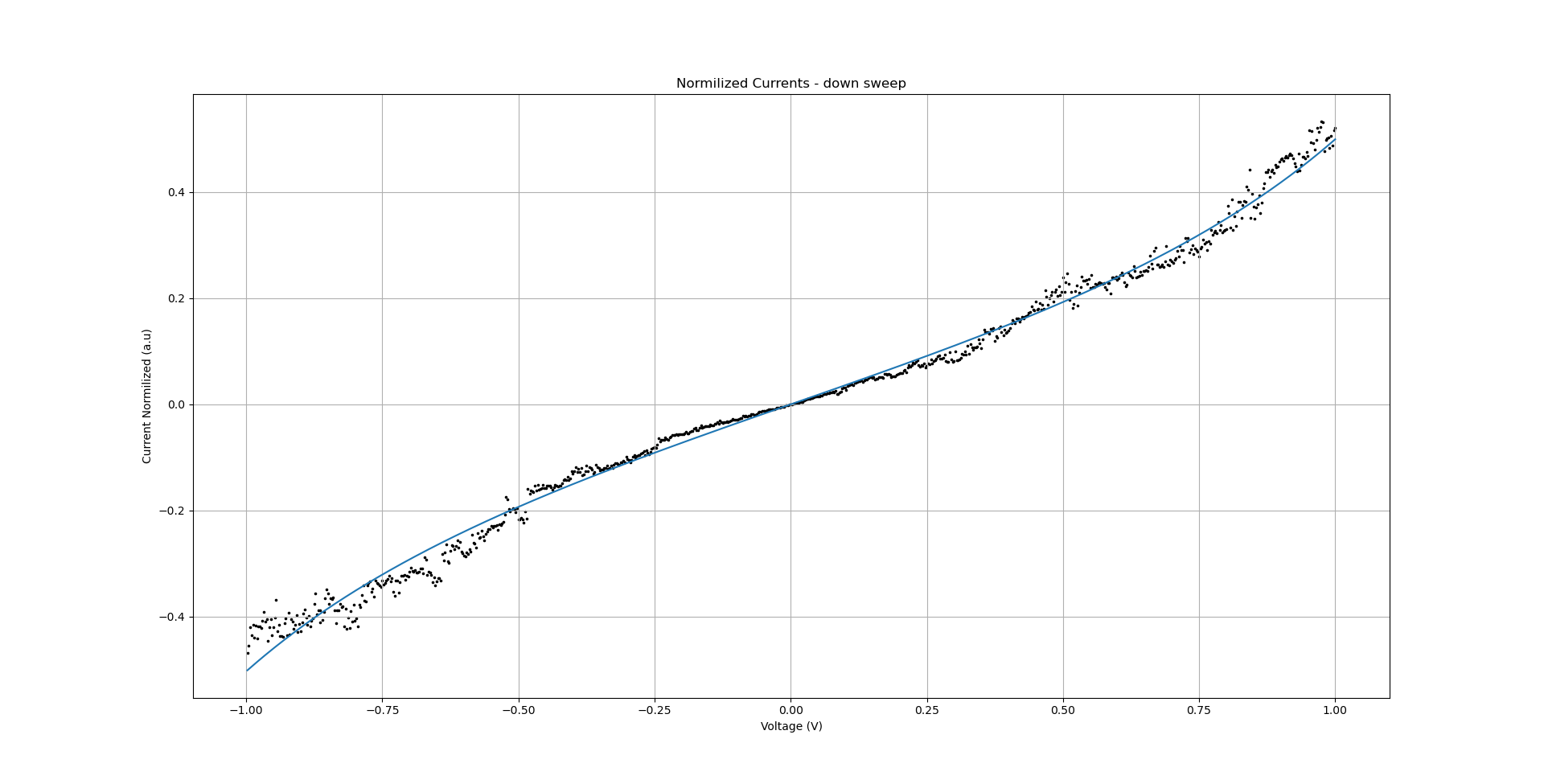}
		\subcaption[]{}
		\label{fig:cat2c}
	\end{minipage}
	\caption{}
	\label{fig:cat2}
\end{figure}

\clearpage

\subsubsection{Measurements with hysteresis and  with poor fitting (origin and tails)}

\begin{figure}[!htb]
	\centering
	\begin{minipage}{.7\textwidth}
		\centering
		\includegraphics[width=1\textwidth]{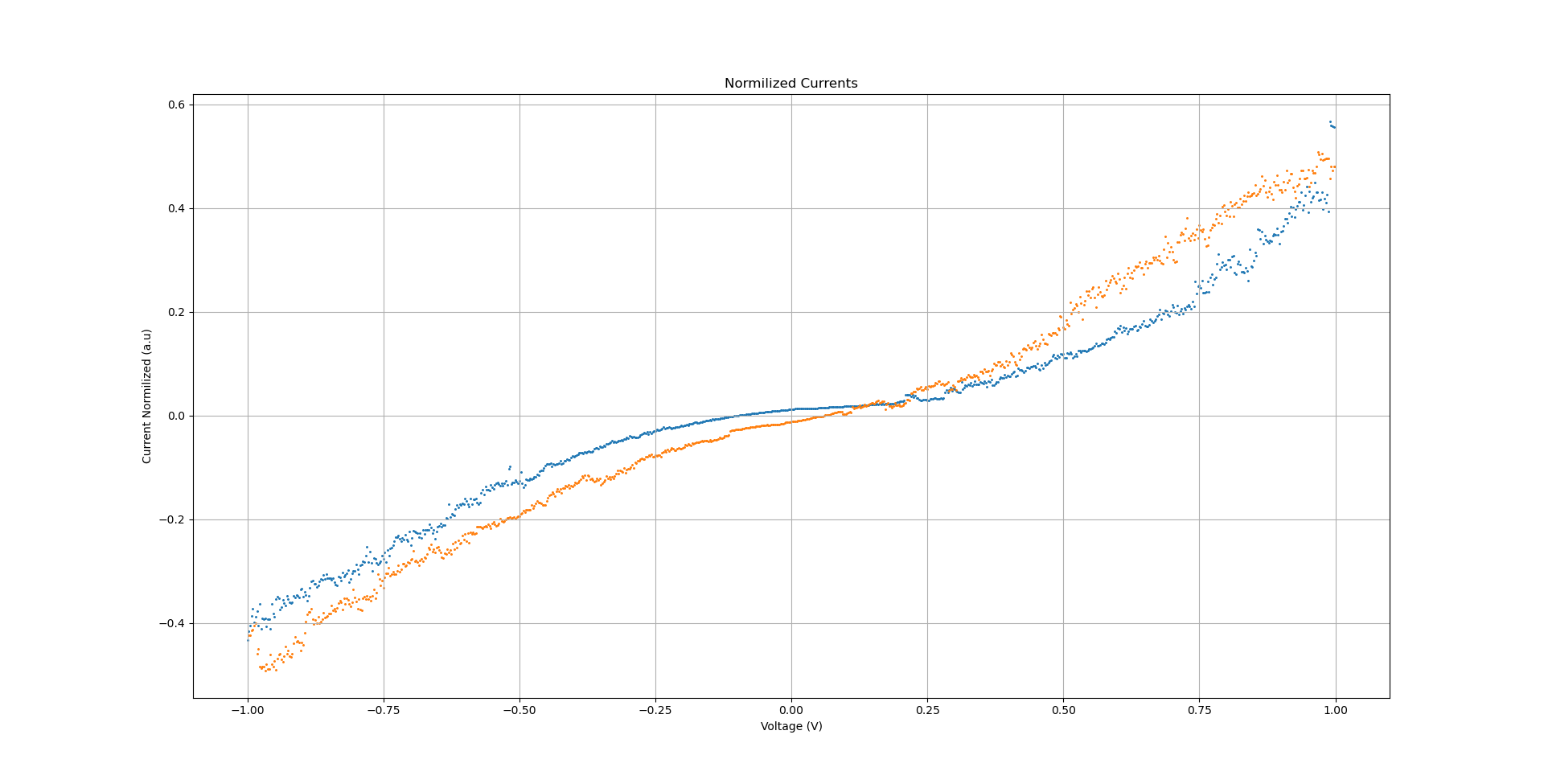}
		\subcaption[]{}
		\label{fig:cat7a}
	\end{minipage}
	\begin{minipage}{.7\textwidth}
		\centering
		\includegraphics[width=1\textwidth]{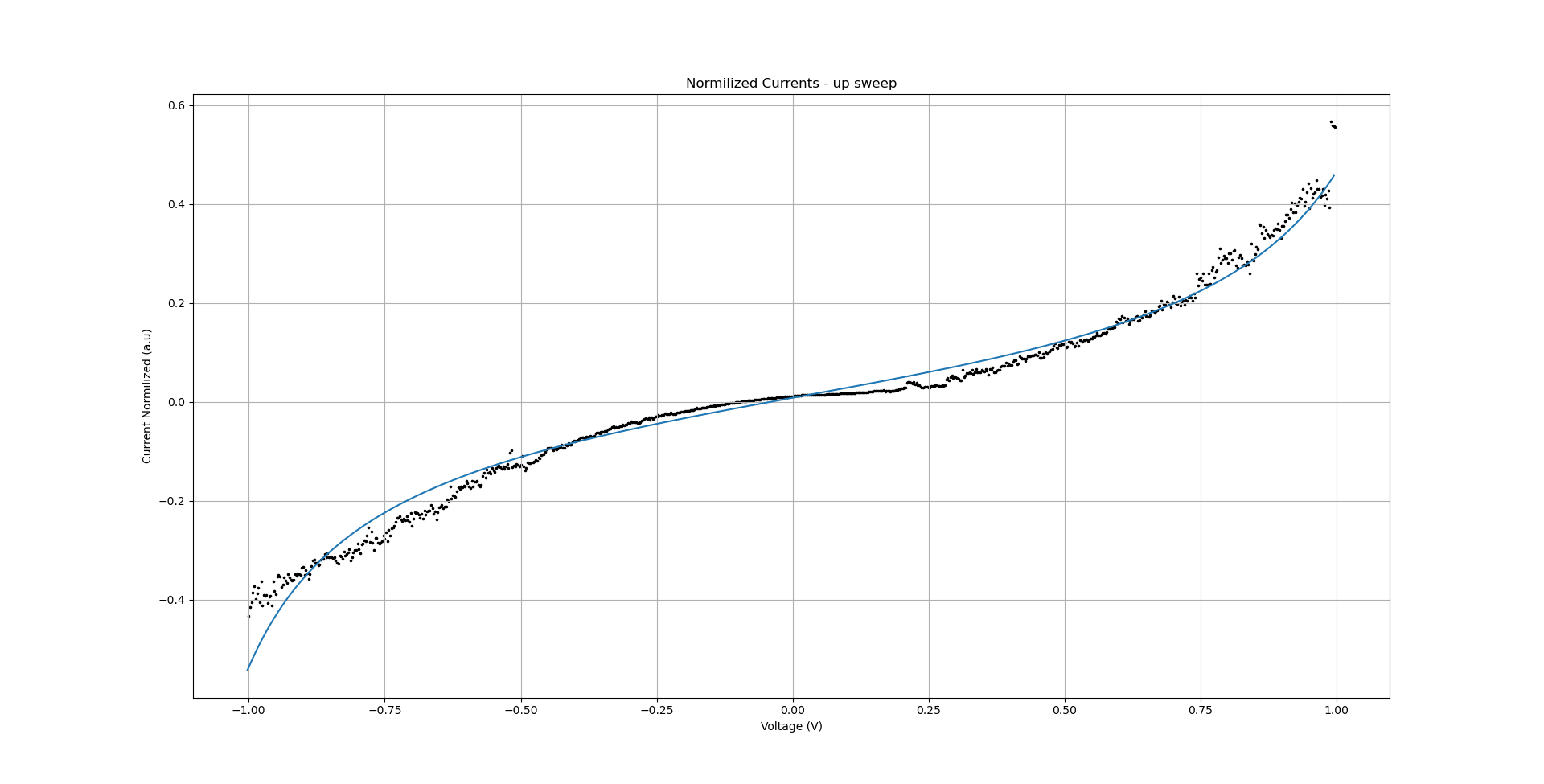}
		\subcaption[]{}
		\label{fig:cat7b}
	\end{minipage}
	\begin{minipage}{.7\textwidth}
		\centering
		\includegraphics[width=1\textwidth]{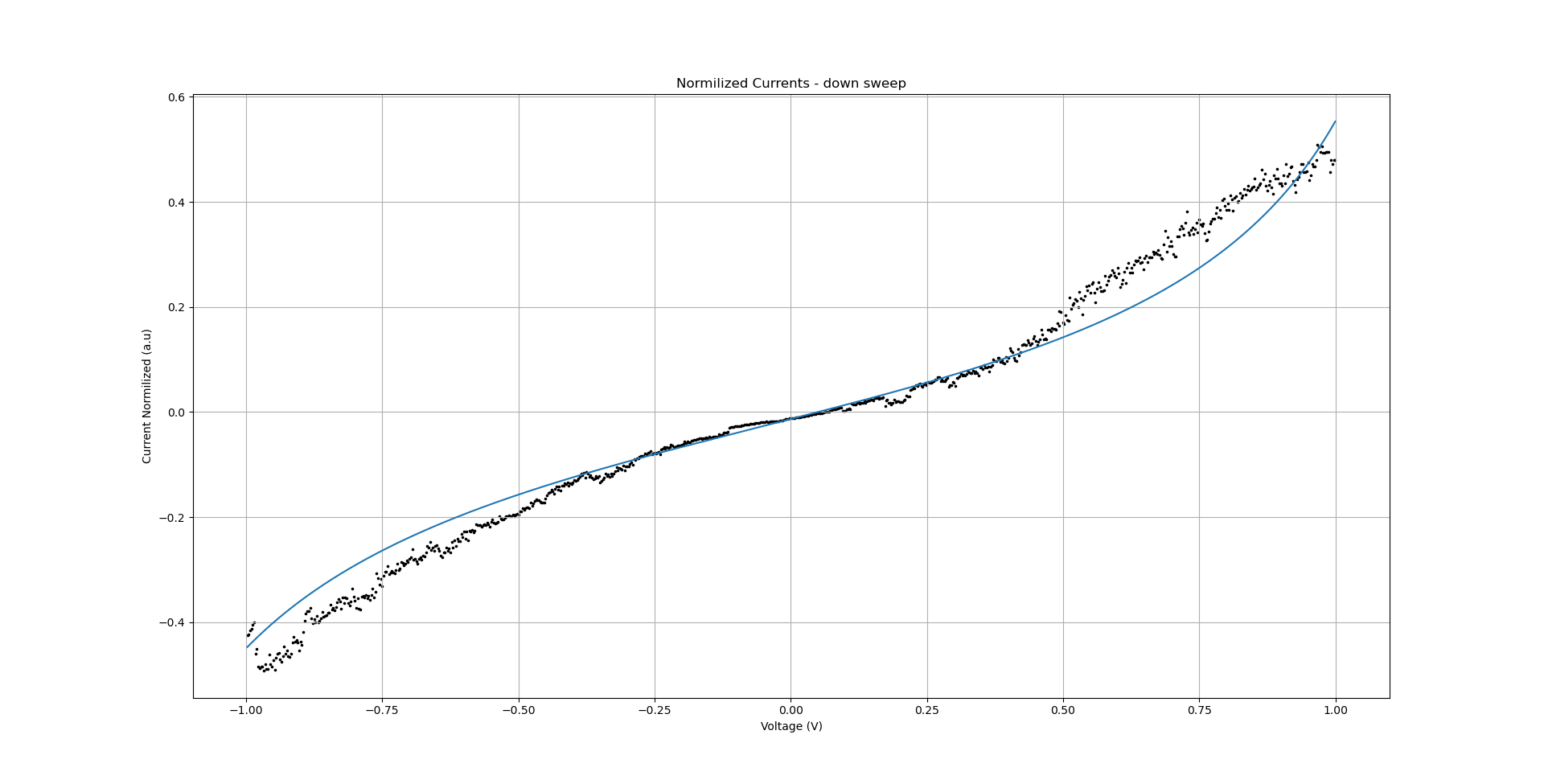}
		\subcaption[]{}
		\label{fig:cat7c}
	\end{minipage}
	\caption{}
	\label{fig:cat7}
\end{figure}

\clearpage

\subsubsection{Measurements with hysteresis and  with poor fitting (origin)}

\begin{figure}[!htb]
	\centering
	\begin{minipage}{.7\textwidth}
		\centering
		\includegraphics[width=1\textwidth]{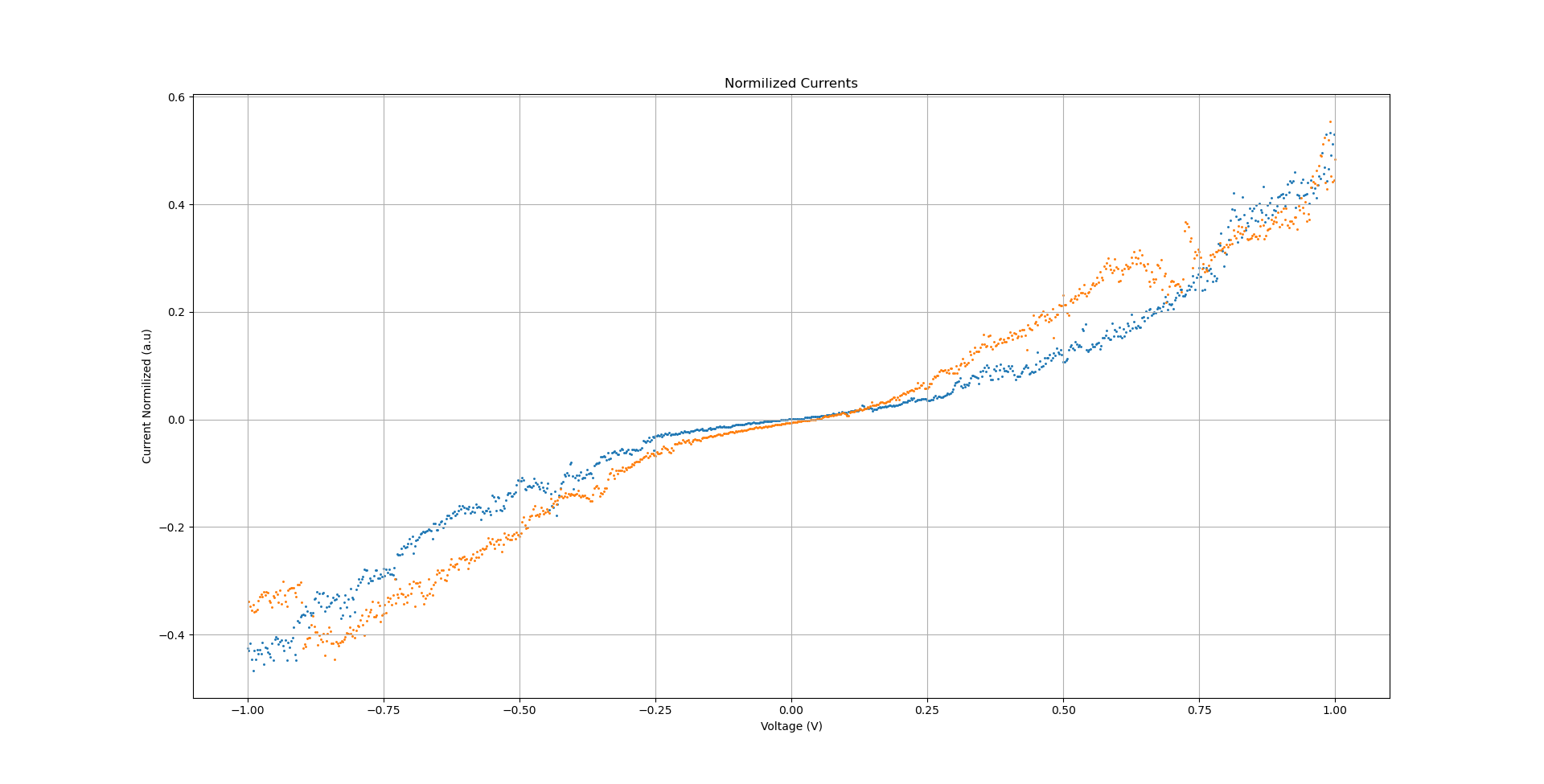}
		\subcaption[]{}
		\label{fig:cat8a}
	\end{minipage}
	\begin{minipage}{.7\textwidth}
		\centering
		\includegraphics[width=1\textwidth]{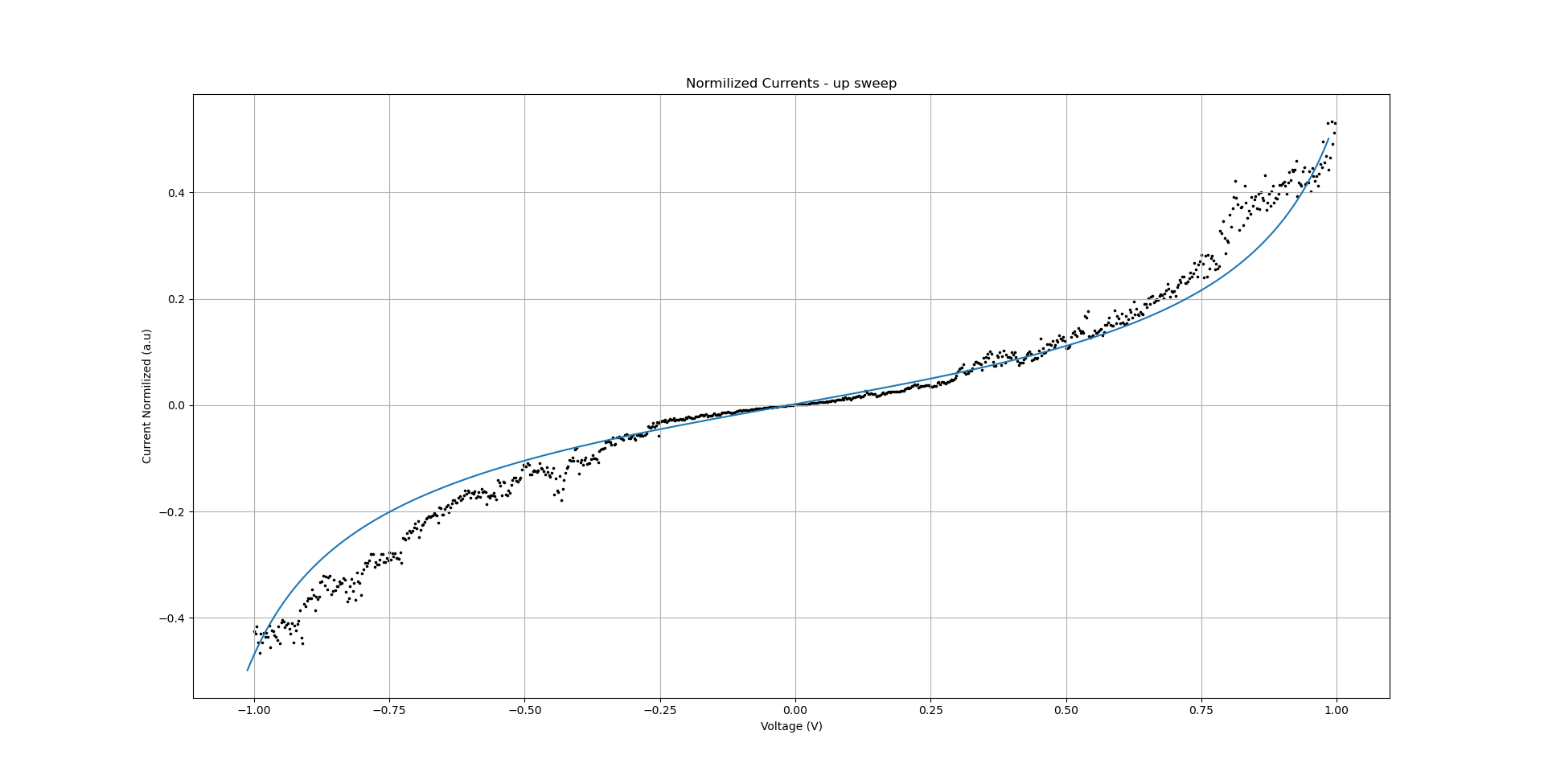}
		\subcaption[]{}
		\label{fig:cat8b}
	\end{minipage}
	\begin{minipage}{.7\textwidth}
		\centering
		\includegraphics[width=1\textwidth]{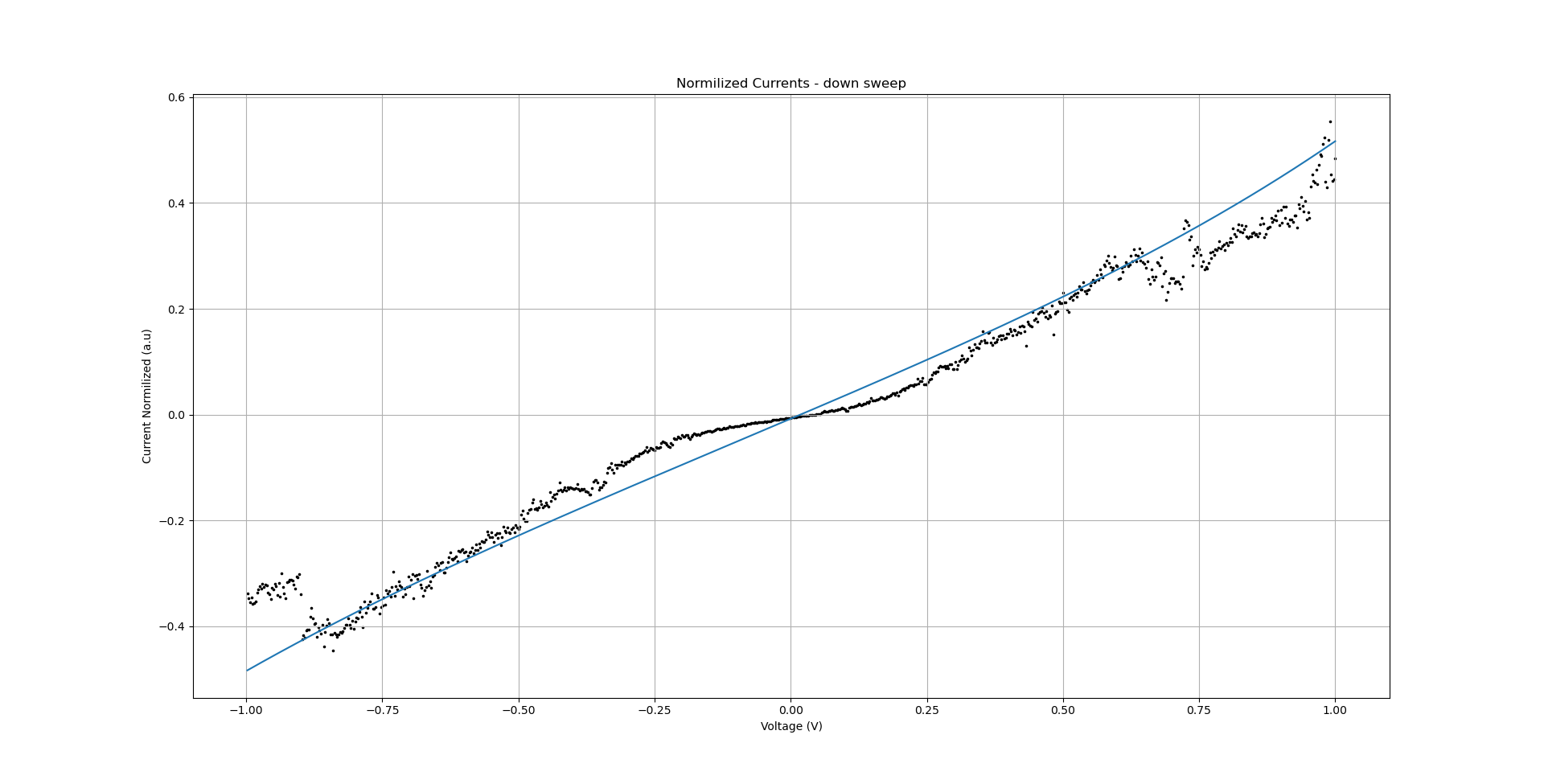}
		\subcaption[]{}
		\label{fig:cat8c}
	\end{minipage}
	\caption{}
	\label{fig:cat8}
\end{figure}

\clearpage

\subsubsection{Measurements with hysteresis and  with poor fitting (tails)}

\begin{figure}[!htb]
	\centering
	\begin{minipage}{.7\textwidth}
		\centering
		\includegraphics[width=1\textwidth]{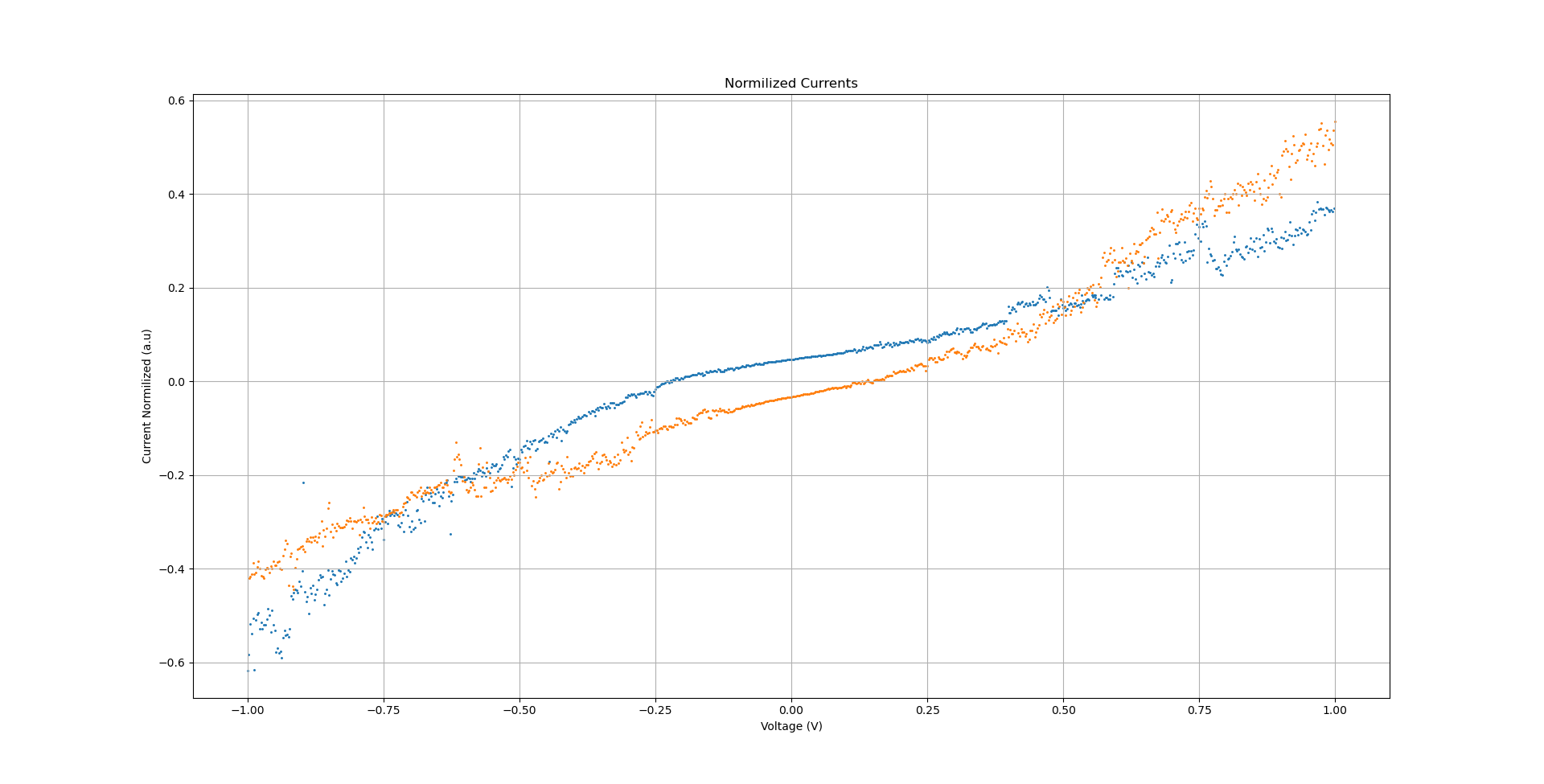}
		\subcaption[]{}
		\label{fig:cat9a}
	\end{minipage}
	\begin{minipage}{.7\textwidth}
		\centering
		\includegraphics[width=1\textwidth]{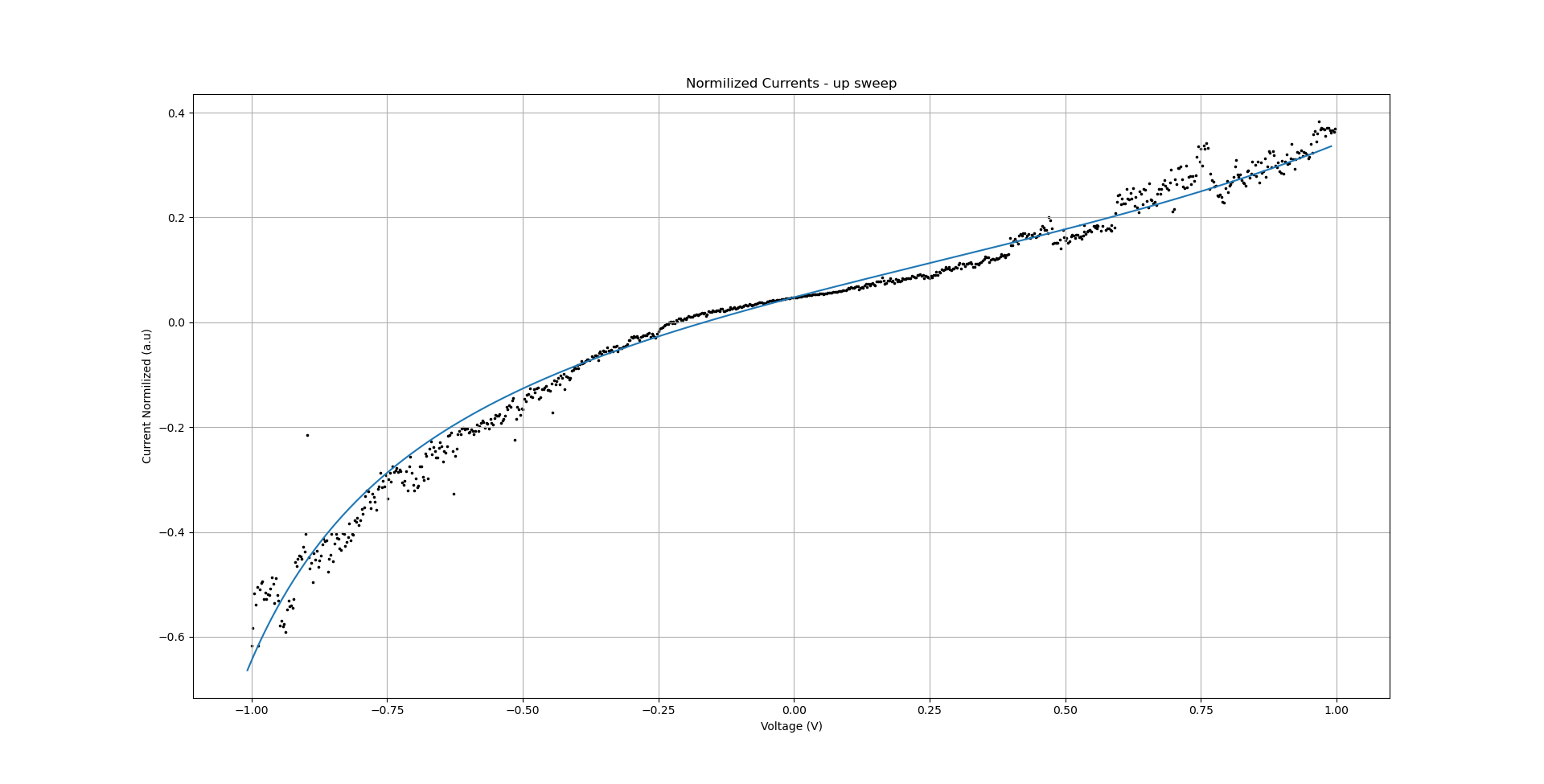}
		\subcaption[]{}
		\label{fig:cat9b}
	\end{minipage}
	\begin{minipage}{.7\textwidth}
		\centering
		\includegraphics[width=1\textwidth]{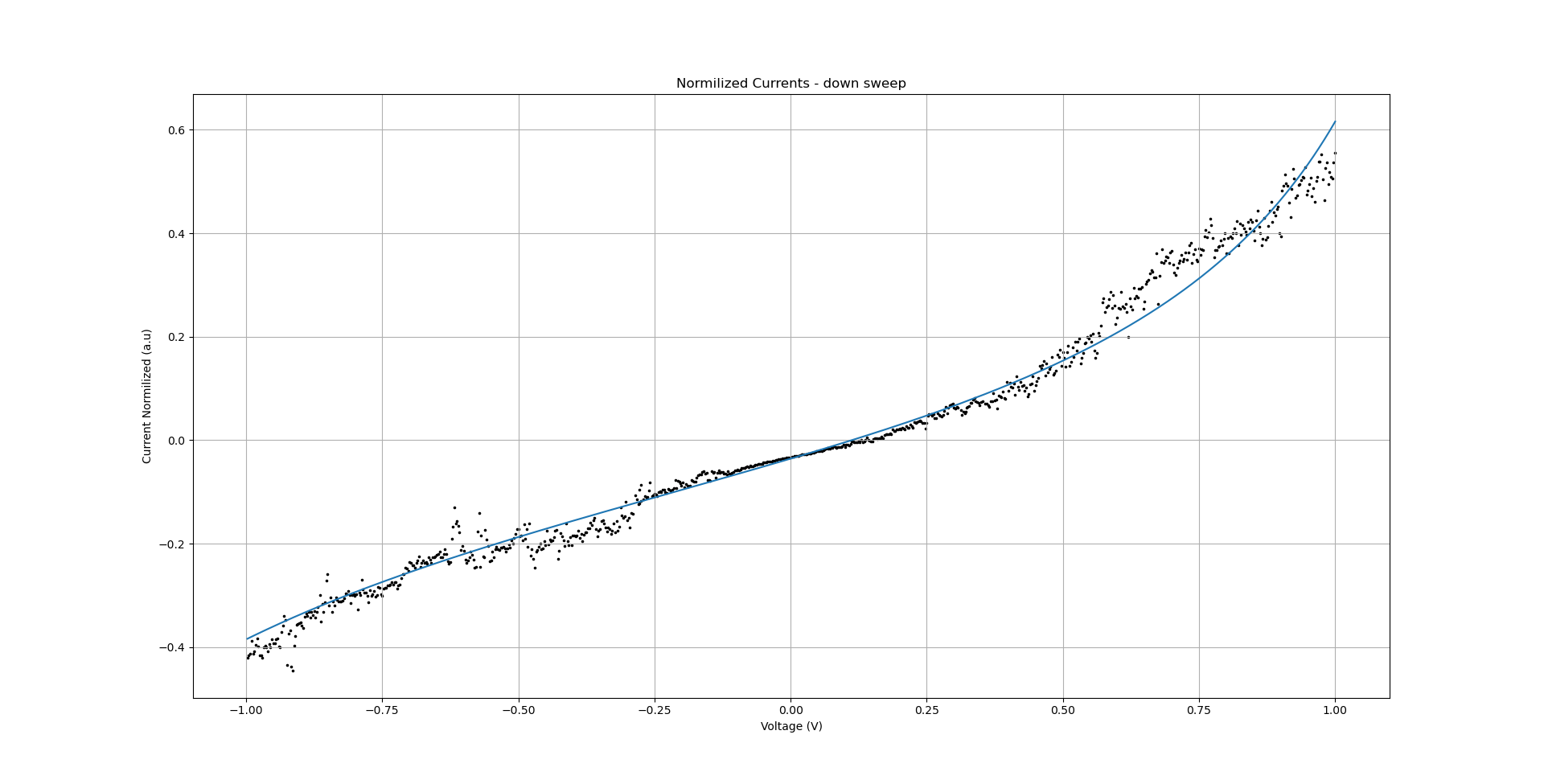}
		\subcaption[]{}
		\label{fig:cat9c}
	\end{minipage}
	\caption{}
	\label{fig:cat9}
\end{figure}

\clearpage

\section{Measurements without Fitting}

\begin{figure}[!htb]
	\centering
	\begin{minipage}{.7\textwidth}
		\centering
		\includegraphics[width=1\textwidth]{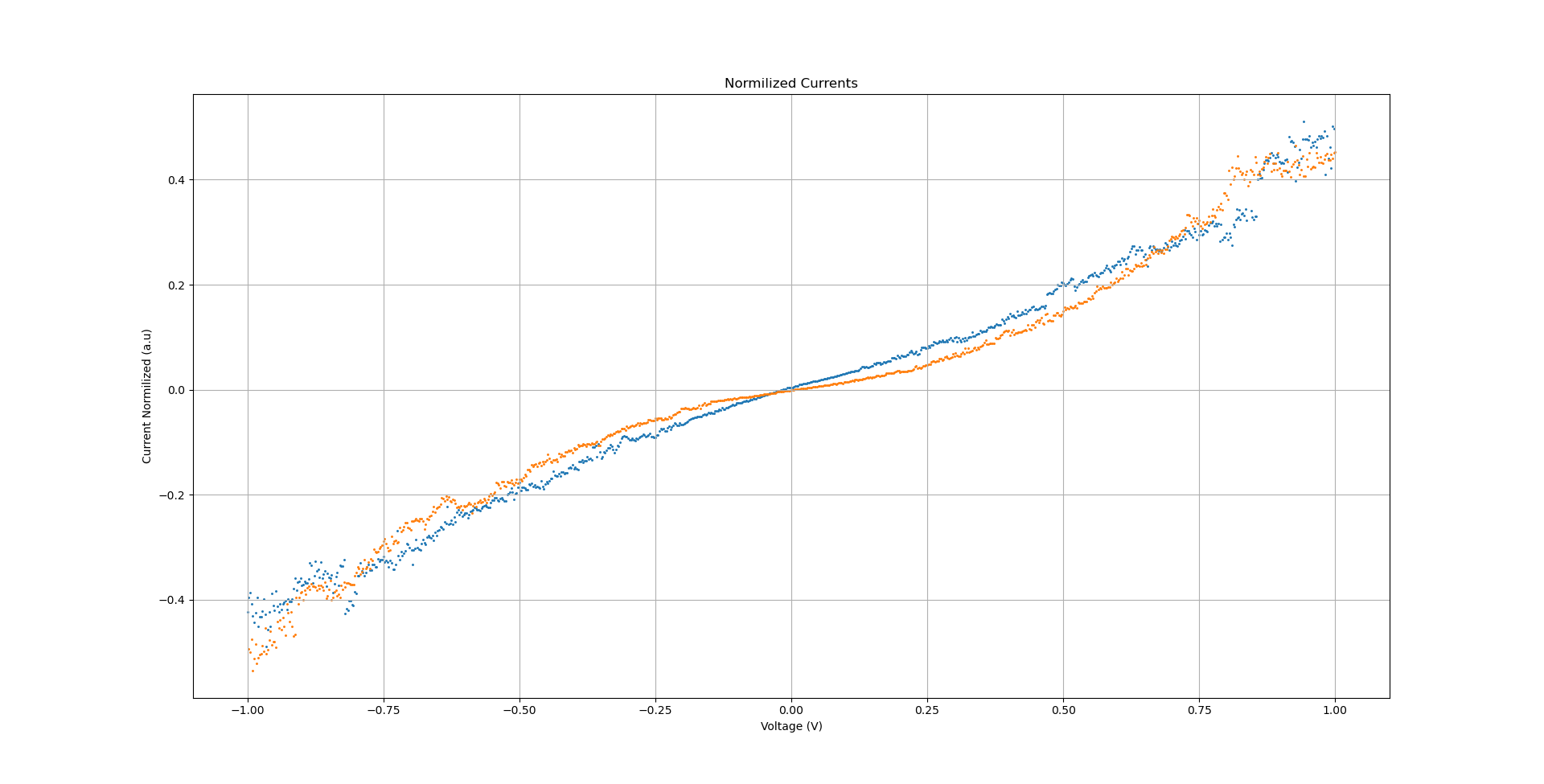}
		\subcaption[]{}
		\label{fig:cat4a}
	\end{minipage}
	\begin{minipage}{.7\textwidth}
		\centering
		\includegraphics[width=1\textwidth]{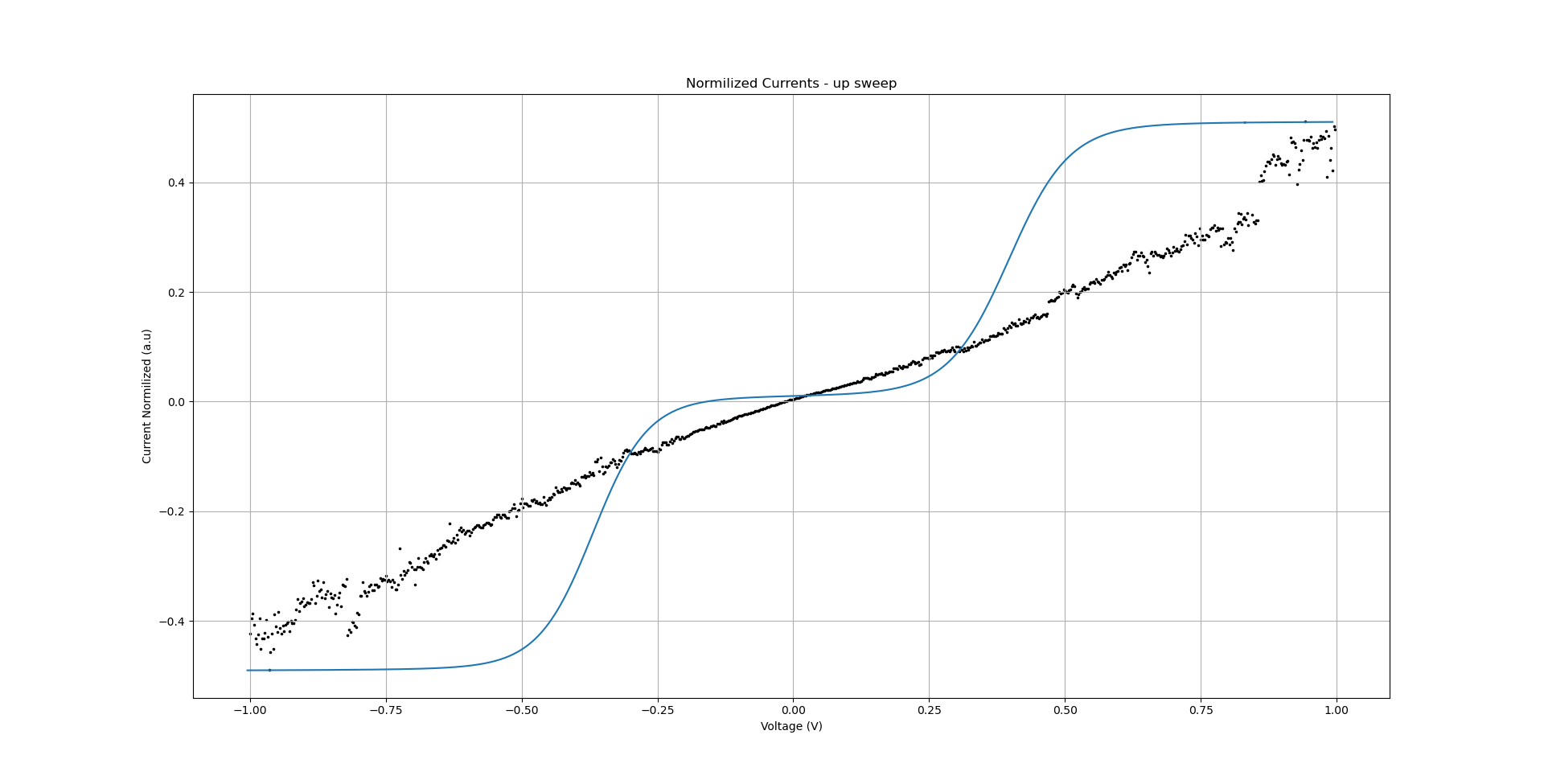}
		\subcaption[]{}
		\label{fig:cat4b}
	\end{minipage}
	\begin{minipage}{.7\textwidth}
		\centering
		\includegraphics[width=1\textwidth]{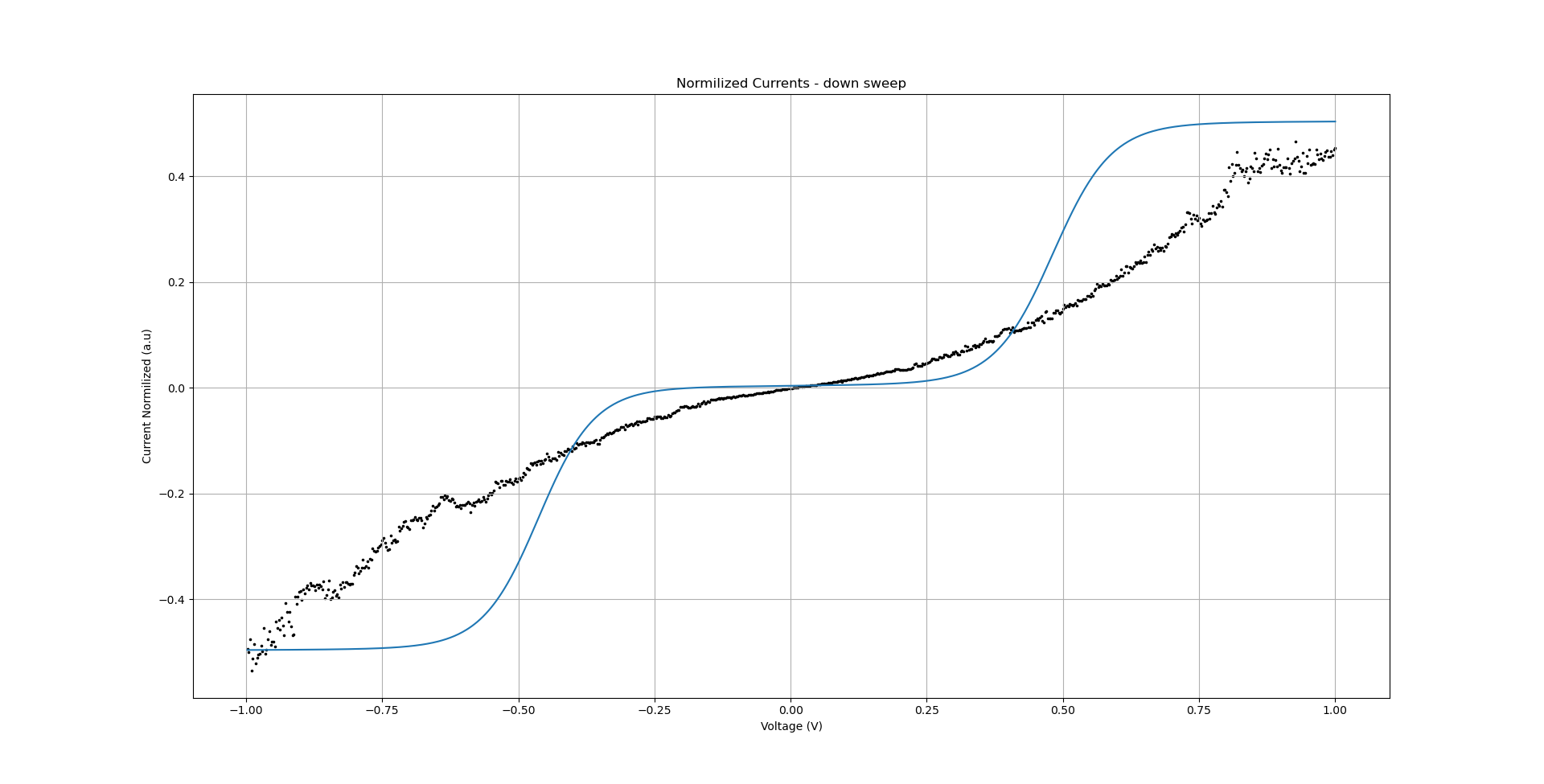}
		\subcaption[]{}
		\label{fig:cat4c}
	\end{minipage}
	\caption{}
	\label{fig:cat4}
\end{figure}
	\chapter{Our best fits where the SLM fails}
\label{app:somefits}

\begin{figure}[!htb]
	\centering
	\begin{minipage}{1\textwidth}
		\centering
		\includegraphics[width=1\textwidth]{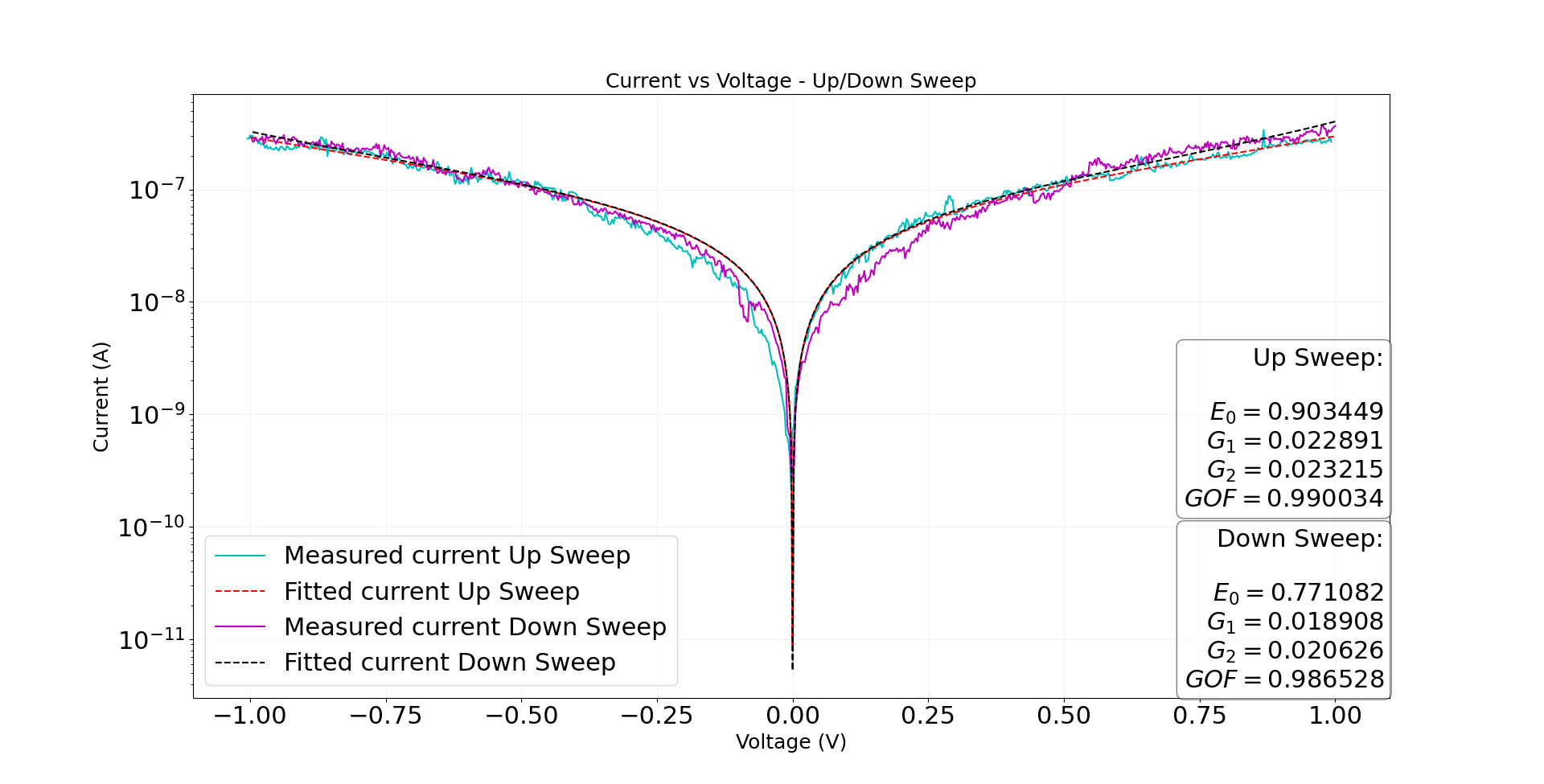}
	\end{minipage}
\end{figure}
\begin{figure}[!htb]	
	\begin{minipage}{1\textwidth}
		\centering
		\includegraphics[width=1\textwidth]{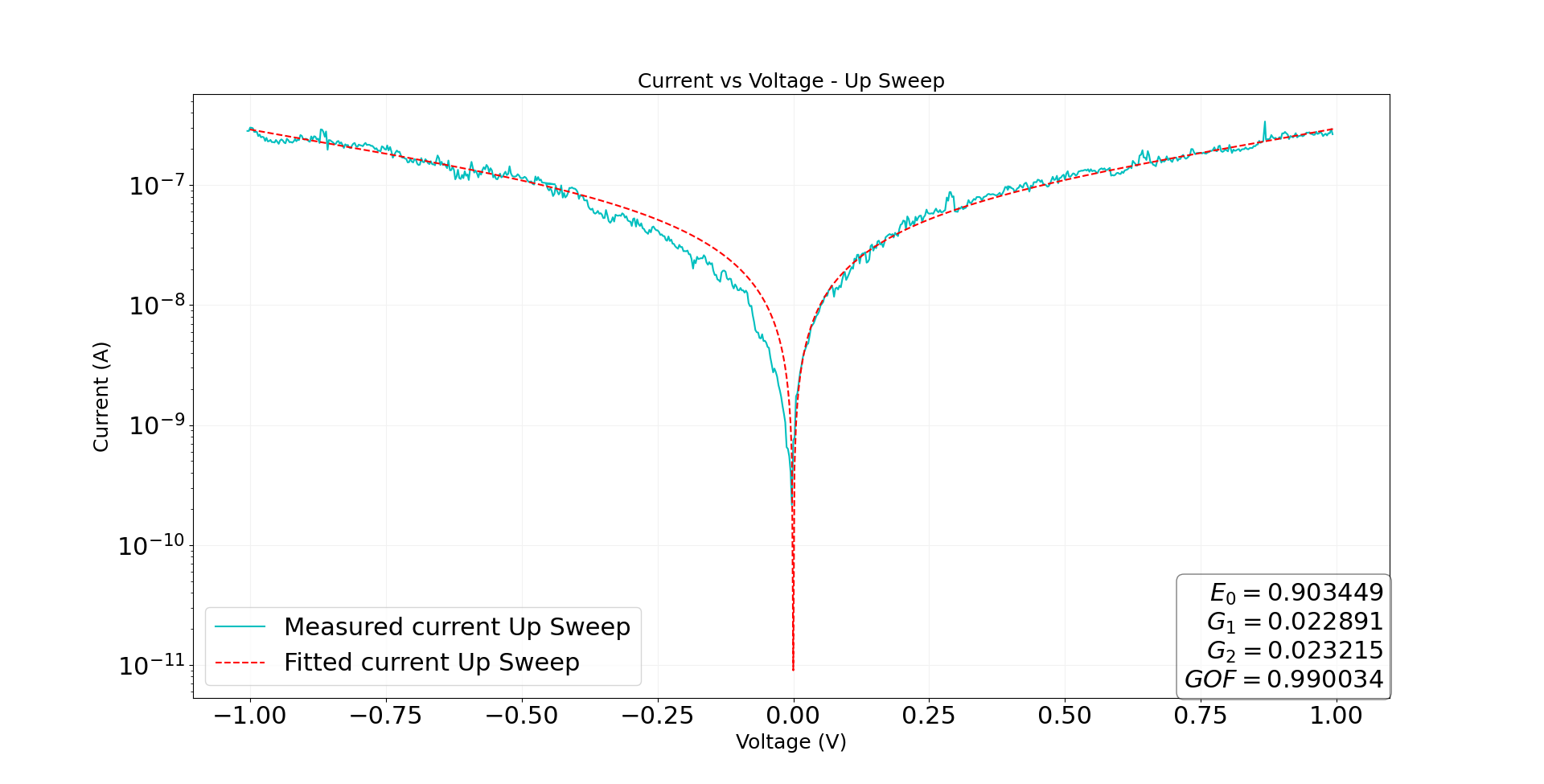}
	\end{minipage}
	\begin{minipage}{1\textwidth}
	\centering
	\includegraphics[width=1\textwidth]{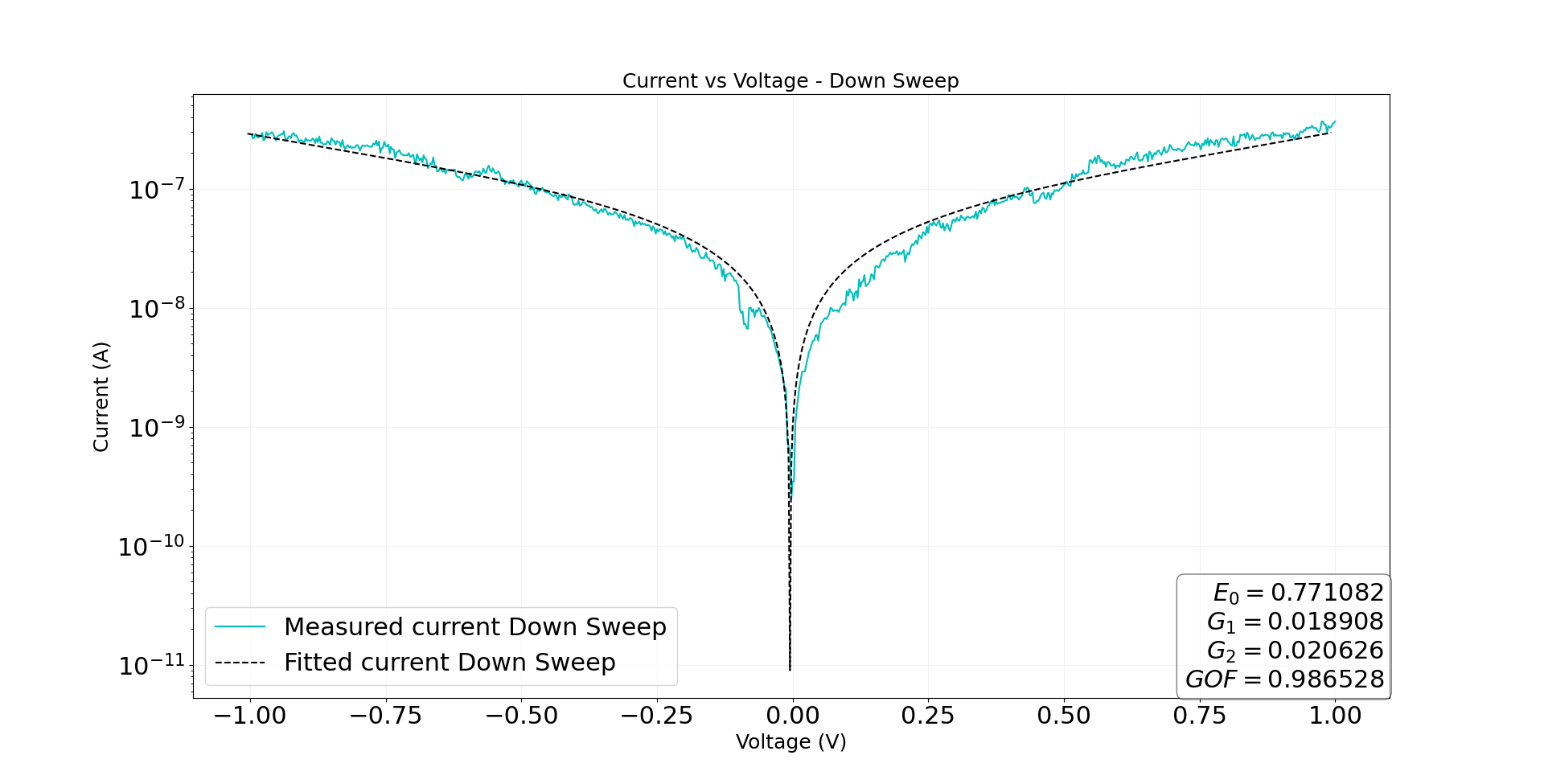}
	\end{minipage}
	\caption{}
\end{figure}

\begin{figure}[h]
	\centering
	\begin{minipage}{1\textwidth}
		\centering
		\includegraphics[width=1\textwidth]{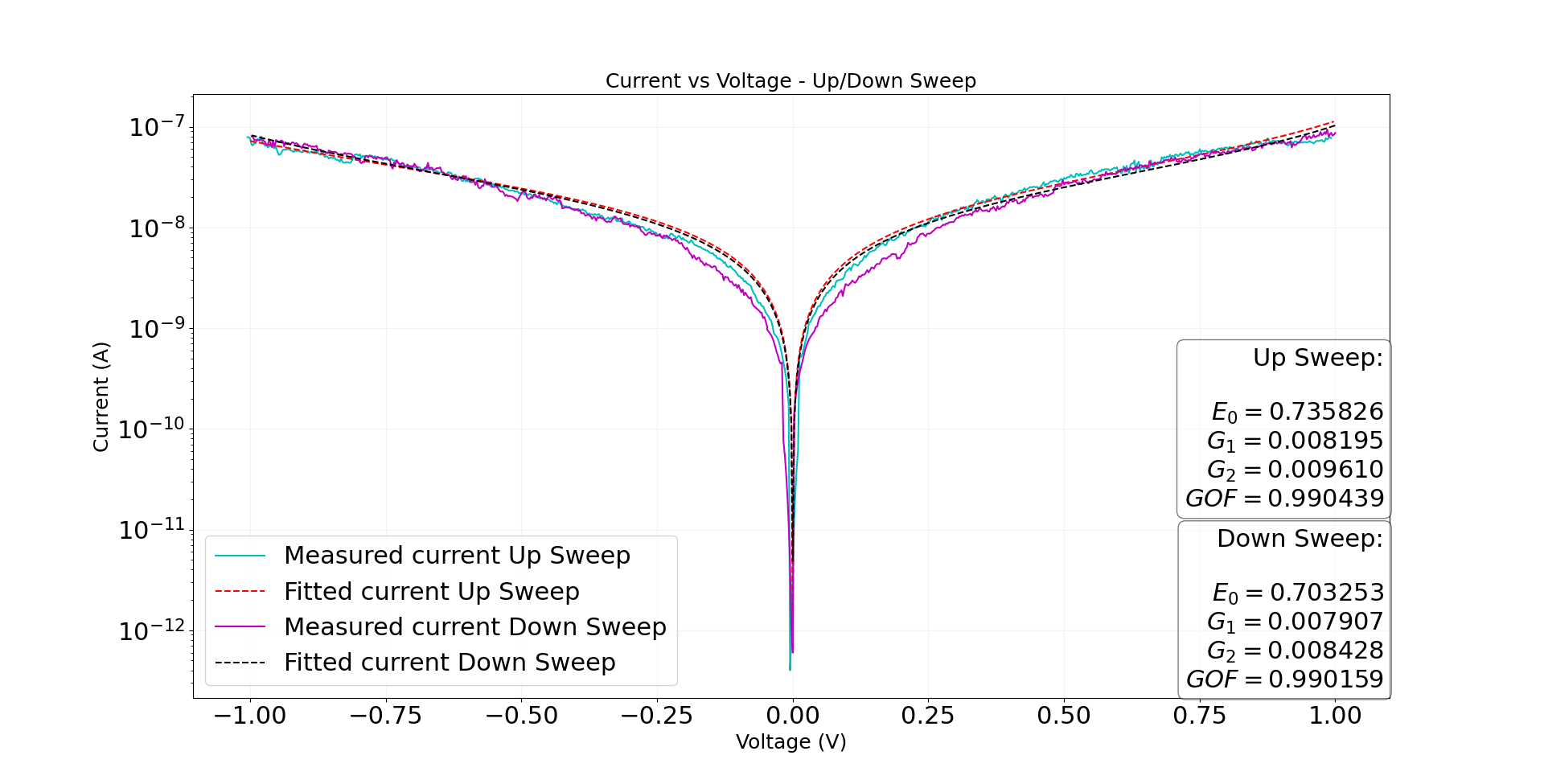}
	\end{minipage}	
	\begin{minipage}{1\textwidth}
		\centering
		\includegraphics[width=1\textwidth]{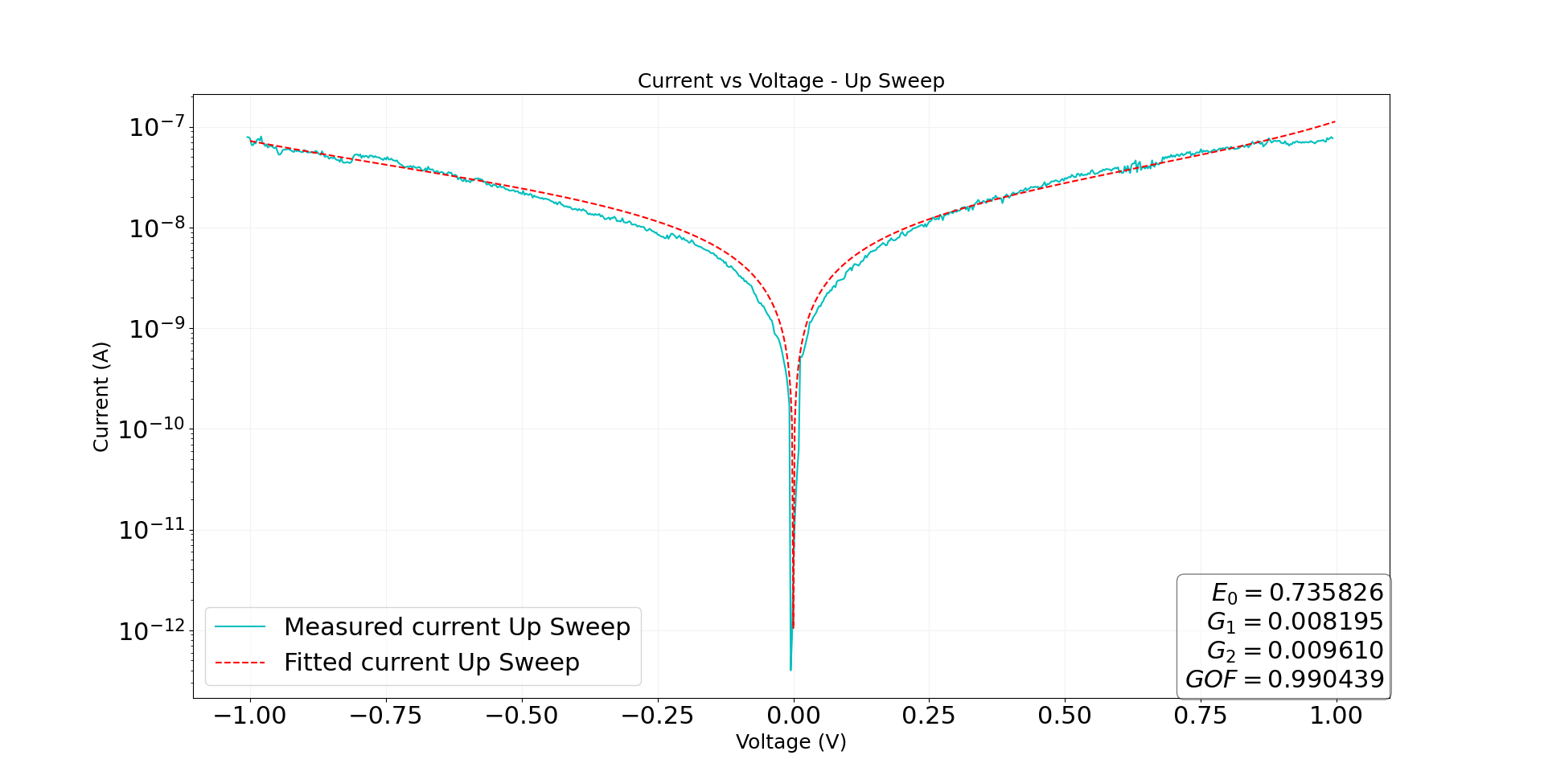}
	\end{minipage}
	\begin{minipage}{1\textwidth}
		\centering
		\includegraphics[width=1\textwidth]{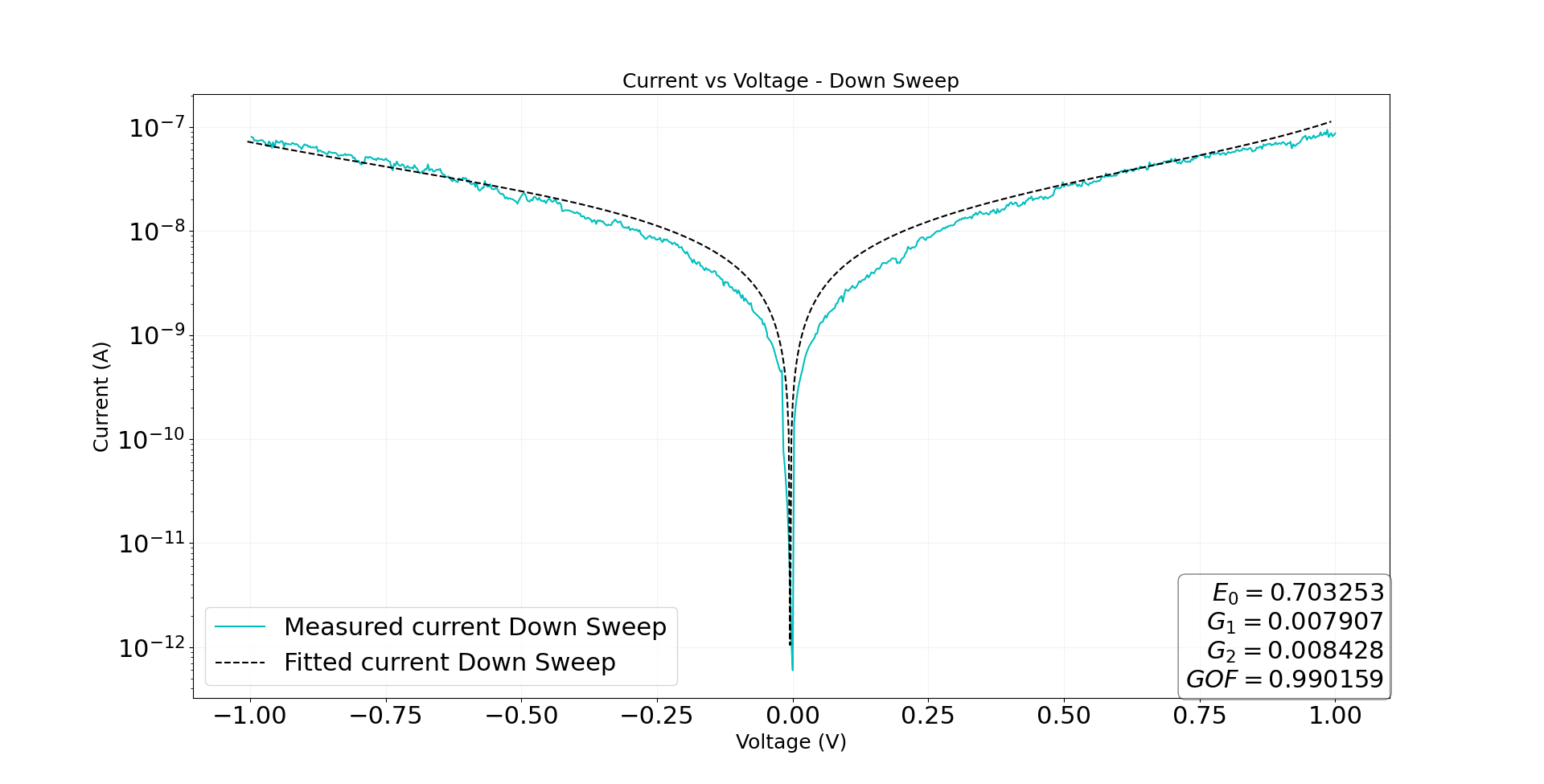}
	\end{minipage}
	\caption{}
\end{figure}

\begin{figure}[h]
	\centering
	\begin{minipage}{1\textwidth}
		\centering
		\includegraphics[width=1\textwidth]{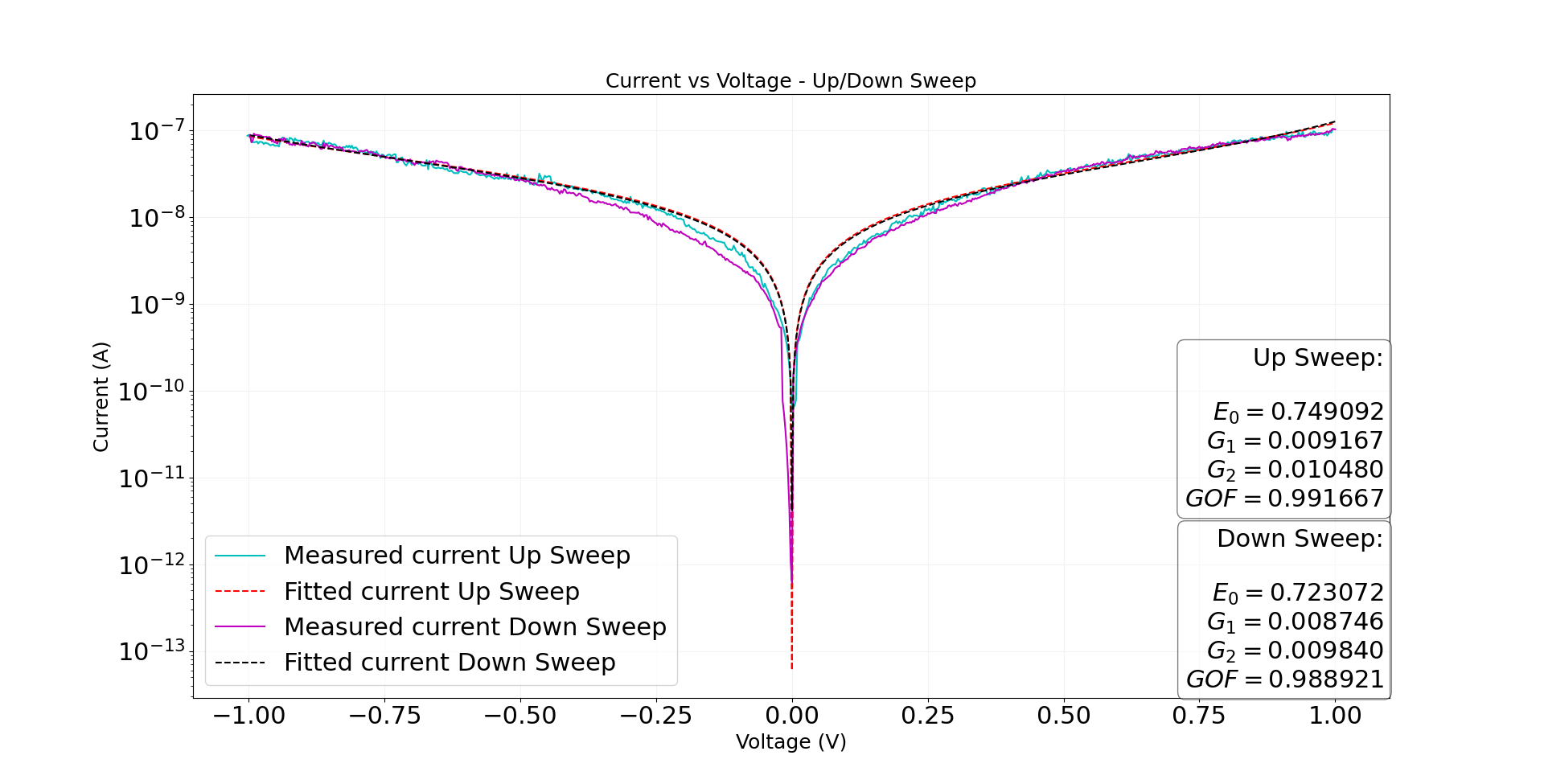}
	\end{minipage}	
	\begin{minipage}{1\textwidth}
		\centering
		\includegraphics[width=1\textwidth]{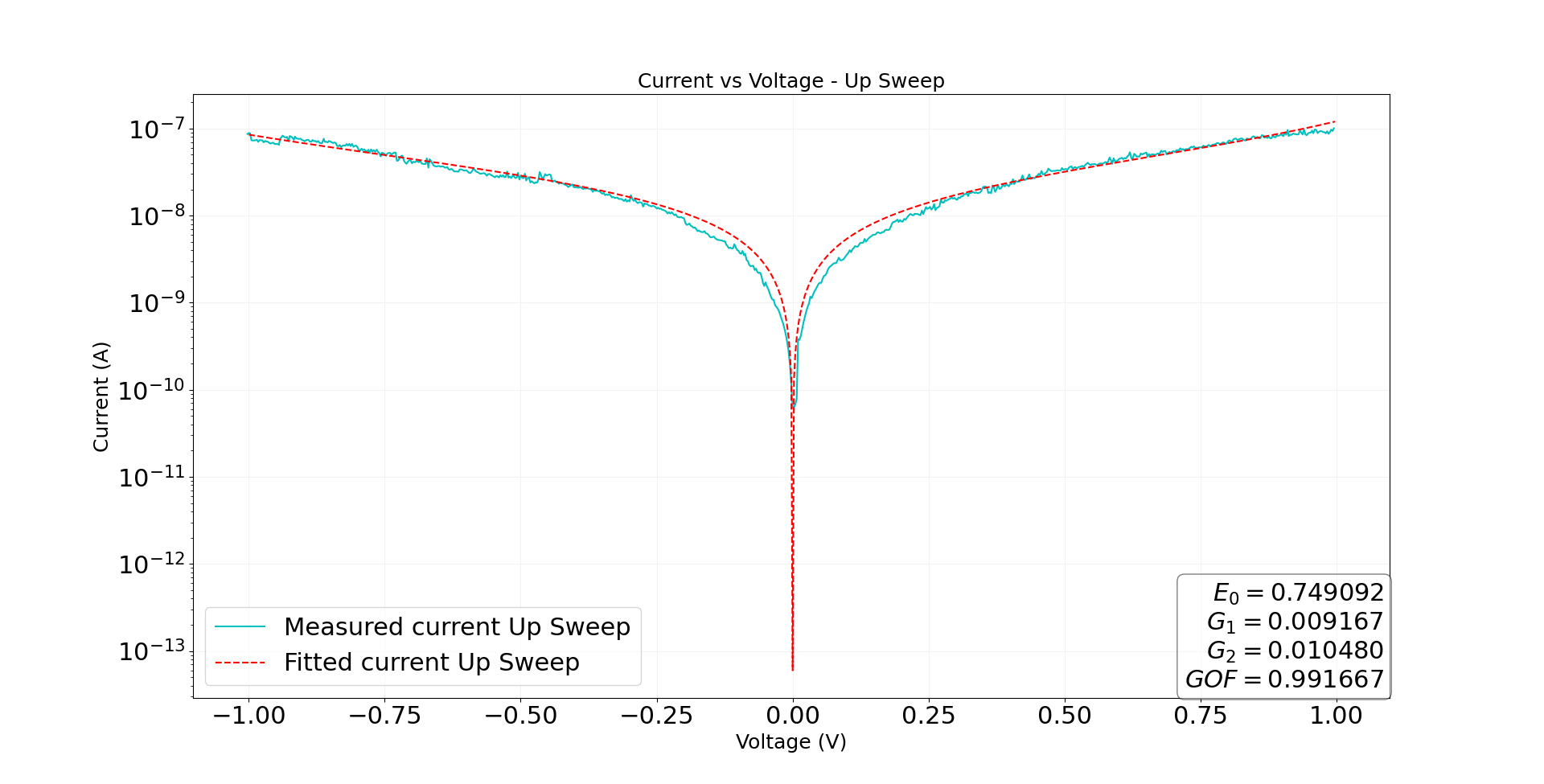}
	\end{minipage}
	\begin{minipage}{1\textwidth}
		\centering
		\includegraphics[width=1\textwidth]{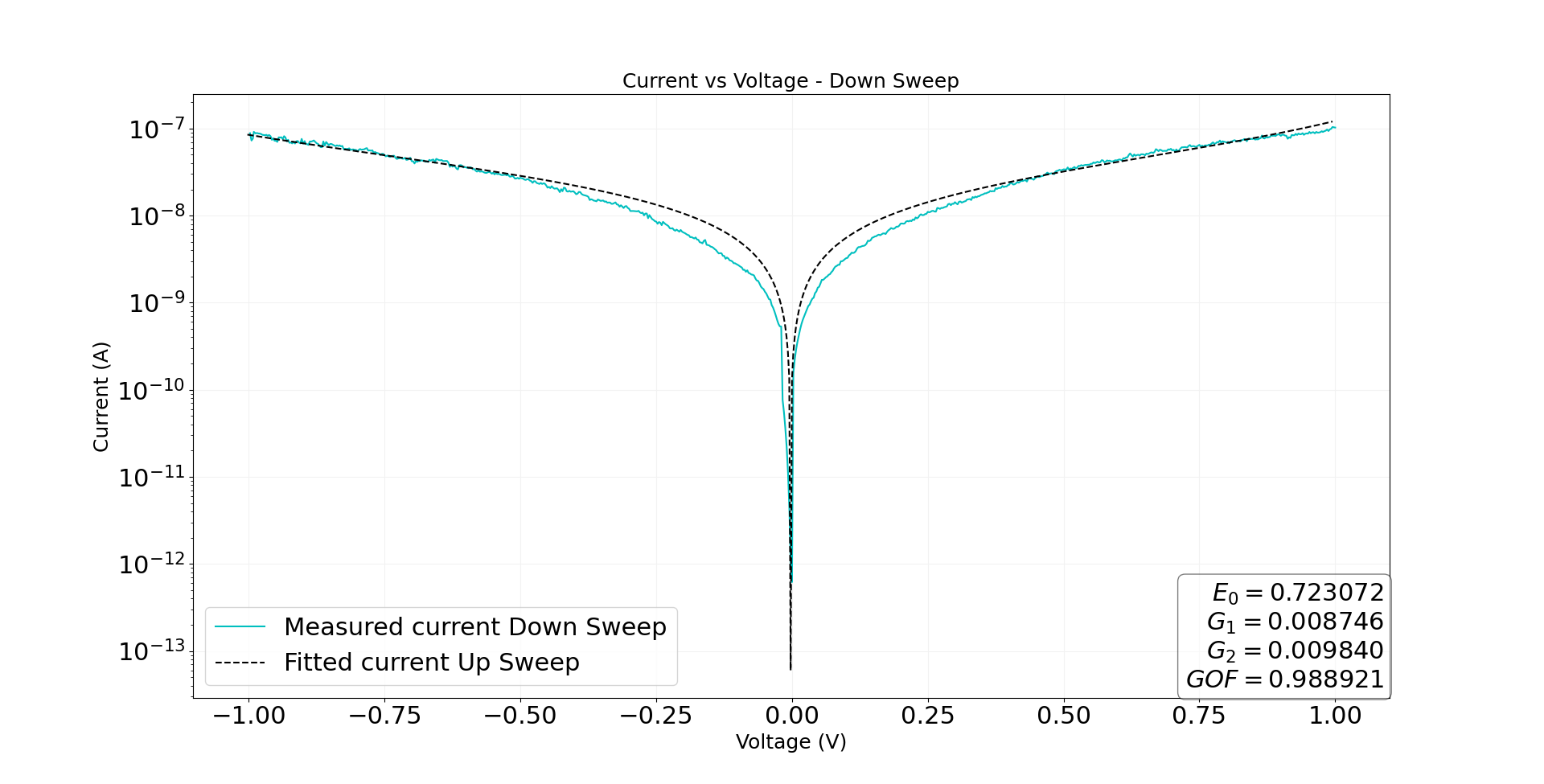}
	\end{minipage}
	\caption{}
\end{figure}

\begin{figure}[h]
	\centering
	\begin{minipage}{1\textwidth}
		\centering
		\includegraphics[width=1\textwidth]{iv_000144_ud.png}
	\end{minipage}	
	\begin{minipage}{1\textwidth}
		\centering
		\includegraphics[width=1\textwidth]{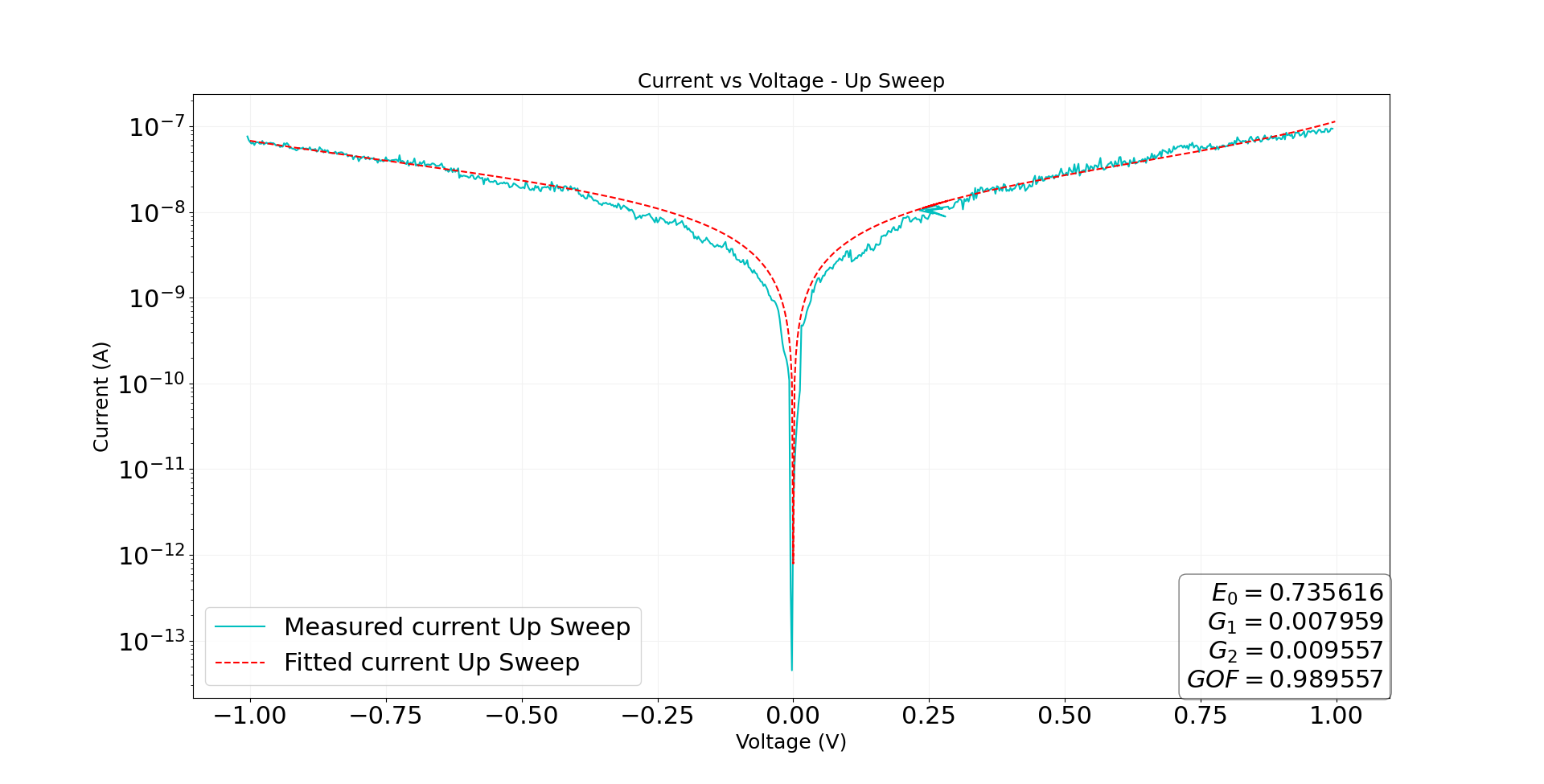}
	\end{minipage}
	\begin{minipage}{1\textwidth}
		\centering
		\includegraphics[width=1\textwidth]{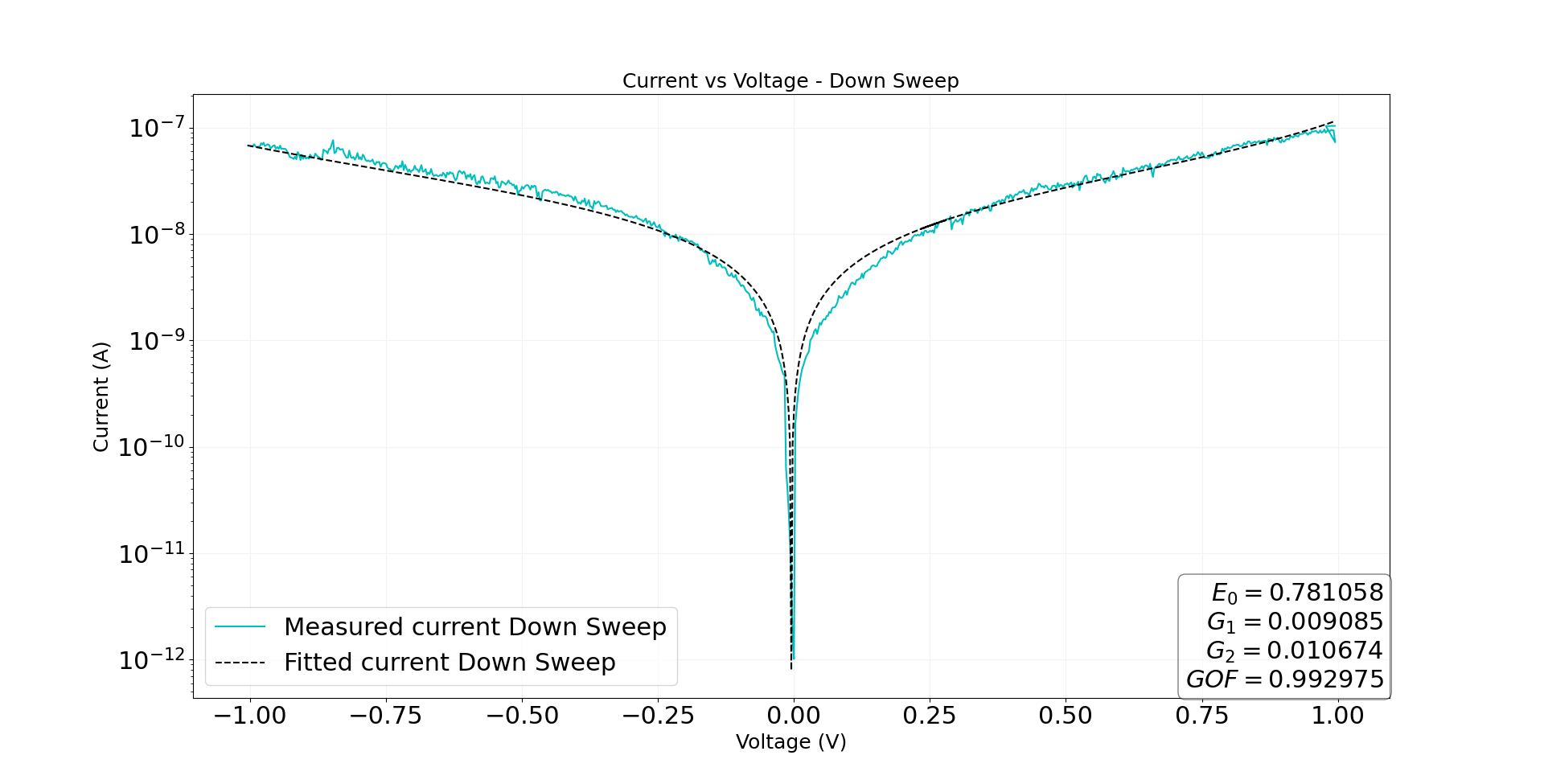}
	\end{minipage}
	\caption{}
\end{figure}

\begin{figure}[h]
	\centering
	\begin{minipage}{1\textwidth}
		\centering
		\includegraphics[width=1\textwidth]{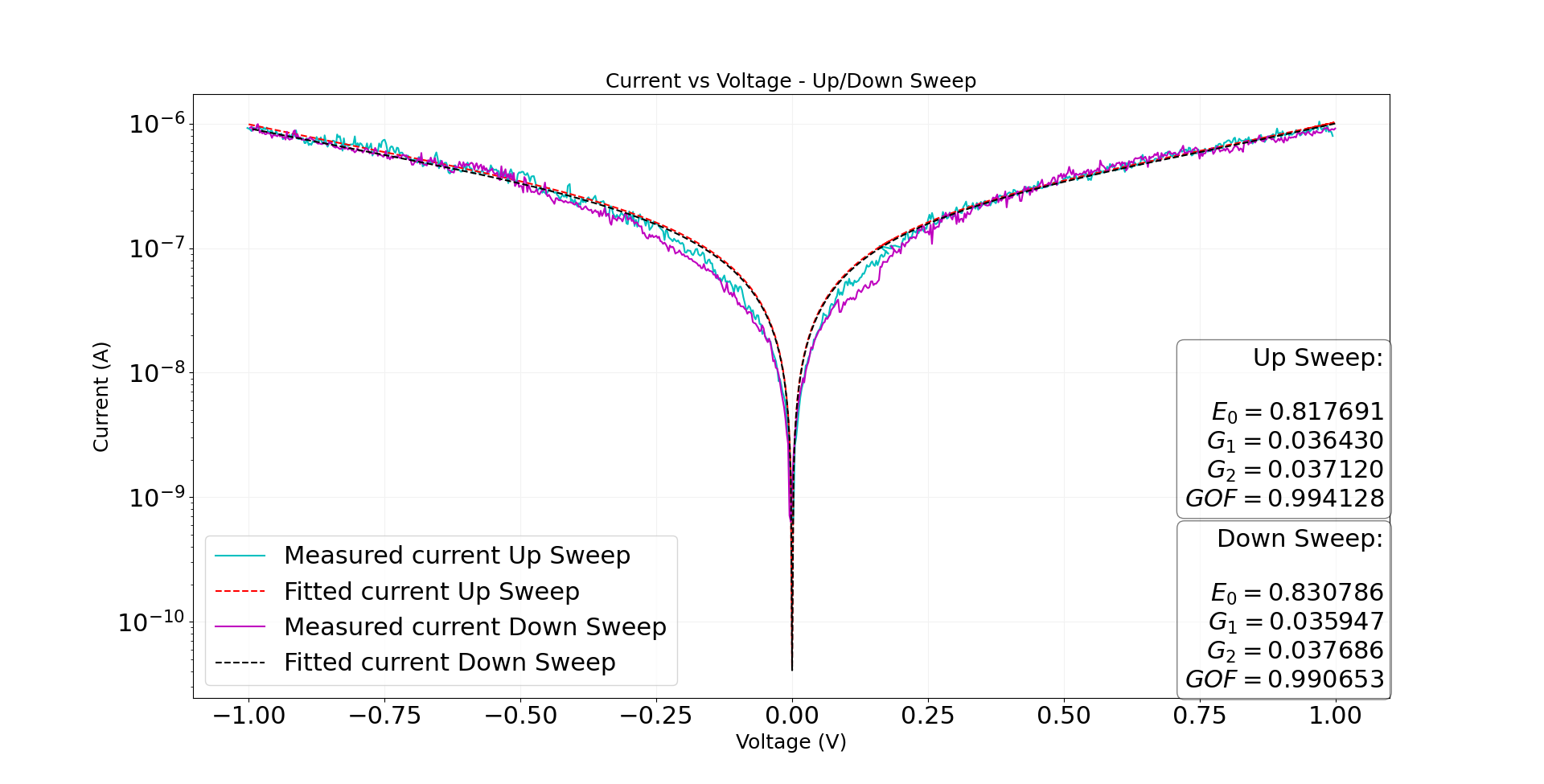}
	\end{minipage}	
	\begin{minipage}{1\textwidth}
		\centering
		\includegraphics[width=1\textwidth]{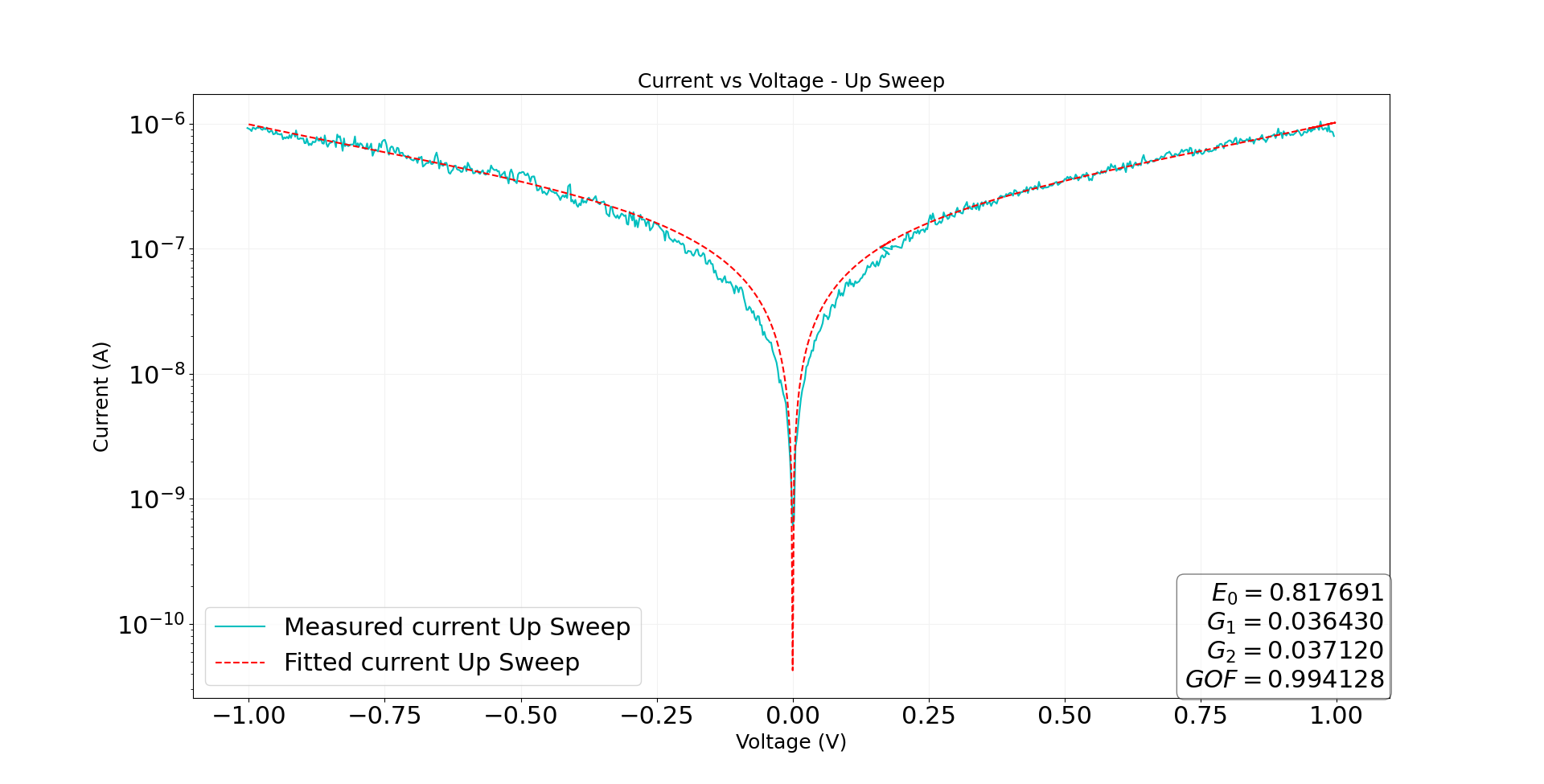}
	\end{minipage}
	\begin{minipage}{1\textwidth}
		\centering
		\includegraphics[width=1\textwidth]{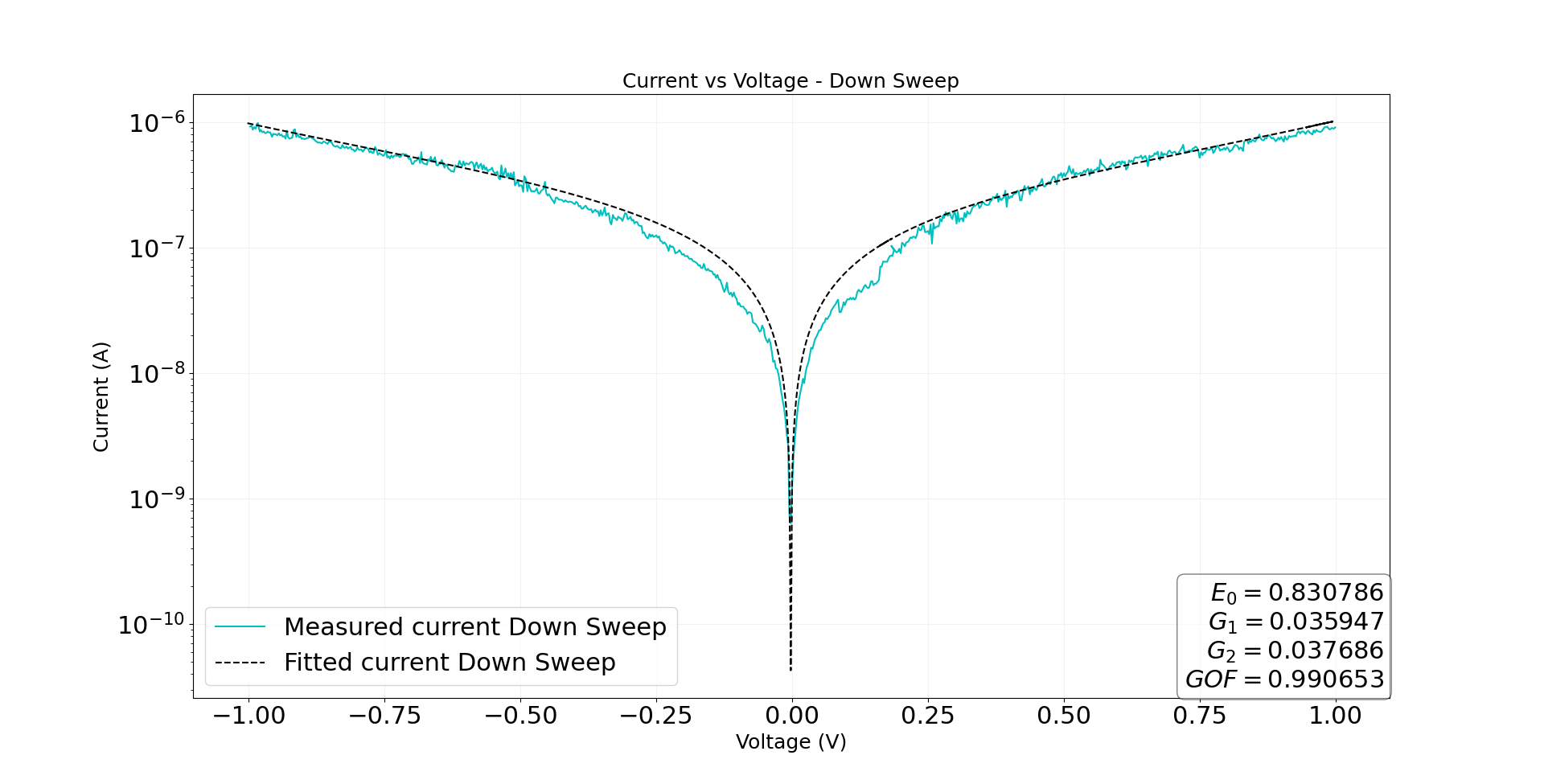}
	\end{minipage}
	\caption{}
\end{figure}

\begin{figure}[h]
	\centering
	\begin{minipage}{1\textwidth}
		\centering
		\includegraphics[width=1\textwidth]{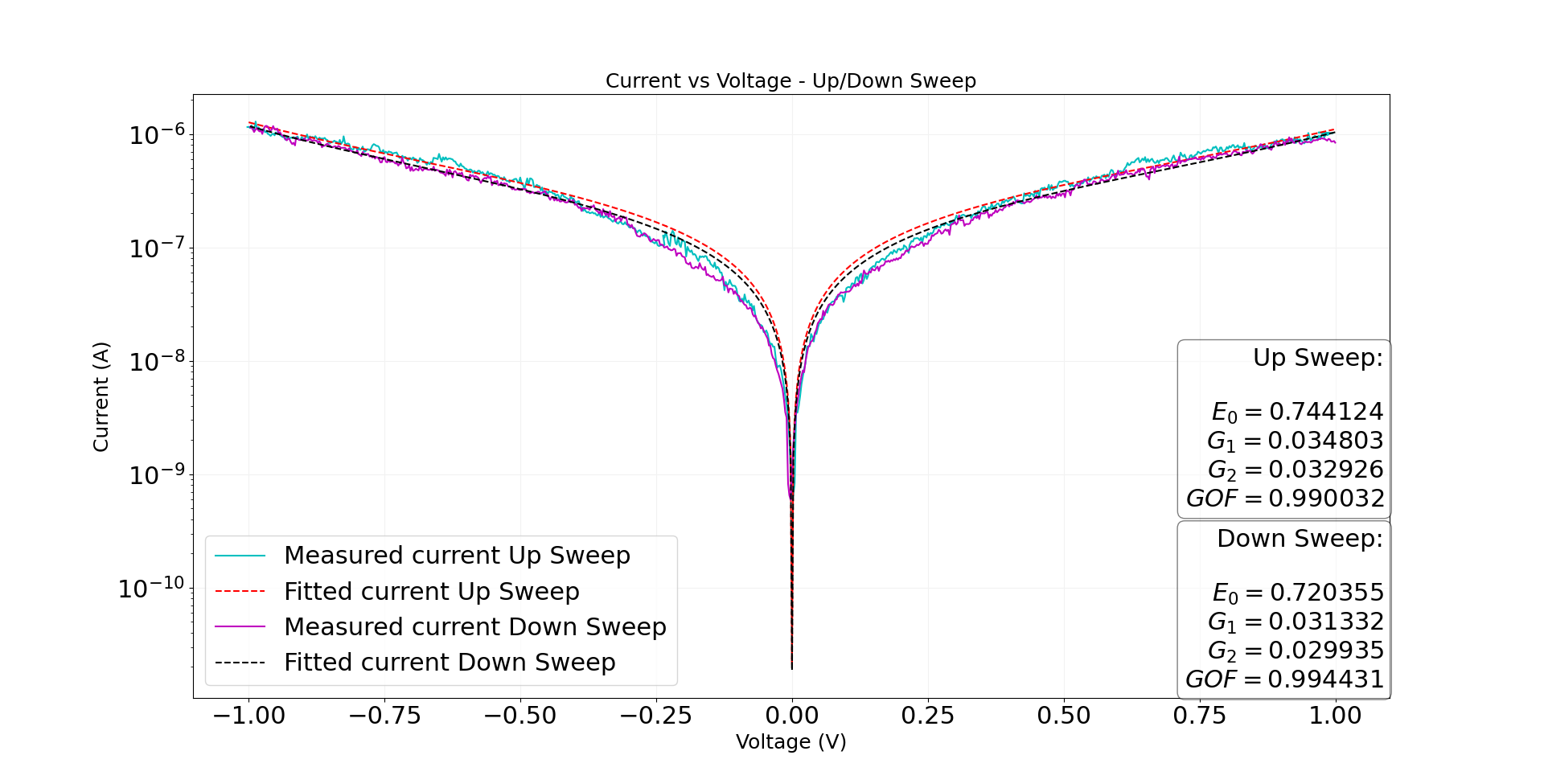}
	\end{minipage}	
	\begin{minipage}{1\textwidth}
		\centering
		\includegraphics[width=1\textwidth]{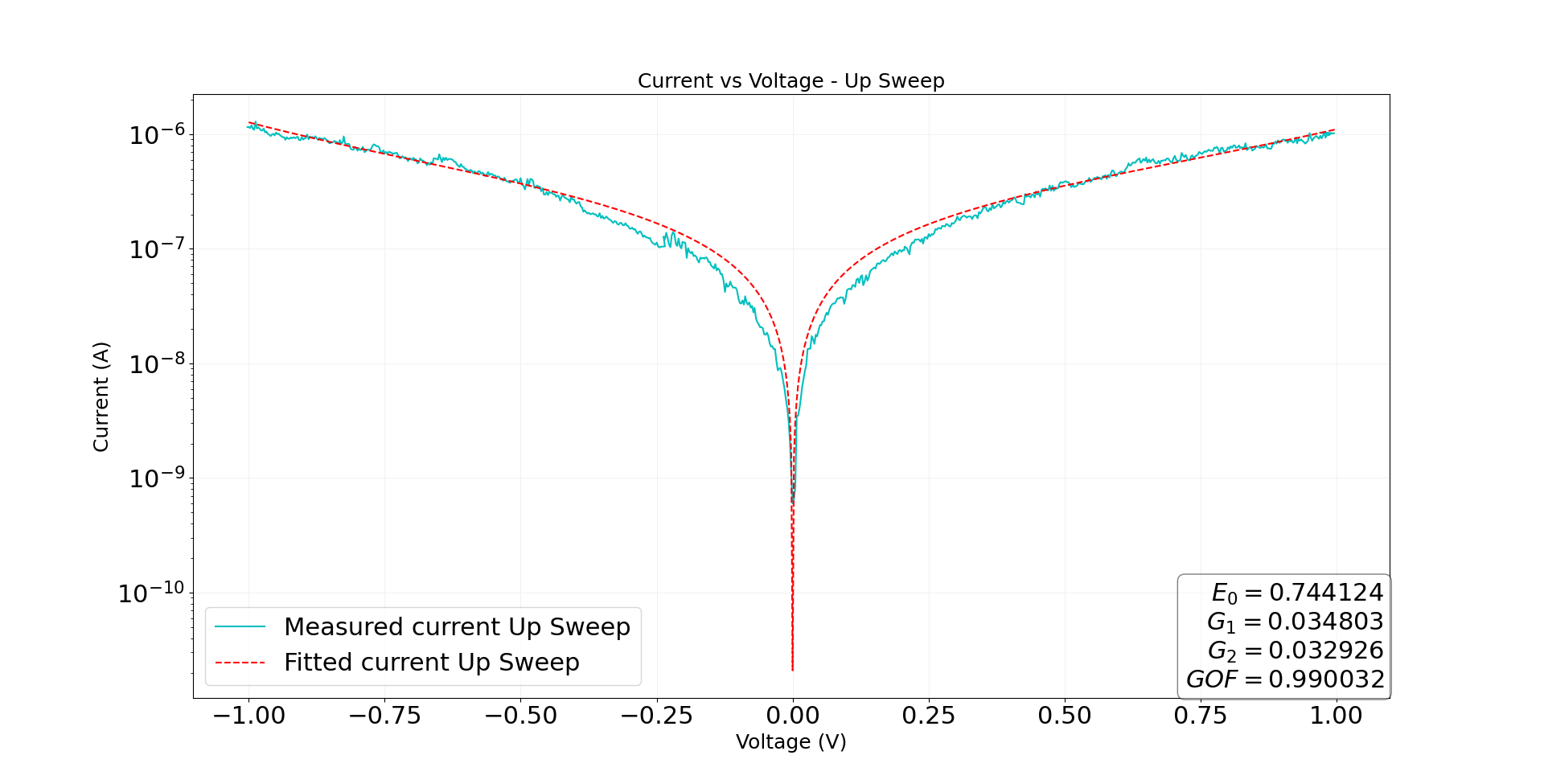}
	\end{minipage}
	\begin{minipage}{1\textwidth}
		\centering
		\includegraphics[width=1\textwidth]{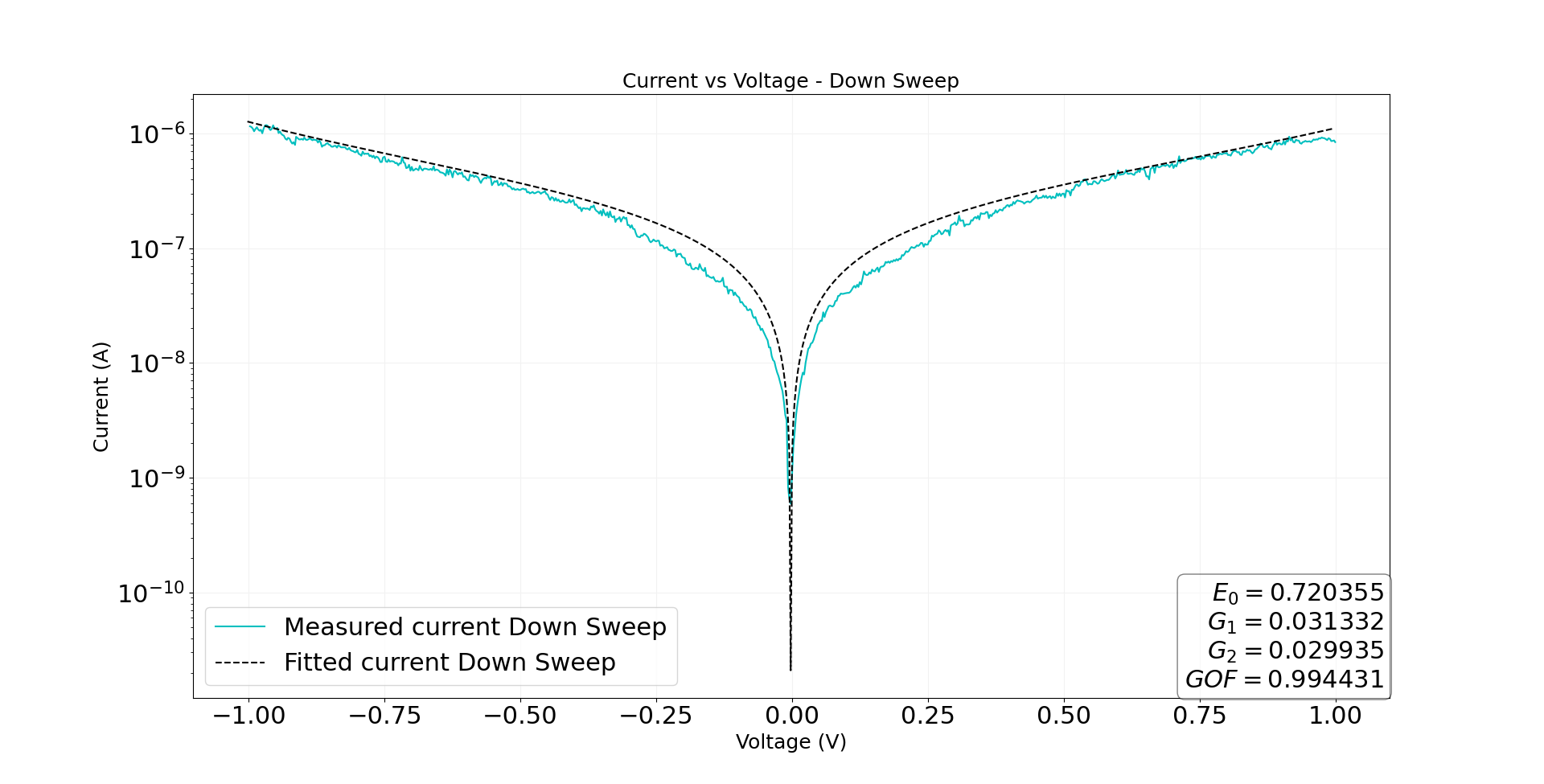}
	\end{minipage}
	\caption{}
\end{figure}
	\bibliographystyle{plain}
	%\bibliography{bibliography_mcbj.bib}

\end{document}